\documentclass[12pt]{report}
\usepackage{graphicx}
\usepackage{amssymb}
\usepackage{epstopdf}
\usepackage{xspace}
\usepackage{xcolor}
\usepackage{multicol}
\usepackage{floatrow}
\usepackage{float}
\usepackage{dcolumn}% Align table columns on decimal point
\usepackage{bm}% bold math
\usepackage[dvips]{epsfig} 
\usepackage[font=footnotesize,labelfont=bf]{caption}

\usepackage[backend=bibtex8, style=ieee, maxbibnames=3, arxiv=abs]   
           {biblatex}
\NewBibliographyString{submitted}
\DefineBibliographyStrings{english}{%
  submitted = {submitted for publication},
}

\usepackage{xspace}
\newcommand{\degree}{\ensuremath{^\circ}\xspace}

\newcommand{\htp}{\ensuremath{\mathrm{H}_2^+}\xspace}

\usepackage{etoolbox}
\makeatletter
\patchcmd{\@makechapterhead}{50\p@}{-10pt}{}{}
\patchcmd{\@makeschapterhead}{50\p@}{-10pt}{}{}
\pretocmd{\subsection}{\addtocontents{toc}{\protect\addvspace{-2\p@}}}{}{}
\pretocmd{\subsubsection}{\addtocontents{toc}{\protect\addvspace{-2\p@}}}{}{}
\makeatother

\DeclareFontShape{OT1}{cmr}{bx}{sc}{<-> cmbcsc10}{}

\newcommand{\opal}{\textsc{OPAL}}
\newcommand{\opalcycl}{\textsc{OPAL-cycl}}

\newcommand{\figref}[1]{Figure~\ref{#1}}
\newcommand{\tabref}[1]{Table~\ref{#1}}

\newcommand{\secref}[1]{Section~\ref{#1}}

\newcommand {\RM}[1]{\mathrm{#1}}

\newcommand {\bs}[1]{\mathbf #1}

\renewcommand{\epsilon}{\varepsilon}

\usepackage{suffix}

\newcommand{\INLINEBOX}[2]{%
   \begin{center}%
    \fcolorbox{#1!60!black}{#1}{%
      \addtolength{\linewidth}{-0.6cm}%  fixed value, works for normal article text
      \begin{minipage}{\linewidth} #2 \end{minipage}%
    }%
   \end{center}\vspace{1pt}%
}

\newcommand{\MARGINBOX}[1]{%
  \mbox{}%
  \marginpar%
   [\tiny\raggedleft\hspace{10pt}#1]%
   {\tiny\raggedright\hspace{0pt}#1}%
}

\newcommand{\TODO}[2][]{\MARGINBOX{\textcolor{red!80!black}{\emph{ToDo (#1):}} #2}}
\WithSuffix\newcommand\TODO*[2][]{\INLINEBOX{red!20!white}{\emph{ToDo (#1):} #2}}

\newcommand{\FIXME}[2][]{\MARGINBOX{\textcolor{blue!80!black}{\emph{FixMe (#1):}} #2}}
\WithSuffix\newcommand\FIXME*[2][]{\INLINEBOX{blue!20!white}{\emph{FixMe (#1):} #2}}

\newcommand{\NOTE}[2][]{\MARGINBOX{\textcolor{green!80!black}{\emph{Note (#1):}} #2}}
\WithSuffix\newcommand\NOTE*[2][]{\INLINEBOX{green!20!white}{\emph{Note (#1):} #2}}

\begin{document}

~~~~

\thispagestyle{empty}

~~

\vspace{-0.5in}

\begin{center}
{\Large \bf IsoDAR@KamLAND:\\ \vspace{0.1in}  A Conceptual Design Report for the
  Technical Facility}
\end{center}

%\vspace{0.5in}

{ \it Abstract:}  
This conceptual design report describes the technical facility for 
the IsoDAR electron-antineutrino source at KamLAND. The IsoDAR source 
will allow an impressive program of neutrino oscillation and 
electroweak physics to be performed at KamLAND. This report provides 
information on the physics case,  the conceptual design for the 
subsystems, alternatives designs considered, specifics of installation 
at KamLAND, and identified needs for future development. We discuss 
the risks we have identified and our approach to mitigating those 
risks with this design. A substantial portion of the conceptual design 
is based on three years of experimental efforts and on industry 
experience. This report also includes information on the conventional 
facilities.

%\newpage
%\thispagestyle{empty}

\vspace{0.05in}

\noindent M. Abs$^{13}$, 
A.~Adelmann$^{20}$,
 J.R~Alonso$^{16}$, 
S.~Axani$^{16}$, 
W.A.~Barletta$^{5,16}$, 
R.~Barlow$^{11}$, 
L.~Bartoszek$^{2}$, 
A.~Bungau$^{11}$,
L.~Calabretta$^{14}$,  
A.~Calanna$^{14}$, 
D.~Campo$^{14}$, 
G.~Castro$^{14}$, 
L.~Celona$^{14}$, 
G.H.~Collin$^{16}$, 
J.M.~Conrad$^{16}$,
S.~Gammino$^{14}$, 
R.~Johnson$^{3}$,  
G.~Karagiorgi$^{17}$, 
S.~Kayser$^{16}$, 
W.~Kleeven$^{13}$,
A.~Kolano$^{11}$, 
F.~Labrecque$^3$, 
W.A.~Loinaz$^1$, 
J.~Minervini$^{16}$, 
M.H.~Moulai$^{16}$,
H.~Okuno$^{21}$,  
H.~Owen$^{8,17}$, 
V.~Papavassiliou$^{19}$,   
M.H.~Shaevitz$^9$, 
I.~Shimizu$^{22}$, 
T.M.~Shokair$^{15}$, 
K.F.~Sorensen$^{10}$, 
J.~Spitz$^{18}$, 
M.~Toups$^{16}$, 
M.~Vagins$^{6}$, 
K.~Van Bibber$^{4}$, 
M.O.~Wascko$^{12}$, 
D.~Winklehner$^{16}$, 
L.A~ Winslow$^{16}$, 
J.J.~Yang$^7$

{\footnotesize
\begin{center}
$^1$Amherst College, Amherst MA, US\\
$^2$Bartoszek Engineering, Aurora IL, US\\
$^3$Best Cyclotron System, Inc., Springfiled VA, US\\
$^4$University of California, Berkeley, CA, US\\
$^5$University of California, Los Angeles, CA, US\\
$^6$University of California, Irvine, CA, US\\
$^7$China Institute of Atomic Energy, Beijing CN\\
$^8$The Cockcroft Institute Daresbury Laboratory, Daresbury, UK\\
$^9$Columbia University, New York NY, US\\
$^{10}$ FLIBE Energy Inc., Huntsville AL, US\\
$^{11}$University of Huddersfield, Huddersfield, UK\\
$^{12}$ Imperial College London,  London, UK \\
$^{13}$ Ion Beam Applications, S.A., Ottignies-Louvain-la-Neuve, BE\\
$^{14}$INFN Laboratori Nazionale del Sud, Catania, IT \\ 
$^{15}$ Lawrence Livermore National Laboratory, Tracy CA, US\\
$^{16}$Massachusetts Institute of Technology, Cambridge MA, US\\
$^{17}$ University of Manchester, Manchester UK\\
$^{18}$ University of Michigan, Ann Arbor MI, US\\
$^{19}$New Mexico State University, Las Cruces NM, US\\
$^{20}$Paul Scherrer Institute, Villigen, CH\\
$^{21}$RIKEN Nishina Center for Accelerator-based Science, Wako,  JP\\
$^{22}$Tohoku University, Sendai, JP\\
\end{center}
}

\noindent {\it \footnotesize Corresponding Authors: Janet Conrad (conrad@mit.edu) and Mike Shaevitz (shaevitz@nevis.columbia.edu)}

\newpage 

\tableofcontents

\clearpage
\chapter{Introduction}

The IsoDAR experimental program will perform unique
searches for sterile neutrino oscillations and non-standard
neutrino interactions.  The experiment is being designed to definitively address these physics
topics using a well-understood, high-intensity $^8$Li $\beta$-decay-at-rest antineutrino source
coupled
with a massive detector such as KamLAND that has good inverse-beta-decay identification
capabilities with high efficiency.   The high statistics data sample will allow excellent sensitivity
to these new physics signatures and allow study of %JOSE added "of"
any signals that may be
detected.  With respect to sterile neutrinos, accurately mapping out the oscillation wave within the detector will conclusively
test if an observed  signal is associated with the recently published
sterile neutrino anomalies
that are of great interest in the community
\cite{abazajian:sterile_whitepaper}. The oscillation wave can be used to 
determine the number of extra sterile neutrino flavors by differentiating the oscillatory behavior of a (3+1) versus a (3+2)
oscillation model \cite{bungau:daedalus}.  The high intensity also will allow the
study of antineutrino-electron scattering.
This is a very clean and well-understood interaction that can be used to search for indications
of non-standard neutrino interactions proposed in many extensions of
the standard model \cite{conrad:neutrino_scattering}.

The IsoDAR neutrino source \cite{bungau:daedalus} consists of 
an accelerator producing 60 MeV protons that impinge on a
$^9$Be target, producing neutrons. IsoDAR can use the same cyclotron
design as the injector cyclotron for the two-cyclotron DAE$\delta$ALUS system \cite{abs:daedalus}. The protons  enter
a surrounding $\ge$99.99\% isotopically pure $^7$Li sleeve, where neutron
capture results in $^8$Li; this isotope undergoes
$\beta$ decay at rest to produce an isotropic
$\bar \nu_e$ flux with an average energy of $\sim$6.4 MeV and
an endpoint of $\sim$13 MeV.
The $\bar \nu_e$ will interact in a scintillator detector via inverse beta decay (IBD),
$\bar \nu_e +p \rightarrow e^+ + n$, which is easily tagged through
prompt-light plus neutron-capture coincidence.
When
paired with KamLAND \cite{abe:kamland}, the experiment is capable 
to observe
$8.2\times 10^5$ reconstructed IBD events in five years.
With this data set, IsoDAR can provide a 5$\sigma$ test of sterile neutrino
oscillation models, allow precision  measurement
of $\bar \nu_e$-$e$
scattering, and search for production and decay of exotic particles.

This document describes a conceptual design for the Technical
Facility that forms the IsoDAR neutrino source
at the KamLAND.
The purpose of this Conceptual Design is to establish the feasibility
of the IsoDAR experiment. Hence, for each subsystem, the design is
worked out in sufficient detail to allow estimates of cost, weight,
overall dimensions, power consumption and other utility needs. 
Decisions have been made about major design features, but
alternatives that remain possible are noted.  We present the rationale
for making these choices,  supported by relevant experimental data
(especially on the front-end design) 
and industrial experience (especially with the cyclotron design).

The conceptual design for the conventional facilities, describing the housing
and utility-support for this technical facility, will be the subject of a separate
report. However, to provide context, an overview is presented here.

The outline of the document is as follows:
\begin{description}
\item{\bf Chapter 2} gives an
outline of the physics reach of a such an experiment, using realistic efficiencies, resolutions,
and background rates based on information from the KamLAND collaboration.
(Table \ref{parSterile} lists the neutrino source and detector parameters used in the
physics calculations and studies.)  
\item{\bf Chapter 3} gives general information on cyclotrons
and why these accelerators are a good choice for a high-intensity neutrino source.
The chapter contrasts the choices made for the IsoDAR  system with other options.
\item {\bf Chapter 4} gives conceptual design solutions for the
Technical Facility for the IsoDAR system at the KamLAND site.  
The subsections are:
\begin{enumerate}
\item The front end, which includes the ion source and low energy beam
  transport (LEBT);
\item The cyclotron design specifics; 
\item The medium energy beam transport (MEBT) system which transport;
the beam from the cyclotron cavern to the targeting cavern;
\item The target engineering and target production and shielding calculations;
\item Radiation protection throughout the system;
\item Controls and Data Acquisition;
\item The interfaces to the conventional facilities.
\end{enumerate}
In each subsection, the open questions involved in the
subsystem are identified. Also, in each subsection, risks
for each subsection are considered.
\item{\bf  Chapter 5} summarizes the conventional facility
  requirements.  The subsections are:
\begin{enumerate}
\item Space constraints and civil construction;
\item Utilities
\end{enumerate} 
\item{\bf  Chapter 6} describes the simulations involved in the design
  of the Technical Facility.
\end{description}

\begin{table}
{\footnotesize
\begin{center} 
      \begin{tabular}{|l|l|} \hline
        Accelerator  & 60~MeV/amu of H$_2^+$  \\  
        Beam Current  & 10~mA of protons on target  \\  
        Beam Power (CW)  & 600~kW  \\  
        Duty cycle  & 90\%  \\  
        Protons/(year of live time)  & 1.97$\times 10^{24}$ \\ 
        Run period  & 5~years  \\  
        Live time  & 5~years$\times$0.90=4.5 years  \\  
        Target   & $^9$Be with FLiBe sleeve (99.995\% pure $^7$Li)  \\  
        Sleeve diameter and length & 100~cm and 190~cm \\ 
        $\overline{\nu}$ source  & $^8$Li $\beta$ decay (6.4 MeV mean
energy flux)  \\  
        Fraction of $^8$Li produced in target &10\% \\  
        $\overline{\nu}$ flux during 4.5~years of live time  &
1.3$\times 10^{23}$~$\overline{\nu}_e$ \\  
        $\overline{\nu}$ flux uncertainty  & 5\% (shape-only is also
considered) \\  \hline
        Detector  & KamLAND   \\  
        Distance between face of target and center of detector & 16.1~m
  \\  
        Fiducial mass  & 897~metric tons   \\  
        Fiducial radius  & 6.5~m   \\  
        Total detector radius  & 13~m   \\ 
        Detection efficiency  & 92\%   \\  
        Vertex resolution  & 12~cm/$\sqrt{E~\mathrm{(MeV)}}$ \\ 
        Energy resolution  & 6.4\%/$\sqrt{E~\mathrm{(MeV)}}$ \\  
        Visible energy threshold (IBD and $\overline{\nu}_e$-electron) & 3 MeV  \\  
        IBD event total & 8.2$\times 10^5$ \\  
        $\overline{\nu}_e$-electron event total &  2600 \\ \hline 
        Expected $\bar\nu_e$ disappearance sensitivity  & $\sin^22\theta_{new}>0.005$ @ $\Delta m^2=1$eV$^2$ \\ 
          Expected $\sin^2\theta_W$ 1$\sigma$ precision & 3.2\% \\ \hline
      \end{tabular}
\end{center}
\caption{\label{parSterile}Summary of IsoDAR and KamLAND detector
  parameters used in physics calculations,  and expected sensitivities
to new physics.}}
\end{table}

\clearpage

\chapter{Scientific Goals}

IsoDAR has two primary science goals:  sterile neutrino searches using inverse beta decay events \cite{bungau:daedalus} and nonstandard interaction searches using elastic scattering events \cite{conrad:neutrino_scattering}. A range of other studies that 
will provide valuable measurements are also under development, but
have not yet been published.  An example is spectrum endpoint 
studies discussed below.  

\section{Sterile Neutrino Searches}
% word for word from Pontocorvo paper  
Searches for light sterile neutrinos with mass $\sim$1~eV are motivated by observed
anomalies in several experiments. Intriguingly, these results come from a wide range of experiments covering neutrinos, anti-neutrinos, different flavors, and different energies. Short baseline accelerator
neutrino oscillation experiments
\cite{aguilar:neutrino_oscillations,
      aguilar-arevalo:neutrino_oscillations}, 
short baseline reactor experiments
\cite{mueller:reactor_antineutrinos,
      mention:anomaly}, 
and even the radioactive source experiments, which were originally intended as calibrations for the chemical solar neutrino experiments\cite{abdurashitov:sage, 
      kaether:gallex}, 
have all observed anomalies that can be interpreted as due to one 
or more sterile neutrinos. 

To understand these anomalies in terms of the $\nu$SM for neutrino
oscillations standard model ($\nu$SM), the observations must be compared to the data from the
large range of experiments with null results  
\cite{armbruster:numu_oscillations, 
      astier:numu_oscillations, 
      stockdale:numu_oscillations, 
      dydak:numu_oscillations}, 
and then to a model. These ``global fits'' are most often to models
with one or more sterile neutrinos added to the
oscillation probability calculation \cite{sorel:sterile_neutrinos}. 
The extended models are referred to as ``3+1", ``3+2", or ``3+3"
neutrino models depending on the number of additional sterile 
neutrinos. The global fits tend to prefer 3+2 and 3+3 models with 
$CP$ violation;
3+1 models are very hard to reconcile between
the experiments with signals and those with null results \cite{conrad:sterile_fits}.

The diversity of experiments showing these anomalous results has motivated a number of proposals to address them. Suggestions range from repeating the source experiments, to specially designed reactor antineutrino experiments, to accelerator-based ones. Many of these proposals, however, do not have sufficient
sensitivity to make a definitive $>5\sigma$ statement about the
existence of sterile neutrinos in all of the relevant parameter space. The experiments that are designed to make a definitive measurement are based on pion or isotope DAR sources. Notably, the full DAE$\delta$ALUS complex
could be used to generate a pion DAR beam for such a
measurement.  However, the IsoDAR concept calls for simply using the DAE$\delta$ALUS injector cyclotron to generate an isotope DAR source. Such a complex situated next to a kiloton-scale scintillator detector such as
KamLAND would enable a definitive search for sterile
neutrinos by observing a deficit of antineutrinos as a function of the distance $L$ and antineutrino energy $E$ across the detector---the definitive
signature of neutrino oscillation. This is the concept behind the
IsoDAR proposal~\cite{bungau:isodar_target}.

The proposed IsoDAR target is to be placed adjacent to the KamLAND detector.
The antineutrinos propagate 9.5~m through a combination of rock, outer muon
veto, and buffer liquid to the active scintillator volume of KamLAND. The scintillator is contained 
in a nylon balloon 6.5~m in radius bringing the total distance from target to detector center to
16.1~m. The antineutrinos are then detected via the IBD interaction. This interaction has a well known cross section with an
uncertainty of 0.2\%~\cite{vogel:ibd}, and creates a distinctive
coincidence signal between a prompt positron signal, 
$E_{vis}=E_{\bar{\nu}_{e}} - 0.78$~MeV, and a delayed neutron 
capture giving a 2.2~MeV gamma ray within $\sim$200~$\mu$s. 

KamLAND was designed to efficiently detect IBD. A standard analysis has a 92\%
efficiency for identifying IBD events~\cite{gando:kamland1}. In IsoDAR's
nominal 5 year run, $8.2\times10^{5}$ IBD events are expected. The
largest background comes from the 100~reactor antineutrino IBD events
detected by KamLAND per year~\cite{abe:kamland}. The reactor
antineutrino rate is dependent on the operation of the nuclear
reactors in Japan which has been significantly lower in 2012 and
2013~\cite{gando:kamland2}. The sterile neutrino analysis uses an
energy threshold of 3~MeV. Due to the very effective background rejection
provided by the IBD delayed coincidence tag, this
threshold enables use of the full KamLAND fiducial volume, R$<$6.5~m
and 897~metric tons, with negligible backgrounds to the reactor antineutrinos signal.

\begin{figure}[tb]
\begin{center}
\includegraphics[scale=.7]{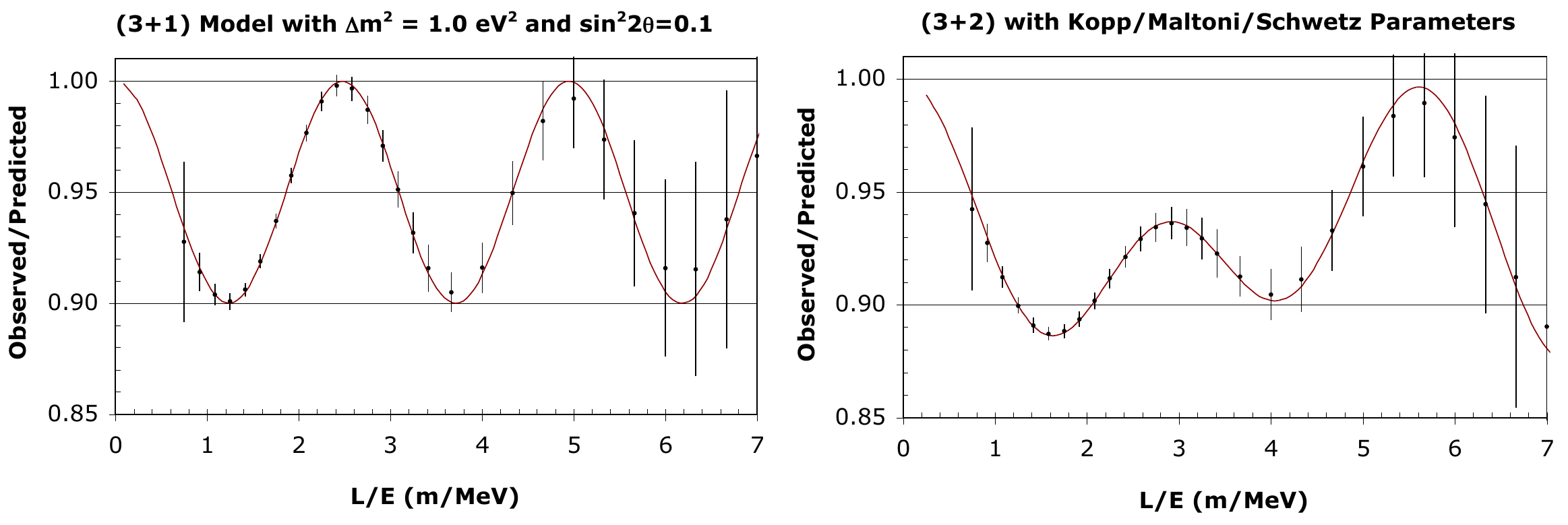}
\end{center}
\vspace{-.5cm}
\caption{\footnotesize The $L/E$ dependence of sample data sets for 5 years of running for 3+1 (left) and 3+2
  (right) oscillation scenarios.   The solid curve is the oscillation probability with no smearing in the reconstructed position and energy
  and the data points with error bars are from simulated events including smearing.
\label{waves}}
\end{figure}

The sterile neutrino analysis makes use of neutrino oscillations's $L/E$ signature. Therefore, the energy and vertex 
resolutions are essential in determining sensitivity. The KamLAND detector has a vertex reconstruction resolution 
of $12{~\rm cm}/\sqrt{E(\rm MeV)}$ and an energy resolution of  $6.4\%/\sqrt{E(\rm MeV)}$~\cite{gando:kamland1}.  
Sample data sets for reasonable 3+1 and 3+2 sterile models are shown in Figure~\ref{waves} for the nominal detector
 parameters, summarized in Table~\ref{parSterile}. In most currently favored oscillation scenarios, the $L/E$ signal is observable. Furthermore, separation of the various 3+N models may be possible as exemplified by Figure~\ref{waves} (right).

\begin{figure}
\begin{center}
\includegraphics[width=0.6\columnwidth]{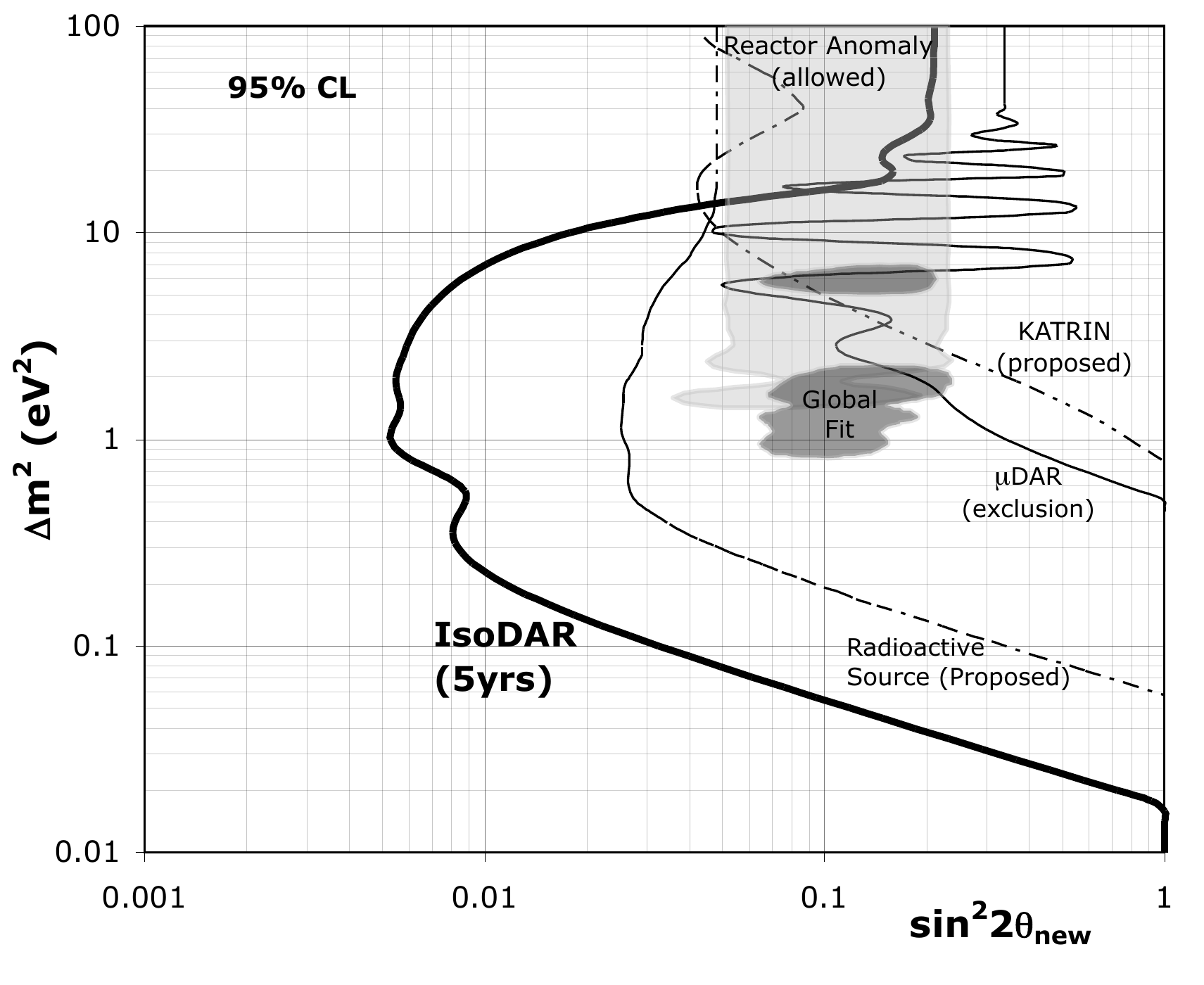} 
\vspace{-.2cm}
\caption{\footnotesize Sensitivity of IsoDAR in a nominal 5 year run, in comparison
  with other experiments.  The existing $\mu$DAR~\cite{conrad:karmen_lsnd} exclusion curve and
Reactor+Gallium~\cite{mention:reactor_anomaly} allowed region are
also shown.  With
respect to future experiments, the expected sensitivities from a
propose PetaBequerel source experiment~\cite{cribier:pbq} and from
KATRIN~\cite{formaggio:katrin}, now under construction, are also indicated. Reprinted from Ref.~\cite{bungau:daedalus}.\label{sensitivySterile}}
\end{center}
\end{figure}

To understand the sensitivity relative to other proposals,
the IsoDAR 95\% CL is compared to other electron antineutrino disappearance experiments in the two
neutrino oscillation parameter space in
Figure~\ref{sensitivySterile}. In just five years of running, IsoDAR
could rule out the entire global 3+1 allowed region, $\sin^2{2\theta_{new}}=0.067$ and $\Delta
m^{2}=1$~eV$^{2}$ at 20$\sigma$. This is the most definitive
measurement among the proposals in the most probable parameter space
of $\Delta m^{2}$ between 1-10~eV$^{2}$.

\section{Precision Electroweak Tests of the Standard Model}
% word for word from Pontocorvo paper 

In addition to the $8.2\times10^5$ IBD interactions, the IsoDAR
neutrino source~\cite{bungau:daedalus}, when combined with the KamLAND
detector~\cite{abe:kamland}, could collect the largest sample of low-energy $\bar
\nu_e$-electron scatters (ES) that has been observed to date \cite{conrad:neutrino_scattering}. Approximately 2600 ES events would be collected above a 3~MeV visible energy threshold over a 5 year run, and both the total rate and the visible energy can be measured. This can be compared to the samples from  the 
Irvine experiment (458 events from 1.5 to 3~MeV
\cite{reines:nu-e_scattering}); 
TEXONO (414~events from 3 to 8~MeV~\cite{deniz:nu-e_scattering}); 
Rovno (41~events from 0.6 to 2~MeV~\cite{derbin:nu-e_scattering}); and 
MUNU (68 events from 0.7 to 2~MeV~\cite{daraktchieva:munu}).  

In the Standard Model, the ES differential cross section is given by

\begin{equation}
\frac{d\sigma}{dT} = \frac{2 G_F^2 m_e}{\pi}
\left[ 
g^2_R + g^2_L\left(1 - \frac{T}{E_\nu}\right)^2 - g_R g_L \frac{m_e T}{E^2_\nu}
\right],
\label{glgrxs}
\end{equation}

\noindent where $T\in\left[0,\frac{2E_{\nu}^2}{m_e+2E_{\nu}}\right]$ is electron recoil energy, $E_\nu$ is the energy of the incoming $\bar\nu_e$, and the weak coupling constants $g_R$ and $g_L$ are given at tree level by $g_R = \sin^2\theta_W$ and $g_L=\frac{1}{2}+\sin^2\theta_W$.  Eq.~\ref{glgrxs} can also be expressed in terms of the vector and axial weak coupling constants, $g_V$ and $g_A$, using the relations $g_R=\frac{1}{2}(g_V-g_A)$ and $g_L=\frac{1}{2}(g_V+g_A)$.

The ES cross section can therefore be used to measure the weak couplings, $g_V$ and $g_A$, as well as $\sin^2\theta_W$, a fundamental parameter of the Standard Model as described in Ref.~\cite{nakamura:pdg}.  Although $\sin^2\theta_W$ has been determined to high precision~\cite{baak:electroweak}, there is a longstanding discrepancy~\cite{nakamura:pdg} between the value obtained by $e^+e^-$ collider experiments and the value obtained by NuTeV, a precision neutrino-quark scattering experiment
\cite{zeller:nu-nucleus_scattering}.  
Despite having lower statistics than the NuTeV, IsoDAR would measure $\sin^2\theta_W$ using the purely leptonic ES interaction, which does not involve any nuclear dependence.  This could therefore shed some light on the value of $\sin^2\theta_W$ measured by neutrino scattering experiments.

The ES cross section is also sensitive to new physics in the neutrino sector arising from nonstandard interactions (NSIs), which are included in the theory via dimension six, four-fermion effective operators.  NSIs give rise to weak coupling corrections and modify the Standard Model ES cross section given in Eq.~\ref{glgrxs} to
\begin{equation}\label{epsilonxsec}
\frac{d\sigma}{dT}= \frac{2 G_F^2 m_e}{\pi} [ (\tilde g_R^2+\sum_{\alpha \neq e}|\epsilon_{\alpha e}^{e R}|^2)+(\tilde g_L^2+\sum_{\alpha \neq e}|\epsilon_{\alpha e}^{e L}|^2)\left(1-{\frac{T}{E_{\nu}}} \right)^2-(\tilde g_R \tilde g_L+ \sum_{\alpha \neq e}|\epsilon_{\alpha e}^{e R}||\epsilon_{\alpha e}^{e L}|)m_e {\frac{T}{E^2_{\nu}}}],
\end{equation}
\noindent where $\tilde g_R= g_R+\epsilon_{e e}^{e R}$ and $\tilde g_L=g_L+\epsilon_{e e}^{e L}$. The NSI parameters $\epsilon_{e\mu}^{e L,R}$ and $\epsilon_{e\tau}^{e L,R}$ are associated with flavor-changing-neutral currents, whereas $\epsilon_{ee}^{e L,R}$ are called non-universal parameters.  We can estimate IsoDAR's sensitivity to these parameters by fitting Eq.~\ref{epsilonxsec} to the measured ES cross section, assuming the Standard Model value for $\sin^2\theta_W$.  In general, lepton flavor violating processes are tightly constrained so we focus only on IsoDAR's sensitivity to the two non-universal parameters $\epsilon_{ee}^{e L,R}$.

The ES interaction used for these electroweak tests of the Standard Model is very different from the IBD interaction used for the sterile neutrino search. The IBD signal consists of a delayed coincidence of a positron and a 2.2 MeV neutron capture $\gamma$, whereas the ES signal consists of isolated events in the detector.  Another difference is that at IsoDAR energies, the IBD cross section is several orders of magnitude larger than the ES cross section.  In fact, if just 1\% of IBD events are mis-identified as ES events, they will be the single largest background.  On the other hand, as was suggested in Ref.~\cite{conrad:sin_theta_w}, the IBD signal can also be used to reduce the normalization uncertainty on the ES signal to about 0.7\% .  A final difference is that the incoming $\bar\nu_e$ energy for IBD interactions in KamLAND can be inferred from the visible energy on an event-by-event basis, while the incoming $\bar\nu_e$ energy for ES interactions in KamLAND cannot.  Therefore, the differential ES cross section is measured in visible energy bins corresponding to the kinetic energy of the recoil electron, and the dependence on the incoming $\bar\nu_e$ energy is integrated out according to the IsoDAR flux.

The backgrounds to the ES signal can be grouped into beam-related backgrounds, which are dominated by mis-identified IBD events, and non-beam backgrounds, arising from solar neutrino interactions, muon spallation, and environmental sources. We adopt a strategy similar to the one outlined in Ref.~\cite{abe:solar_neutrinos} to reduce the non-beam backgrounds.  First, a cosmic muon veto is applied to reduce the background due to radioactive light isotopes produced in muon spallation inside the detector. This reduces the live time by 62.4\%.  Next, a visible energy threshold of 3 MeV is employed to reduce the background from environmental sources which pile up at low energies.  Finally, a fiducial radius of 5 m is used to reduce the background from external gamma rays emanating from the rock or stainless steel surrounding the detector.  To reduce the beam-related backgrounds, an IBD veto is employed to reject any ES candidate that is within 2 ms of a subsequent event with visible energy $>1.8~\mathrm{MeV}$ in a 6 m fiducial radius.  The IBD veto is estimated to have an efficiency of $99.75\% \pm 0.02 \%$, where the uncertainty is assumed to come from the statistical uncertainty on measuring the IBD selection efficiency with 50,000 AmBe calibration source events.

\begin{figure}[tb ]
	\begin{center}		
		\includegraphics[width=0.7\textwidth]{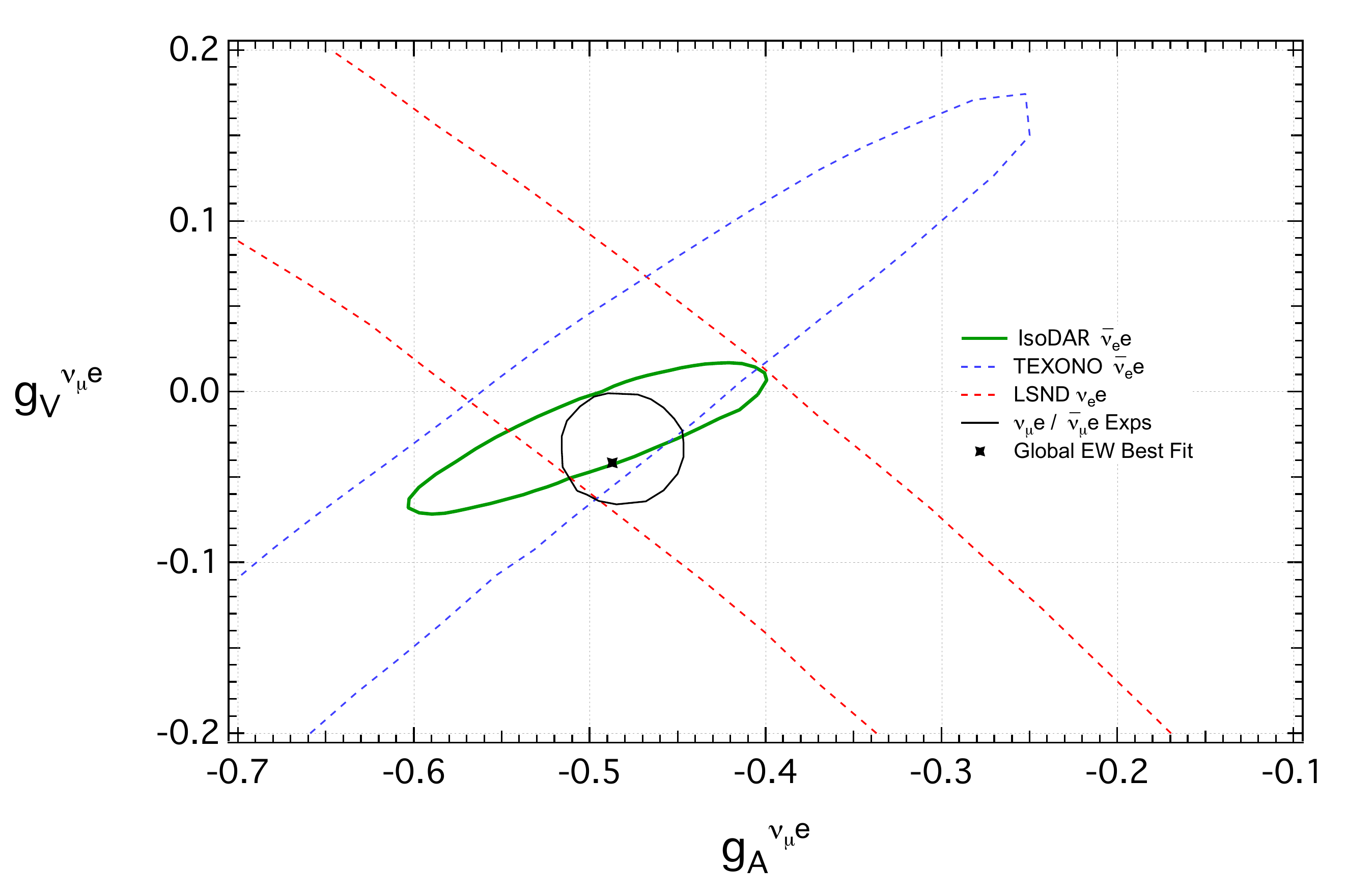}
	\end{center}
	\caption{\footnotesize\label{gV_gA} IsoDAR's sensitivity to $g_V$ and $g_A$ along with allowed regions from other neutrino scattering experiments and the electroweak global best fit point taken from Ref.~\cite{nakamura:pdg}. The IsoDAR, LSND, and TEXONO contours are all at $1\sigma$ and are all plotted in terms of $g_{V,A}^{\nu_\mu e}= g_{V,A}^{\nu_e e}-1$ to compare with $\nu_\mu$ scattering data. The $\nu_\mu e/\bar{\nu}_\mu e$ contour is at 90\% C.L.}
	\end{figure}
\begin{figure}[tbh ]
	\begin{center}		
		\includegraphics[width=0.49\textwidth]{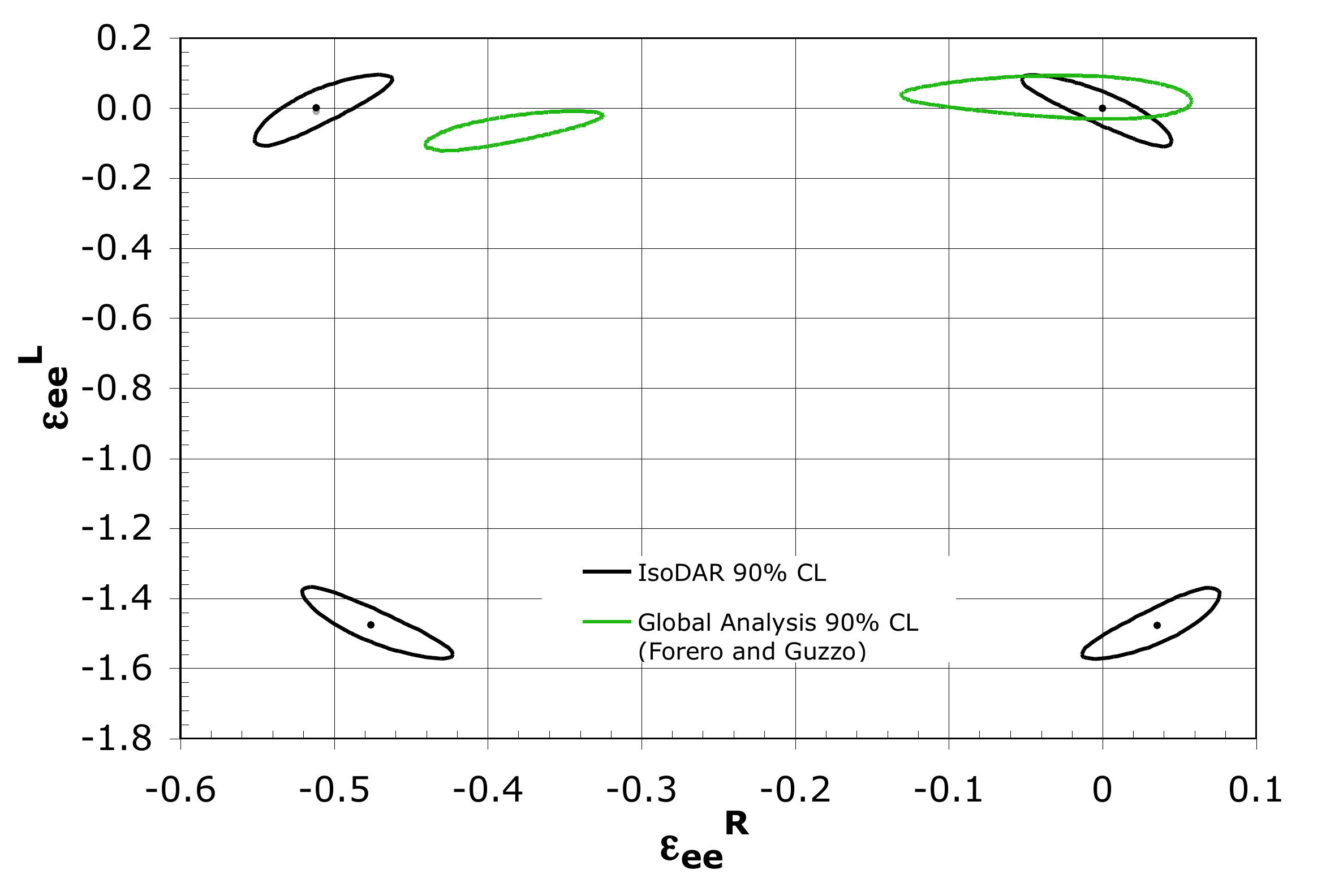}
		\includegraphics[width=0.49\textwidth]{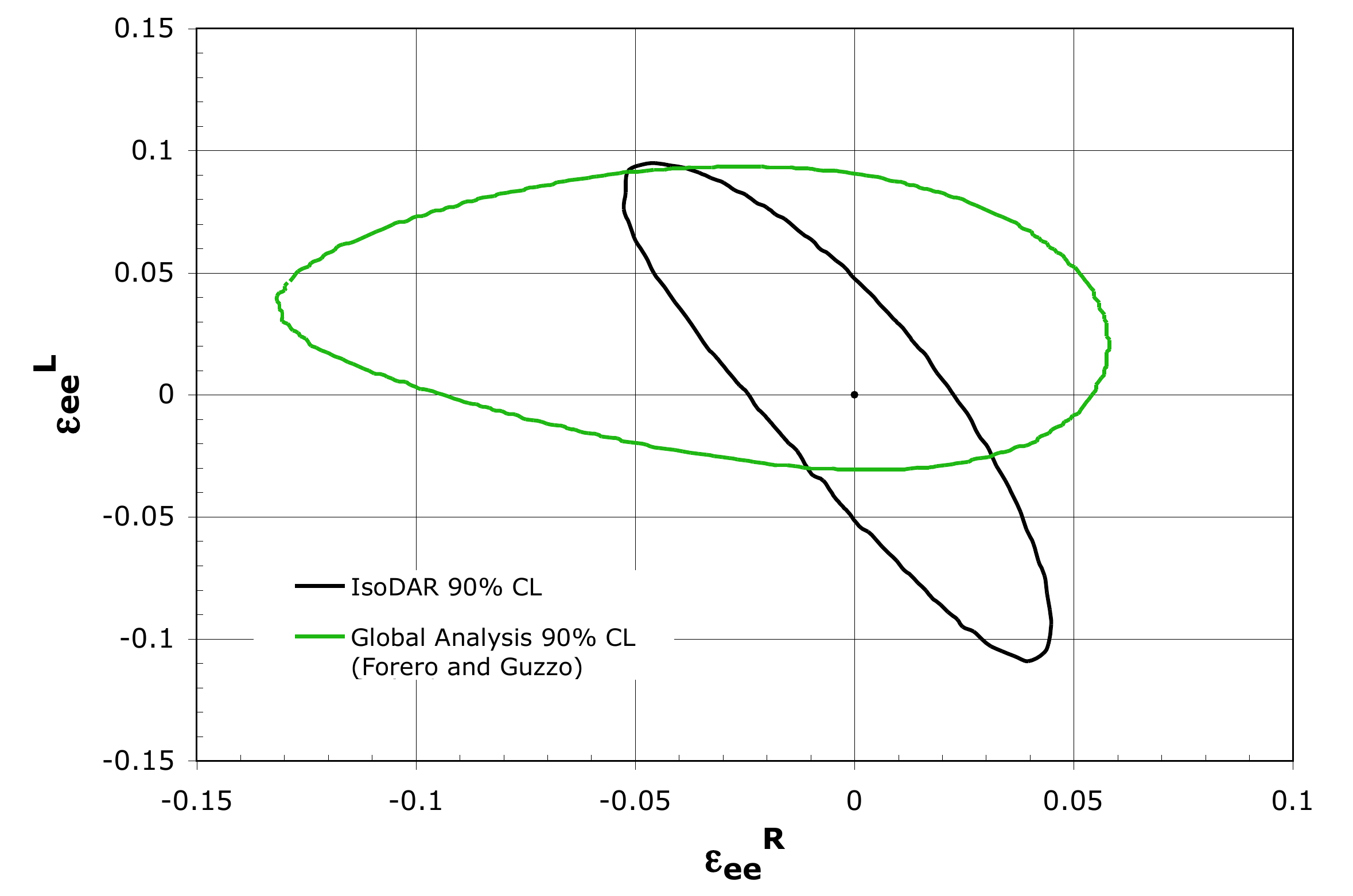}
	\end{center}
	\caption{\footnotesize \label{NSI} (Left) IsoDAR's sensitivity to $\epsilon_{e e}^{e L}$ and $\epsilon_{e e}^{e R}$. The current global allowed region, based on Ref.~\cite{forero:nu-e_nsi} is also shown. (Right) A zoomed-in version of the left plot emphasizing the region near $\epsilon_{ee}^{e  L}$ $\sim0$ and $\epsilon_{ee}^{e  R}$ $\sim0$.}
\end{figure}

A total of 2584 signal events, 705 IBD mis-ID background events and
2870 non-beam background events are expected
assuming a nominal 5 year IsoDAR run with a 90\% duty factor.
We assume that the energy spectrum of the non-beam backgrounds can be
measured with 4.5 years of KamLAND data before the IsoDAR source turns
on.  The energy spectrum of the beam-related background 
(mis-identified IBD events)
can be extracted from beam-on data with a dedicated delayed
coincidence selection. 
Given these assumptions, a combined fit to the rate and spectral shape 
of the differential ES cross section results in a 3.2\% measurement of $\sin^2\theta_W$ 
corresponding to an uncertainty of $\delta\sin^2\theta_W = $ 0.0076 (0.0057 statistics only).  
A 50\% background reduction results in a 2.5\% measurement of $\sin^2\theta_W$ 
with an uncertainty of 0.0059 (0.0048 statistics only) and a 100\% 
background reduction results in a 1.7\% measurement of $\sin^2\theta_W$ 
with an uncertainty of 0.0040 (0.0037 statistics only). 

%Given these assumptions, recasting the IsoDAR
%measurement as $\sin^2\theta_W$, results in a 3.2\% measurement from
%a combined fit to the rate and spectral ``shape'' of the differential
%ES cross section, where the error in $\sin^2\theta_W$ is 0.0076
%(0.0057 statistics-only).   For a 50\% background reduction, these
%values become 2.5\% and 0.0059 (0.0048), respectively, 
%and for the case of a 100\% background reduction become 1.7\% and
%0.0040 (0.0037), respectively.

To compare the sensitivity of IsoDAR with that of other experiments, the fits to the ES cross section can also be done in terms of $g_V$ and $g_A$. Fig~\ref{gV_gA} shows the IsoDAR 1$\sigma$ contour in  the $g_V$ versus $g_A$ plane as well as contours from other experiments. IsoDAR would be the most sensitive $\nu_e e/\bar\nu_e e$ experiment to date and could test the consistency of $\nu_e e/\bar\nu_e e$ couplings with $\nu_\mu e/\bar\nu_\mu e$ couplings.

Finally, we can also estimate IsoDAR's sensitivity to the non-universal NSI parameters $\epsilon_{ee}^{e L}$ and $\epsilon_{ee}^{e R}$, assuming the Standard Model value for $\sin^2\theta_W=0.238$.  The results are shown in Fig.~\ref{NSI} along with the current global allowed region~\cite{forero:nu-e_nsi}.  In the region around $\epsilon_{ee}^{e L}$ and $\epsilon_{ee}^{e R}\sim0$, the IsoDAR 90\% confidence interval significantly improves the global picture.

\section{Use of the Spectrum End Points for Physics and 
Calibration}
% new to this document

In addition to the rich physics provided by the IsoDAR beam, the $^8$Li
neutrino beam can be used as a calibration source for the reactor,
geo- and solar- neutrino analyses. Similar to the decays of $^{12}$B
from muon spallation, the beam provides an isotropic source of
positrons and neutrons from the IBD interactions that can be used to calibrate the
energy scale. These events will not suffer from the electronics
effects that appear in the muon spallation analyses.

The measured electron and positron spectra may be interesting to $^8$B
solar neutrino analyses like those done by SuperK and SNO.  The $^8$Li
decay proceeds through the same wide excited state of $^4$Be as $^8$B.
These measurements would be a good test of the conversion procedure
from accelerator based beta and alpha spectrum $^8$B experiments to
the $^8$B neutrino spectrum.

Studies have also shown that the sterile neutrino oscillation analysis has the statistical
power to determine the $^8$Li endpoint along with identifying the sterile neutrino
effects.  Preliminary estimates indicate that the endpoint can be determined with
an uncertainty of around 0.015 MeV using this type of analysis.

\clearpage
\chapter{General Considerations About Cyclotrons}

\section{Introduction to Cyclotrons \label{introcyc}}
% some text is directly lifted from various proposals (esp accelerator
% proposals).

Cyclotrons fall into three types: classical,  synchrocyclotrons and
isochronous cyclotrons.   We are proposing an isochronous cyclotron,
which is the type typically used for high intensity beams.  This type operates
at a fixed frequency, while increasing the magnetic field with
radius in order to 
compensate for relativistic effects.  

There are multiple ways to inject ions into a cyclotron.  In our case,
the cyclotron is fed by an ion source through a  Low Energy Beam
Transport (LEBT) system.  The ion source is held at high voltage (in
our case, 70 kV), in a
cage,  and the LEBT is at ground.   Multiple types of ions will leave
a source; the ion of interest, in our case, H$_2^+$, is selected in the LEBT using magnets
and strategically placed collimators and dumps.

\begin{figure}[t]
\centering
\includegraphics[width=3.5in]{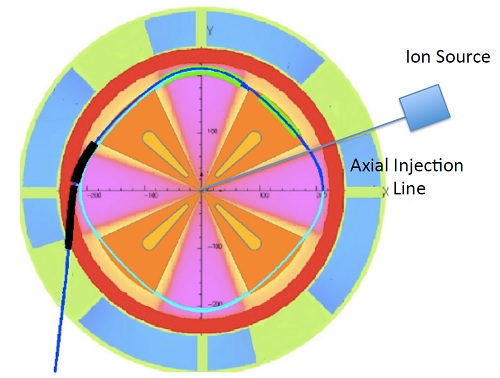}
\caption{\footnotesize  Schematic of the IsoDAR cyclotron. The beam is
  injected in the center.  The hills
  of the magnet are indicated in magenta.   Orange RF cavities are
  overlaid in the valley regions (yellow).  The coil is red.  The
  magnetic field outside of the coil is in the opposite direction of
  the interior region, and
  is indicated in blue and green on this figure.    The green, which is a weaker
  magnitude field, is at the position of gaps in the steel.    The
  last turn of the ions in the cyclotron is shown in cyan.  The beam
  is kicked outward and directed with magnetic channels to the
  extraction pipe.
\label{injcyc}}
\end{figure}

The cyclotron will accelerate  the H$_2^+$ ions to full energy, in our
case 60  MeV/amu, operating in continuous wave (CW) mode.
The beam is injected at the center of the cyclotron via a ``spiral inflector'' that
directs the beam from its axial direction into the median plane of the
cyclotron.  Then, an RF system accelerates the particles and the magnetic
field causes the ions to follow spiral orbits (``turns'') with radii
increasing with energy.   The separation between turns grows smaller
with increasing energy.    Many cyclotrons, including IsoDAR, use a compact
resistive magnet.  Having a constant RF frequency
allows particles of all energies to be accelerated simultaneously,
vastly increasing total beam current.  For vertical focusing and
isochronicity, the pole faces are cut into four sets of wedges,
alternating high-field ``hill'' regions,
where the gap between the poles is small, with low-field “valleys” where the poles are very far apart.
RF cavities and vacuum pumps are placed in the valley
regions.     On the last turn, the beam can be extracted with an
electrostatic deflector that diverts the beam into specialized magnetic
regions (``magnetic channels'') designed to direct the beam to an
extraction pipe.    A schematic of the IsoDAR cyclotron with the
features described above is 
shown in Fig.~\ref{injcyc}.     

\section{Why Choose a Cyclotron as the IsoDAR Driver?}

\begin{figure}[t]
\centering
\includegraphics[width=4in]{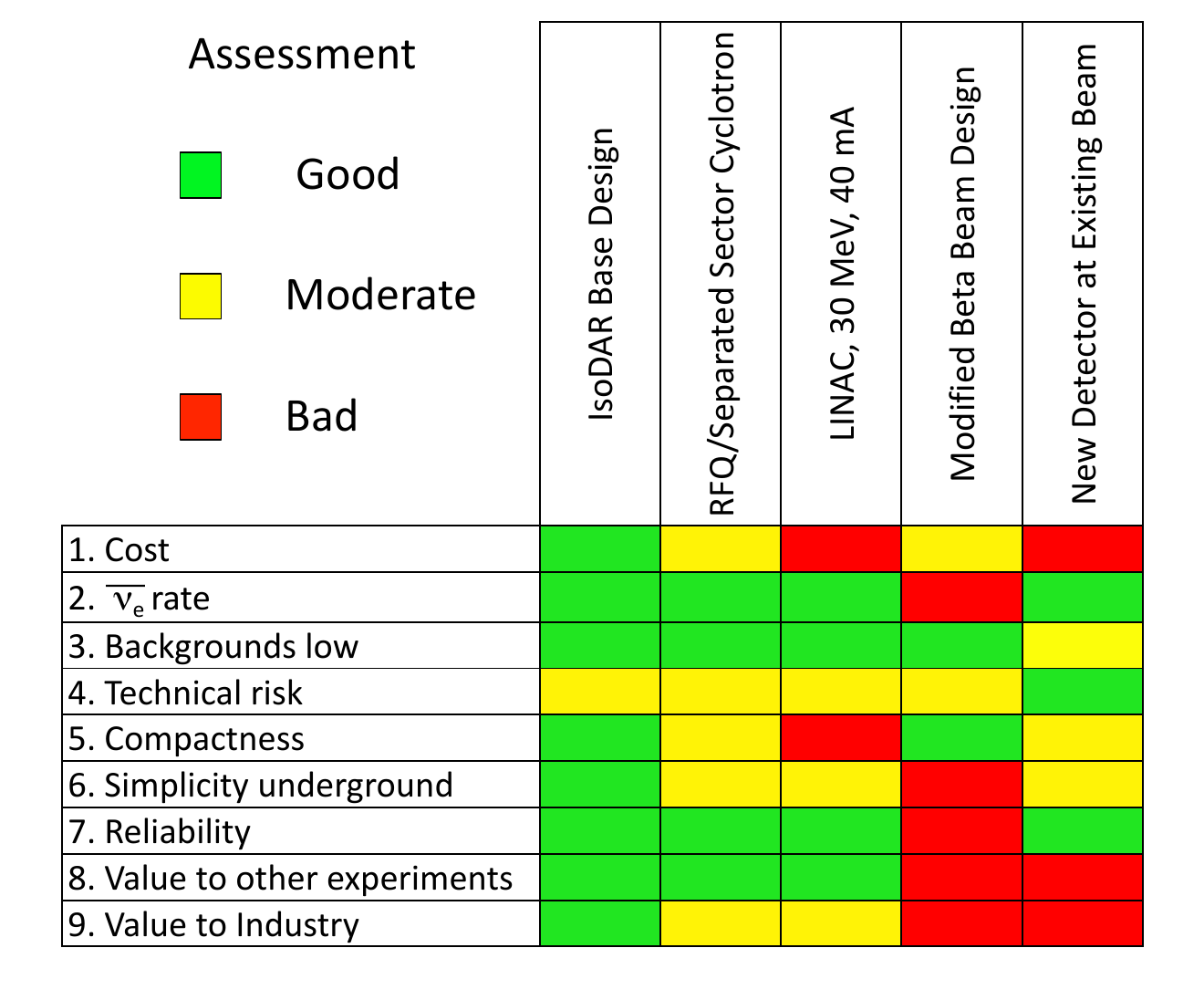}
\caption{\footnotesize  Conclusions of our study on cost-effectiveness
  of solutions for IsoDAR \cite{adelmann:isodar}.
\label{effective}}
\end{figure}

IsoDAR requires the highest possible flux of neutrons at an energy low enough for efficient capture in
 the $^7$Li target.  Neutron production is possible over a very wide
 range of energies, from D-T (deuterium-tritium) generators 
 at a few hundred kilovolts, to spallation sources at GeV proton
 energies.    We have performed a study (see
 Ref.~\cite{adelmann:isodar}) of the most cost-effective
 choice of driver for IsoDAR,  taking into account the rates required,  the
 spaces likely to be available at KamLAND and other underground facilities,  and
 reasonable power and water requirements.
A summary of the different designs in the study, and the parameters
considered for cost-effectiveness, is shown in Fig.~\ref{effective}.
This  indicated  the most efficient and economical production occurring at a proton energy of about 60 MeV, where the ratio 
of neutrons produced per incident proton is about 0.1 \cite{adelmann:isodar}.  At lower
energies the ratio of neutrons produced per incident proton is reduced, so substantially higher proton currents 
are needed, while at higher energies, where this ratio is higher, the size, cost and complexity of the 
accelerator system become very large.  As a result, we have selected
60 MeV protons as the source-of-choice driver for 
IsoDAR.

Since we brought out this study, the suggestion has been made to use
advanced D-T generators as neutron sources.  Great progress has been
made in efficiency of these generators \cite{sengbusch:n_generators},
with most work being done for
D-D (deuterium-deuterium) that do not have the environmental and licensing risks of using
tritium.  Using gaseous targets, a vendor is quoting a flux of
1--$5\times 10^{11}$
neutrons/second for a 50 milliamp deuteron beam accelerated to 300
keV. This is extrapolated to 1--$5\times 10^{13}$ if tritium is
used as a fuel.  While impressive performance, this is still a factor
of 20 below the $1\times 10^{15}$ neutrons/second produced by the
IsoDAR system.  Unless multiple generators are deployed, in a 5-10
year run,  observation of a clear oscillation wave, with differentiation
between 3+1 and 3+2 models, will not be attainable.     Also, the
rates are too low to allow for the ES scattering program.
Integration of the $^7$Li sleeve with the D-T generator for best utilization of the neutrons 
is also complicated. 
The generator target is gaseous, with a complex differential pumping system
required because of the windowless connection to the source.  
The source of neutrons is thus distributed over a distance of 0.5 to 1 m, 
%JOSE The generator target would be located between the electrodes of the 300 kV HV platform 
% making it difficult to place the sleeve close to the source of neutrons.    
making efficient coupling of the sleeve to the neutron source difficult.
It is for these reasons that we have decided to continue with our
scenario of targeting 60 MeV protons.

\section{Comparison to Cyclotrons ``On the Market''}

The IsoDAR cyclotron is at the technological edge of cyclotrons being
produced today, in terms of intensity.  Industrial interest in IsoDAR is
demonstrated by
their no-cost collaborations with us.\footnote{IBA regularly participates 
in our yearly Erice Workshops, and Best Cyclotron Systems, Inc. 
continues to support our ion source and cyclotron injection tests at their development laboratories.}
In this section, we consider
what is available now and is likely to be available in the near future.

High intensity cyclotrons in the energy range of interest to IsoDAR
are used to produce a constantly expanding tool set
of diagnostic and therapeutic isotopes
\cite{henning:medical_cyclotrons, 
      norenberg:isotope_needs, 
      aprahamian:nsac_isotopes1, 
      aprahamian:nsac_isotopes2,
      moschini:cyclotron_isotopes}. 
Need for these isotopes prompted a DOE
cost/benefit study for a 70 MeV, 2 mA proton machine by the
JUPITER Corporation \cite{jupiter:cyclotron_cost}.  
Ion Beam Applications (IBA) Cyclotron 
Solutions \cite{iba:cyclone70} and Best Cyclotrons, Inc. 
\cite{best:70p} are already
selling  70 MeV proton machines running  at 750 $\mu$A. 
CIAE is building a 100 MeV proton cyclotron
\cite{zhang:ciae_cyclotrons}.    
We expect that CIAE will commercialize this machine in the future.

Cyclotrons are also sold for scientific research. 
Laboratori Nazionali di Legnaro recently purchased a 70 MeV, high intensity 
cyclotron for the SPES (Selective Production of Exotic Species) project \cite{infn-lnl:spes}
from Best Cyclotron Systems, Inc. 
This cyclotron is currently being installed and in spring 2016 the high power 
commissioning will start.
The accelerator science goal is to eventually deliver 1 mA of protons.

All of these cyclotrons use H$^-$ ions, and are limited in intensity
by beam losses during acceleration, which is proportional to the beam
intensity.    The H$^-$ ions, which are fragile, with
only 0.7 eV of binding, are lost primarily due to two
sources: 1) electromagnetic (Lorentz) stripping of the second electron and 2)
interaction with residual gases.     Lorentz stripping limits the
highest magnetic fields allowable in the cyclotron, which in turn
limits the lowest extraction radius.     Since the cost of the cyclotron rises
with radius,   commercial cyclotrons balance acceptable losses versus
size.       The residual gases come
%JOSE changed two instances of "gasses" to "gases"
from two sources:  air which remains after pump-down, and pressure 
bursts from beam hitting the cyclotron vacuum chamber.   Commercial
cyclotrons also balance cost of the vacuum system versus acceptable losses.

While the losses due to H$^-$ are a problem, these ions do have several
advantages.     Design of the ion source to provide sufficient beam to
supply 1 mA is straightforward.    Also, the beam can
be extracted very cleanly using a stripping foil.     
The alternative which has been less competitive in commercial markets
is to use proton machines.  These must use an
electrostatic septum for extraction, yielding much higher losses and
activation.     Thus, H$^-$ has, so far, been the ion of choice for
the isotope industry.

If this technology were pushed to higher currents,  several issues
would arise.     The losses described above, which would already be
unmanageable, would be compounded  by  ``space charge'' effects, where
the beam repels itself, making the bunch very large.  This leads to
additional losses from 
beam-beam interactions in the final,
narrowly-separated turns.  Lastly,   H$^-$ ion sources would need
significant improvements to intensity.

\section{Choice of H$_2^+$ Ion \label{sec:intro_h2p}}

The breakthrough which allows the IsoDAR cyclotron to deliver 10 mA of
protons on target is directly related to the choice of ion:  H$_2^+$,  two protons bound by a single electron,
with binding energy of 2.75 eV.      This eliminates the problem of
Lorentz stripping in a compact cyclotron.    The problem of beam-gas interactions is addressed 
by building our machine with a factor of
two better vacuum than the industry standard, which is
within the range of existing machines.     

An important ``pro'' of H$_2^+$ acceleration is that it reduces
space-charge effects.   A simple way to see this is to note that for
every two protons injected at the center, there is only $+1$ electric
charge.   Thus, we have 5 mA of H$_2^+$ while we provide 10 mA of
protons to the target.
A measure of the space charge effect is the 
generalized perveance ($K=(qI)/(2\pi\epsilon_0 m\gamma^3 \beta^3)$). 
The K value of our 5 mA H$_2^+$ beam injected at 70 keV (35 keV/amu) 
is similar to that of 2 mA of protons (or H$^-$) injected at 30 keV \cite{mitsumoto:cyclotron}.  
As this H$^-$ performance has been demonstrated in commercial cyclotrons today, 
we expect that the required H$_2^+$ current should be achievable.   

One ``con'' of the choice of H$_2^+$ is that high intensity ion
sources of this type are still under development.  However, a Berkeley
multi-cusp ion H$_2^+$ source, built in 1983 and described in Ref~\cite{ehlers:multicusp1}, came 
within 20\% of our design goal for IsoDAR of 50 mA.   For this reason, we are 
confident  we will reach the necessary current.     In fact, it should
be noted that 50 mA is not the requirement for the experiment---5 mA of
accelerated H$_2^+$ beam is the requirement.
The total current which is accelerated depends upon the current from
the ion source and the combined
buncher, inflection, and RF capture efficiencies. If we attain 20\% combined efficiencies, which is the best
achieved at PSI \cite{adelmann:private},  then for IsoDAR we will require
only 25 mA from the ion source.    
On the opposite end of the scale,  if we achieve 10\% combined efficiencies, which is comparable to the best efficiency obtained
in commercial high-current compact cyclotrons
\cite{kleeven:private}, we require 50 mA for IsoDAR running.  Our
ongoing  experiments \cite{alonso:vis1}
will allow us to measure our combined efficiencies with the IsoDAR design.

A second con of the choice of this ion is the rigidity of the H$_2^+$.  This requires that
the spiral inflector on the central axis be larger than for the typical H$^-$ machine, 
with greater gaps; compensated appropriately for axial 
magnetic fields.    Our recent experimental work has been dedicated to
demonstrating the successful design of the larger spiral inflector and
benchmarking the model with measurements, and by means of numerical simulations.
The rigidity also leads to a slightly larger machine (4 m pole diameter)
than the usual commercial cyclotron (3 m pole diameter).

With \htp, we have two options for extraction. The first is by stripping, similar
to the extraction of $\mathrm{H}^-$. 
Instead of bending towards the outside, the protons emerging from the stripper
would bend sharply inwards, 
%and correct placement of the foil in the hill/valley
%magnet configuration can bring the extraction orbit to the outside of the machine
%approximately 320$^\circ$ from the stripper location.
and depending on the placement of the foil in the hill/valley magnetic 
configuration one can bring the beam out in a number
of different fashions.  Selection of the extraction path would rely
on simulations calculating optimum beam size and minimum beam loss.
The second, and our preferred option, is to use 
an electromagnetic septum, whereby we 
%deliver 5 mA of \htp, which is equivalent to 10 mA of protons, to the target.
bring the \htp ion intactly to the outside of the cyclotron.  
It will be stripped close to the extraction point so the 5 mA of \htp becomes
10 mA of protons which are sent through the transport line and delivered to the target.
This allows us to use the same extraction design as for the
DAE$\delta$ALUS injector cyclotron.    Unlike the case of proton
machines,  we can protect our electromagnetic septum with an upstream
``pre-septum'' %JOSE addition
%stripping foil that removes about 50 $\mu$A of beam that would otherwise strike the septum, 
%and thus avoiding major activation issues for this septum.
%The protons produced in this foil bend inwards, missing the septum.  They
%can be directed to an isotope production target, 
%which is of interest to our collaborators.     
%The extraction of the intact H$_2^+$ is as well  of
%interest to our industrial collaborators, should this machine be
%commercialized,  because the beam can be subdivided using foils
%downstream, peeling off $\sim 1$ mA at a time to various isotope
%targets.   Therefore, this very powerful machine can still be paired with
%today's targets, making its commercialization more attractive.  This
%is not an option if a proton or H$^-$ machine were to be used.
narrow stripping foil, and thus avoid major activation and thermal damage
issues for the septum.  It is well known that high intensity
beams form halos, where a small fraction of the beam is found at distances 
substantially greater than the extent of the main beam bunch.   
The beam can be cleaned with collimators at low energies, 
however this halo regrows within a few turns, and using collimators at high energy 
exacerbates the neutron production and machine activation problem.  Using \htp and the narrow 
stripper foil provides a method for converting this inevitable beam loss for beams of our intensity 
into a ``controlled" loss, as the stripped particles, that would damage the septum, are bent inwards 
avoiding the septum, and loop around to exit the machine cleanly into a well-shielded dump.  
This option for halo control is not available for proton machines, where septum extraction is
 necessary and a narrow pre-septum foil would not prevent the proton halo from striking the
  extraction septum.  Prospects even exist for using the 60 MeV halo particles stripped by the 
  pre-septum foil, around 50 microamps, for producing radioisotopes inside the beam dump.
  
\section{Potentially Running Deuterons}

A secondary design, involving
acceleration of deuterons rather than H$_2^+$, is also under study.
As this maintains the same $q/m$ ratio,  other elements of the accelerator can remain the
same in this case.  However,  it may be possible to reach the
necessary rates of $^8$Li with ions of less energy (40 MeV/amu) if deuterons are
used,  leading to a more compact cyclotron design.  This is attractive
for installation.   
The issues with running deuterons are related to beam losses,
%that, because of the high-velocity neutrons, cause more serious activation of the machine and
%beamlines.  Thus, this alternative needs careful study.   as it is well-known that activation from beam loss is substantially greater for deuterons 
due to the high-velocity neutrons produced when deuterons intercept any material.   
Note too, that the narrow pre-septum stripper used for halo cleaning is also not an option 
for deuterons, further increasing the difficulty of achieving clean high-intensity beams.  
Beam accelerated in a deuteron machine must then be $\mathrm{D}^-$, 
and extraction accomplished using a stripping foil.
Beam loss in a deuteron cyclotron must be very carefully studied, 
with thorough beam simulation calculations.  
We have not yet performed these simulations.

\clearpage

\chapter{Conceptual Design for the IsoDAR System at KamLAND}
The IsoDAR neutrino source at KamLAND consists of four major systems:
\begin{enumerate}
\item The front-end (ion source and Low Energy Beam Transport - LEBT),
\item the cyclotron and peripheral systems, 
\item the beam transport from the cyclotron to the target/neutrino source, and
\item the target/source including shielding. 
\end{enumerate}
The conceptual design for these four items will be described in the next 
sections, followed by the support systems (radiation protection, controls, and interfaces). Table~\ref{faqs} provides at-a-glance information for some frequently
asked questions about IsoDAR installed at KamLAND.

\begin{figure}[p]
\begin{center}
\includegraphics[width=5.5in]{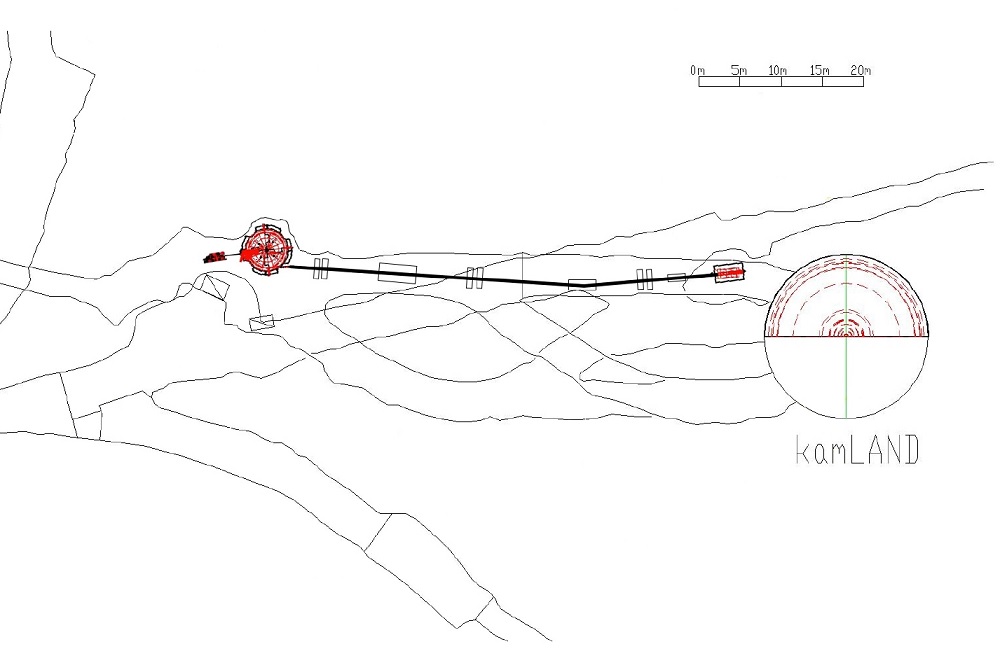}
\end{center}
%\vspace{-0.5in}
\caption{\footnotesize  Conceptual design, plan view of the IsoDAR system installed at KamLAND.
\label{kamplan}}
%\vspace{-0.25in}
\end{figure}

\begin{figure}[p]
\begin{center}
\includegraphics[width=5.5in]{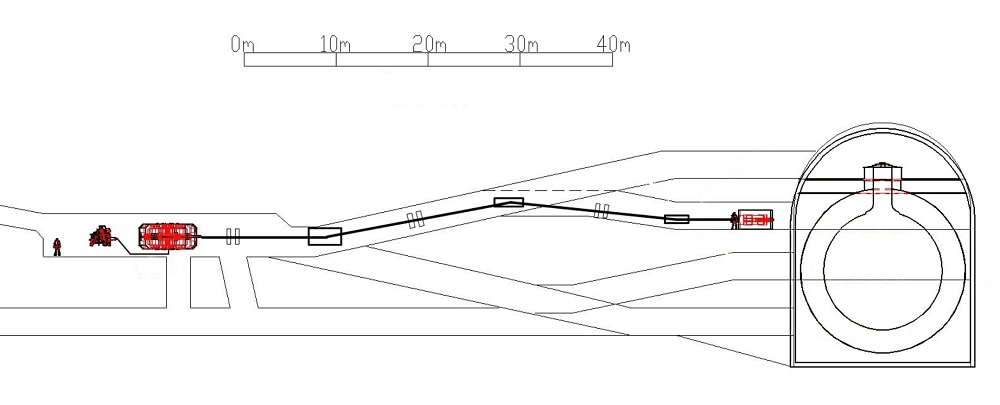}
\end{center}
%\vspace{-1.0in}
\caption{\footnotesize  Conception design, elevation view of the IsoDAR system installed at KamLAND.
\label{kamelev}}
\end{figure}

\begin{table}[tb]
\begin{center}
{\footnotesize
\begin{tabular}{|lcc|}
\hline
Question: What is the... & Design value & See... \\ \hline 
...total power requirement? & 3.4
MW & Sec. \ref{power}\\ 
...water requirement? & 5000 liters/minute & Sec. \ref{water}\\
...wait-time after beam-off before access to the detector? &
$< 30$ minutes
& Sec.~\ref{interlock} \\
...long term activation level? & $<$0.1 Bq/gm &  
Secs.~\ref{sec:shielding_calculations}, \ref{environ} \\
...conceptual layout in the KamLAND space? & &  Figs.~\ref{kamplan},
\ref{kamelev}\\ \hline
\end{tabular} \caption{\footnotesize Answers to some frequently asked
  questions about installation at KamLAND \label{faqs}}}
\end{center}
\end{table}

Figs.~\ref{kamplan} and \ref{kamelev} show the conceptual layout of the cyclotron 
and the target in the KamLAND space.   The cyclotron is located in an area currently holding an unused water-treatment plant, the target is in the back of a control room, whose functions have recently been transferred to an above-ground location.  The important constraint is to locate the target as close as allowed by background considerations. 
The beam transport between the cyclotron and the target is sketched, basically following the passageway (``drift'') between the two caverns.
A better description of the proposed transport line is given in section 4.3.
Because the weight of the cyclotron is 450 metric tons,  it will be necessary to have mine
engineers involved in the final engineering lay-out, to ensure rock stability in the areas below the cyclotron.

\clearpage
\section{Front End Design Details\label{sec:frontend}}

\subsection{Ion Source \label{ion_source_design}}

As mentioned in Section \ref{sec:intro_h2p}, being able to deliver the required 
\htp ion current of 5 mA on target depends on two factors. One is the combined
beam transport and RF capture efficiency (essentially how much of the beam produced
in the ion source reaches the target). This will be discussed in detail in Section
\ref{sec:lebt_design}. The other factor is the amount of \htp produced in the ion
source to begin with. The expected transport and RF capture efficiency for a 
conventional LEBT design is estimated to be between 10\% and 20\%. This includes 
losses in the LEBT from collimation as well as improved RF capture through 
pre-bunching the beam before entering the cyclotron. The $10-20\%$ efficiency leads to 
a requirement of 25 to 50 mA of \htp DC from the ion source. This is high for
state-of-the-art high current ion sources (which are usually optimized for either 
protons or H$^-$), but not unheard of. Based on literature, we have identified two 
possible candidates:
\begin{enumerate}
\item Off-resonance Electron Cyclotron Resonance (ECR) ion sources 
      that use a flat-field configuration. This type of ion source 
      is usually used for high proton current, but is often reported
      with a high parasitic \htp fraction. An example is the 
      Versatile Ion Source (VIS) \cite{miracoli:vis1}, built by INFN
      Catania.
\item Short multicusp ion sources without a magnetic filter. These 
      ion sources are typically used for proton or H$^-$ production 
      and utilize a number of techniques to suppress unwanted \htp. 
      By omitting these, the \htp fraction can be improved. In 
      addition, the mean free path of \htp in the ion source 
      plasma is on the order of a few cm. It is thus beneficial
      to have the bulk of ion production close to the extraction 
      aperture to avoid dissociation of already formed \htp.
      A prototype ion source at LBNL reported \htp current densities 
      at extraction of about 50 mA/cm$^2$ \cite{ehlers:multicusp1}.
\end{enumerate}
At this point, we are considering both ion source types. In the summer 
of 2014, we did preliminary tests of \htp ion beam generation, transport and injection 
into a test cyclotron in collaboration with Best Cyclotron Systems, Inc. in Vancouver.
For those tests, we borrowed the VIS from Catania. It was possible to extract 
$\approx 15$ mA, but we concluded that there are upgrade-possibilities that will 
allow us to reach a DC \htp beam current of $25-35$ mA 
(cf. Section \ref{sec:vis_tests}). 
Based on this result, we have decided to build and test a multicusp ion source similar 
to the LBNL prototype (cf. Section \ref{sec:miso_tests}).   We
identify the multicusp ion source as our primary design, with the VIS
as the alternative option.

\subsubsection{Tests of the ECR Ion Source VIS \label{sec:vis_tests}}

\begin{figure}[t]
\centering
	\includegraphics[width=1.0\textwidth]{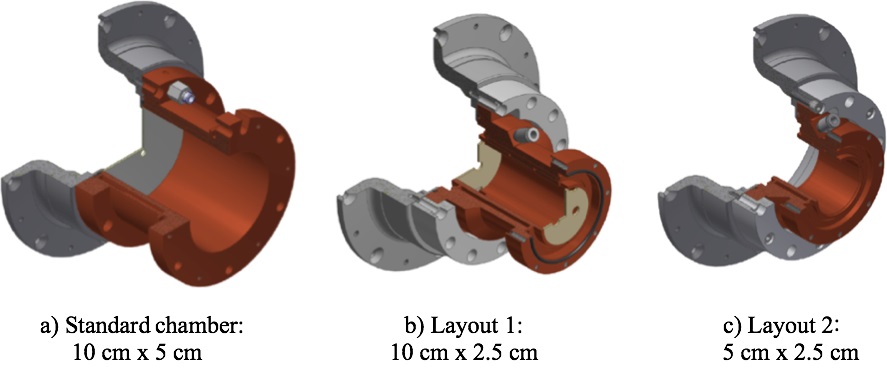}
	\caption{The three plasma chamber designs of the VIS. Courtesy of INFN Catania.}
	\label{fig:vis_chambers}
\end{figure}

In summer 2014, the collaboration, led by MIT and INFN,
performed beam extraction, transport and injection tests using the VIS. This
was done in Vancouver, at one of the laboratories of Best Cyclotron Systems, Inc.
who provided a 1 MeV test cyclotron with a large spiral inflector (design by
INFN Catania). The results pertaining to the performance of the VIS are reviewed
briefly here. Two papers reporting on the tests in detail are in preparation  
(one by the INFN ion source group \cite{castro:vis2} and one by MIT
\cite{winklehner:bcs_tests}).

The VIS is an ECR ion source. Electrons are heated by electron cyclotron resonance
using 2.45 GHz microwaves introduced into the plasma chamber through a waveguide
and a magnetic field provided by two permanent magnet solenoid rings. The magnetic 
field corresponding to electron resonance at 2.45 GHz is 875 Gauss. In the initial configuration (see 
Figure \ref{fig:vis_chambers} a), the plasma chamber was still optimized for the 
production of protons. It was possible to extract a maximum of 12 mA of \htp 
with a 1:1 ratio of protons to \htp using a high source pressure 
($\approx 1.2\cdot10^{-5}$ mbar at the exit) and low microwave power (300 W). 
The Catania ion source group then devised two alternative designs of the plasma 
chamber which should improve the \htp fraction by reducing the path length from \htp
creation to extraction. These can be seen in Figure \ref{fig:vis_chambers} b) and c).
Tests with Layout 1 yielded an improved \htp current of $\approx 15$ mA. 
Layout 2 will be tested in Catania during the coming year and further improvement is
expected.

Current can be increased by increasing the size of the extraction aperture. 
Emittance measurements showed that we are well within the acceptance of the cyclotron
spiral inflector and can increase the extraction aperture of the VIS from
8 mm diameter to 10 or even 12 mm. This would yield another increase in current of 
50-120\%, overall leading to an estimated current increase to $25 - 35$ mA.

\subsubsection{Tests of the Multicusp Ion Source MIST-1 \label{sec:miso_tests}}
\begin{figure}[t]
\centering
\begin{minipage}{.4\textwidth}
  \includegraphics[height=0.25\textheight]
                  {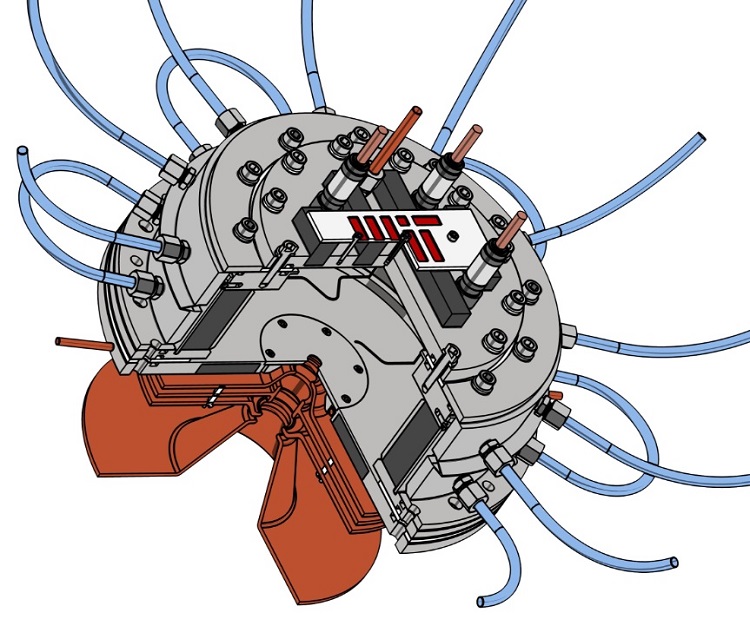}
\end{minipage}
\hspace{.05\textwidth}
\begin{minipage}{.5\textwidth}
  \includegraphics[height=0.25\textheight]
                  {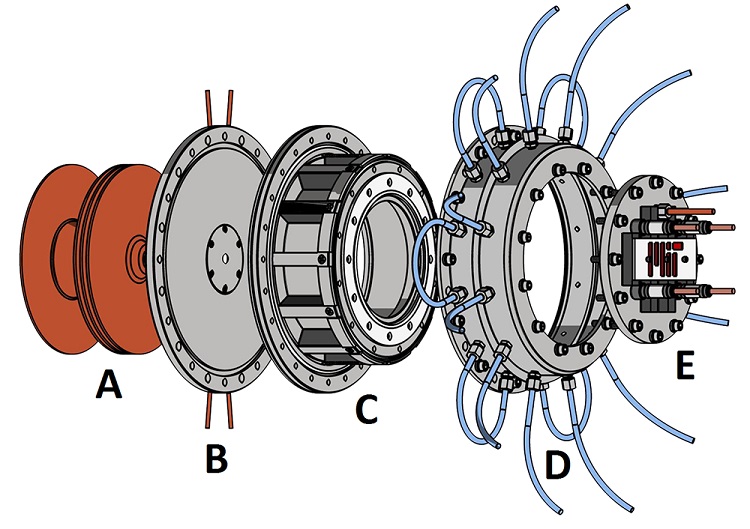}
\end{minipage}
\caption{CAD renderings of MIST-1. Left: Quarter-Section view, 
         Right: The major pieces pulled apart along the axial 
         direction. From left to right: A) Preliminary extraction 
         system, B) front plate with exchangeable extraction hole,
         C) plasma chamber with permanent magnets, D) Water cooling
         sleeve around plasma chamber, E) Exchangeable back-plate.
         Note: The ceramics insulator around the extraction 
         electrodes has been omitted for better visibility.
         \label{fig:miso_cad}}
\end{figure}

Because the VIS tests were only marginally successful, 
a decision was made to build a prototype multicusp
ion source at MIT (called MIST-1).   
While literature indicates this is expected to
produce more current, construction of a prototype allows concrete
comparison of the
performance and long time stability of both types of ion sources.
The major design choices with respect to the production of \htp are:
\vspace{-\topsep}
\begin{itemize}
\item Short Chamber to minimize losses through dissociation.
\item Magnetic configuration with field a free region near the extraction aperture.
\end{itemize}
\vspace{-\topsep}
The design of MIST-1 is well underway \cite{axani:mist1}. 
The main parameters are listed in
\tabref{tab:mist}. A detailed CAD model has been produced (see Figure \ref{fig:miso_cad})
and a full set of fabrication drawings has been generated from it. 
Funding has been secured and fabrication of the parts is currently ongoing with assembly in mid-October 2015 and first commissioning tests at the end of October 2015. 
A publication with the design details and test results will 
be available in winter 2015.

\begin{table}[b]
	\caption{MIST-1 - Main Ion Source Parameters.\label{tab:mist}}
	\centering
    \renewcommand{\arraystretch}{1.25}
		\begin{tabular}{ll}
            \hline
		    \textbf{Design Parameter} & \textbf{Value} \\
            \hline \hline
            Plasma Chamber Dimensions (diameter, length) & 150 mm, 
            80 mm \\
            Permanent Magnet Material        & Sm$_2$Co$_{17}$ \\
            Extraction Voltage               & 15-22 kV  \\
            Heating Method                   & Tungsten Filament \\
            \hline
		\end{tabular}
\end{table}

\subsubsection{Risk Assessment and Mitigation \label{sec:ion_source_risk}}
\textbf{Risk: Ion Source Long Time Stability.} As IsoDAR is designed to run continuously
for 5 years, the long time stability is a major concern for the ion source. Hands-on
maintenance of the ion source is going to be limited, because of the close proximity of
the neutron producing cyclotron. This is mainly a risk for multicusp-type ion sources driven by filaments or internal RF antennas.

\emph{\textbf{Mitigation:} We do not expect this to be a risk for a ECR-type ion
source, which have demonstrated outstanding long-term stability in the past.
For the multicusp-type ion source, we will develop alternative heating methods (external antenna and 2.45 GHz microwaves). MIST-1 is being designed with an exchangeable 
back plate for quick filament replacement and easy further development of alternative
heating methods. Another option is dual source operation utilizing a switching magnet.
We estimate ~50k\$ for source development beyond the ongoing project}

\textbf{Risk: Ion source maximum \htp current is too low}. We already know we can 
deliver 15 mA of \htp by using the VIS. With future upgrades and larger extraction
aperture, we are certain to reach at least 25 mA. Being able to run IsoDAR then depends strongly on LEBT and bunching (see next section). The performance of the multicusp
option will be benchmarked soon. Papers suggest 40 mA from a 10 mm diameter aperture 
should be possible.

\emph{\textbf{Mitigation:} As mentioned earlier, the real current needed depends on 
the bunching and injection efficiencies. The stretch goal of the ion source is to
deliver the current necessary to run IsoDAR while operating with the lowest possible 
combined injection efficiency (50 mA), however, a more realistic expectation is that 
ion source and buncher will both work at an average efficiency thus relaxing the ion
source requirement to about 35 mA. Moreover, the geometric emittance of the beam depends
on the beam energy. The experiments at BCS have shown that we can inject at 60 keV beam
energy. If we increase the beam energy to 80 keV (still within the limits of the spiral inflector) the reduced geometrical emittance will allow opening the source aperture
even more, thus generating more beam current. Finally, we are investigating the 
possibility of a direct ion source-RFQ-cyclotron injection scheme (see next section),
which could increase the combined injection/RF capture efficiency to $> 40\%$.}

\clearpage
\subsection{Low Energy Beam Transport (LEBT) \label{sec:lebt_design}}
The Low Energy Beam Transport line (LEBT) mainly serves the following purposes:
\vspace{-\topsep}
\begin{itemize}
\item Transport the beam from the ion source to the accelerator.
\item Preserve beam quality (emittance).
\item Filter out unwanted ion species created in the source (protons and 
      $\mathrm{H}_3^+$).
\item Bunch the beam to increase the cyclotron RF capture efficiency.
\end{itemize}
\vspace{-\topsep}
In addition, the LEBT can provide space for beam diagnostics.
Based on the findings during the tests in Vancouver, we have identified two
options for the LEBT:
\vspace{-\topsep}
\begin{enumerate}
\item A `conventional' LEBT consisting of a dipole analyzing magnet, several 
      focusing solenoids and a two-gap buncher. This option has the larger footprint,
      but also leaves space for diagnostics. The placement of the ion source
      with respect to distance from the cyclotron is somewhat flexible.
\item An RFQ-based LEBT in which the ion source is closely coupled to a 
      Radio-Frequency Quadrupole (RFQ) which is re-entrant to the cyclotron
      yoke. This would be a very compact system with a high bunching efficiency,
      but the ion source would be very close to the cyclotron making hands-on
%      maintenance more difficult. While the RFQ-based LEBT offers a number of (Jose suggestion)
	maintenance more difficult. While RFQ-based axial injection into a compact cyclotron offers a number of 
      advantages (as discussed in Section \ref{sec:rfq_lebt}, it is clearly in need 
      of additional R\&D, as it has never been realized before. 
\end{enumerate}
\vspace{-\topsep}
Until an experiment demonstrating injection of \htp into a cyclotron
using the RFQ is performed, the RFQ is considered an alternative design and
the conventional LEBT will be the primary design.

Two possibilities for the ion source and two possibilities for the LEBT 
make a total of 4 possible combinations. We will focus on the
two options that make more sense at this point, although the other two 
combinations are not excluded. The two options are:
\vspace{-\topsep}
\begin{enumerate}
\item \textbf{Multicusp ion source with conventional LEBT.} 
      At this point, we expect the multicusp ion source to 
      deliver the better performance, but with the drawback 
      of limited lifetime. We have a switching magnet and 
      thus could envision the use of two sources to 
      essentially double the lifetime. This is our primary design.
\item \textbf{ECR ion source with RFQ LEBT.} ECR ion sources 
      have virtually infinite lifetime, but so far delivered 
      only 15 mA. This means better bunching is necessary, 
      which is something the RFQ-based LEBT would provide.  
\end{enumerate}
\vspace{-\topsep}

%Layout Options 1 and 2 are shown in Figures \ref{fig:conv_lebt} 
%and \ref{fig:rfq_lebt}, respectively and are discussed in more 
%detail in the following subsections. 

Layout Options 1 is shown in Figure \ref{fig:conv_lebt} and 
both options are discussed in more detail in the following 
subsections. 

\clearpage
\subsubsection{Option 1: Conventional LEBT \label{conv_lebt}}

\begin{figure}[!t]
\centering
\includegraphics[width=1.0\textwidth]
                {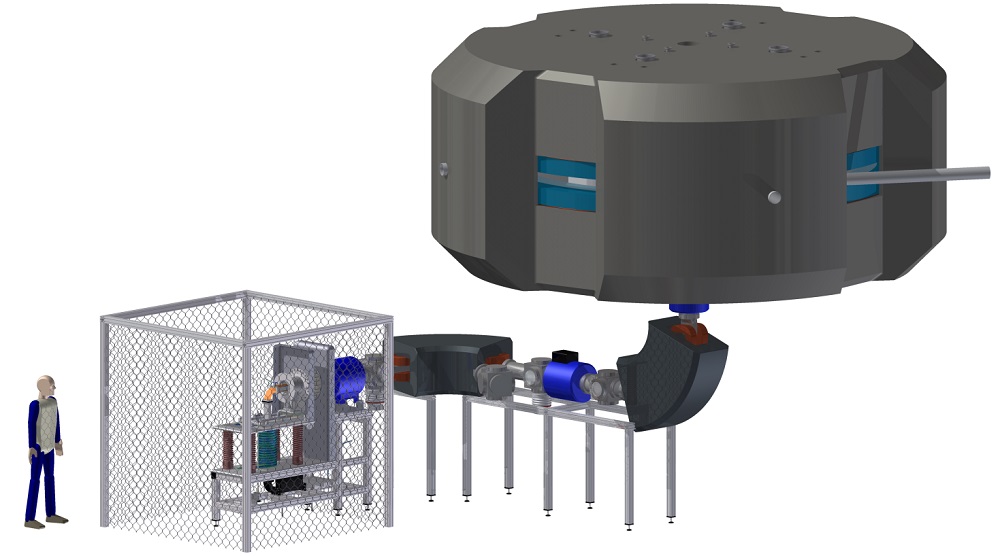}
	\caption{\footnotesize Relative size of the ion source, the
	         conventional LEBT and the cyclotron. The VIS was used
	         as a placeholder for MIST-1 for now. We expect the high
	         voltage platform and cage to be of similar design and 
	         size. 
	         \label{fig:conv_lebt}}
\end{figure}

In this option, the ion beam is extracted at 80 kV from the multicusp ion 
source and transported through a system of ion beam focusing elements. 
The layout can be seen in Figure \ref{fig:conv_lebt}. A dipole analyzing 
magnet placed close to the ion source separates protons and $\mathrm{H}_3^+$ from
the \htp ions. Solenoid magnets placed right after the ion source, between the
two bending magnets and close (possibly slightly re-entrant) to the yoke of the
cyclotron will provide final beam focusing into the spiral inflector (cf. Section
\ref{sec:injection}). Diagnostic boxes will provide space for
beam current and quality monitoring, and a two-gap buncher, operating at the 
frequency of the cyclotron, which will bunch the beam for improved RF capture 
efficiency. According to simulations (cf. Section \ref{sec:lebt_sim}) LEBT 
transmission is expected to be on the order of 95\% assuming high space charge 
compensation.

The length of this LEBT is on the order of 5 m. Depending on the availability of 
space in the Kamioka mine, the ion source and LEBT may be put in a trench underneath 
the cyclotron. A longer LEBT can be designed if shielding between the source and 
the cyclotron becomes necessary for hands-on maintenance of the ion source.

\clearpage
\subsubsection{Option 2: RFQ LEBT \label{sec:rfq_lebt}}

%\begin{figure}[!t]
%\centering
%    \includegraphics[width=1.0\textwidth]
%                    {source_rfq_cyclotron.png}
%    \caption{\footnotesize Relative size of the ion source, the RFQ LEBT and
%	         the cyclotron. PLACEHOLDER! This should be updated to VIS with RFQ.
%             \label{fig:rfq_lebt}}
%\end{figure}

% The RFQ is a linear accelerator that can focus, bunch, and accelerate a continuous 
The RFQ is a linear beam transport structure that can focus, bunch, and accelerate a continuous 
beam of charged particles at low energies with high bunching and transmission
efficiencies. The basic principle is that of voltage modulated in the radio frequency
range applied to a set of four rods (or vanes). Wiggles on the rods provide a
longitudinal electric field component that accelerates the ions. With properly tuned
RF frequency and vane voltage, the \htp ions are continuously focused, bunched and
accelerated in the RFQ. The RFQ would almost completely sit inside 
the cyclotron yoke with the ion source being only a slight
protrusion on the bottom (or top) of the machine.

This method has several advantages over a conventional LEBT design: 
\vspace{-\topsep}
\begin{itemize}
\item The bunching efficiency is higher than for the previously considered two-gap 
      buncher and thus the overall injection efficiency is higher. This relaxes the 
      constraints on the \htp current required from the ion source. 
\item The overall length of the LEBT can be reduced significantly.
\item The RFQ can also accelerate the ions. This enables the ion source platform
      high voltage to be reduced from 80 kV to 15 kV, making underground 
      installation easier. 
\end{itemize}
\vspace{-\topsep}

\begin{table}[!b]
    \caption{\footnotesize Preliminary IsoDAR RFQ-Linac Injector Parameters.
             (\textsuperscript{a} The PARMTEQ 
              study will soon be updated to 
              the actual cyclotron frequency of 32.8 MHz. However, is
              not expected to significantly change 
              the preliminary findings.) \label{tab:rfq_params}}
    \centering
    \renewcommand{\arraystretch}{1.25}
    \begin{tabular}{ll}
        \hline
        \textbf{Parameter} & \textbf{Value} \\
        \hline\hline
            Operating frequency\textsuperscript{a} & 33.2 MHz \\
            Injection energy & 15 keV \\
            Final beam energy & 80 keV \\
            Design input current & 10 mA \\
            Current limit & 22 mA \\
            Transmission at 10 mA & 99\% \\
            Inp. trans. emittance (6-rms, norm.) & 0.5 $\pi$-mm-mrad \\
            Nominal vane voltage & 43 kV \\
            Bore radius & 1.27 cm \\
            Maximum vane modulation & 1.94 \\
            Structure length & 1.09 m \\
            Peak RF field surface gradient & 4.66 MV/m \\
            Structure RF power & 0.82 kW \\
            Beam power & 0.64 kW \\
            Total input RF power & 1.46 kW \\
        \hline
    \end{tabular}
\end{table}

The idea of using an RFQ re-entrant to the cyclotron yoke, to axially inject ions 
into a cyclotron was first presented in 1981 \cite{hamm:rfq1}, but has not yet been 
realized. A preliminary feasibility study for use in IsoDAR was conducted
in collaboration with the author of the original 
paper \cite{winklehner:rfq1}. Here the established RFQ design
code PARMTEQ \cite{crandall:parmteq1} was used. The parameters of the preliminary 
design are listed in Table \ref{tab:rfq_params}. The results showed a total phase 
width of the beam of 90\degree, but 60\% of the beam is within $\pm10\degree$. 
The energy spread of the beam at 80 keV has a FWHM value of 5 keV but has a ``halo" 
that extends out to $\pm15\%$. This would lead to a cyclotron RF acceptance of 
60\%. Assuming a reduction to 40\% (still twice the maximum acceptance compared to 
the conventional LEBT) due to de-bunching while going through the spiral inflector, 
the ion beam current required from the source would be on the order of 13 mA.

The ion source would be closely coupled to the RFQ and matched using a segmented
einzel lens as part of the multi-electrode extraction system. 
%Jose added
The RFQ is a very good q/A filter; protons and $\mathrm{H}_3^+$ ions are dumped onto the vane surfaces in no more than 10 cm, beyond this only \htp ions survive.  Approximately 150 watts will be deposited on the vane surfaces, but this load will be small compared to the heating from RF currents, so the cooling system for the vanes will handle this readily.
%end Jose addition
%A conceptual design is depicted in Figure \ref{fig:rfq_lebt}.  
The compactness of the design is attractive for underground 
installation.

%more Jose suggestions
The 5 keV energy spread implies that the beam will begin losing its tightly bunched nature as soon as it leaves the RFQ.  For this reason it is necessary to have the RFQ exit close to the spiral inflector, to minimize the distance to the first accelerating gap in the cyclotron.  This minimum distance, from RFQ exit through spiral inflector to the dee gap is estimated to be no more than 40 cm, which is the basis for the 60$\%$ capture efficiency quoted above.
%end Jose suggestion.

From this conceptual design study, we find that using an RFQ indeed is a possibility 
for IsoDAR, albeit with additional research and development necessary.

%The typical acceptance of an unbunched beam into the cyclotron RF bucket is on the 
%order of 5\%, and with a 2-gap multiharmonic buncher, this acceptance can be increased 
%to 10-20\%.

%RFQ's are very attractive for low energy ion accelerators, e.g. for
%applications with high current beams or in combination with sources such as an ECR,
%because the source can be close to ground potential and is easy to operate and to 
%service. Because the basic RFQ concept can be implemented over a wide range of
%frequencies, voltages, and physical dimensions, it is an ideal structure to use for
%bunching an intense ion beam for injection into a cyclotron 

\subsubsection{Risk Assessment and Risk Mitigation\label{sec:lebt_risk}}
\textbf{Risk: New RFQ Technology}. Direct axial injection into
a compact cyclotron has never been demonstrated. It is possible
that there are limitations, that have not been thought of, that will
prove to be show-stoppers. 

\emph{\textbf{Mitigation:} A prototype machine will be built and
thoroughly tested before the RFQ direct injection LEBT will become
the primary design option. As a fallback, the conventional LEBT
design has proven to be well understood and reliable. The focus in
this case will shift to increasing the ion source performance
further.}

\textbf{Risk: RFQ Power Estimate} The RFQ costs depend strongly 
on the RF power necessary to drive the rods/vanes, the length of 
the RFQ, the engineering complexity involved in machining the 
vanes/rods and designing the vacuum vessel.
     
\emph{\textbf{Mitigation:} We will work with RF engineers with intimate knowledge of RFQ design to obtain accurate RF system cost estimates.}

\textbf{Risk: Beam spread between RFQ exit and spiral inflector.}
The conceptual design studies undertaken so far showed that the 
beam spreads transversally and longitudinally after exiting the RFQ. 
To an extent, where it might become unusable.

\emph{\textbf{Mitigation:} In the next design iteration, we will have 
much more accurate simulations of the transport from the RFQ exit to 
the spiral inflector entrance and of the spiral inflector and acceleration
through the first few turns of the cyclotron using the PARMTEQ output beam. Additional focusing elements can be put in between RFQ and spiral inflector.}

\subsection{Front End - Current Status and Future Work}
Regarding the IsoDAR front end (ion source and LEBT) we are currently 
working on the following items:
\vspace{-\topsep}
\begin{itemize}
\item Design and construction of a multicusp ion source optimized for \htp. 
      This project is planned to conclude in fall 2015 with the commissioning
      of the ion source and publication of a paper about the first test.
\item Design studies and simulation of an RFQ-based LEBT. We are further 
      investigating the use of an RFQ as LEBT and are refining the PARMTEQ
      model as well as simulations of the beam transport from the exit of the
      RFQ through the spiral inflector and the first turns of the cyclotron
      using OPAL (cf. Chapter \ref{sec:simulations}). This will help us confirm
      the preliminary results regarding feasibility and come up with more 
      accurate cost estimates. This is likely an ongoing effort that will 
      lead into a full design study of a RFQ-based LEBT for direct cyclotron
      injection.
\end{itemize}
\vspace{-\topsep}
In order to arrive at a preliminary design report (PDR) level, the following 
issues need to be addressed:
\vspace{-\topsep}
\begin{itemize}
\item Further development of the multicusp ion source with respect to plasma
      heating methods are important. The filament was chosen for now because
      it is the simplest method, but it is limited by lifetime. An external
      RF antenna or 2.45 GHz microwave heating will be investigated. The 
      conceptual design of the ion source includes an exchangeable back plate
      to facilitate this.
\item The availability of space at the installation site in the Kamioka mine 
      needs to be included in the LEBT design.
\item As part of the preliminary design phase, both options listed in the LEBT 
      section (Section \ref{sec:lebt_design}), need to be carefully
      evaluated. For this, a full design study of an RFQ-based LEBT for direct cyclotron
      injection, including construction of a prototype,  is necessary.
\end{itemize}
\clearpage
\section{Cyclotron Design Details \label{sec:cyclotron}}

\begin{figure}[!t]
\centering
\includegraphics[width=5.in]{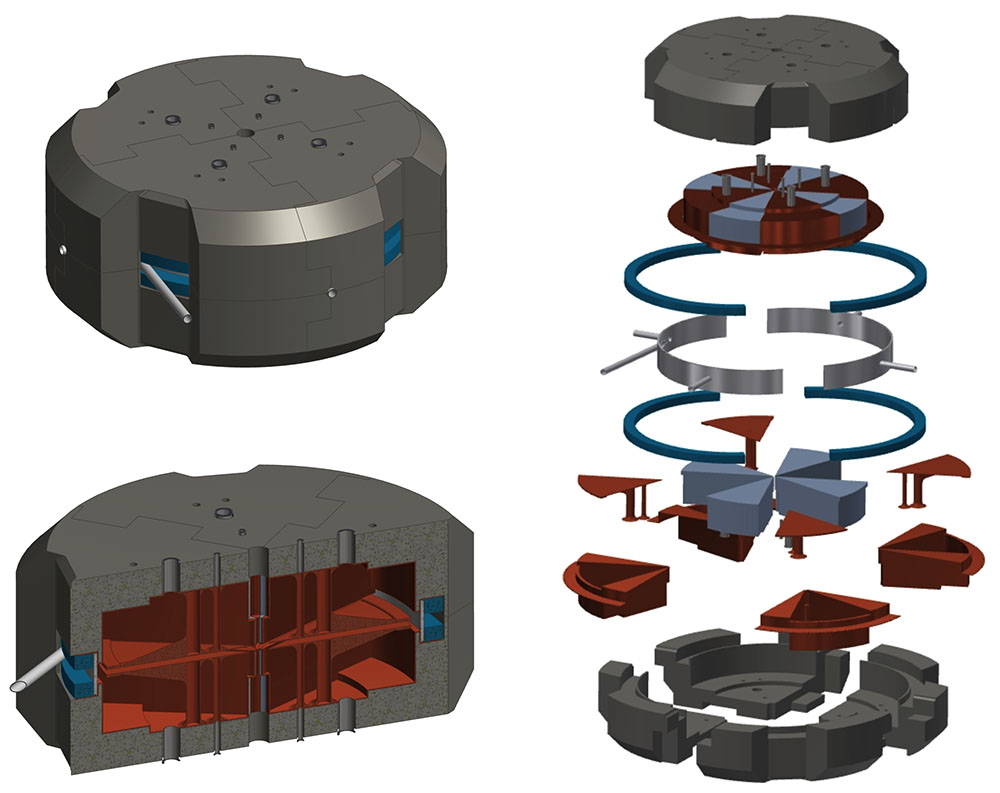}
\caption{\footnotesize Views of the cyclotron. Top Left: Oblique view of the
  cyclotron. One sees the vacuum jacket and the extraction line.
  Bottom Left: Cutaway view of the cyclotron. One sees the coil, the
  iron and the Dees. Right: Exploded view of the cyclotron showing all 
  components. Each component is subdivided in small pieces that can be 
  transport into the Kamioka mine.
\label{cycloviews}}
\end{figure}

In this section we describe the details of the cyclotron design. 
Beam dynamics in the cyclotron, from injection to extraction, are discussed
in Chapter 6, where the beam is traced from the ion source through to the target.

\begin{table}[!b]
\begin{center}
\renewcommand{\arraystretch}{1.25}
{%\footnotesize
\begin{tabular}{|lr|lr|}
\hline
Design element & Design value & Design Element & Design value \\ \hline 
$E_{max}$ & 60 MeV/amu	& $E_{inj}$ & 35 keV/amu \\
$R_{ext}$ &	1.99 m &
$R_{inj}$ &55 mm  \\
$<B>$ @ $R_{ext}$ &1.16 T	 &	
$<B>$ @ $R_{inj}$ &	0.97 T  \\
Sectors		& 4		& 	
Hill width	&	25.5 - 36.5 deg \\
Valley gap	& 1800 mm	&
Pole gap	& 80 - 100 mm  \\
Outer Diameter & 6.2 m	& 
Full height & 2.7 m  \\
Cavities	& 4	&
Cavity type	& $\lambda/2$, double gap  \\
Harmonic &	 4th		&
rf frequency	& 32.8 MHz  \\
Acc. Voltage	& 70 - 240 kV	 &
Power/cavity &	$ 310$ kW  \\
$\Delta E$/turn	 &1.7 MeV &	
Turns &95  \\
$\Delta R$ /turn @ $R_{ext}$	& $>20$ mm~$^\dagger$	&
$\Delta R$/turn @ $R_{inj}$ & $>56$ mm  \\
Coil size & 200x250 mm$^2$ &
Current density	 & 3.167 A/mm$^2$  \\
Iron weight & 450 tons	&
Vacuum  & $< 10^{-7}$ mbar \\
\hline
\multicolumn{4}{l}{ {$^\dagger$}Additional shift from first harmonic precession}\\ 
\end{tabular} \caption{\footnotesize Details of the proposed cyclotron design.\label{cyclodesigntable}}}
\end{center}
\end{table}

Table~\ref{cyclodesigntable} describes the specifics of the cyclotron design.
The cyclotron design, developed at
INFN-Catania, will accelerate 5 mA of \htp ions up to 60 MeV/amu. 
This current is an order of magnitude higher than currently available
in commercial cyclotrons, giving it a potential application as a
production source of medical isotopes \cite{alonso:isotopes}. 
IsoDAR will be a four sector machine, with a pole radius of 220 cm
and a large vertical gap of 10 cm. The hill angular width is 25.5\degree in the central region and increases up to 36.5\degree in 
the extraction region. There are two normal-conducting coils, 
each of which has an inner radius of 223 cm and a size of 200 $\times$ 250 mm$^2$; 
the current density is 3.167 A/mm$^2$. 
Located outside the coils is the outer steel yoke, which accounts for the bulk of the 
450 metric ton weight of the cyclotron and for its 6.2 m outer diameter.
%The external yoke completes the system, 
%resulting in a 6.2 meter outer radius and the bulk of the mass of the cyclotron. 
The average magnetic field
varies between 1.05 and  1.2 T, while the minimum and maximum
values are 0.28 T in the valley and 2.11 T in the hill section. 

Injection will be done using a spiral inflector placed at the center of the cyclotron. The axial hole from the yoke surface ($z = 1350$ mm) to a distance of 450 mm from the median plane has a diameter of 280 mm. There is enough room to install quadrupoles and solenoids inside this axial hole to optimize the injection efficiency. Also in the region nearest to the median plane, from $z = 450$ mm to $z = 80$ mm, the axial hole has a generous diameter of 140 mm which easily allows installation of a large spiral inflector.

The outer parts of the poles have a special shape to allow a smooth crossing of 
the $\nu_r=1$ resonance, which increases the inter-turn separation for extraction 
efficiency to close to 100\% using two electrostatic deflectors.
Four RF double-gap cavities, based on a design for a commercial cyclotron, and tuned for 4th harmonic operation, are placed in the four magnet valleys.
The $\lambda$/2 cavities produce an accelerating voltage
that rises from 70 kV at inner radii to 240 kV at the outer radius.  

\begin{figure}[!t]
\centering
\includegraphics[width=3.5in]{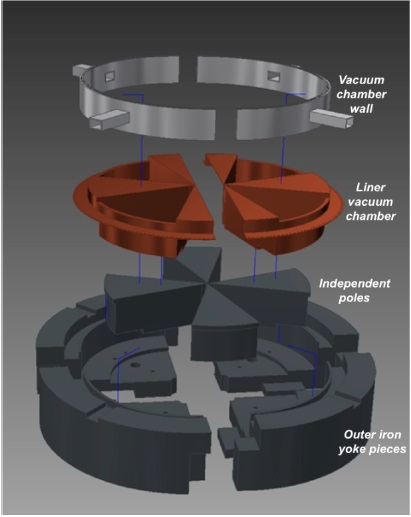}
\caption{\footnotesize  Cyclotron split into parts.
\label{cyclosplitparts}}
\vspace{0.2in}
\end{figure}

Support systems for the accelerator and target include the
water-cooling systems; electrical power-converters for driving the
magnets, ion-source components, and RF systems;  ventilation systems;
safety and access control systems; and overall systems management and
control computers.  These will be located in close proximity to the
technical components, although the human control points can be located
remotely.  

%Installation of IsoDAR in the Kamioka mine adds two major constraints: the 
%limitations of maximum size (2.5 m) and weight (12-50 tons), to allow for
%handling of each component/part at its final location on site (cave). 
%Jose added:
We are fortunate in that access to the underground site is ``drive-in", i.e. does not require 
a hoist or conveyance (elevator) to reach the level of the experiment.  
Nonetheless, size and weight restrictions still exist.  The maximum weight of a piece
to be transported is set by the load limit of a bridge on the secondary road leading from the
main highway to the adit (mine entrance).  This limit is currently 20 tons (should confirm).
The size limit is determined by the height of the access tunnel: four meters at the most
constricted part. 
The cyclotron, as well as all other parts of the IsoDAR system, must be segmented into
pieces that do not exceed these limits.

This creates several challenges:

\vspace{-\topsep}
\begin{itemize}
\item Many pieces, such as the main poles and coils for the cyclotron are usually made in one
piece.  These will need to be broken apart into smaller pieces.
\item In the case of the coil, either the coil must be split into halves, or it must be wound underground.
\item The steel pieces will need to be reassembled to the micron-level accuracy needed
to ensure accuracy of the magnetic field.  
\item The vacuum liner, shaped around the hills and valleys of the cyclotron magnet will
also need to be brought in in pieces and welded in place.
\item A substantial lay-down area will be needed for staging of cyclotron parts, and for welding
and assembling individual components, and performing the actual assembly of the cyclotron.
\item Rigging equipment in the staging area will be needed, with hoisting capacity
of over 100 tons.
\end{itemize}
\vspace{-\topsep}

A plan has been developed for the cyclotron segmentation, 
figures \ref{cycloviews} and \ref{cyclosplitparts} show the main components of the cyclotron subdivided in 
small pieces that can be transported into the Kamioka mine.

\subsection{Cyclotron Injection \label{sec:injection}}
\begin{figure}[!t]
    \centering
    \includegraphics[width=0.5\textwidth]{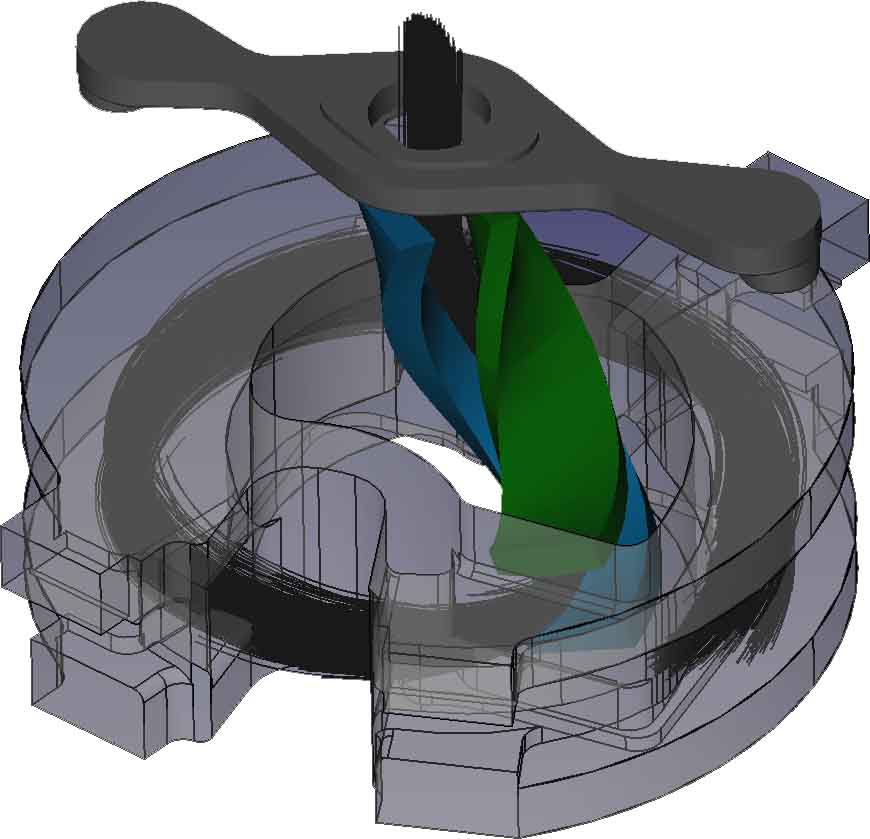}
    \caption{CAD rendering of the spiral inflector and particle trajectories for the BCS tests. 
             The positive and negative electrostatic deflectors are shown in 
             blue and green respectively. The cyclotron magnetic field in this 
             image is directed vertically upwards. Particles enter the spiral 
             inflector via the rectangular grounded collimator (solid gray) and 
             are guided into the cyclotron mid-plane by means of the cyclotron 
             magnetic field and the electrostatic potential between the 
             electrodes. The copper housing (transparent gray) isolates the 
             spiral inflector from the RF fields driving the
             cyclotron. 
             The recesses on both sides of the housing (one of them visible on 
             the lower left) provide space for the tips of the dees.}
    \label{fig:bcs_inflector_opal}
\end{figure} 
The injection system design is a crucial component to achieve the highest 
beam current.   Our design balances injecting the  \htp molecule at high 
energy against avoiding
the use of complex high-voltage platforms.
Higher energy injection mitigates space charge and to reduces 
geometrical (non-normalized) emittance.
An energy of 35 keV/amu was selected because ion sources like VIS or MIST-1
can work with an extraction voltage of 70 kV. 
In case of an RFQ-based LEBT, the ion source voltage can be much lower and
the RFQ can pre-accelerate the beam to the necessary 35 keV/amu.
As in commercial cyclotrons, an axial injection system based on a Spiral 
Inflector (SI) will be used to bend the beam from the axial direction to the 
median plane. This concept is visualized in \figref{fig:bcs_inflector_opal}
using trajectories calculated with OPAL (cf. \secref{sec:simulations}).
The exit point of the spiral inflector is also the starting point of 
acceleration in the Central Region (CR) of the cyclotron. 
The central region and the spiral inflector have to be designed carefully 
and together because they have strong interplay. This procedure has already 
been performed and tested for the injection tests during the summers of 2013 
and 2014 at the Best Cyclotron Systems development laboratory  
(BCS) in Vancouver. In the following subsection,
we will describe the design process of the central region for the BCS test 
stand cyclotron and point out the main differences of the full IsoDAR machine.

\subsubsection{Spiral Inflector Design}

The axial injection of the ionized beam into a cyclotron is 
realized using an electrostatic device called a spiral inflector, which consists
of two curved electrode deflectors (see Figure~\ref{fig:bcs_inflector_opal}). 
The 
electrostatic potential between these electrodes is able to bend the beam 90\degree
from the axial line to the median plane of the cyclotron. The helical trajectory of the beam is determined both by the shape of the electrodes (electric field) and the
magnetic field of the cyclotron as the beam is bent into the median plane. The applied voltage depends on the velocity of the beam as well as the rigidity of the ions. 

\begin{figure}[!t]
    \centering
    \includegraphics[width=0.6\textwidth]{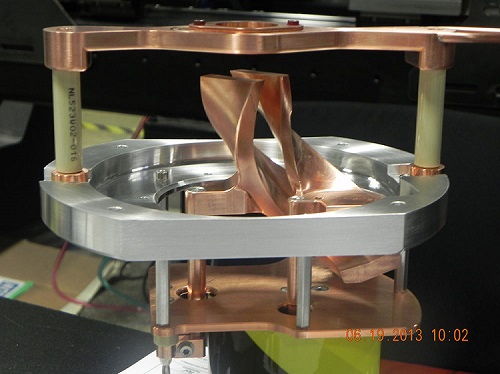}
    \caption{Photograph of the spiral inflector used during the tests at Best Cyclotron Systems.}
    \label{fig:BCS_inflector_photo}
\end{figure} 

The spiral inflector tested at BCS (shown in the photograph in \figref{fig:BCS_inflector_photo}) 
was designed to mimic the conditions required 
for the IsoDAR inflector, its primary defining characteristic being the 15 mm gap
between the electrodes. Compared to other similar designs \cite{jongen:cyclotron1},
this is rather large because it has to take into account the larger beam size due 
to space-charge effects. 
Due to the large distance between the electrodes, the SI occupies a volume in which the magnetic field components’ variation is not negligible, as it is assumed in the analytical treatment of the spiral inflector. This effect has to be carefully taken into account in order to shape correctly the spiral inflector electrodes and to avoid the introduction of a high energy-spread during the beam transport through the device.
The spiral inflector has a copper housing that surrounds the electrodes to minimize the interaction between the electric fields generated by the dees and the
electrostatic fields between the electrodes. 
To minimize beam striking the uncooled electrodes, a water-cooled, grounded, rectangular collimator shields the spiral inflector entrance. 

The preliminary design of the spiral inflector called for a nominal voltage of $\pm11$ kV and a tilt angle of 16\degree for optimal beam injection at 60 keV. The total height of the device was approximately 80 mm.  These parameters were found by using a MATLAB code based on an analytic theory for spiral inflectors
\cite{bellomo:axial_injection, toprek:spiral_inflector}. The electrode shape was
calculated using VectorFields OPERA \cite{opera:online}.
This design was later modified to account for some effects due to fringe fields on 
the electrodes. One of these modifications was to reduce the overall length of each
electrode by 5 mm.

First turn acceleration is achieved by the dee tips extending into recesses in the
spiral inflector housing. The shapes of the dee tips and the spiral inflector 
housing have been designed to guide the particles from the spiral inflector exit 
to the acceleration region while providing the necessary energy gain and beam
focusing.

The final OPERA simulation 
%, shown in \figref{fig:BCS_spiral_inflector_4turn_opal} 
(neglecting space-charge effects) of the spiral inflector was able 
to transport a 60 keV beam with a normalized rms emittance of 
0.62 $\pi$-mm-mrad at a transmission efficiency of 100$\%$.
This was corroborated by the tests performed at BCS \cite{winklehner:bcs_tests}.
Details to the simulations of the spiral inflector, both without and with 
space-charge, can be found in \secref{sec:simulations}.

\begin{figure}[!t]
\centering
\includegraphics[width=0.5\textwidth]
                {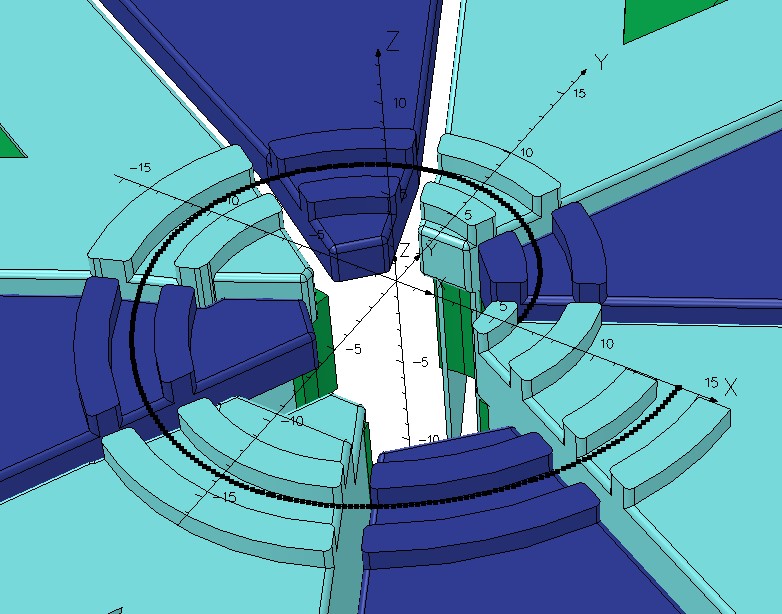}
	\caption{Schematic view of the IsoDAR central region. The back/forward 
	         path integration is used to placing the protrusions on dees and 
	         liners. 
	         \label{fig:IsoDAR_CentralRegion}}
\end{figure}

The main differences between the BCS test stand design and the IsoDAR
design are:
\begin{itemize}
  \item The IsoDAR cyclotron will be equipped with 4 accelerating RF 
        cavities, while the test cyclotron used only 2 cavities.
        \figref{fig:IsoDAR_CentralRegion} shows a first pass at the 
        first-turn design trajectory for IsoDAR.
  \item The harmonic mode (and the frequency) will be decreased from  
        6\textsuperscript{th} harmonic (49.2 MHz) to 
        4\textsuperscript{th} harmonic (32.8 MHz).
  \item The injection energy will be increased from 30 keV/amu to 
        35 keV/amu.
  \item The spiral inflector gap will be increased from 15 mm to 16.5 mm
        to accommodate a larger beam.
\end{itemize}

The main reason for reducing the harmonic mode is to relax the 
conditions for isochronism of the magnetic field, but in terms of 
injection, it has the added advantage of increased phase acceptance
and better vertical focusing.
The increased beam energy will slightly reduce space charge effects and 
also decreases the geometric emittance by 7\%.

\subsubsection{Risk Assessment \& Mitigation \label{sec:inflector_risk}}
\emph{Note: The risk of beam spread between RFQ and spiral inflector in case of the RFQ based LEBT has already been discussed in \secref{sec:lebt_risk}.}

\textbf{Risk: High voltage on the spiral inflector electrodes.} The high 
voltage (nominally $\pm 12.7$ kV) on the spiral inflector electrodes can 
lead to arcing from one electrode to the other or to ground, thereby 
creating unstable injection conditions.

\emph{Mitigation: During the BCS tests we were able to run the spiral 
inflector electrodes as high as $\pm13.5$ kV without beam using 
conventional polishing and cleaning methods for the electrodes.
Electro-polishing and conditioning in an inert gas environment will
improve this even more.
Beam striking the electrodes could potentially induce arcs through 
electron emission. 
This can be avoided by presenting a well-defined beam to the spiral 
inflector to assure close to 100\% transmission.
Beam dynamics at the end of the LEBT and through the spiral inflector will
be studied with the utmost care.
We have excellent simulation tools to predict beam behavior and benchmarks
from the BCS tests.}

\clearpage
\subsection{Cyclotron Main Magnet}

\begin{figure}[t]
\centering
\includegraphics[width=3.in]{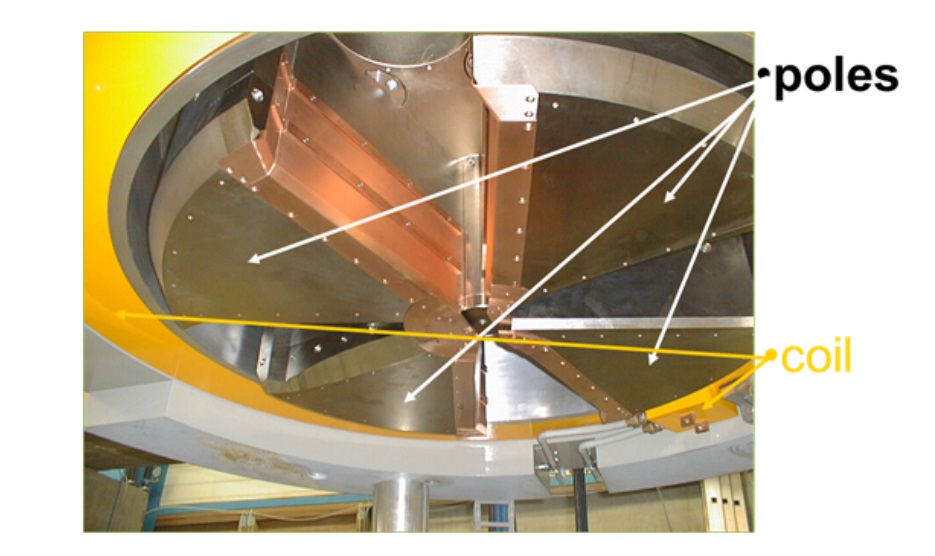}~
\includegraphics[width=3.in]{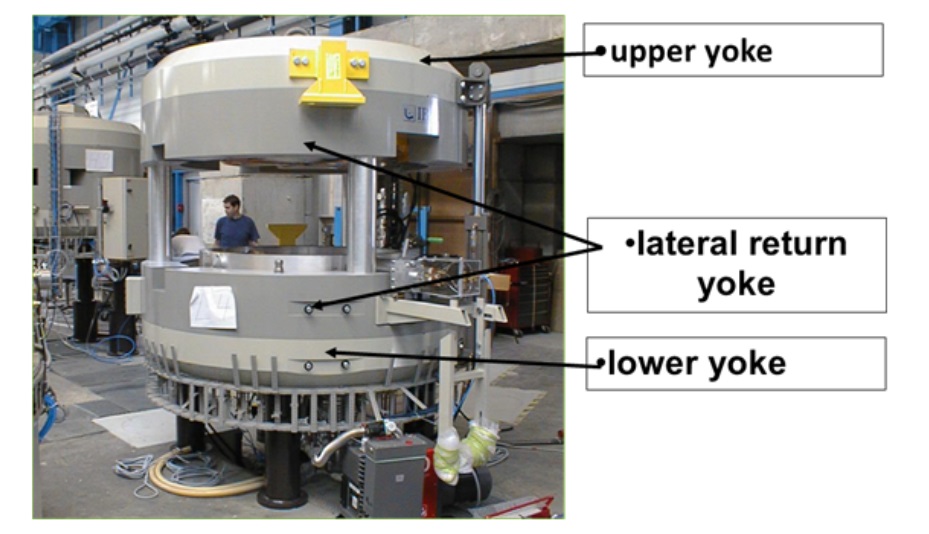}
\caption{{\footnotesize  Left:  Photograph of a similar cyclotron
    showing the upper
  poles and coil;  Right: showing the yoke.} 
\label{photopolescoilyoke}}
\vspace{0.2in}
\end{figure}

Fig.~\ref{photopolescoilyoke} identifies the parts of the magnet that
will be discussed below. 
The magnet is the heart of the accelerator and is characterized by
following features:
\vspace{-\topsep}
\begin{itemize}
\itemsep-0.1em
\item A strong and well-controlled energizing device: the coil and its power supply.
\item A large assembly of magnetic iron pieces (yoke, poles, flux return, ...), where accuracy in geometry, assembly reproducibility, and material composition require high
industry standards
\item A series of dismountable devices to allow magnetic field tuning : typically the radial shim edges of the pole tips.
\item An ``opening system" to allow access to the core of the equipment: the yoke lifting system.
\item A measurement system and post-processing models to adjust the field map as requested by the beam optics: the mapping system.
\end{itemize}
In order to synchronize the beam trajectories with the maximum of the electrical accelerating field, and, so,
optimize the energy transfer from the RF to the beam, the magnetic field has to be shaped very accurately all over the
whole accelerating region, from the early turns up to extraction. 
%Jose
Well developed and thoroughly benchmarked design codes are used to determine the theoretical magnetic field value at every
coordinate point of the entire beam path.  
Not only isotonicity (ensuring orbits at all energies take exactly the same time), but also focusing 
and resonance crossings must be taken into consideration. (Some of these points are discussed
in the Simulations chapter.)
The fabrication of the magnet parts must be done with extreme precision, and there is 
%end Jose
need for excellent agreement between the real (measured) average magnetic field in magnet gap compared
to the calculated one (up to $1\times 10^{-5}$ accuracy level). The
field values are affected by how the coil is
energized, as well as by local geometrical details (mechanical tolerances) and local material (iron) composition.

\subsubsection{Magnet Coil and Power Supply \label{sec:coil}}

The magnet is energized via 
a pair of resistive coils surrounding the poles. For a ``conventional" cyclotron of our specifications, the coil parameters would be: 225 cm inner radius, 
250 cm outer radius, and 20 cm height.
The coils would be made of square section copper insulated
conductors, with a cooling channel in their center (central hole) in
which cooled de-ionized water circulates during operation. These conductors are wound and molded with resin forming the two 
coils (one on each side of the machine median plane). 

The coils for the IsoDAR
cyclotron are a challenge due to the coil size that will not fit through the
Kamioka mine tunnels.
Either the coil must be split into two pieces (which is our primary design) or
the coil must be wound underground (an alternative design).
A precedent for such a split-coil design is the TRIUMF cyclotron, in which
the 15-meter diameter coil is actually constructed from six parts.\cite{TRIUMF}.

\begin{figure}[t]
\centering
\includegraphics[width=3.in]{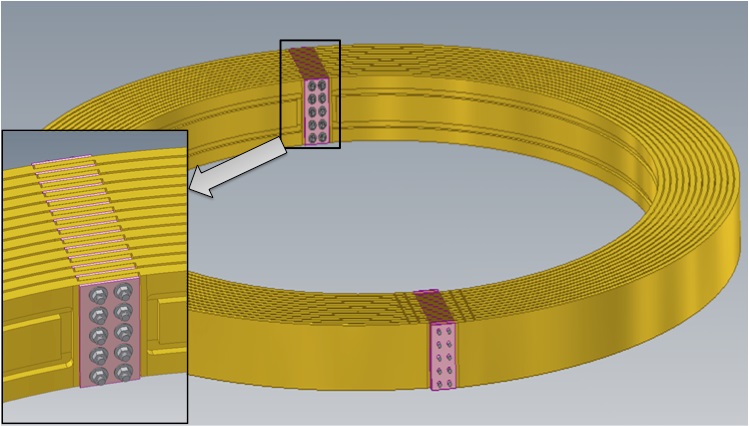}
\caption{{\footnotesize Split coil design.} 
\label{splitcoil}}
\vspace{0.2in}
\end{figure}

{\bf Split coil characteristics}

Figure \ref{splitcoil} shows schematically how a split-coil assembly might be constructed. 
This concept would consist of aluminum or copper sheets bent into half-circles, and overlapped 
at the junction points.  Insulated sheets between turns prevent shorts.  The assemblies are tightly
clamped at the junction points.  Open cooling channels are milled into the sheets; the insulator seal must
be good enough to prevent leakage.
This concept would be implemented with as few turns as possible, to minimize the number of joints.
The required ampere-turns are then provided by a very high-current, low-voltage driver.  The power supply needed is 30 volts, 12 kilo-amps (360 kilowatts).

The design of the mechanical joint interface:
\vspace{-\topsep}
\begin{itemize}
	\item Must ensure that parts can be assembled to required accuracy.
	\item Guarantee insulation between conductors turns, and conductor continuity 
	      to ensure magnet performance and stability over time.
	\item Ensure the lowest resistivity at conductor interfaces.
	\item Guarantee the cooling circuit continuity inside the conductors without 
	      any leaks.
	\item Handle the high mechanical stress between parts due to high currents in 
	      the conductors in magnetic ambiance that generates large 
	      forces while main coil global stiffness must remain very high to avoid 
	      any non linear and unpredictable effects on magnetic field.
\end{itemize}

{\bf Winding the coil underground}

A detailed cost analysis that includes what would in all probability be a less-expensive power supply,
might conclude that winding the coils underground in the cyclotron vault would be a better option.

It would require transporting and assembling a coil-winding machine, bringing in the rolls of conductor and insulation (that would almost surely fit in through the tunnel), designing and constructing a collapsible potting fixture and mold
for the wound coils, providing the appropriate epoxy-mixing, vacuum and ventilation capabilities. 
And, very importantly, substantial space would be required for this process.  

Scarce staging-area space could be efficiently utilized by performing the coil-winding process prior to bringing
in any of the steel pieces or other parts of the cyclotron.  
The coil winder could be dismantled and removed then to make room for the cyclotron parts.  The coils would be stored away from the construction site until needed for the assembly of the cyclotron.

\subsubsection{Yoke, Flux Return and Poles}

\begin{figure}[t]
\centering
\includegraphics[width=0.25\textwidth]{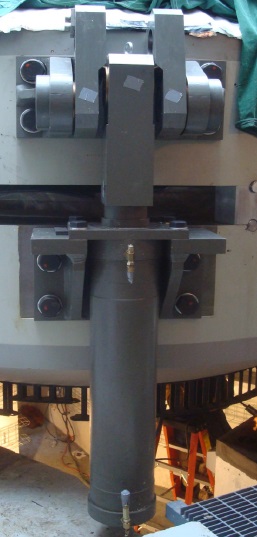}
\caption{{\footnotesize  A hydraulic lifting system installed on a
    similar cyclotron.}}
\vspace{0.2in}
\end{figure}

The typical approach to  large magnet steel construction is to
employ the casting manufacturing process because it allows
reaching a ‘near net shape’ with the raw material and reduces the
number of interface between parts. 
However, in the ISODAR case,  although the accelerator itself is a ``large machine", parts size and weight handling restrictions of
components clearly lead to prefer the "small machines" industrialization process for the magnet (meaning made out of forged and laminated steel pieces).
Total magnet weight is estimated to $\sim$500 tons but said restrictions lead to cut it into approximately 42 different pieces.
The IBA design and cost estimation is based on a magnet split in 42 parts, out of laminated iron sheets, finished by classical
machining. This offers high and reproducible iron quality but requires more 
machining and assembly activities and also more attention to all interfaces 
mechanical design and adjustments.
Per IBA experience, laminated iron sheets exceeding 250-300 mm thickness don't meet classical accelerators material composition and homogeneity requirements. 
Then, starting from standard size slices of 2.5 m $\times$ 2.5 m $\times$ 230 mm
(approx weight : 10 tons) and achieving a 80\% filling factor (20\% lost material during machining) we can envisage 8
useful tons per slice.  

Beyond the machining of the 42 parts described above, final machining of the each fully assembled half yoke will
probably be required to achieve global tolerances (gap and poles
position, etc.). This requires a 
large capacity milling machine (~6.2 m outer diameter and 250 tons).
We assume such a mill can be located.

% Move this comment into the cost book:
% and include
%the commensurate high hourly rates in the cost estimate.

Handling the parts during and after the manufacturing process and to
the final site location will require design and construction of dedicated
handling frames.   By using multi-purpose and/or reusable frames we
could achieve that job with a set of 25 frames.

The yoke is designed so that the upper poles of the magnets can be
opened to access its median plane.  This is required for many purposes:
maintenance, manufacturing and testing phases, upgrades, etc.
Generally this functionality is achieved by either big hydraulic or screw jacks with the appropriate level of control,
regulation and safety. This equipment has to be installed prior to
magnetic mapping, discussed below,  not only for practical reasons
(installation of mapping tool, pole edges modifications) but also because it causes some magnetic field perturbations
in the beam chamber that need to be taken it into account to tune the
magnetic field.

\paragraph*{Options for Yoke Lifting System}

The primary design for IsoDAR is based on a hydraulic lifting
system.    
% An alternative approach to easily handle those heavy yokes with high accuracy 
% uses screw jacks instead of hydraulic.   
Screw jacks can be used as an alternative, but are more complex to control and drive.
Due to the fact that machine will be unique and 
used by mainly well trained people, we kept the standard hydraulic device.

\subsubsection{Magnetic Field Mapping \label{sec:mapping}}

\begin{figure}[t]
\centering
\includegraphics[width=0.6\textwidth]{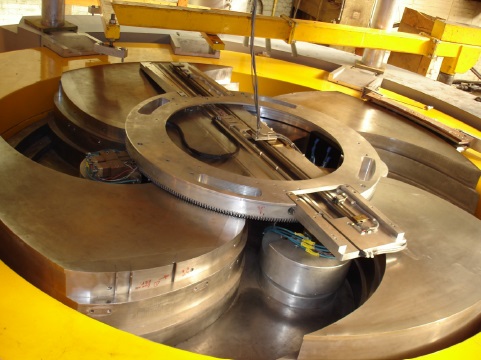}
\caption{{\footnotesize A mapping system installed in a similar
    cyclotron at IBA. (Image courtesy of IBA.) \label{mapsys}}}
%\vspace{0.2in}
\end{figure}

Careful magnetic field mapping to a high level of
accuracy is crucial for producing a successful cyclotron.  
In fact, taking into account limited capabilities of manufacturing processes 
and the segmented structure of this magnet, the field mapping of the magnet 
is mandatory.
Measurement equipment and tools for such a purpose are specific to the
magnet design and do not exist ``off-the-shelf.''  
It is necessary to design and manufacture the measuring equipment itself.
Fig.~\ref{mapsys} shows the type of hardware that is constructed for
the purpose. It typically consists of a large mechanical structure
holding a set of probes (Hall probe, search coils or a both) that is able, 
through a set of motors, drivers, controls and feedback systems, to measure the 
field in every desired positions inside the accelerating gap and extraction area
of the magnet.

Developing a mapping tool requires care as this tool accuracy will define performance accuracy of the measured magnet itself. Many 
precautions are mandatory concerning calibrations, mechanical stiffness, thermal 
stability and tracking as well as the use of perfect "non-magnetic" materials (to 
avoid measures perturbations and unexpected distortions). 
The assembly and fine tuning of the device in Lab and {\it in situ} will require approximately 6 man months and the mapping campaign itself on the equipment will 
require 24 man months including post-processing. We assume 1 engineer, 1 technician, 
and 2 physicists are involved. At this point, no device for magnetic
field radial components estimation has been evaluated. Even though
this field is, in principle, 
not present, in practice radial fields can be destructive for 
the beam optics. Such measurements are similar but this  metrology
device can be tricky to design as one needs to measure a very tiny but potentially 
very destructive radial field component mixed in a very high vertical field component
ambiance (measured by nominal mapping process).

Based on measurement results, physicists can virtually track the beam along its path, by computing the equation
of motion that takes all the acting forces (electromagnetism, space
charge, etc.) and relativistic effects into account.  Based on this, 
they can suggest new pole-geometry adjustments necessary to keep the beam at appropriate positions
and shape and guarantee its final characteristics at extraction.
This is a standard procedure used in the cyclotron industry.

\subsubsection{Risk Assessment \& Risk Mitigation \label{sec:magnet_risk}}

\textbf{Risk: Manufacturing Imperfections.} Because no manufacturing process is perfect, we must measure the local magnetic field developed by the accelerator magnet 
in an iterative process. During the mapping iterations, physicists compare 
measurements results with expected calculated (theoretical) magnetic field map values 
and propose local magnet iron shape adjustments, typically on pole
tips.   Difficulty of the process is a risk.

\emph{Mitigation: For that reason, pole tips, or at least the edges of the pole 
tips (boundary valley/hill) are to be dis-mountable to allow re-machining and 
multiple tuning iterations.
This mapping process is looped until final expected map is obtained, allowing, theoretically, the appropriate beam acceleration and extraction.
In IsoDAR case, the machine must be dismounted after surface tests in order to be 
shipped to its final site. This will require re-measuring the magnetic field after
re-assembly and final welding (to seal the internal part of the pole) on site, during a second (shorter) mapping campaign.}

\textbf{Risk: Split Coil Design.} The split coil is an engineering challenge in 
IsoDAR project for several reasons. First is the need for accurate mechanical 
interface. Second,  the consequence of the  choice of low turns and high current on 
the required power supply to feed coils is out of the usual industrial experience.   
For comparison, proton therapy machines typically require a \mbox{30 V / 750 A} 
(22.5 kW), with highly regulated equipment.

\emph{Mitigation: An alternative is to produce the coil underground,
  in which case the coil can be a single, unbroken piece.  
We discuss this alternative above.  }

\textbf{Risk: Mapping Tool.} 
Developing a mapping tool requires care as this tool accuracy will define performance accuracy of the measured magnet itself.
 
\emph{Mitigation:  Many 
precautions are mandatory concerning calibrations, mechanical stiffness, thermal 
stability and tracking as well as the use of perfect "non-magnetic" materials (to 
avoid measures perturbations and unexpected distortions). }

\subsection{Cyclotron RF System}
Table~\ref{tab:rf_design} provides a detailed description of the RF system.
The system for IsoDAR cyclotron is divided into subsystems
as follows:
\vspace{-\topsep}
\begin{itemize}
\item The RF cavities allowing a fast acceleration to the final particle energy and keeping a large inter-turn
separation. The cavities are triangular shaped and are located in the magnet valleys. There are 4
accelerating electrodes (Dees).
\item The RF tuning system that will compensate for thermal drift of
  cavities and maintain a stable resonating frequency.
\item The RF amplifiers that provide the necessary power to create the required RF electric field inside the cavity and the power to accelerate the beam.
\item The RF couplers that inject the RF power from the amplifiers into the cavities
\item The Low Level RF system (LLRF) that controls and regulates the RF amplitude on the Dee and drives the tuning
system.
\end{itemize}
\begin{table}[!b]
	\caption{Details of the RF design.
	         \label{tab:rf_design}}
	\centering
    \renewcommand{\arraystretch}{1.25}
		\begin{tabular}{ll}
            \hline
		    \textbf{RF System Component} & \textbf{Design Value} \\
            \hline \hline
			Resonance Frequency & 32.8 MHz\\
			Dee voltage in the central region & $<80$ kV \\
			Dee voltage at extraction radius & $>200$ kV\\
			Dee radial extension & 2 m \\
			Dee angular extension & 38$^\circ$\\
			Cavity height & 1800 mm \\
			Number of Dee stems & 4 per Dee (2 up, 2 down)\\
			Number of Dee’s & 4\\
			Acceleration harmonic & 4th\\
			Power dissipated per cavity & 100 kW\\
			Cavity Q-factor & 9540 \\
			\hline
		\end{tabular} 
\end{table}
We consider these in the Sections below.

\subsubsection{RF Cavities (Dees, Stems and Liners)}

\begin{figure}[!t]
\centering
	\includegraphics[width=0.6\textwidth]{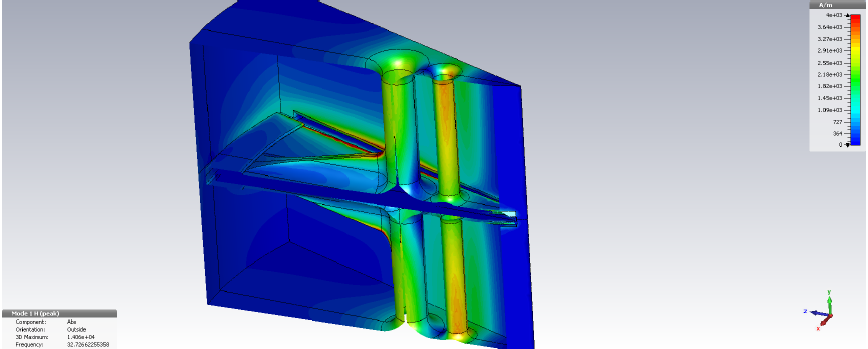}
	\caption{{\footnotesize  RF Dee -- color scheme represent current density} 
			 \label{RFdee}}
%\vspace{0.2in}
\end{figure}

The RF cavities of the cyclotron have been calculated using CST microwave 
studio \cite{cst:microwave_studio} in order to define the shape of the Dee
and the Dee stems.
The optimization work is intended to reach the right resonance frequency, the Dee voltage Law, together with the
minimal power dissipation and hence the highest quality factor (Q-factor).
The field maps allow also the calculation of the couplers and voltage
pick-up but this last calculation is yet to be performed.

The Dees and cavity walls are all made of OFHC copper that combines good electrical conductivity and good solder
ability.
The Dees are fabricated out of a 20 mm thick copper plate and are made
rigid by a supporting arm.
The four round stems support the Dee and provide the right inductance in order to get the required resonance
frequency. Their size and location allow reaching the required Dee voltage profile versus radius.
Each Dee will be equipped with cryopanels to efficiently pump the 
median plane.
One Dee is equipped with one electrostatic deflector for
extraction.  (This can be seen in Fig. \ref{injcyc}.) This fact has not been taken into
account in the RF simulations to date.

The liner, in addition to being the electrical ground surface for the RF system, also 
provides the vacuum integrity for the cyclotron.  It is shown in Fig. \ref{fig:vacliner1}.
It is an important and delicate part of
this structure for multiple reasons:
\vspace{-\topsep}
\begin{itemize}
\item A single liner covering the entire pole exceeds the maximum
  dimension that can be transported into the Kamioka tunnels.
Consequently this part is divided into four separated pieces and
re-soldered in place. Copper is not
easily soldered due to its high thermal conductivity.
\item The liner should be vacuum-tight since the choice has been made to limit the high vacuum region to the most
limited volume. (Maintaining the vacuum level in this machine is critical due to the high
stripping cross section of \htp with the residual gas.)
\item The preferred manufacturing process of this cavity is electroforming. At least 10 mm of copper should be
accumulated to maintain the required rigidity of the whole structure. A second choice is the assembly by
soldering of many different panels,  but the mechanical tolerances may
be difficult to maintain with this
process. Electroforming is a conventional manufacturing process that
is widely used in cyclotron manufacturing
\end{itemize}

\subsubsection{RF Tuning System}

The tuning system is a movable plate or loop that corrects the resonance frequency drift mainly caused by RF heating.
The system here is based on four separated tuning plates (one per cavity) and located in the median plane. The plate
is accurately displaced by an electrical motor via a ball screw.
Similar systems are already used many commercial cyclotrons.

\subsubsection{RF Amplifiers}

Traditionally a large part of the cyclotron RF budget is dedicated to the RF power source.
This is especially true for this cyclotron, as it requires a high Dee voltage and accelerates 600 kW of beam power. This
means that the total RF power exceeds 1 MW.   Here we assume
1200 kW of RF power, which corresponds to a 50$\%$ efficiency of RF power to beam power.  
Since the four cavities are driven separately we are speaking about 4 distinct amplifier chains of 300 kW each.

The traditional way to produce such a high RF power is still today the use of vacuum tubes (triodes or tetrodes) and
this is especially true in the accelerator world where the load constraints are more severe than simple antennas.
On the other hand solid state technology made a significant step forward in the past years in terms of power density
and robustness. Today, the last generation of LDMOS from NXP or Freescale allows to make solid state amplifiers at
nearly competitive price with respect to vacuum tubes and that are also very reliable.

\subsubsection{RF Lines and Couplers}

The RF power is fed to the cavities through EIA 6 1/8" solid copper
coax line. 
There are basically two ways to couple RF to the cavities:
\vspace{-\topsep}
\begin{itemize}
\item Capacitively in the median plane
\item Inductively in the bottom (or top) of the cavity liner.
\end{itemize}
The choice between these two alternatives is mainly driven by the
available space.
Our primary choice is capacitive coupling.  This is preferred because the coupler is less prone to multipacting. This is also magnified
by the fact that the coupling antenna could be easily DC biased.
The usual experience in this field is that both systems can work
and couplers up to 800 kW at 107 MHz have been
successfully built and operated.
Four couplers are needed in this case but this moderates (300 kW) the power to be injected via one coupler.

\subsubsection{LLRF Electronics}

The LLRF (Low Level RF) controls the RF amplitude and phase of the signal sent to the amplifiers. It also controls the tuning
mechanism in order to maintain constant resonant frequencies in the cavities.
The amplitude and phase control is based on feed-back signals coming from the cavities and amplifiers.
In order to avoid the multipactor effect the LLRF manages also a pulsing mode that provides for an easy start-up and
reduces the power dissipated in the amplifiers. One LLRF system per cavity is planned.

An additional electronic element is needed to keep a stable phase difference between the different cavities.
%This unit is called here the CPREU (cavity phase regulation electronic unit).

\subsubsection{Risk Assessment \& Risk Mitigation \label{sec:rf_risk_cost}}

\textbf{Risk: Design and Manufacturing Process.}

\emph{Mitigation: A full scale RF cavity will be built and tested before the start of the construction of the other three cavities.}

\subsection{Vacuum System and Pumping \label{vacuum}}

The vacuum requirement for the cyclotron is at least $5\cdot10^{-8}$ mbar, calculated to keep losses due to dissociation of the \htp ions from gas collisions to less than 1\% over the entire acceleration cycle.
This level is achievable if careful attention is paid to materials and construction techniques.  Surfaces likely to be struck by beam particles need to be adequately cleaned and treated to minimize gas desorption that can lead to pressure bursts and total beam loss.  

The vacuum system comprises two elements: the vacuum enclosure and the pumps.  The enclosure is addressed here, the pumps specified are listed at the end of this section.
The vacuum enclosure includes the surfaces of the pole faces, either lined or not, and an outer ring that, in addition to providing the vacuum enclosure, serves as a precision spacer to keep the pole gap dimensions at designed values.

The preferred option for the pole-face sections is to construct a liner that is attached to the steel, so the liner itself does not need to support atmospheric stresses over the entire inner diameter of the cyclotron.
One option considered was the
IBA solution for the Cyclone 230 and Cyclone 70xp \cite{iba:cyclone70}. The vacuum chambers of these  
two cyclotrons are made of centrifuge cast aluminum. This technique has a good track record for vacuum chambers of
these two machines, for diameters about 2 to 2.5 m. 

\begin{figure}[!t]
\begin{center}
\includegraphics[width=5.5in]{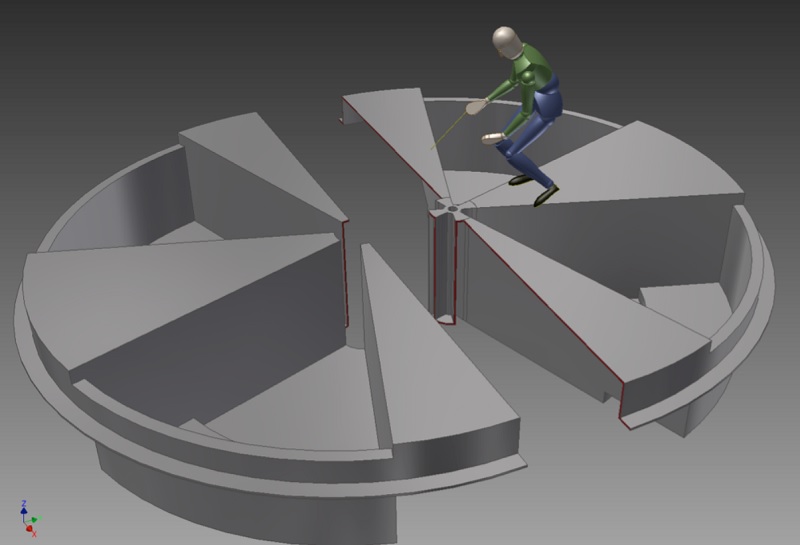}
\end{center}
%\vspace{-1.0in}
\caption{\footnotesize  Concept for assembly of the vacuum liner.  Pieces capable of fitting through the access tunnels (drifts) are brought in to the cyclotron staging area.  Seams still to be welded are outlined in red.  (INFN-LNS, Catania)
\label{fig:vacliner1}}
\end{figure}

However, the maximum diameter of the IsoDAR chamber, 4.5 m, is larger than the Kamioka tunnel size, so such a vacuum chamber would require being brought in in pieces and welded into its final size in the cyclotron assembly staging area.  In addition, for accommodation of the RF dees and pumping, it is better to have the liner follow the contours of the hills and valleys.  For such a structure, use of non-magnetic stainless steel is preferred.  Surfaces close to the RF dees may need to be copper-plated to ensure good electrical properties.
A procedure for welding this liner has been developed, taking into account that this must be accomplished underground, at the cyclotron assembly site.  This is shown schematically in Fig. ~\ref{fig:vacliner1}

The alternative design executes vacuum-tight welds of the steel pole
pieces, that also must be brought into the assembly site in pieces,
and not have a separate liner.  
Achieving vacuum-tight welds in these steel pieces, while preserving the 
dimensional accuracy required for the magnetic field map, may be more of a challenge than 
the preparation of a liner, however it does simplify the overall structure of the vacuum system.

In both cases surface preparation and cleaning are very important.  Cleaning will be more of a challenge for the welded steel option, but adequate procedures can be developed for either option.  
The recommended surface preparation is to sputter a layer of titanium nitride.  This can either be done on the separate pieces at the fabrication plant, or through a discharge tube placed on a cart and moved around the mid plane of the assembled cyclotron.  It will be more important to ensure good coating for the steel surfaces in the absence of a liner.

A critical item in the vacuum interface design is the feedthroughs for the high-power RF.  With many hundreds of kilowatts of RF power being delivered to the dee structures (beam power is 600 kW), the electrical and thermal characteristics of these insulated feedthroughs must receive very close attention.

For either option (liner, or no-liner) a ring at the mid-plane must be provided, with upper and lower o-ring interfaces, to allow the splitting of the magnet halves for assembly and service.  This ring, also of diameter larger than the access-drift opening, must also be brought in, in halves, and welded into a ring at the installation site.  It must include ports for the extraction line, radial probes and other diagnostics.

For the pumping system, the IsoDAR cyclotron will be equipped with
\vspace{-\topsep}
\begin{itemize}
\item 8 cryo-pumps (4500 l/s) with their ancillaries and interfaces, and
\item 2 turbo-molecular pumps (300 l/s) to bring the pressure down to 
      a level where cryo-pumps become effective.
\item cryogenic panel(s) inside the Dee structures. 
\end{itemize}

\subsubsection{Risk Assessment \&  Mitigation\label{sec:vacuum_risk_cost}}

\textbf{Risk: Achieving required vacuum level in presence of beam.}  The calculated required vacuum is $5\cdot10^{-8}$ mbar.  Before the beam is turned on, the pumping plan provided will reach this with little difficulty, assuming no vacuum leaks.  However, introduction of beam, with gas-desorption from surfaces hit by stray beam particles, could raise the pressure above the acceptable level.

\emph{Mitigation: Beam losses from other sources must be minimized.  In addition, options for further surface cleaning and treatment, as well as possible addition of more pumping capacity, should be developed.}

\clearpage
\subsection{Extraction}

\begin{figure}[!t]
\begin{center}
\includegraphics[width=5.5in]{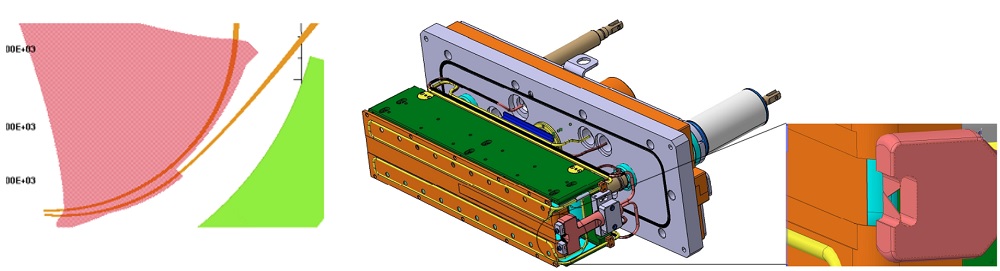}
\end{center}
%\vspace{-1.0in}
\caption{\footnotesize  Extraction elements for the IBA C70 cyclotron.  The last two particle orbits are shown schematically on the left.  Details of the electrostatic deflector, fitting in the notch of the (pink) pole are shown in the center section, while the expansion on the right shows the cooled pre-septum designed to protect the septum from beam that would strike it.
%Extraction elements for the IBA C70 cyclotron.  The entrance to the electrostatic septum is detailed in the expanded insert.  A water-cooled pre-septum (rose-colored) more narrowly defines the entrance channel for the extracted beam.  The turn prior to the extraction turn skims by on the inside of the (orange) septum, while the extracted turn enters into the strong electrical field of the electrostatic deflector (represented by the teal area behind the pre-septum).  Beam particles that would strike (and cause damage to) the thin septum electrode are instead stopped on the pre-septum foil.  The graphic on the left illustrates a notch in the main magnet pole to reduce the field seen by the beam after passing through the electrostatic septum, to increase the separation of this beam from the circulating beam.  The green piece is a steel compensator to correct for fringe fields from the edge of the pole that would defocus the emerging beam.
\label{fig:C70_extractor}}
\end{figure}

Extracting an \htp beam can be accomplished by stripping (liberating two protons that, depending on the placement of the stripper, have clean orbits that exit the machine), or by a conventional electrostatic septum-based channel.

We do not choose stripping extraction due to the high current that
will limit the lifetime of the stripping foil.   The average lifetime
of strippers used in H$^-$ cyclotrons is $\sim$20 mA hour.  This would
indicate a need for replacing the foil many times per day.
Moreover, the baseline design of an electrostatic septum system rather
than stripping foil is consistent with follow-on use of the IsoDAR
cyclotron design for in the future DAE$\delta$ALUS project.
This larger cyclotron requires injection of \htp ions, so these must
be transported as intact ions from the injector cyclotron.

%The baseline design option will be the electrostatic septum system, because of the desire to consider the IsoDAR cyclotron as the injector for the larger cyclotron called for in the future DAE$\delta$ALUS project.  This larger cyclotron requires injection of \htp ions, so these must be transported intactly from the injector cyclotron.

The extraction channel will include two electrostatic septa, followed by one or more magnetic channels that will bring the beam to the outside of the cyclotron steel (recall figure \ref{injcyc}).  The engineering design of these elements is quite mature, the parameters required for effective operation are well-known.  The only real difference between the existing state of the art and the IsoDAR design is the introduction of the narrow pre-septum stripper detailed below.  

Figure \ref{fig:C70_extractor} shows a notch in one of the magnet poles (pink) where the first
 electrostatic septum is located.  
The turn prior to the extraction turn skims by on the inside of the septum, 
while the extracted turn enters into the strong electrical field of the electrostatic deflector.  
The track of this orbit misses the edge of the pole, experiencing less bending, 
and so increasing the separation for good extraction.  
The fringing field of the pole is compensated by the green steel piece seen in the figure, 
to prevent defocusing of the beam.  
The detail in the middle of the figure shows mechanical details of the 
electrostatic deflector assembly.  
The orange piece is the thin septum electrode, typically less than 1 mm thick.  
The teal-colored area demarks the region of the strong bending electric field.  
The expansion at the right of the figure shows the cooled pre-septum, 
designed to mask the septum and intercept beam that would strike the septum.  
The ``v'' notch increases the surface area for beam deposition 
and improves conductive cooling efficiency.  
Beam in the extracted turn passes between the ``v" notched foil and the base of the harp holding the foil, and enters the main electrostatic deflector channel.
This extraction system for the C70 cyclotron is designed to handle a maximum of 1 mA.

%Elements of a typical extraction setup is shown in Fig.  ~\ref{fig:C70_extractor}, the configuration used by IBA for extracting positive-ion beams from their C70 cyclotron.  While the location of the first electrostatic septum is different from ours, the basic configuration and elements are similar.

%Shown is the electrostatic septum assembly, the inset shows the water-cooled pre-septum (rose-colored) designed to protect the very thin septum.  Note the ``v" shaped foil, to reduce the heat load of intercepted beam on its edge.  The graphic at the left shows the magnet pole with a notch so that beam in the extraction channel see less beam.  The green steel piece is a compensator to minimize optical distortion of the beam shape.  This extraction system is designed for the IBA C70, designed to deliver at most 1 mA of beam.

Key to efficient ``classical" extraction is having clean beam separation at the location of the first extraction septum.  While achieving a suitably high orbit separation is not a problem -- calculations 
and simulations presented in Chapter 6 indicate turn separation is expected to be about 14 mm, the critical number is the ratio of this orbit separation and the transverse size of the beam bunch at the location of the extraction septum.  Because of the very high beam current (a factor of five over the above-mentioned C70), even if the principal envelope of the beam is less than the turn separation, there will be halo and tails that come in the space between the orbits at the location of the septum.  Even if the septum is very thin, a few tenths of a mm, for instance, the power deposited on the front edge of the septum will cause serious thermal damage and erosion.

A possible mitigation of this problem is to introduce a narrow 
stripper, of width approximating the septum electrode.  Placing this 
stripper just upstream of the pre-septum, beam that would strike the 
septum will be stripped into protons which will be bent inside, 
missing the septum assembly, and transported to a suitably placed beam dump at
the end of the proton orbits.  
These strippers will intercept a very small fraction of the total beam, so lifetime of the strips 
should not be a major issue, but nevertheless a magazine of these stripper ``strings" 
will be needed to ensure sufficient operating time between service intervals to change the 
stripper magazines.

%The quality of the extracted beam heavily relies on the design of the
%extraction system.   Beam transmission can be
%affected by the overall losses due to out-gassing of surfaces where beam is lost.

%The IsoDAR primary design has extraction using electrostatic
%deflectors, in accord with the plan of IsoDAR as an intermediate
%step for the DAE$\delta$ALUS project.  Electrostatic deflectors
%suffer from losses on the septum due to insufficient turn separation,
%and so cooling is required.  Nevertheless, septum damage and
%significant beam loss are risks.
%To mitigate this risk, IBA has found that an upstream string of stripping foils to intercept
%the beam that would strike 
%the electromagnetic septum is feasible.   This will allow removal of a limited amount of
%particles through striking the string while avoiding losses on the deflector septum. 
%The stripped fraction will be transported out of the vacuum chamber through a 
%dedicated exit beam line.
%With this stripping string, the extraction system is therefore composed of:
%\vspace{-\topsep}
%\begin{itemize}
%\item  1 stripping string with adjustable position to limit losses on the first electrostatic deflector;
%\item 2 electrostatic deflectors to steer the  beam out of the magnetic field;
%\item 1 gradient corrector to correct the beam optics in the extraction channel.
%\end{itemize}

\subsubsection{Risk Assessment \& Mitigation \label{sec:extraction_risk}}
\textbf{Risk: Inadequate turn separation.} If beam bunches are not radially separated by a sufficient amount, unacceptable beam loss will occur on the thin stripper rendering its lifetime to be inadequately short.

\emph{Mitigation: Understanding beam dynamics sufficiently well that this does not happen, and having adequate beam diagnostics to be able to tune the beam for maximum turn separation.}

\textbf{Risk: Placement of the thin stripper.} It is not obvious where the optimum location is for the thin stripper.  It must be such that stripped protons do not strike the septum, but also that halo particles do not miss the stripper and strike the septum.

\emph{Mitigation: Adequate beam dynamics simulations, including tracing of particle orbits.}

\clearpage
\subsection{Cyclotron Design - Current Status and Future Work}
%The present design of the cyclotron is preliminary. In the next phase, a more accurate 
%design (physics as well as engineering) has to be produced. Here we shortly describe the main items to be addressed:

We have presented a preliminary design for the IsoDAR cyclotron. In the next phase we must produce a more complete physics and engineering design.  The following is a specific list of topics that must receive more work.
\vspace{-\topsep}
\begin{itemize}
\itemsep-0.05em
\item An engineering design of each component of the pole and yoke and the construction, assembling and dismounting procedure are necessary.
\item The transport of the cyclotron pieces into Kamioka mine has to be properly 
evaluated.
\item The assembling procedure and the assembling tools have to be designed.
\item To reduce the power consumption of the coil of about 20\% the pole width could be increased from the present maximum value of 36.5\degree up to $44\degree-45\degree$.
This could change the dimensions, and consequently the weight, of the return yoke, which has to be fixed soon according to the limitations of the entrance space and the available tools.
%\item A more realistic model of RF cavities working in $4^{\mathrm{th}}$ harmonic 
%(32.8 MHz) instead of the $6^{\mathrm{th}}$ harmonic (49.2 MHz) has been proposed
%recently. 
\item The profile of voltage vs. radius has changed and the maximum achievable voltage 
has to be carefully evaluated to optimize the power consumption and the inter-turn
separation between the orbits. 
%\item The new model of the RF cavity has a capacitive coupling instead of the inductive loop of the previous model. This need to remove some hole from the valley region and check if there is enough room in the median plane at azimuthal position of the RF cavities;
\item Examine the dimensions of the axial bore to not preclude the use of an RFQ-based LEBT.  This bore may need to be enlarged, but the impact on the main magnetic field in the central region must be evaluated.
\item A new run of beam dynamic simulation including space charge effect taking account of all the changes here suggested is also mandatory.
\item The beam dynamic simulations are also necessary to establish the correlation between the central region with the extraction channel and the optimization of the beam separation at the azimuthal position of the electrostatic deflector.
\item  Completing the design of the vacuum chamber includes fixing the numerous openings 
for the coupler, the trimming capacitors of the RF cavities, the radial probes, the 
high voltage feed-throughs for the electrostatic deflector, the beam extraction channel,
and the vacuum pumps.
\item The procedure of dismantling the cyclotron has also to be studied in consideration of decommissioning. The proposed solution of welding the sub-components of the pole and the vacuum chamber has to be evaluated also in the perspective of decommissioning.
\end{itemize}
\clearpage
\section{Beam Transport from Cyclotron to Target (MEBT)
         \label{sec:mebt}}

The cyclotron and target are separated by a distance of approximately 51 m.  
It is the task of the beam transport system to take the beam this distance with minimal losses and present it to the target with a spreading technique that distributes the beam over the face of the target for optimal distribution and thermal management of the 600 kW of beam power. 

This section will focus on the overall configuration and components in the transport line, the beam optics will be addressed in Chapter 6.  

\begin{figure}[!t]
\begin{center}
\includegraphics[width=5.5in]{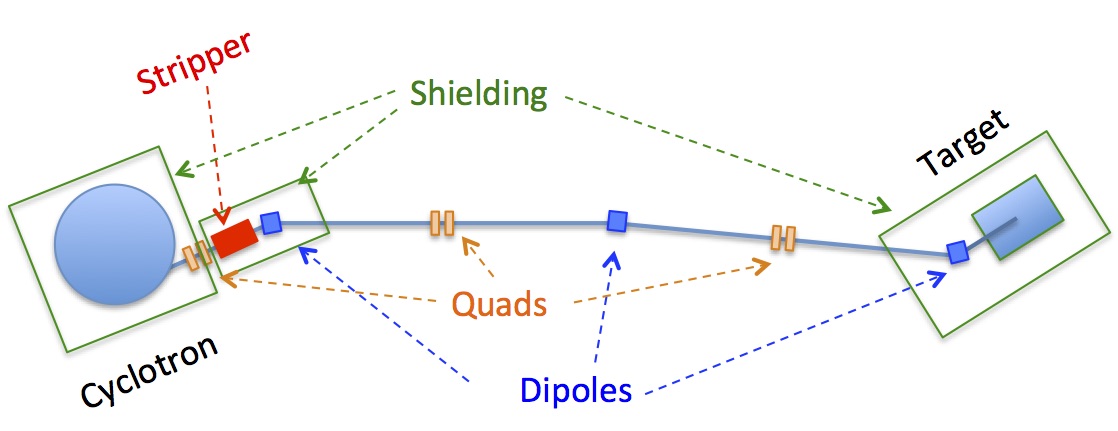}
\end{center}
%\vspace{-1.0in}
\caption{\footnotesize  Schematic of beam line elements transporting the beam from the cyclotron to the target.  Shielding enclosures are indicated for location and approximate shape, but detailed calculations will be needed to determine final dimensions and configuration.
\label{fig:BeamLineSchem}}
\end{figure}

\subsection{Overview}

	Transport is performed by the usual combination of dipole magnets to bend, and quadrupole magnets to focus the beam.  
	A schematic of the beam line layout is shown in Fig. \ref{fig:BeamLineSchem}, and will be described in more detail below.
The beam is transported through a vacuum pipe of sufficient diameter (at least 10 cm) to minimize beam loss through scraping, but not so large as to require oversized magnets. 
A series of roughing and turbo pumps provide the necessary vacuum to minimize losses due to gas scattering.
 
 Collimators and scrapers at appropriate locations, placed in adequate shielding enclosures, control the points where beam is purified to avoid losses in lightly-shielded areas.  Extensive simulations of beam halo growth and other potential beam-loss mechanisms will be required to determine the optimal placement of these devices.
 
 Beam diagnostic boxes are also included at appropriate locations along the beam path.
 These are used for tuning dipoles, quads and steering magnets, 
 and measuring beam profile and intensity.
 
 Just prior to the target, a 30$^\circ$ bending magnet prevents neutrons back-streaming from the target face from traveling down the pipe to the cyclotron, and provides space for a re-entrant neutron dump beyond the magnet in line with the target face.
 This magnet and the back-streaming dump are enclosed in an extension of the main target neutron shielding.
 
 %Between this magnet and the target, a wobbler magnet system sweeps the slightly-defocused beam over the face of the target to distribute the heat load.  
 Beam is spread over the face of the target so as to minimize the power density at any given point.  
 This can either be done by wobbler magnets, or optical elements 
 (that might consist of quadrupole, sextupole and/or octupole magnets).  
 The option offering the best control is wobbler magnets, as the pattern over which 
 the beam is swept is quite arbitrary.  
 However, as a relatively small beam (sigma of a few cm) is being swept, 
 the sweep rate must be high enough to prevent deposition of beam power at a point 
 on the target at a rate greater than what can be dissipated by the cooling systems.  
 Calculations will need to be done to establish that the required sweep rate is 
 within the capabilities of magnets of sufficient power to bend the beam.
  
 \subsection{Stripping} 	
 As the beam emerges from the cyclotron as \htp, two options need to be considered for transport: a) stripping the beam close to the cyclotron, to transport protons, 
 or b) transporting the intact \htp ions to the target.
 The advantage of stripping close to the cyclotron is that transporting protons requires smaller magnets, and avoids gas stripping of the fragile \htp ions, which introduces an undesirable source of neutrons along the length of the beam line, unless the beam line vacuum is extremely good.  Our estimates, based on \htp beam loss measurements at Catania \cite{Calabretta:h2+beamloss}, are that the beam line pressure must be substantially below 10$^{-8}$ torr.
 
 The preferred option is a), stripping the beam as close to the cyclotron as possible.
 
 Again several options exist for stripping this high-power beam:  a)  carbon-based foils, 
 or b) windowless gas strippers.  
 
 The best experience with carbon foils at this energy range exists for stripping of internal  $\mathrm{H}^-$ in isotope-producing cyclotrons.  The lifetime most often quoted is about 2 mA-hours, for 1-to-2 mA beams.  The foils used are usually 200 $\mu$g/cm$^2$.  It should be noted that the two electrons stripped from the  $\mathrm{H}^-$, referred to as ``convoy electrons'', usually are bent back into the foil (the foils are in the strong cyclotron magnetic field), where after repeated passes they lose all their energy. In fact the electrons deposit over ten times more energy in the foil than the protons do \cite{Kim:carbon-foil-lifetimes}.  
 
 In our case, recent experiments by Calabretta \cite{Calabretta:stripping} indicate that 60 $\mu$g/cm$^2$ is adequate to strip 99.5\% of the 60 MeV/amu  \htp ions, and if one placed several foils separated by a few cm in the beam line, the remaining fraction of \htp could be made completely negligible.  
 In addition, as these foils are not in the cyclotron magnetic field, the convoy electrons will not spiral back into the foil.  One can place small permanent magnets behind each foil to bend the electrons sufficiently that they do not hit the next foil.  As a result, the power deposition rate in the foils will be a small fraction of that experienced in the $\mathrm{H}^-$ cyclotrons, and the factor of five higher beam current should still not be a problem for the thermal lifetimes of the foils.
 
 Remaining is to assess the lifetime of the foils due to radiation damage.
 For this, we are working with PSI to perform a test in their transport line between the 72 MeV Injector II cyclotron and their main ring.  Their 2.2 mA beam is lower current, however ours will probably be larger in size, so power densities could be made equivalent, making this a meaningful test for us.
 
This set of stripper foils would be located in a well-shielded enclosure, suitable to slow and stop neutrons produced from nuclear interactions in the foils.  There is adequate space, too, to place carousels of foils that can be remotely actuated so spent foils can be replaced as needed.

 Should the lifetime of the foils in the PSI tests prove to be shorter than desirable, new materials are becoming available that may prove more resilient.  Graphene is being explored as a possible window or stripper material for high-current beams \cite{wang:graphene_window}.  Estimates are that multi-megawatt beams can be transported through graphene foils because of the excellent heat conductivity of this material.  However, 
 experimental measurements of strength, radiation tolerance and lifetimes must still be conducted. 
 
 Should foil lifetimes prove to be too short, another option is to use a helium-gas stripper.  Such a device has been built at RIKEN \cite{imao:gas_stripper}, and is used for stripping high-current uranium beams.  The gas cell itself is about 50 cm long, and operates with 50 torr of helium.  An elaborate set of differentially-pumped ante-chambers, serviced by Roots blowers, recycles 96\% of the helium flowing out of the 1 cm orifices required for the RIKEN beam.  The overall length of the device is almost three meters.
 
 A possible method for shortening the gas stripper and simplifying the differential-pumping structure is to  use plasma windows in the apertures.  This concept, developed at Brookhaven National Lab, has been built at RIKEN as well \cite{hershcovitch:plasma_window}, but has not yet been put into a beam line. Bench tests indicate that this window can significantly restrict the gas flow from the stripper cell, reducing (but not eliminating) the requirements for differential pumping stages.

As mentioned above, the baseline assumption is that the beam is stripped right after the cyclotron, however except for magnet strengths, the beam transport is essentially the same for either ion species.

\subsection{Basic Transport Plan}

	Beam is transported through three sets of quadrupole doublets, and three dipole bends, shown schematically in Fig. \ref{fig:BeamLineSchem}.  The dipoles serve to clean up the beam and allow the beam to follow the natural (not straight nor level) path of the drift (passage way) between the cyclotron and the target.  
	
The first quadrupole doublet will be located three meters from the extraction point, about as close to the cyclotron steel as one can get.  These magnets will focus the \htp beam through the stripper and collimator sections.

{\bf Collimator/scraper and stripper:}  Five meters from the cyclotron, the beam will pass through the stripper foils and will be converted with extremely high efficiency to protons.  
	Immediately following the stripper section will be a collimator-scraper system to intercept 
	large-amplitude particles (halo) and protons scattered in the stripping foils.  
	The strippers and collimator will be housed in a thick steel and concrete shield, calculated to attenuate neutron fluxes 
	from beam intercepted to acceptable levels at the shielding surface. 
	
{\bf Dipole 1:} Immediately following the stripper/collimator assembly will be the first dipole.  This will provide a bend -- approximately 20$^\circ$ -- sufficient to separate any \htp that has not been stripped from the protons .  A beam dump will be built into the shallow-angle port of this dipole  to capture any remaining \htp ions.

{\bf Transport and Dipole 2:} Protons are focused by the second quadrupole doublet located 18 meters from the cyclotron exit, and gently bent by the second dipole to follow the cavern path.
Note, this will be a compound bend because required deflection is both in vertical and horizontal planes. 
Such bends require careful attention to try to make beam envelopes in horizontal and vertical planes equal to prevent emittance dilution of the beam.
Further downstream, beam is focused by the third set of quadrupoles, and is presented to the third dipole, which plays an important role.

\begin{figure}[!t]
\begin{center}
\includegraphics[width=5.5in]{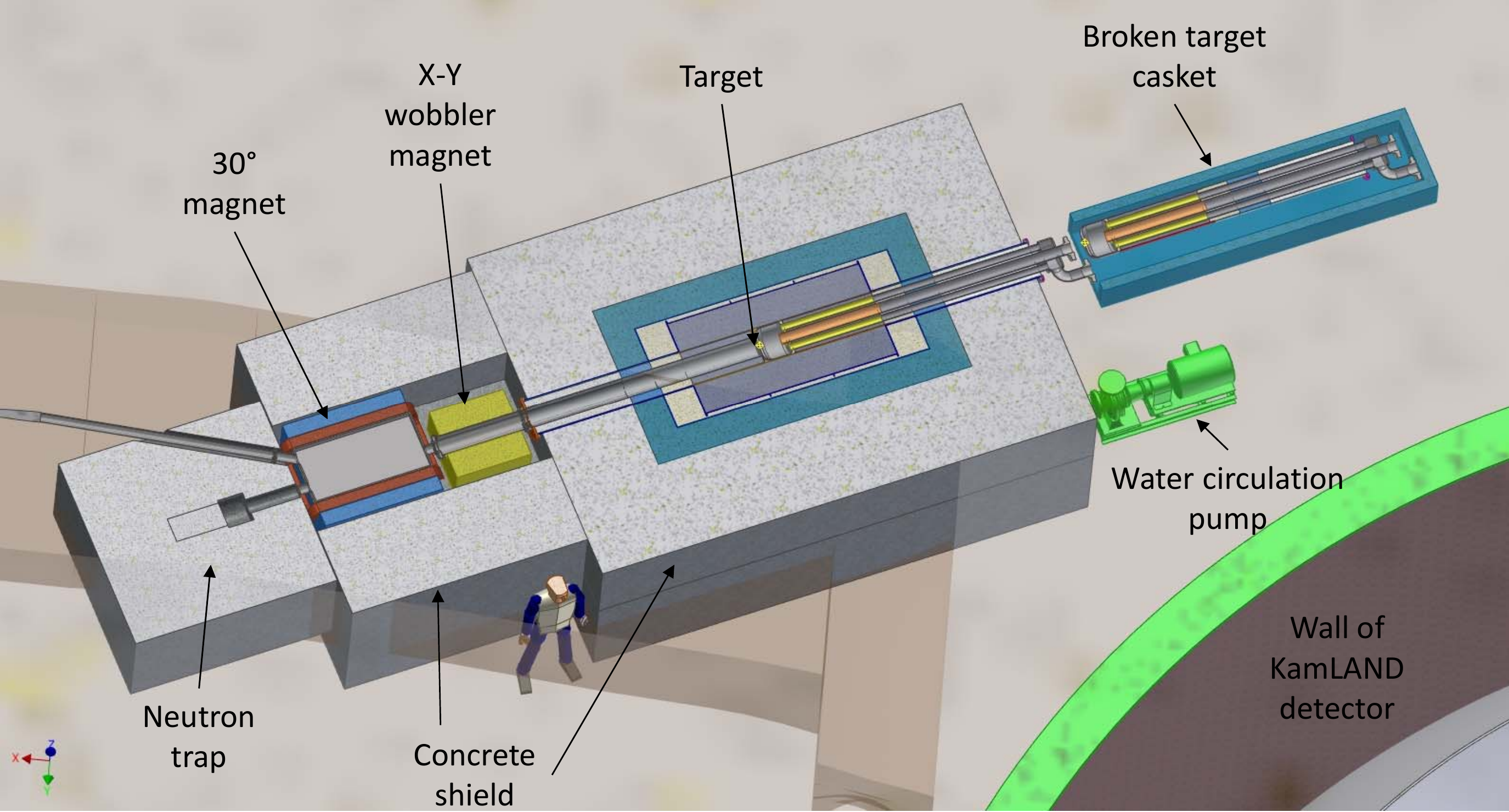}
\end{center}
%\vspace{-1.0in}
\caption{\footnotesize  Schematic of beam-line elements just prior to the target.  The 30$^\circ$ dipole  and combined-function Wobbler magnets are enclosed in additional concrete shielding to contain backscattered neutrons.  A re-entrant stop traps fast neutrons directly on the beam axis  that pass through these magnets.  Also shown in this figure are the casket system to be employed for changing targets, and the main circulating pump for the RAW (RadioActive Water) target-cooling loop.  These systems are described in following sections.
\label{30-degree}}
\end{figure}

{\bf Dipole 3:}  This dipole, located 3 meters upstream of the target will bend the beam through approximately 30$^\circ$ and direct it to the target.  The steel of this dipole will be very close to the opening in the target shield, and so provide shielding for neutrons diffusing in the general backward direction.  
However, there will be a very large flux of neutrons back-streaming directly from the face of the target.  The dipole gap and vacuum chamber must be designed so as to transport these neutrons through the magnet gap and into a large dump located directly behind the dipole.  This geometry is shown schematically in Figure ~\ref{30-degree}.

{\bf Wobbler:} The beam reaching the target will be slightly defocused, so that it's inherent size
 will be approximately 3 to 4 cm in diameter.  
It is necessary to spread the beam over the 20 cm diameter of the beryllium target in order to distribute the 600 kW of beam evenly.
 This is accomplished with a wobbler magnet.  
 The current concept employs a combined-function dipole magnet with two orthogonal coil sets \cite{anferov:scanning_magnet}.
 
 These are two coil pairs, one creating a horizontal field, the other a vertical field, that are driven sinusoidally 90$^\circ$ out of phase with each other.  
 The effect of this is to paint the beam in a circular Lissajous figure.  The yoke of the magnet should be made from steel laminations to be capable of being driven at a sufficiently-high frequency (probably a few hundred Hertz).  To further ensure uniform painting of the beam on the target, the amplitude of the driving sine wave can also be modulated.  This strategy, of painting a relatively small beam with a sophisticated driving function can serve to preserve the sharp edges of the beam at the extremities of the target, and provide adequate flexibility in ensuring the optimal heat deposition on the target, matching the characteristics of the target cooling system.
 
 Thermal heat-dissipation calculations need to be performed to assess the temperature fluctuations on the target face.  If the sweeping rate of the Wobbler is not adequately high it is possible that unacceptable thermal stresses will occur on the beryllium target as the beam heat source may be concentrated for too long a period in one location.
 
 Many applications of this beam-spreading philosophy exist, from radiotherapy with charged particle beams to ion-beam lithography.  One must be careful in planning the driving functions, to ensure that frequencies are different from any natural frequency likely to be present in the power supplies, to prevent a resonant buildup of beam intensity creating hot-spots on the target.  

	This magnet will be located very near to the upstream end of the target shielding.  It will receive a very large radiation dose, so must be made from rad-hard materials:  ceramic insulated coils, no epoxy or organics, and be cooled with the same RAW (``RadioActive Water") water system used to cool the target.
	
	It should be noted that beam-shaping optical elements using multipole magnets could also be used to spread the beam, so there would be no time-variation of the beam on any portion of the target.  However this comes at the cost of less control over the beam shape, and a higher degree of complexity of the magnet elements involved.

\subsection{Beamline Instrumentation and Vacuum}

	As much as possible, non-intercepting instrumentation must be used, including current transformers for beam intensity and 4-plate capacitive pickups to establish beam centering and steering.
	Note that the beam emerging from the cyclotron is tightly bunched, extending only over a few degrees of the 360$^\circ$ RF phase.  While the small energy spread causes the beam to de-bunch as it propagates down the beam line, it still retains most of its bunched nature when it reaches the target, so sensitivity of these pickups will be very high.
	  These instruments would be active continuously, and interlocked for rapid beam-tripping should beam stray outside of acceptable limits.  Beam size monitoring may require temporary insertion of wire harps or scraper plates, to adjust quadrupole settings. 
	Instrumentation stations will be located along the transport line, particularly close to the entrance of quadrupoles, and at dipole magnet entrance and exit points.  Of particular importance is tracking the beam centerline before and through the last 30$^\circ$ magnet.  A set of pickups downstream of the Wobbler, if possible inside the target beam pipe will be needed to ensure centering of the beam on the target face.

	Beam loss monitors located along the transport line, in the form of proportional ionization chambers that measure environmental radiation levels, will ensure that neutron levels generated by beam losses are within acceptable levels.
	
	Vacuum pumping is provided by distributed turbo pumps, located every few meters along the length of the line.  The density of pumps will depend on the species being transported:  with a more stringent requirement for higher vacuum should the beam be unstripped \htp. Pumping needs to be concentrated close to the areas where beam loss will be expected, such as dipole magnets, and intercepting beam monitors.    In particular, a high-capacity station will be needed just prior to the last 30$^\circ$ magnet.  This station will provide the pumping for the areas downstream, including the vacuum in the target, which includes the beam pipe downstream of the wobbler and the vacuum lining around the target ``torpedo."  The vacuum level in this area need not be high, so the poor conductivity around the target will not be a problem.  The area is evacuated only to eliminate the need for a window just upstream of the target.  Vacuum seals on either side of the target, described in the following section, are metallic Jetseal flanges.  
	
	Gate valves are provided to isolate sections of the beam line.  In particular, these are needed in close proximity to the cyclotron extraction point, around the gas stripper, and close to the last pumping station close to the target.  This last valve must be very fast-acting, probably cartridge-driven, to protect the beam line and upstream components from a possible target rupture.

\subsection{Beam loss control}

	While we desire to transport all of the beam extracted from the cyclotron to the target, it cannot be done with 100$\%$ efficiency.  The resulting beam losses can be divided into two categories:  controlled, and uncontrolled.  It will be necessary to deposit on a collimator particles outside of the principal beam envelope, whether from halo on the extracted beam, or from scattered or unstripped ions emerging from the stripper.  This beam loss is generated close to the cyclotron, inside the large shielding surrounding these components, so can be classified as ``controlled'' loss.  

	Uncontrolled loss occurs at other areas in the beam line, and generates neutron doses that activate beam line components and reach the rock walls.   A rule of thumb used in high-intensity accelerator transport lines, is that uncontrolled beam losses should be kept to less than 1 watt/meter of beam line length.  This figure of merit, which is pretty independent of beam energy, is designed to keep activation of beam line components below levels that would not allow hands-on maintenance.  This power loss corresponds to about 15 nA (at 60 MeV).  For a 10 mA beam this corresponds to about one part in 10$^{ 6}$ of allowed uncontrolled beam loss per meter of beam line.  

	The principal sources of beam loss will be halo generation and gas scattering.  Careful beam dynamics studies will be needed to estimate the effect of beam halo generation in this beam line, and any necessary mitigation by addition of more downstream shielded collimators.  Experience with high-power beams has been that passing the beam through a collimator/scraper will clean the halo, however a few meters downstream the halo has regenerated.  This has in fact been modeled with simulation codes taking into account strong space-charge forces of these high-intensity beams.  These same codes can be used to track halo formation and develop a suitable strategy for placement of collimators and shielding blocks.
	
	Gas scattering is more easily calculated, and establishes the required vacuum levels in the transport line.  Preliminary calculations indicate that the pressure in the transport line should be below $10^{-7}$ torr for protons, or at least one or two orders of magnitude better for \htp.  This beam loss, however, occurs along the length of the line, so controlling losses requires excellent vacuum along the entire length of the beam line.
	
\subsection{Current Status and Future Work}

The design concepts described above come from extensive experience in designing beam transport lines, and are at a suitable level for this CDR.  The engineering layouts, specification of the magnet length, bores (for quadrupoles) or gaps (for dipoles) and strengths, and materials will be determined in follow-on studies.

Evaluation of stripper options must take place, including possible experiments, to assess effectiveness, viability and lifetimes of different stripping techniques.  Experiments of stripper-foil efficiency for \htp ions have recently been conducted at LNS-Catania, and we are planning on radiation-lifetime measurements at PSI in the coming months.

Careful simulations of beam-halo generation must be performed, and from these simulations of halo growth and re-growth following collimation, a suitable strategy for collimation and scraping must be developed. These collimators, with suitable shielding enclosures, will prevent neutron flux reaching the rock walls from uncontrolled beam loss.

Simulations of beam loss from gas scattering, particularly for assessing the option of not stripping the \htp ions, must be performed, to determine locations where shielding may be required, again to mitigate neutron flux to the rock walls.

A maintenance plan must be developed, to service or replace activated components.  Of particular importance are the stripper and downstream cleaning collimators, and the areas adjacent to the target, including the last dipole and wobbler magnets.  

Target-changing strategies, discussed in the following section, must take into account the impact of possibly moving spent targets in their shielding caskets, and new targets, through the MEBT area.  
 It may be necessary to move some of the beam-line elements out of the way to allow these elements to pass by.

\subsection{Risk Assessment \& Mitigation \label{sec:mebt_risk}}

\textbf{Risk: Unacceptably-short stripper-foil lifetime.}

\emph{Mitigation:  Explore an optimized gas-stripper technology.}

\textbf{Risk: Uncontrolled beam loss - gas scattering.}

\emph{Mitigation: Ensure adequate vacuum in the beam line.}

\textbf{Risk: Uncontrolled beam loss - halo.} 

\emph{Mitigation: Beam-dynamics simulations with appropriate codes }

\textbf{Risk: Neutron containment from the target back-streaming along the beam path }

\emph{Mitigation: Proper design and placement of last beam-line elements, and a neutron dump, that interpose adequate material thickness to attenuate the neutrons. }

\
\clearpage
\section{Target and Shielding}

The protons impinge on a $^{9}$Be target,  producing neutrons
which are moderated in the cooling water and enter a surrounding sleeve where they are captured by $^7$Li to produce $^{8}$Li.  
The cylindrical sleeve contains 99.995\% isotopically-pure $^{7}$Li.  The following subsections describe:
\begin{enumerate}
\item The target and shielding design specifications.
\item The results of simulation justifying the design.
\item Our approach to acquiring and handling the sleeve material, containing fluorine, lithium and beryllium, called FLiBe.
\item The preliminary mechanical design of a target prototype.
\end{enumerate}
With respect to the specifications of the $^7$Li used in the target, we note that our original studies indicated that  99.99\% pure enrichment provides the needed
production rate \cite{bungau:daedalus}.    For this CDR, we have chosen the 99.995\% specification for the $^7$Li due to its availability for use in the advanced reactor
industry.  In the third section, below, we note that our approach to acquiring the FLiBe gives us flexibility on the specification.

\subsection{Target and Shielding Design Details}

\subsubsection{Design Specifics}

\begin{figure}[t]
\centering
\includegraphics[width=4.5in]{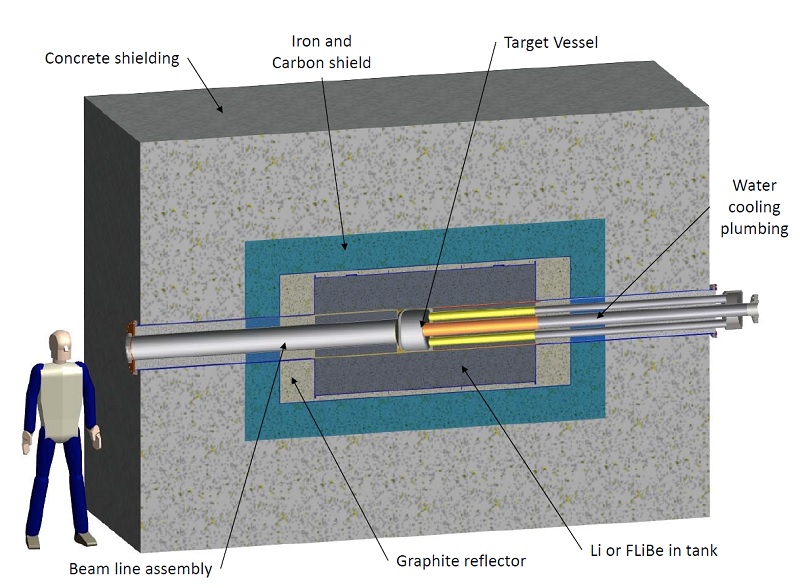}
\caption{\footnotesize Cross section of the target and shielding assembly. 
\label{targ}}
\end{figure}

The components of the current design are: the beryllium target, the water moderator/coolant, the $^7$Li sleeve, the graphite reflector, a carbon-enriched steel shell, all surrounded by a concrete shield. The target is centered in the sleeve and in the block of graphite reflector. A uniform wobbled
beam spreads the beam over the 20 cm target face. The wobbler magnets will be installed upstream of the target shielding concrete leaving only a hole large enough for the beam pipe through the shielding. A cut-away of the assembly
is shown in Fig.~\ref{targ}.

\begin{table}[b]
\begin{center}
\begin{tabular}{ l l}
\hline
\hline
Parameter & Value \\
\hline
Thickness of beryllium target & 1.7 cm \\
Distance to Bragg peak in beryllium & 2.04 cm\\
Diameter of target & 20 cm   \\
Total power on target & 600 kW \\
Power deposited in the beryllium & 300 kW\\
Power per unit area in the beryllium & 950 W/cm$^2$ \\
\hline
\hline
\end{tabular}
\caption{\footnotesize \label{targdes}Target Design Parameters.}
\end{center}
\end{table}

\begin{figure}[t]
\centering
\includegraphics[width=4.5in]{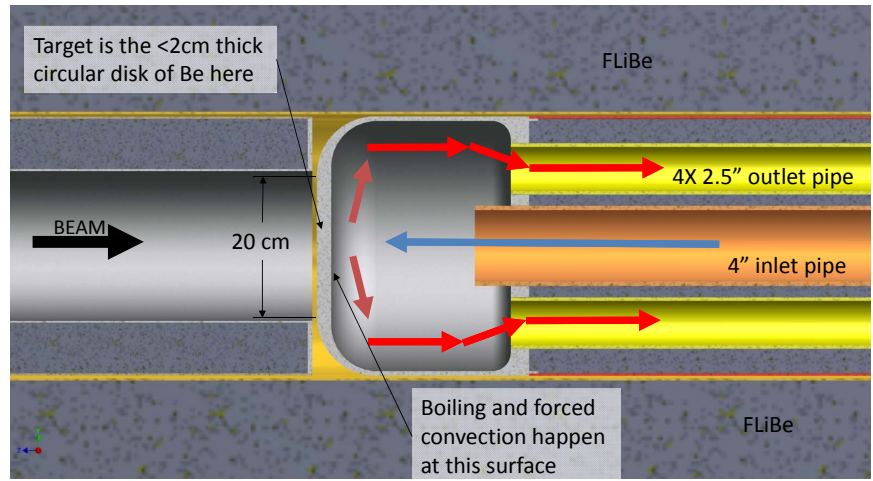}
\caption{\footnotesize Detail of the target region showing the direction of cooling water flow
\label{waterflow}}
\end{figure}

As seen in Fig.~\ref{waterflow}, the target consists of a formed hollow cylinder of beryllium with the target proper being a 1.7 cm thick end face of the cylinder. Water at high pressure is introduced along the central axis, and strikes the back surface of the target, providing the primary means of dissipating the 600 kW of beam power. Water is conducted away from the target volume by tubes located radially around the entrance pipe.  The target thickness is carefully calculated to optimize the production of neutrons and minimize heat deposited in the beryllium.   Protons of 60 MeV in Be have a Bragg peak at 2.04 cm. Thus the 1.7 cm thick Be target intercepts the high energy portion of the proton trajectory, where neutron production is high, while placing the stopping point of the protons, where energy deposition rate is greatest, within the volume of cooling water. 
%Jose
The peak power-density deposited on the target is diluted by the wobbler that uniformly spreads the beam over the full target face.  The wobbler frequency, around 60 Hz, is high enough in relation to thermal time constants to ensure distribution of the heat load.  Note, too, that the megahertz beam-bunch structure will, for the purposes of thermal diffusion, appear as a purely steady-state heat load.
 %Beam is spread over the 20 cm diameter face of the target by means of a wobbler magnet that bends the beam in a circular pattern at about a 60 Hz rate.  This decreases the peak power density deposited on the target itself.

The beryllium target vessel is surrounded by the sleeve, a cylinder with outer diameter of 1 meter and length of 1.9 meters.
%Jose added
The sleeve contains the material that absorbs the neutrons to produce the $^8$Li.  This toroidal sleeve will have an inner cylindrical wall of beryllium, to facilitate neutron transmission from the target, and outer walls of stainless or other material metallurgically compatible with the Li-containing material housed inside.
  The most convenient lithium-containing material to use is called FLiBe, a mixture of lithium fluoride and beryllium fluoride, a material with the enrichment in $^7$Li at the required level (or even higher).  Obtaining and directly handling FLiBe, as well as chemical hazards associated with employing FLiBe in the target, are discussed in Sec.~\ref{sec:flibe}.   For the discussion of the design below, it is important to note that tritium is a by-product of interactions in the sleeve and so we introduce measures to safely pump out and and store the tritium gas from the sleeve in the design.

\begin{figure}[!p]
\centering
\begin{minipage}{.45\linewidth}
  \includegraphics[width=1.0\linewidth]{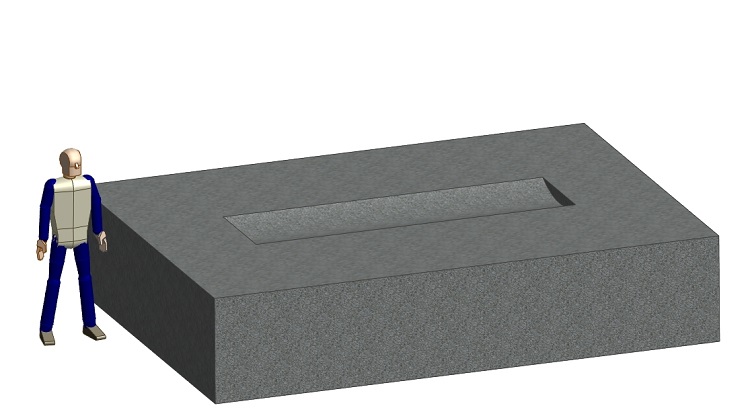}
\end{minipage}
\hspace{.05\linewidth}
\begin{minipage}{.45\linewidth}
  \includegraphics[width=1.0\linewidth]{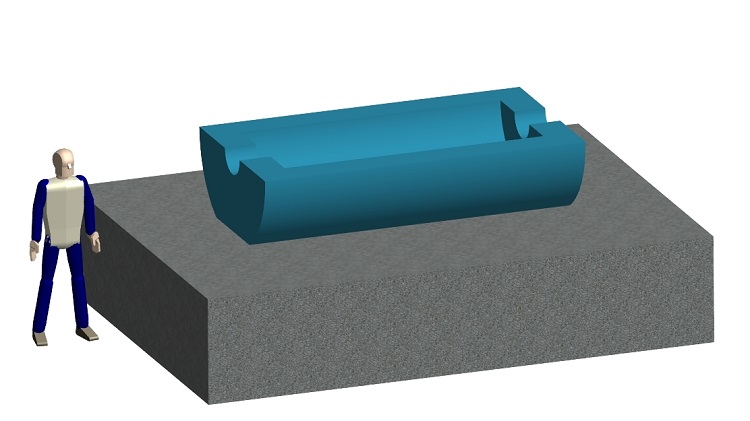}
\end{minipage}
\begin{minipage}{.45\linewidth}
  \includegraphics[width=1.0\linewidth]{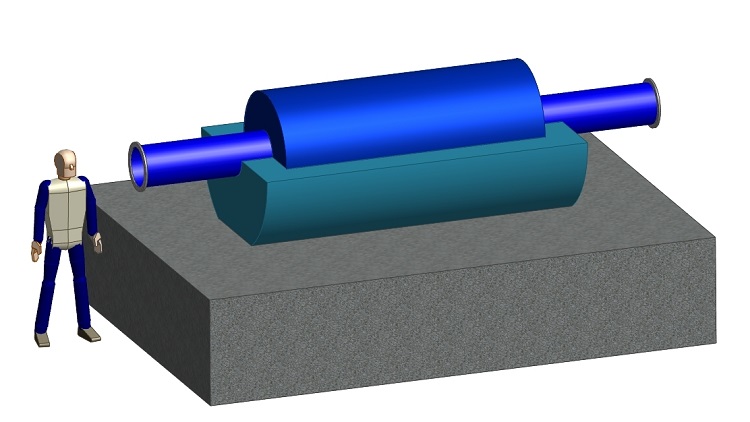}
\end{minipage}
\hspace{.05\linewidth}
\begin{minipage}{.45\linewidth}
  \includegraphics[width=1.0\linewidth]{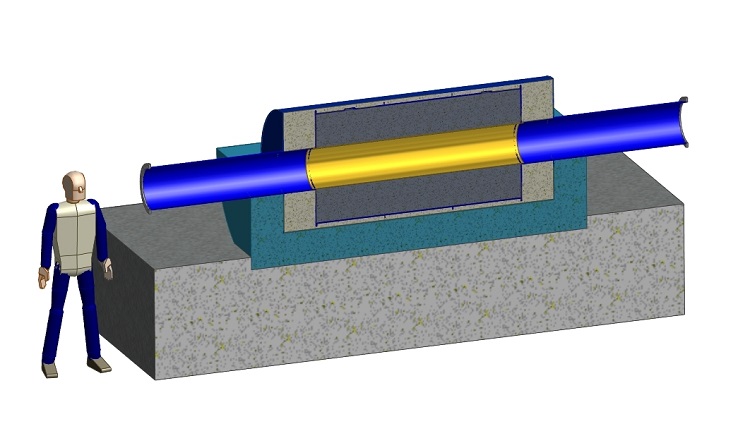}
\end{minipage}
\begin{minipage}{.45\linewidth}
  \includegraphics[width=1.0\linewidth]{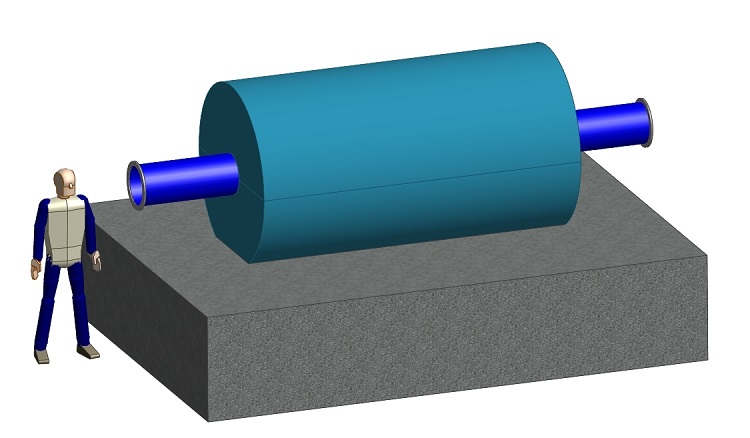}
\end{minipage}
\hspace{.05\linewidth}
\begin{minipage}{.45\linewidth}
  \includegraphics[width=1.0\linewidth]{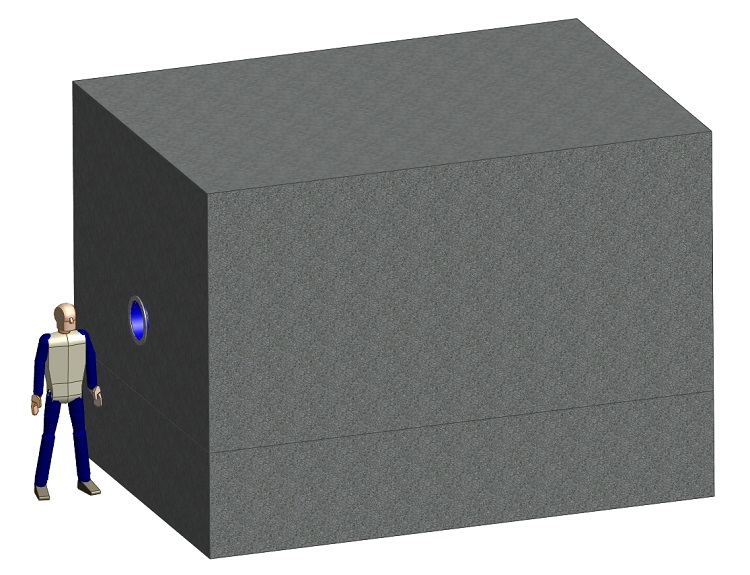}
\end{minipage}
\caption{\footnotesize {\it Top Left:} First, stacking of concrete blocks for the shielding; 
{\it Top Right:} Second, placing the bottom half of the steel shielding cylinder on the concrete pad.  
{\it Middle Left:} Third, placing the vessel containing the graphite reflector and the FLiBe into the steel cradle; 
{\it  Middle Right:} Section view for  third step, cut through the vessel containing the graphite reflector and the FLiBe on the steel cradle;
{\it Bottom Left}: Fourth, placing the upper half of the steel shielding on the target.  
{\it Bottom Right:}  Fifth, stack the upper portion of the concrete shielding blocks.  
\label{IsoDAR_target-08-0to5}}
\end{figure}

\begin{figure}[!p]
	\centering
  %\setlength{\fboxsep}{-\fboxrule}
  %\fbox{%
    \begin{minipage}[b]{1.0\textwidth}
    \centering
      \begin{tabular}{c}
        \includegraphics[width=0.45\textwidth]{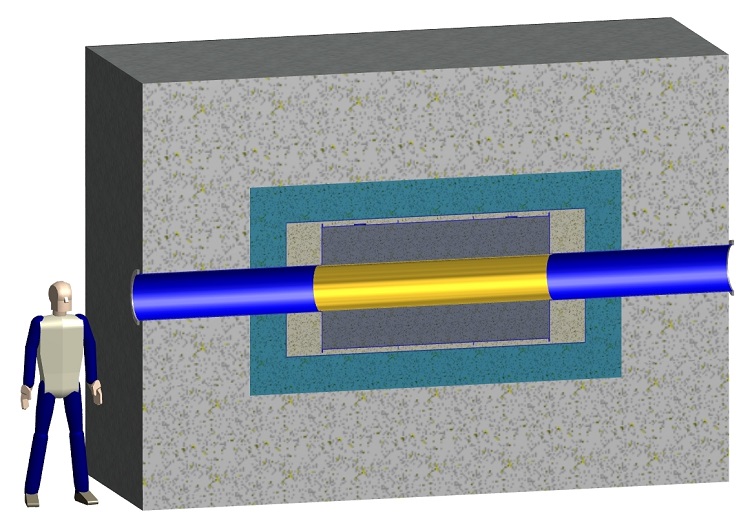}
        \\ 
        \includegraphics[width=0.45\textwidth]{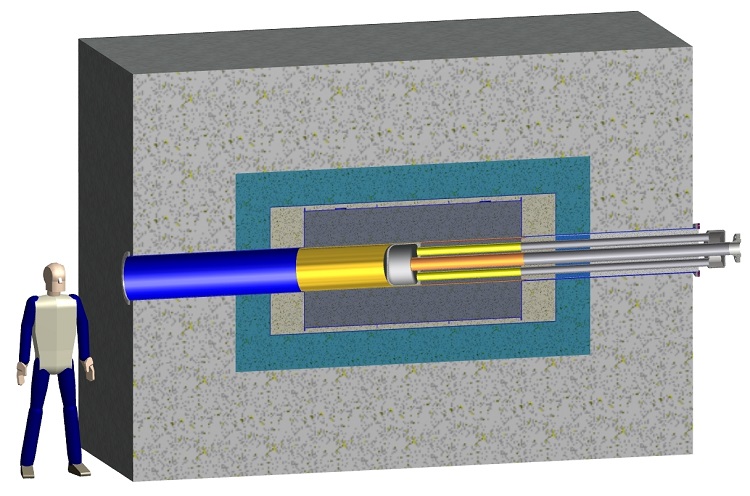}
        \\ 
        \includegraphics[width=0.45\textwidth]{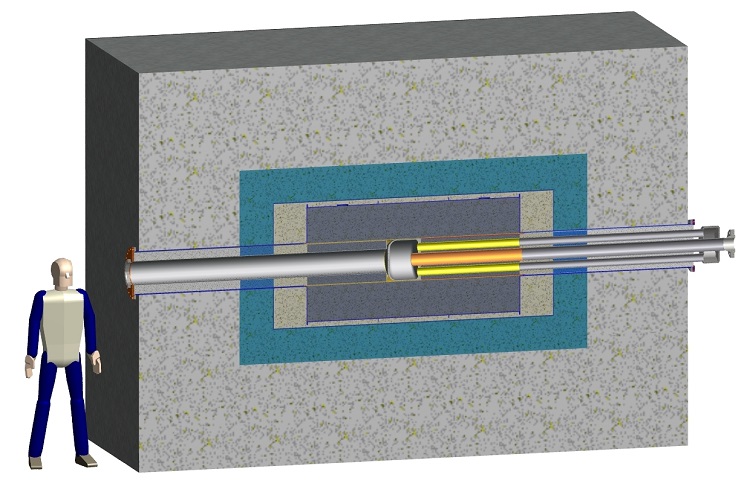}
        \\ 
	\end{tabular}
      \caption{\footnotesize {\it Top:}  Section view through the target without the target or beam tube modules in place;  
{\it Middle:} Section view showing the target module and its plumbing in place.  A complete remote handling and target replacement system has not been worked out.
{\it Bottom:}  Section view showing the beam pipe module and the target module in place.
\label{IsoDAR_target-08-6to8}}
    \end{minipage}
\end{figure}

The sleeve is surrounded by a 5 cm graphite neutron reflector, and then by shielding to limit neutrons escaping to the rock walls of the cavern.  The required thickness of the target shielding is discussed below.
Assembly of the target may be accomplished conceptually by the
sequence of pictures shown in Figs.~\ref{IsoDAR_target-08-0to5} and \ref{IsoDAR_target-08-6to8}.  The sequence proceeds as follows.  The bottom layer of concrete blocks is placed to form the cradle shown.  The concrete shielding shown weighs 46 tons (41.7 tonnes).  Second, place the bottom half of the steel
shielding cylinder on the concrete pad.  This steel half-cylinder weighs 20.2 tons (18.3 tonnes).  It may need to be
brought into the mine in pieces and assembled in place,
requiring a craning system above the target.  Interfaces between pieces will have steps to create a radiation labyrinth.  Third, place the sleeve containing the graphite reflector, the FLiBe, and the beam pipe into the steel cradle.  A section view
through the sleeve containing the graphite reflector and the
FLiBe on the steel cradle is included in the figures. 
%Jose
The beam pipe sections, of stainless, are welded to the ends of the FLiBe sleeve to form the vacuum enclosure sealed at the upstream end by the connection to the wobbler, and downstream by the assembly designed to facilitate target changes.
This assembly could be made outside the KamLAND mine and brought in as a single piece weighing approximately 7,275 lbs (3,299 kg). Fourth, place the upper half of the steel shielding on
the target.  This shell also weighs 20.2 tons, (18.3 tonnes).
Fifth, the upper portion of the concrete shielding blocks are stacked and secured in place.  The new concrete shown weighs 95.3 tons (86.4 tonnes).  Note that the parameters of the maximum part size and weight that can be transported to the mine and through the mine to the target location are not yet defined in detail.

\clearpage

%\begin{figure}[p]
%	\centering
%  %\setlength{\fboxsep}{-\fboxrule}
%  %\fbox{%
%    \begin{minipage}[b]{1.0\textwidth}
%    \centering
%      \begin{tabular}{cc}
%        \includegraphics[width=0.45\textwidth]{IsoDAR_target-08-000l.jpg}
%        & \includegraphics[width=0.45\textwidth]{IsoDAR_target-08-001l.jpg}
%        \\ 
%        \includegraphics[width=0.45\textwidth]{IsoDAR_target-08-002l.jpg}
%        & \includegraphics[width=0.45\textwidth]{IsoDAR_target-08-003l.jpg}
%        \\ 
%        \includegraphics[width=0.45\textwidth]{IsoDAR_target-08-004l.jpg}
%        & \includegraphics[width=0.45\textwidth]{IsoDAR_target-08-005l.jpg}
%        \\ 
%	\end{tabular}
%     \caption{\footnotesize {\it Top Left:} First, stacking of concrete blocks for the shielding; 
%{\it Top Right:} Second, placing the bottom half of the steel shielding cylinder on the concrete pad.  
%{\it Middle Left:} Third, placing the vessel containing the graphite reflector and the FLiBe into the steel cradle; 
%{\it  Middle Right:} Section view for  third step, cut through the vessel containing the graphite reflector and the FLiBe on the steel cradle;
%{\it Bottom Left}: Fourth, placing the upper half of the steel shielding on the target.  
%{\it Bottom Right:}  Fifth, stack the upper portion of the concrete shielding blocks.  
%\label{IsoDAR_target-08-0to5}}
%    \end{minipage}%
%  %}%
%  %\fbox{%
% 
%  %}
%\end{figure}

\clearpage

\clearpage

\begin{figure}[!t]
\centering
\includegraphics[width=4.5in]{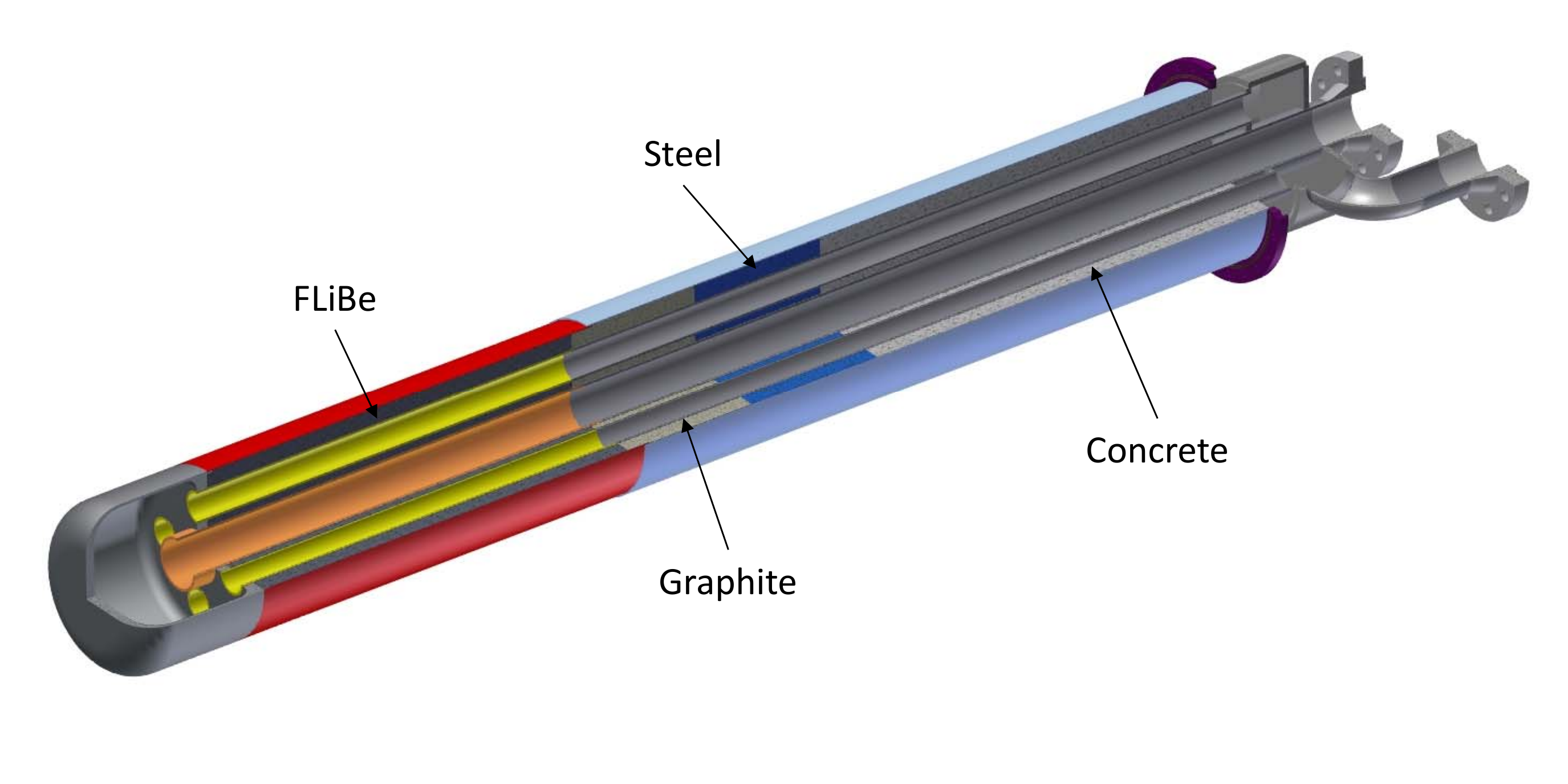}
\caption{\footnotesize 3/4 section view of the target ``torpedo".  The outer shell parts are:  grey - beryllium (integral target piece), red - beryllium, light blue - stainless steel, purple - vacuum flange.
\label{torpedo}}
\end{figure}

Fig.~\ref{torpedo} shows a 3/4 section view through the target ''torpedo".  The beryllium target vessel is attached to one 4" inlet pipe and four 2.5" outlet pipes.  Preliminary calculations show that the heat deposition cannot be removed from the Be target by forced convection or by boiling at one atmosphere alone.  Both forced convection and boiling are required to remove the total amount of heat.  If we can paint the full 20 cm diameter of the target with the beam, then we only need to remove 0.95 kW/cm$^2$ from the inside face of the beryllium target.  This is much less than the state of the art, which is ~6 kW/cm$^2$ with submerged jet impingement cooling.

The water velocity in the inlet jet must be kept to less than 15 m/sec to avoid erosion of the Be surface.  Preliminary calculations assumed that the water velocity was 12 m/sec, which leads to 1,545 gallons per minute (5,848 l/m) of water flow through the target.  Since about half the power is deposited directly in the water behind the target, carrying away 300 kW at this flow rate only increases the water temperature by a 0.74C difference between the inlet and outlet.  CFD calculations will be necessary to determine how well heat is extracted from the inner surface of the beryllium target and how sub-cooled the water needs to be.  The sub-cooling allows more energy to be carried away by the transition to boiling, but requires the cooling loop to be run at elevated pressure, possibly several hundred psi.  The cyclotron will have to be interlocked such that any loss of water pump power in the primary cooling loop immediately stops the beam.  The CFD will also show whether the inner shape of the Be vessel is correct to shape the water flow from inlet to exit ports and if there are any places on the vessel with inadequate heat extraction.  The combination of phase change and forced convection with the possibility of cavitation induced erosion means that the CFD will at best inform the design of the first prototype.  Only physical testing will determine the actual target lifetime.

\subsubsection{Target Maintenance Strategies}

The beryllium target will probably suffer metal crystal lattice degradation from beam hitting on the vacuum side and erosion from cavitation due to the boiling and forced convection on the water side.  These processes can lead to failure of water containment in the target vessel.  There can't be a window between the target and the beam vacuum (because of lack of cooling for the window,) so any water leaking from the target vessel goes directly into the beam pipe vacuum space.  Diagnostics will be in position to sense water leaks and shut a fast gate valve in the beam line before much water can leak toward the cyclotron.  In the event of a water leak from the target vessel, the target ''torpedo" will be exchanged for a new one.

Remote handling procedures will need to be developed as the broken target will be too hot for humans to be around if unshielded.  Techniques such as those used in neutrino beam target halls at Fermilab will be adapted to safely remove and transport broken targets.  

\begin{figure}[!t]
\centering
\includegraphics[width=4.5in]{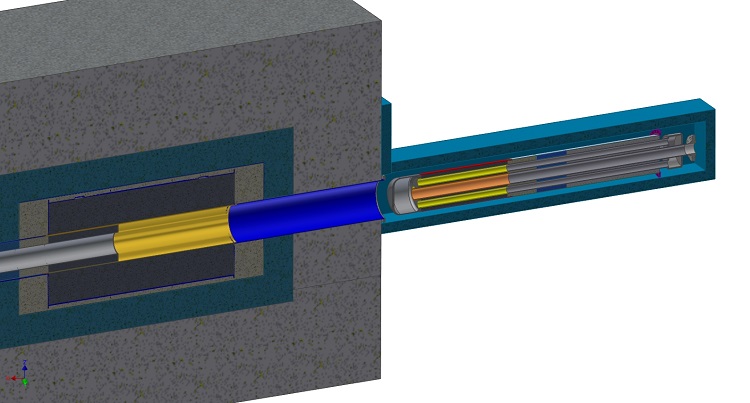}
\caption{\footnotesize Elevation section view showing target torpedo extracted from the beam pipe into steel casket.
\label{target_repl-001}}
\end{figure}

Given the magnets and neutron shielding at the upstream end of the target, the most logical place to remove the torpedo from is the downstream end of the target shield.  Neutron shielding at the downstream end is not shown in the pictures, but it is assumed that some concrete blocks will be necessary.  These can be moved aside on a rail system.  The water cooling plumbing at the downstream end is removed and then the vacuum seal between the FLiBe tank and the torpedo is disassembled.  A casket is brought into place at the downstream end, and the torpedo can be pulled into the steel casket as shown in Fig.~\ref{target_repl-001}.  Details of the sliding, actuation of movement and alignment need to be worked out.  The door of the casket is lowered into place and the casket can be taken to storage.  A 4 inch thick steel casket weighs 8.75 tons (7.93 tonnes).  With a target inside, (1,436 lbs, or 651 kg), the weight is 9.5 tons (8.6 tonnes).  These will have to be moved on a motorized rail system.

\begin{figure}[!t]
\centering
\includegraphics[width=4.5in]{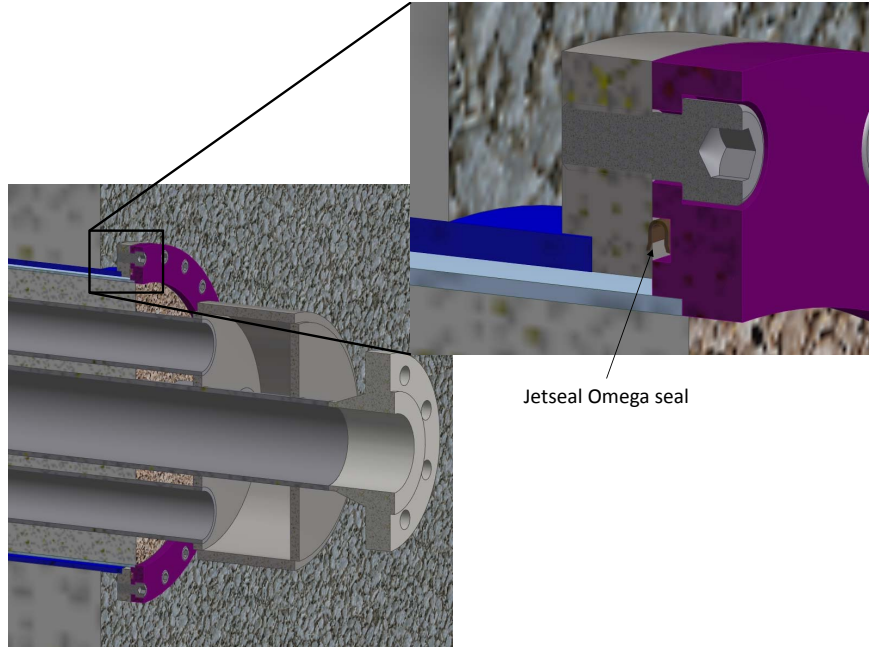}
\caption{\footnotesize Detail of ``omega" seal from Jetseal, Inc.  Jetseal is one of several possible all-metal seals that can be used to seal the target torpedo to the beam pipe.
\label{target_seal-001}}
\end{figure}

The details of the vertical vacuum seal are shown in Fig.~\ref{target_seal-001}.  The advantage of an omega seal in a vertical sealing groove is that there is an easy way to lock the seal into the groove using a metal tab.  This method was used in the EXO experiment to retain a large Jetseal seal in a copper cryostat.

The cavity in the FLiBe tank will also be radioactive once used, so the replacement operation to install a new target will be the reverse of extracting a broken target.  The new target will be inserted into its cavity by pushing it out of a casket.  The purpose of the casket holding the new target is that the casket will be a shield from radiation coming from the cavity.

\subsubsection{How Target Design Affects The Physics}

\begin{figure}[t]
\centering
\includegraphics[width=5.5in]{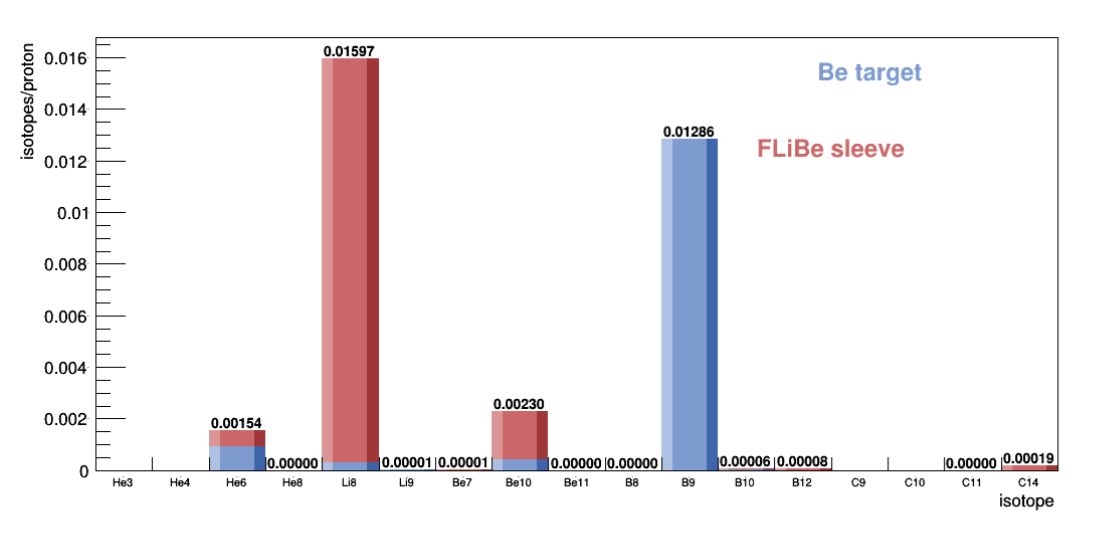}
\caption{\footnotesize The isotope production rates per incident 60 MeV proton.  The production 
is separated by color into the two sources: Be target and FLiBe sleeve.
\label{isotopes}}
\end{figure}

The isotope production per incident 60 MeV proton is shown in Fig.~\ref{isotopes}.  The antineutrino $\bar\nu_e$  flux of interest for the
IsoDAR measurements is from the $\beta$-decays of the produced $^8$Li isotopes.   The current design produces about 0.016 $^8$Li isotopes
per incident proton and is what has been used for the sensitivity studies shown in Chapter 2.  

\begin{figure}[t]
\centering
\includegraphics[width=5.5in]{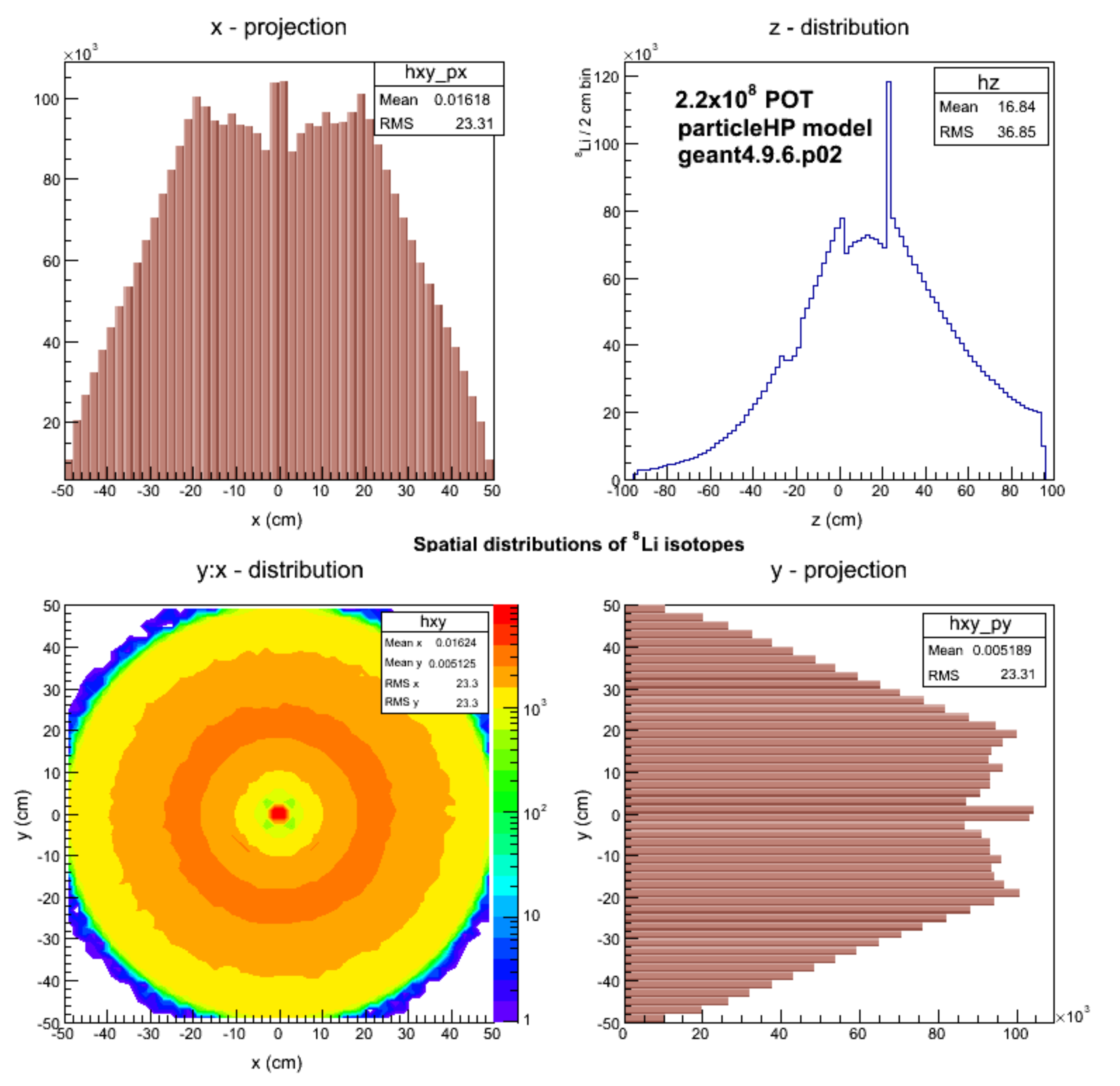}
\caption{\footnotesize Distributions for the antineutrino $\bar\nu_e$ production points in the IsoDAR target.  
\label{isotopes_dist}}
\end{figure}

As seen from Fig.~\ref{isotopes}, the dominant 
source of $^8$Li isotopes is from the outer FLiBe sleeve, which has an outer radius of 50 cm and a length of 190 cm in the current design.  Thus, the
antineutrino $\bar\nu_e$ production points will be distributed throughout the FLiBe sleeve.  Fig.~\ref{isotopes_dist} shows the production point
distributions.  The x and y distributions have an $\sigma = 23$ cm and for the z distribution $\sigma = 37$ cm.  

The spread in these production points can
affect the oscillation measurements by introducing an uncertainty in the antineutrino flight path.  For the studies in Chapter 2, the antineutrino
production points have been assumed to come randomly from a $(1 \times 1 \times 1)$ $m^3$ region centered on the target.  This distribution is
consistent with the actual distribution shown in Fig.~\ref{isotopes_dist}.  Studies have shown
that the additional smearing from the antineutrino production point does not affect the oscillation measurements very much as shown in Fig.~\ref{waves}.

\subsubsection{Current Status and Future Work}
{\bf Target Cooling:}  A conceptual design for the neutrino production target has been
developed, which meets the required antineutrino rates to accomplish
the physics with a 60 MeV 10 mA proton beam.  This beam is expected to
deposit of order 300 kW of heat into the beryllium target structure
and an additional 300 kW of heat in the water directly.  This heat
must be removed by a combination of forced convection and boiling at
the surface behind the target using cooling water.  The next step in
this design sequence is to perform detailed simulations of the cooling
and material integrity backed up by prototype studies.

The future work would involve simulating the behavior and efficacy of
the Be target cooling system and design a prototype to study the
removal of the 600 kW expected to be deposited in the Be target.  To
remove the heat at this level, one needs to use boiling water at the
inside surface of the Be target plus forced convection.  A numerical
analysis to determine the cooling parameters will be necessary.  In
addition, the effects of the large water flow on the target need to be
simulated and prototyped, especially with respect to the erosion of
the Be material.

The parts of this study include:
\vspace{-\topsep}
\begin{enumerate}
\item 	A detailed Computational Fluid Dynamic (CFD) analysis needs to
  be developed and used to test the full system.
\begin{enumerate}
\item	Calculate the energy deposition in the target and water
\item	Do CFD analysis to determine the effect of static pressure
\item	Do CFD analysis varying the nozzle to target spacing
\item	Do CFD analysis varying the nozzle diameter and number of nozzles
\item	Determine the water pressure and flow velocity
\end{enumerate}
\item	A prototype Target Be vessel with external utilities and a
  simulated heat source needs to be designed and used to validate the
  CFD simulation results.
\begin{enumerate}
\item	Required tests to validate the simulation need to be developed.
\item	A full prototype system that can accomplish these tests needs to be designed.
\end{enumerate}
\end{enumerate}
\vspace{-\topsep}
%{\bf Target Maintenance:}  With a high degree of certainty, the target ''torpedo" will need to be changed during the course of the experiment.  A plan must be developed so this can be accomplished, without exposing maintenance personnel to the very high radiation levels that will be present on the target.  The best option appears to be that the spent target should be removed from the back of the shielding and placed into a shielded casket.  Remote robotic handlers need to be developed that can disconnect the water circuits and retract the torpedo into the casket.  There needs to be sufficient space alongside the target shielding to transport the casket to the front of the target location, then out by the cyclotron vault, and eventually out to the surface, where the casket can be transported to a suitable storage area for radioactive material.

\subsubsection{Risk Assessment \&  Risk Mitigation\label{sec:target_eng_risk}}

\textbf{Risk: Target Lifetime.} The beryllium target will be subject
to a difficult thermal and radiation environments.  The lifetime of
the target are difficult to estimate.  

\emph{Mitigation: Modeling, tests, and suitable inventory of
  replacements for the target:  As mentioned already, experimental tests on beryllium erosion from high-pressure water will help in establishing wear rates and possible modes of target failure.  At the start of running, at least one spare target will be kept on hand, and based on lifetime of the first target, enough lead time provided to ensure there is always at least one spare in inventory.}

\subsection{Target Shielding Considerations\label{sec:shielding_calculations}}
The proposed site for the IsoDAR target is currently configured as a
control room. Once the present structure is removed, the space in the
tunnel has approximate cross section dimensions of 2.25 m floor to
ceiling and 3.5 m side to side. As discussed above, this space must
contain the target assembly, sleeve and graphite reflector surrounding
the sleeve. The remaining space is available for neutron shielding
which will be a combination of iron, concrete with high loading of
boron and other neutron-absorbing materials. 
A preliminary design for the shielding has been developed through
simulations and calculations. The calculations indicate that the space
for neutron shielding will need to be larger than the present cavern, particularly in the
vertical direction. Below, we describe the basis for our simulation for 
this design along with a discussion of the additional studies needed to
complete the specifications of the shielding dimensions. 

%We need consistency of usage:  Geant4 or GEANT4.  I've changed what I noticed to all caps... JA
\subsubsection{Monte Carlo validation with experimental data}\label{sec:ccc}
GEANT4~\cite{agostinelli:geant4} provides an extensive set of 
hadronic physics models for
energies up to 10 - 15 GeV, both for the intra-nuclear cascade region 
and for modelling of evaporation. To cover all combinations of incident particle type, energy and target material,
various models are combined into physics packages called physics lists
in order to address the full spectrum of hadronic collisions. There are many different models 
(data based, parametrized and theory-driven) and each one uses 
different approximations and has its own applicable energy
range. Monte Carlo codes usually come with their own physics models and the
user is offered default selections. However, due to the vast range 
of applications, GEANT4 will not give the users any default physics
models, the users themselves have to work out what models to use for what processes. 
One of the best hadronic models available in GEANT4 is the Liege 
intra-nuclear cascade model coupled with the independent evaporation/fission code
ABLA, which has been validated against experimental data for
spallation processes in many different heavy elements
\cite{heikkinen:geant4}.
However, the INCL-ABLA validation results presented at the IAEA benchmark for
spallation reactions~\cite{iaea:spallation} show that, for energies lower than
100 MeV, the results of the Liege model are not so good as above this
energy. This is because the model does not have pre-equilibrium:
INCL cascade is directly ``coupled'' to equilibrium de-excitation
handled by ABLA and therefore it does not describe well enough low
energy reactions where nuclear structure effects start to play their
role. Above 100 MeV, INCL-ABLA works well being one of 
the best models available. 

\paragraph{The GEANT4 model}
Recent GEANT4 developments have introduced a new data driven model
called \textit{particle\_hp} for describing the proton inelastic interactions 
in the energy range 0 - 200 MeV. 
The particle high precision package \textit{particle\_hp} uses evaluated nuclear
data bases for inelastic interactions of proton, neutron,
deuteron, triton, $^3$He, alpha and gamma. It is applicable for incident 
particle energies in the range 0 - 200 MeV. The package works with GEANT4
versions \textit{GEANT4.9.5} and later. It includes three physics
lists: for protons  \textit{QGSP\_BIC\_PHP}; for neutrons  \textit{QGSP\_BIC\_NHP}; and one
for all particles, i.e. neutrons, protons, deuterons, tritons, $^3$He and
alpha,  \textit{QGSP\_BIC\_AllHP}. The last two physics lists give the same results
for neutrons as the physics list  \textit{QGSP\_BIC\_HP} from the standard
GEANT4. 

The evaluated nuclear data libraries differ and thus the results of
the Monte Carlo simulations will depend on the library. 
Two databases  \textit{TENDL 2012} and  \textit{ENDF VII} were converted into 
GEANT4 format. They can be found in two versions: with information on
residual nucleus yields and without information on residual nucleus
yields (in this case GEANT4 will not generate any ion). The \textit{ENDF VII} 
library uses experimental data for projectile energies up to 150
MeV. These data are essentially nuclear reaction cross sections
together with the distribution in energy and angle of the secondary 
reaction products. The  \textit{ENDF VII} database for proton
projectiles contains data only for 48 isotopes (including Be). 
Unlike the earlier versions, the  \textit{TENDL 2012} database
contains information not only for proton, but also for deuteron, triton, $^3$He 
and alpha. The \textit{TENDL 2012} library uses some experimental data
and \textit{TALYS} calculations for projectile energies up to 200
MeV. The database can be applied to all target materials but the best
results are obtained for targets with atomic number in the range
12-339. The  \textit{TENDL 2012} database contains information for all
isotopes. 

The predictions of this model rely on the existing tabulated
experimental data. The data on some isotopes were
measured in more detail than on other isotopes in the experimental database. 
Therefore the model predictions will continue to improve
as more experimental data becomes available. This being said, even
with the limited amount of data for protons on beryllium, this model
describes better the inelastic proton interactions for 60 MeV incident
energy than any other theoretical model available. 

\begin{figure}[t]
   \centering
   \includegraphics*[height=75mm, width=90mm]{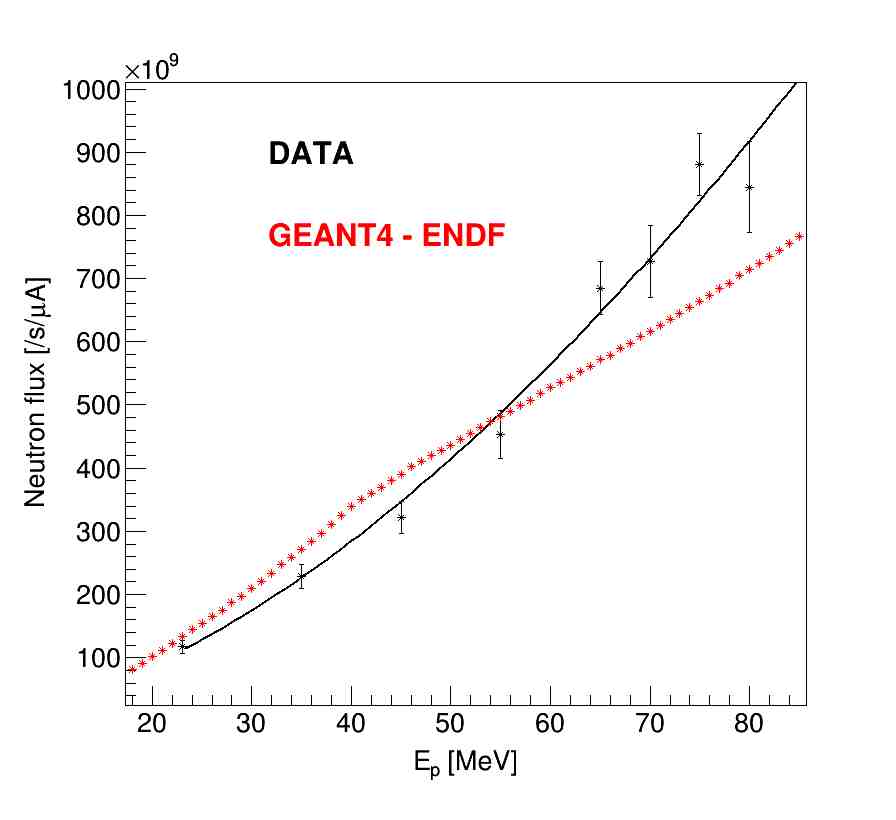} 
   \caption{Validation of the \textit{particle\_hp} physics package
     with experimental data taken from Ref.
     \protect\cite{tilquin:n_flux} for various
     proton energies on a Be target.  }
   \label{fig:Tilquin_ENDF_valid}
\end{figure}

\paragraph{Validation with experimental data and MCNPX}
Simulations using the beryllium target were performed for several proton beam energies,
and were compared to experimental measurements \cite{tilquin:n_flux}. 
In the experiment, the Be
target has a diameter of 40.3 mm and a thickness of 3.7 cm to stop the
incident protons. The target
is surrounded by a vessel filled with a manganese sulphate solution; neutron yield is 
obtained by measuring the production of $^{56}$Mn.
The beam profile at the target position is gaussian
and the full widths are 15 mm horizontally and 30 mm vertically. The total 
number of neutrons per second and per ${\mu}$A emitted by the target calculated by GEANT4
is compared with the measured experimental values and
fig.~\ref{fig:Tilquin_ENDF_valid} shows the results. At about
55 MeV proton energy, there is a good agreement between the GEANT4
predictions and the measured data. Below this value, the simulation
overestimates slightly the experimental data but above 55 MeV the code
predictions underestimates the experimental values. A possible reason 
is that at higher proton energies, a higher fraction of spallation 
neutrons produced in this experiment were captured on $^{55}$Mn present 
in the manganese bath than estimated, due to resonances  at 
higher energies in the neutron capture cross section for $^{55}$Mn. This 
would result in an overestimation of the evaluated total number of
neutrons emitted by the target in this experiment. 

In the IsoDAR experiment 60 MeV protons impact on a beryllium
target and from fig.~\ref{fig:Tilquin_ENDF_valid} we see that the model
predictions for 60 MeV protons match relatively well the experimental data. 

The experimental data on neutron yield from a Beryllium target in the
60-70 MeV proton energy range is rather scarce. The integrated yield
at 70 MeV was measured in Ref~\cite{tilquin:n_flux} while differential yields
for few angles were measured at various beam energies in
Ref~\cite{waterman:n_spectra, harrison:neutrons}. 
New measurements of neutron
yield produced by a 62 MeV proton beam on a thick beryllium target were
performed at Laboratori Nazionali del Sud (LNS) of INFN
using the existing superconducting cyclotron~\cite{osipenko:neutrons}. 
A 62 MeV proton beam with an operating beam current of 30-50 $\mu$A was
extracted from the cyclotron and transported through the beam
transport system to the target. The beryllium target 
had a thickness of 3 cm and a 3.5 cm diameter. This thickness was
chosen to ensure complete absorption of the protons. The neutrons 
produced in the target were measured by the time-of-flight technique. 
Eight neutron detectors were installed around the target at the same 
height with respect to the beam line at different angles and at two 
different distances (150 cm and 75 cm). The electric charge deposited 
by the beam on the target was measured by a digital current integrator 
and used for absolute normalization of the data.

\begin{figure}[!t]
   \centering
   \includegraphics*[height=130mm, width=140mm]{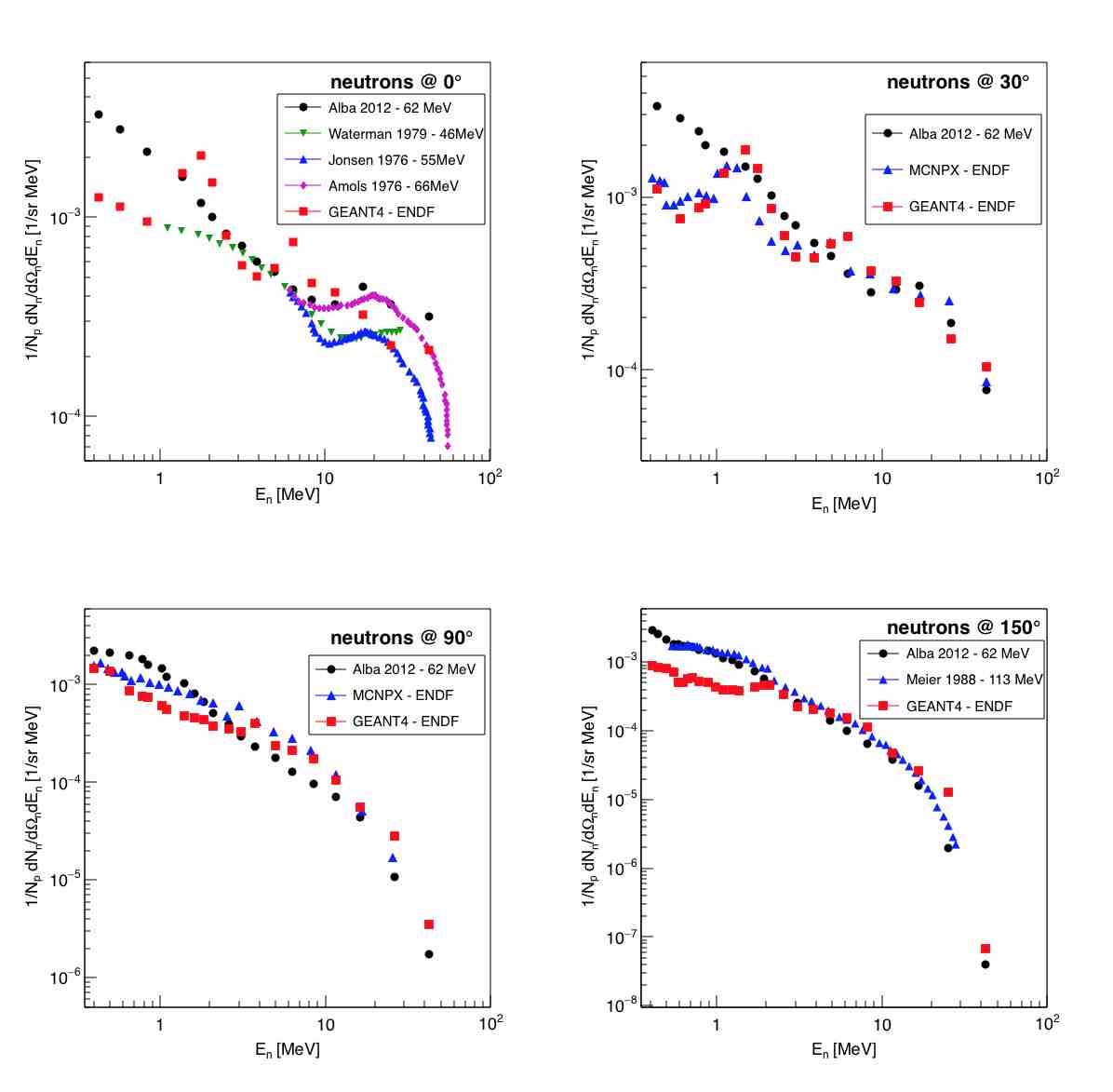} 
   \caption{Comparison of the \textit{particle\_hp} physics package predictions with
     the experimental data taken from this experiment, Ref. \cite{osipenko:neutrons, waterman:n_spectra, harrison:neutrons, johnsen:n_production, amols:neutrons} and
     MCNPX simulations. The neutron yield was measured at
     $0^{\circ}$, $30^{\circ}$,  $90^{\circ}$ and  $150^{\circ}$.}
   \label{fig:data2}
\end{figure}

The target and the detector set up were modeled with GEANT4
simulations. The \textit{particle\_hp} physics package was used to simulate the
neutron yield produced by 62 MeV protons. Fig.~\ref{fig:data2} 
shows the results at $0^{\circ}$, $30^{\circ}$,  $90^{\circ}$ and
$150^{\circ}$. 
The comparison of the simulation with the experimental data taken from this
experiment and also from Refs.
\cite{waterman:n_spectra,
      johnsen:n_production,
      amols:neutrons}
at $0^{\circ}$ for lower beam energies shows a 
significant disagreement in the lower neutron energy range. This disagreement 
demonstrates that at lower beam energy data cannot be extrapolated to 
higher beam energies simply by an overall factor, and the kinematic 
limits have to be taken into account. The comparison shows that for 
low energy neutrons (below 10 MeV) there is 
a disagreement between data and simulation especially at lower
angles. The \textit{particle\_hp} physics package which uses the \textit{ENDF VII}
library for particle cross sections was released in December
2011 and therefore does not contain the experimental results measured in 
Ref.~\cite{osipenko:neutrons} at a later date. However, at larger
angles there 
is a good agreement between these two, even for low energy neutrons 
(above 2 MeV).  At larger angles ($150^{\circ}$) there is a good agreement 
between the measured data in this experiment and ref.
\cite{meier:neutrons} and 
the GEANT4 code predictions for neutron energies above 2 MeV. At 
neutron energies below 1-2 MeV, the GEANT4 predictions lie below data.

The GEANT4 simulations were compared not only with the experimental data
from this experiment but also with MCNPX results using in both simulation codes
the ENDF data library for measured cross sections. The comparison
shows a good agreement between the predictions of the two codes at larger
angles, $30^{\circ}$ and $90^{\circ}$. There is a good agreement 
between the two codes and data  in some kinematic regions, like for 
larger angles and larger neutron energies.
There is also a good agreement between the GEANT4 and 
MCNPX predictions at all angles and for all neutron energies. 

Referring again to the Louvain-la-Neuve experiments \cite{tilquin:n_flux}; these measured the total number of neutrons emitted when 
proton beams having an energy in the range 23 and 80 MeV impact on 
a beryllium target.
% were performed at the Louvain-la-Neuve (LLN) 
%cyclotron in Belgium~\cite{tilquin:n_flux}. 
Eight beams of different
proton energies (23, 35, 45, 55, 65, 70, 75 and 80 MeV) were used
to measure the neutron fluxes produced by the p+Be reaction. The 
beryllium target was surrounded by a large metallic cylinder filled 
with an aqueous manganese sulphate ($MnSO_{4}$) solution. The neutrons 
were slowed down to thermal energies by water and partially captured by 
manganese. The thermal flux is determined by $^{56}Mn$ activation
measurements. The neutrons interact with the Mn but part of
them are lost by reactions on H, O, and S present in the bath or by
self absorption in the target. In the second stage, the beam and the
beryllium target were replaced by a Ra-Be source with known activity
(14.8 GBq). The number of neutrons emitted by the beryllium target
was determined by knowing the neutrons emitted by the Ra-Be source
and by measuring the isotopes produced in the bath with the source and
with the beryllium target when hit by the proton beam. The relative 
errors over the neutron yield measurements vary between 7.3\% and 
9.1\%~\cite{tilquin:n_flux}.

The discrepancies are mainly for neutrons produced with energies below
2 MeV and this is due to the fact that the new experimental data
measured by Alba \textit{et. al}~\cite{osipenko:neutrons} 
in 2012 was not
included yet in the proton ENDF database when it was released. However these low
energy neutrons will not make it through the shielding and therefore
will not pose a problem for rock activation.  

As more experimental data for protons on beryllium becomes available
it will be added to the proton ENDF database increasing the accuracy
of the model predictions.

\subsubsection{Neutron Flux Limit Requirements}
The requirements for the shielding and radiation protection have been
established by Japanese radiation-safety regulations \cite{Uwamino:RIKEN}, which are based on IAEA standards \cite{IAEA}.
To wit, artificially produced radionuclides of any species we are likely to produce
must not exceed an activity level of 
0.1 Bq/g. 

\begin{figure}[!t] 
\centering
\includegraphics[height=75 mm, width=90 mm]{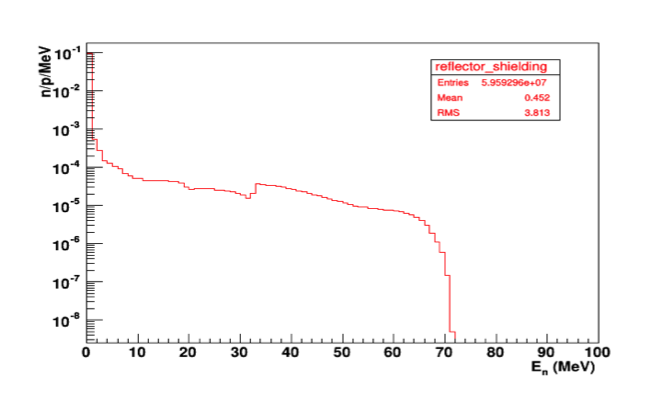}
\caption{The energy spectrum of the neutrons that come out of the
  reflector and enter the shielding.}
\label{fig:001}
\end{figure}

The total neutron flux that is produced in
the target system during the experiment is defined, as an experimental requirement. 
It is straightforward then to calculate the neutron flux at the boundaries of the sleeve and graphite reflector.
Figure ~\ref{fig:001}
shows this neutron flux as it enters into the shielding outside the reflector
One can see that the preponderance of these neutrons are at low energy, however
the higher-energy neutrons are proving to be the most troublesome.
% The neutrons that escape from the beam line components are
%subject to engineering mediation in the sense that beam diagnostics systems
%that measure, monitor and correct both controlled and unexpected beam losses
%during commissioning and running the facility need to be installed. 

%A fraction of the neutrons will escape 
%into the cavern walls during a run causing activation and this must 
%happen inside a Controlled Area defined by the 0.1 Bq/g criterion.

The task at hand is to start with this spectrum, attenuate it with an optimized shielding configuration,
bathe the rock surfaces with the resulting neutron spectrum and
calculate from this the residual activity.  This process is iterated, with further optimization of the shielding, until the requirement of 0.1 Bq/gm is achieved.  The primary tool for this study, again, is GEANT4.

For this study, it is important to understand the rock composition, particularly in regards to trace
elements that may be parents of long-lived isotopes.  Also one must establish the
ground rules of when the residual-activation assay will be performed (how many years after
the completion of the experiment), and what technique will be employed for performing the assay.

In regards to the time the assay is performed, it is not unreasonable to specify this as five years following the completion of the experiment.  This time will be marked as the formal end of the IsoDAR experiment, the time when the area occupied by the experiment is returned to its original owners.  The condition of the area at that time must satisfy the above-stated residual activation requirements.  The IAEA regulations actually state that any area containing artificially-produced radioactivity above the 0.1 Bq/gm level must be in a Controlled Area (for radiation protection). During the course of the experiment, essentially all of the area occupied by the experiment must be in such a Controlled Area.  However, after the beam is shut off, the artificially-produced activities will begin to decay, so the size of the Controlled Area will begin shrinking. The condition for return of the IsoDAR space is that there be no remaining Controlled Areas.

When the experiment is completed, many components will be very radioactive.  It is wise to allow these to cool, for perhaps a year, prior to beginning the disassembly of the equipment.  This decommissioning must proceed at a careful pace, with handling and removal of the still-radioactive pieces taking place in a slow, deliberate manner.  It may take more than a year to complete this process.  A further year or two might be specified before making the final assay.  Note, that the 0.1 Bq/gm is approximately one tenth of the natural background of uranium and thorium in the mine, so that even before the final assay is performed the radiological hazard present in the area will have all but vanished.

For the technique to perform the assay, it is proposed to bring in a large, sensitive germanium gamma detector, and place this in the center of the area to be assayed.  This will measure the integrated residual activity emanating from the walls of the cavern.  Selected locations for this detector should be agreed upon: certainly the point of impact of the beam on the target, as the primary source of neutrons.  But other areas, as well, along the beam line and in the cyclotron vault should be marked as assay points.  The sensitivity of the detector will pick up the presence of artificial isotopes from the  background of uranium, thorium and potassium naturally present in the surrounding rocks.

The following sections will detail the status of the shielding and activation studies to date.  They are by no means complete, as the complex calculations often lag behind the changing configurations of the beam line configurations.  However, even from the incomplete status of the calculations, we have confidence that our methodology is sound, and that the final solutions will provide complete compliance with the regulations.

\subsubsection{Simulating Rock Activation}

\paragraph{Rock analysis}
Radionuclides with half-lives shorter than a day are of no
consequence so the  progenitors of long-lived products
like $^{60}Co$, $^{152}Eu$, $^{154}Eu$, $^{134}Cs$ etc. are the ones
that need to be assessed. Rock samples were collected from various sites in the mine and irradiated 
in the MIT reactor to $10^{18} n/cm^{2}$. The rocks were grouped in
groups of six samples, each set with different fluences, masses,
exposure duration/times etc. labelled: KRS1, KRS2, KRS3, KRS4, KRS5, KRS6. 
The samples were analyzed at MIT and LBNL and the dominant activities after a few weeks in most
samples were $^{46}Sc$ and $^{65}Zn$. The dominant long-lived
activities are $^{152}Eu$ and $^{60}Co$. The conclusions arrived from these studies is that cobalt is present at the level of approximately 30 ppm, and europium at about 1 ppm.  Note that the neutron spectrum at the reactor has essentially no flux above 5 MeV, while there is a possibility of neutrons reaching the rock with energies higher than this.  Of concern then is $^{22}$Na, with a threshold of 13 MeV.  The cross section is low, however as we will see the percentage of sodium in the rock is high.  This will place an emphasis then on designing shielding which drastically reduces the component of high-energy neutrons.

From these initial studies, a goal of neutron attenuation was set:  the 
limit on the
neutron flux out of the shielding must be smaller than 
$10^{-13} n/p/mm^{2}$ in our simulations. 

The KamLAND control room will host the target, the surrounding 
components and the shielding.  The dimensions of the cave are  
 $3.5 \times 2.25  \times 28 $ m. The critical dimension is the height 
because after placing the target system inside the cave the remaining
space to the cavern ceiling/floor ($\approx$ 50 cm) limits the neutron
shielding. The simulation studies have so far suggested that in order to 
maintain the neutron flux below $10^{-13} n/p/mm^{2}$  the room will 
have to be enlarged in the vertical direction. We have been told that excavation of the 
mine is allowed if it can be accomplished without blasting. This option will need to be
explored carefully. There is enough room at the front 
and back end of the target so that shielding material can be added if required. 

\paragraph{Simulations}
KamLAND is located under the peak of Ikenoyama (Ike Mountain,
36.42$^{\circ}$N, 137.31$^{\circ}$E). Various types of rocks are found
in Ikenoyama in unknown quantities, such as Inishi type rocks, skarn
rocks, but also granite and limestone. The Inishi type rock is
characteristic for the Japanese mountains and is made of various
oxides with a high concentration of $SiO_{2}$. The composition
of the rock is given in Table~\ref{tab:Inishi-rock}
\cite{winslow:kamland}. 
Skarns are calcium-bearing silicate rocks that are most often formed
at the contact zone between intrusions of granitic magma bodies and
carbonate sedimentary rocks such as limestone and dolostone. 
A combination of 70\% granite and 30\% limestone defines the skarn-type
rock. The specific gravity for generic skarn is 2.75 $g/cm^{3}$ and
for the Inishi rock is 2.65 $g/cm^{3}$~\cite{abe:kamland_cosmic}.

\begin{table}[!b]
\centering
\caption{\label{tab:Inishi-rock} Chemical composition of the
  Inishi-type rock in elemental percentage
  \protect\cite{tang:muon_simulations}.}
 \begin{tabular}{>{\bfseries}cc>{\bfseries}cc}
\hline\hline
 Compound & Composition ($\%$ ) &Compound & Composition ($\%$ )\\
\hline
\boldsymbol{$SiO_{2}$}  & 60.70 & CaO & 6.00 \\
\boldsymbol{$TiO_{2}$}  & 0.31 & \boldsymbol{$Na_{2}O$}& 6.42 \\
\boldsymbol{$Al_{2}O_{3}$}  & 17.39 & \boldsymbol{$K_{2}O$} & 3.47 \\
\boldsymbol{$Fe_{2}O_{3}$} &  1.10 & \boldsymbol{$P_{2}O_{5}$} & 0.18 \\
FeO & 1.22 & \boldsymbol{$H_{2}O$} & 1.27 \\
MnO & 0.15 & S & 0.01 \\
MgO  & 0.93 & \boldsymbol{$CO_{2}$} & 0.96 \\
\hline\hline
 \end{tabular}
 \end{table}

Preliminary calculations with the GEANT4 code for the Inishi-type rock
composition with an added fraction of Co (30 ppm) have shown that 
the isotopes that are produced are: $^{7}Be$, $^{46}Sc$, $^{44}Ti$,
$^{51}Cr$, $^{54}Mn$, $^{59}Fe$, $^{56}Co$, $^{57}Co$, $^{58}Co$,
$^{60}Co$ and $^{22}Na$. Addition of 1 ppm of europium to the rock composition
 confirmed that
the long-lived $^{152}$Eu and $^{154}$Eu isotopes will also be present
in detectable amounts.
Refined calculations of all the radionuclides produced and their 
activation will be carried out with the GEANT4 code using the
composition of the six rock samples. 

We will return to the rock activation calculations after a discussion of shielding materials
and a first cut at design of the target shielding.

\subsubsection{Shielding Materials Requirements}
The choice of shielding is strongly dependent on neutron energy and an 
efficient shielding must first slow down the fast neutrons by using 
appropriate neutron attenuation materials. Once the neutrons are
slowed down to thermal energies by elastic collisions, then in a 
second stage the thermal neutrons are captured by the absorbers. 
Low-Z materials containing a high fraction of hydrogen (for example
polymers, concrete, water) provide good neutron energy attenuation
(moderation) as a result of elastic scattering of neutrons on protons. At low neutron
energies low-Z materials are more efficient as the cross section
interaction is high and the energy lost in a collision is significant. 
However low density materials emit gamma rays from capture of 
thermal neutrons and therefore the gamma radiation must
be shielded against. At higher neutron energies the interaction cross
section for low-Z material is low and therefore these materials are not effective at slowing
down fast neutrons. High-Z materials with good inelastic scattering
properties such as iron (carbon steel or stainless steel) are
effective for neutrons at higher energies. The slow neutrons are 
captured by absorbers and boron has been largely used in various materials in
addition to concrete because of its high neutron capture cross section.
Therefore good shielding can be built with combinations of both high-Z and
low-Z materials and absorbers to satisfy the requirements. 
There are several factors that must be taken into account when
selecting the shielding materials. Considerations such as
effectiveness, strength, and resistance to damage can
affect radiation protection in many ways. While metals are strong and
resistant to radiation damage, they undergo changes in their
mechanical properties and degrade in time from radiation exposure. On
the other hand, concrete materials are strong, durable and cost effective 
however they are weaker at elevated temperatures and less effective at 
blocking neutrons.   

\subsubsection{Shielding Studies}
The figure of merit in our simulations was the neutron flux recorded at 90 
degrees on a detector sphere surrounding the shielding. Common 
shielding materials are usually borated polyethylene, gadolinium
polyethylene, water and concrete but also lead and steel. 
%Studies have shown that it is required a large thickness of the above shielding
%materials to keep effectively the neutron flux below the desired
%value. Therefore these materials are not optimum solutions for our
%shielding in the confined area in the Kamioka mine. 
One can construct effective shielding enclosures using these materials, 
however there are also newly-developed materials that might
prove effective in overall shielding with less thickness of material.
This would be particularly valuable for our confined environment.

In particular, the Jefferson Laboratory (JLab) has developed two new concrete
materials: a boron carbide aggregate mixture with almost 60$\%$ by weight of boron, 
a very light concrete in which most of the aggregate is shredded plastic.  The latter
is particularly effective for moderating neutrons, while the boron-rich mixture is a good
absorber of the thermalized neutrons.

Studies using an inner layer of plastic concrete as a neutron moderator 
and an outer layer of boron rich concrete as a neutron absorber for the 
available 50 cm from the target system to the cavern ceiling/floor 
showed that the neutron flux was still an order of magnitude higher
than the desired value. 

\begin{figure}[tbp] 
\centering
\includegraphics[width=0.5\textwidth]{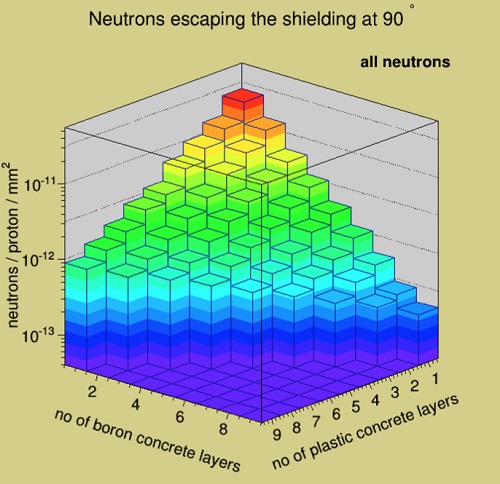}
\caption{Neutron flux at 90 degrees for 100 cm shielding. The
  shielding consists of various layers of plastic concrete and boron
  rich concrete. Each layer is 10 cm thick. The neutron flux at 90
  degrees is above the desired value.}
\label{fig:002}
\end{figure}

\begin{figure}[tbp] 
\centering
\includegraphics[width=1.0\textwidth]{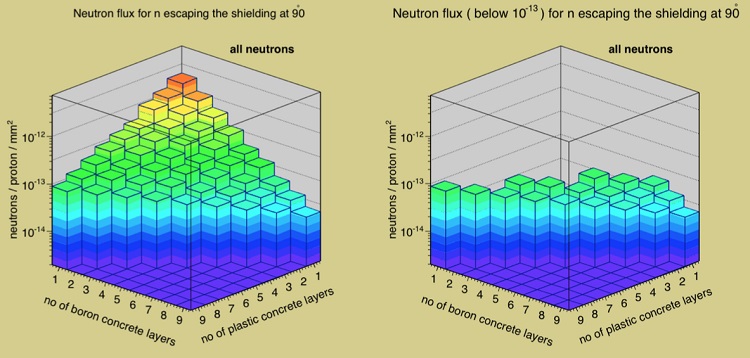}
\caption{Neutron flux at 90 degrees for 120 cm shielding. The
  shielding consists of various layers of plastic concrete and boron
  rich concrete (with each layer being 10 cm thick) and 20 cm
  steel. The neutron flux is below the desired value for several
  material combinations.}
\label{fig:003}
\end{figure}

In the next step a target shielding of various combinations of plastic
concrete and boron rich concrete of total thickness of 100 cm was
considered.  The neutron flux recorded on a detector sphere at 90 
degrees was still above $10^{-13} n/p/mm^{2}$ (fig.~\ref{fig:002}).  
Fast neutrons can still penetrate the 100 cm thick material combination
suggesting that a high-Z material to stop them is required. The inelastic 
scattering on high-Z atoms will reduce the neutron energy to a much lower 
value such that they can be absorbed in the boron rich concrete layer. 
Therefore in addition to the 100 cm shielding thickness of combinations 
of materials developed at JLab, 20 cm of iron (stainless steel) was 
added. In this configuration the inner layer is iron in order to slow down high
energy neutrons to much lower energies, followed by a variable thickness of
plastic concrete to reduce the energy of the low energy neutrons. The outer 
layer is boron rich concrete which absorbs these neutrons. The results 
of the simulations can be seen in fig.~\ref{fig:003}. The neutron flux 
is recorded on a detector sphere of radius 350 cm for various JLab material 
combinations placed on top of the 20 cm inner iron layer. There are several material 
combinations for which the neutron flux is lower than $10^{-13}
n/p/mm^{2}$ in these studies (right plot). The flux is lowest for the combination 10
cm plastic concrete, 90 cm boron rich concrete. 

\begin{figure}[tbp] 
\centering
\includegraphics[width=1.0\textwidth]{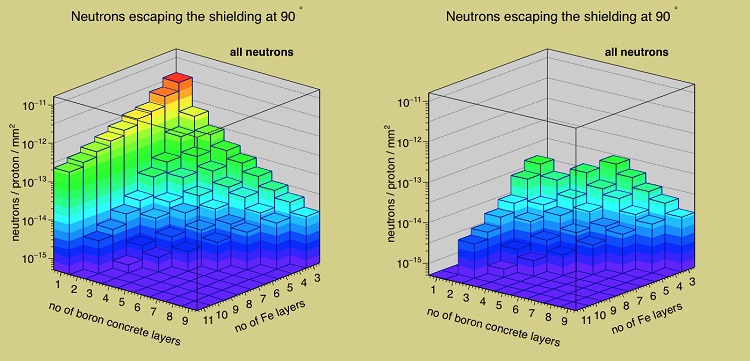}
\caption{Neutron flux at 90 degrees for 120 cm shielding. The
  shielding consists of various layers of iron (stainless steel) and boron
  rich concrete (with each layer being 10 cm thick). The neutron flux
  is below the desired value for several material combinations.}\label{fig:004}
\end{figure}

\begin{figure}[tp!] 
\centering
\includegraphics[width=0.7\textwidth]{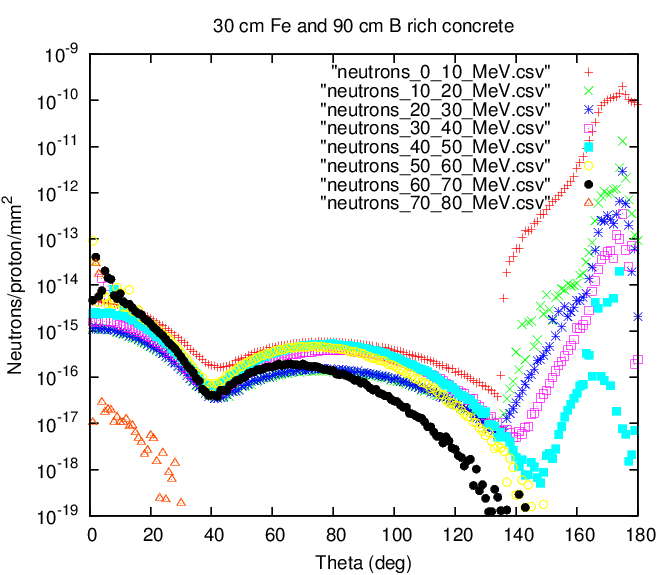}
\caption{The neutron flux detected on a sphere surrounding the
  shielding. The sphere radius is 350 cm. The shielding material
  consists of 30 cm Fe and 90 cm boron rich concrete.}
\label{fig:005}
\end{figure}

However the fact that the best solution was for a minimum thickness 
of plastic concrete suggested that better results could be obtained for 
combinations of steel and boron rich concrete only. Assuming the same 
total shielding thickness of 120 cm, the results are shown in
fig.~\ref{fig:004}. Since steel poses engineering issues on the
reflector and FLiBe sleeve due to its high density, the combination 30
cm steel and 90 cm boron rich concrete was chosen and the total mass
of the target system and shielding becomes 165,331 kg. The neutron
flux for this shielding configuration recorded on the detector sphere
is shown in fig.~\ref{fig:005}. The lower flux values at 40 and 140
degrees correspond to the corners of the concrete shielding block and
the higher values of flux above 140 degrees correspond to neutrons
escaping into the space left for the wobbler magnets in front of the
target.   New GEANT4 calculations with this new configuration have
not yet been performed.The total neutron flux at 90 degrees for all energies is
1.88 $\times 10^{-15} n/p/mm^{2}$.  (Note, the upstream configuration of the beam line, shown 
in Figure \ref{30-degree} was developed in large measure from these results,
showing that attenuation of the back-scattered neutrons was extremely
important.)

\begin{figure}[tbp] 
\centering
\includegraphics[width=1.0\textwidth]{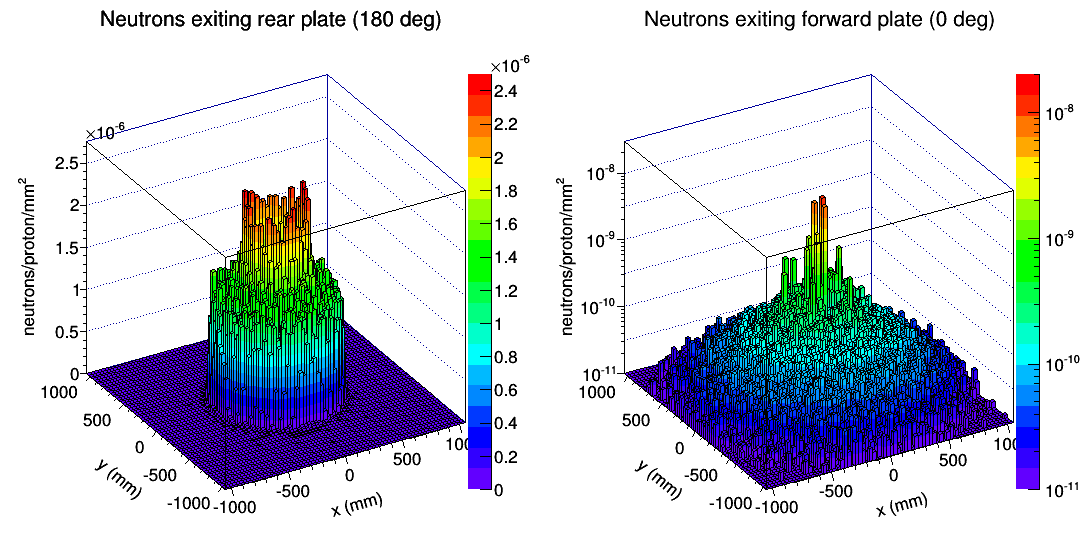}
\caption{flux for 3 Fe and 9 B concrete at 0 and 180 deg.}
\label{fig:007}
\end{figure}

\begin{figure}[tbp] 
\centering
\includegraphics[width=0.8\textwidth]{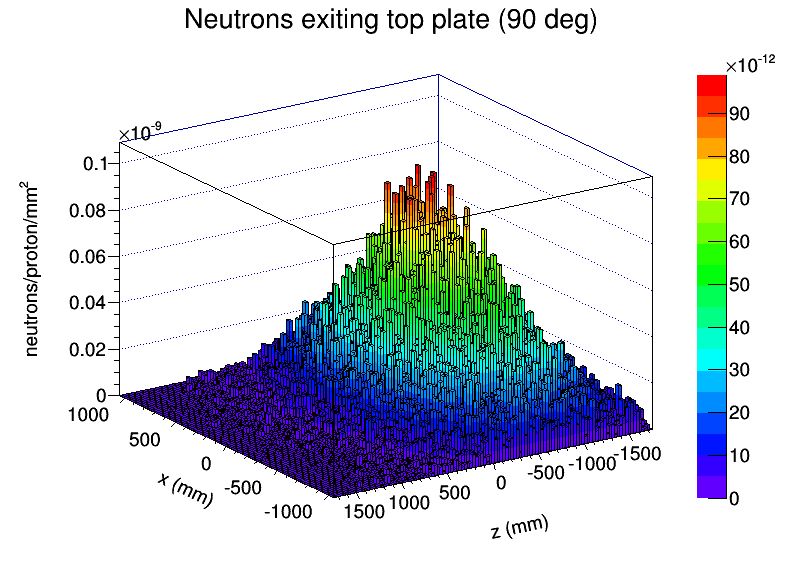}
\caption{flux for 3 Fe and 9 B concrete at 90 deg.}
\label{fig:008}
\end{figure}

For a more accurate neutron flux determination at the surface of the rock,
detector plates were placed on the
shielding block to record the flux at 0, 90 and 180 degrees for the
same shielding configuration. The neutron flux shown detected at the
front and at the back of the target is shown in fig.~\ref{fig:007}.
The neutron contamination in the proton beam pipe is 2.4$\times
10^{-6} n/p/mm^{2}$ while the flux in the space for magnets has a peak
value of 1.4$\times 10^{-6} n/p/mm^{2}$. The flux shown in
fig.~\ref{fig:008} at 90 degrees is forward biased as most of the
neutrons that are backscattered escape in the space left for the
wobbler magnets and therefore have no chance to be scattered in the
sleeve and detected on the first half of the plate. The total neutron
flux is 4$\times 10^{-11} n/p/mm^{2}$.

\clearpage
\subsubsection{Rock activation analysis}

\begin{figure}[t]
   \centering
   \includegraphics*[height=80mm, width=146mm]{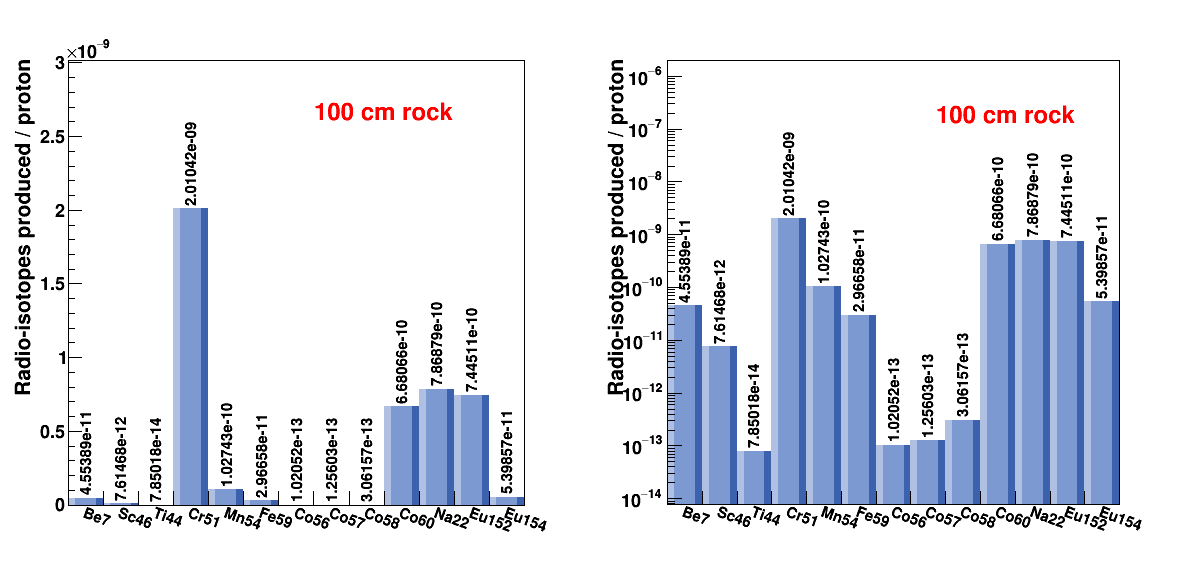} 
   \caption{The isotopes production rate in the   rock on
     both linear and logarithmic scale. The thickness of the rock is
     100 cm.}
   \label{fig:isotope-rate}
\end{figure}

Refined calculations of all the radionuclides produced in the rock and the 
induced activation were carried out with the GEANT4 code. We have used
the target layout described in the previous section. 
From the geometry standpoint we considered for a start a layer of rock 
which extended 1000 mm outside the shielding block in both $X$ 
and $Z$ directions and had 1000 mm thickness. The exact dimensions 
of the rock layer used in our studies are: $ 5512.7 \times 1000 \times 6912.7$
mm. The rock composition was given in Table~\ref{tab:Inishi-rock} 
~\cite{winslow:kamland}. Studies from the MIT reactor indicated that cobalt
and europium are other sources of long-lived isotopes therefore these
elements were added to the chemical composition of the rock shown 
in Table~\ref{tab:Inishi-rock}. The concentration given by reactor 
studies for Co is 30 ppm and for Eu is 1ppm with both concentrations 
being percentages by weight.  

In order to study the radioactive inventory originated from rock
activation, the simulation code was designed to register the production
of any radioactive isotope that decays emitting gamma rays and has a
life time longer than 24 hours. Shorter lived isotopes can pose a
safety concern only if the target area needs to be accessed in emergency
situations, therefore we limited our study to include only isotopes that
live longer than 24 hours. The isotopes that are recorded in our simulations are:  
$^{7}$Be, $^{46}$Sc, $^{44}$Ti, $^{51}$Cr, $^{54}$Mn, $^{59}$Fe, 
$^{56}$Co, $^{57}$Co, $^{58}$Co, $^{60}$Co, $^{65}$Zn, $^{75}$Se, 
$^{84}$Rb, $^{85}$Sr, $^{88}$Y, $^{95}$Zr, $^{94}$Nb, $^{95}$Nb, 
$^{106}$Ru, $^{109}$Cd, $^{111}$In, $^{113}$Sn, $^{125}$Sn, $^{124}$Sb, 
$^{125}$Sb, $^{125}$I, $^{132}$Cs, $^{134}$Cs, $^{137}$Cs, $^{133}$Ba, 
$^{139}$Ce, $^{141}$Ce, $^{144}$Ce, $^{152}$Eu, $^{154}$Eu, 
$^{153}$Gd, $^{160}$Tb, $^{161}$Tb, $^{170}$Tm, $^{169}$Yb, 
$^{172}$Hf, $^{182}$Ta, $^{185}$Os, $^{192}$Ir, $^{198}$Au, 
$^{199}$Au, $^{203}$Hg, $^{210}$Pb, $^{207}$Bi, $^{228}$Th, 
$^{239}$Np, $^{241}$Am, $^{243}$Am. Reactor studies indicated that 
some rock samples from various locations in the mine contain a 
large fraction of $^{22}Na$, so
it was added to this list. 

Simulations have shown that the isotopes that are produced in the
layer of rock described above are: $^{7}Be$, $^{46}Sc$, $^{44}Ti$,
$^{51}Cr$, $^{54}Mn$, $^{59}Fe$, $^{56}Co$, $^{57}Co$, $^{58}Co$, $^{60}Co$,
$^{22}Na$, $^{152}Eu$, $^{154}Eu$. The rate of the isotope production
is shown in Fig.~\ref{fig:isotope-rate}. Although $^{51}Cr$ is
produced in a significant amount, its half life is only 27 days
(Table~\ref{tab:half-life}) while other isotopes produced
at a lower rate ($^{44}Ti$, $^{60}Co$, $^{22}Na$, $^{152}Eu$ and 
$^{154}Eu$) pose a safety concern as their half life is of the order
of several years.

\begin{table}[b]
\centering
\caption{\label{tab:half-life} Half life of the isotopes produced in
  the rock.}
 \begin{tabular}{|>{\bfseries}c|c|>{\bfseries}c|c|}
\hline\hline
% Isotope & $^{7}Be$ &$^{46}Sc$ & $^{44}Ti$ & $^{51}Cr$&
%$^{54}Mn$&$^{59}Fe$&$^{56}Co$&$^{57}Co$&$^{58}Co$&$^{60}Co$&$^{22}Na$&$^{152}Eu$&$^{154}Eu$\\
%\hline
 %Half life &53.12 d& 83.79 d&63 y&27.70 d&312.3 d&44.50 d&77.27 d&271.79 d&70.86 d&5.2 y&2.60 y&13.5 y&8.6 y\\
%\hline
Isotope& half life & Isotope& half life\\
\hline
\boldsymbol{$^{7}Be$}  & 53.12 d &\boldsymbol{$^{57}Co$}  & 271.79 d\\
\boldsymbol{$^{46}Sc$}  & 83.79 d & \boldsymbol{$^{58}Co$}  & 70.86 d\\
\boldsymbol{$^{44}Ti$}  & 63 y & \boldsymbol{$^{60}Co$}  & 5.2 y\\
\boldsymbol{$^{51}Cr$}  & 27.70 d & \boldsymbol{$^{22}Na$}  & 2.60 y \\
\boldsymbol{$^{54}Mn$}  & 312.3 d & \boldsymbol{$^{152}Eu$}  & 13.5 y\\
\boldsymbol{$^{59}Fe$}  & 44.50 d & \boldsymbol{$^{154}Eu$}  & 8.6 y\\
\boldsymbol{$^{56}Co$}  & 77.27 d & & \\
\hline\hline
 \end{tabular}
\setlength{\extrarowheight}{20pt}%
 \end{table}

\begin{figure}[t]
   \centering
   \includegraphics*[height=80mm, width=98mm]{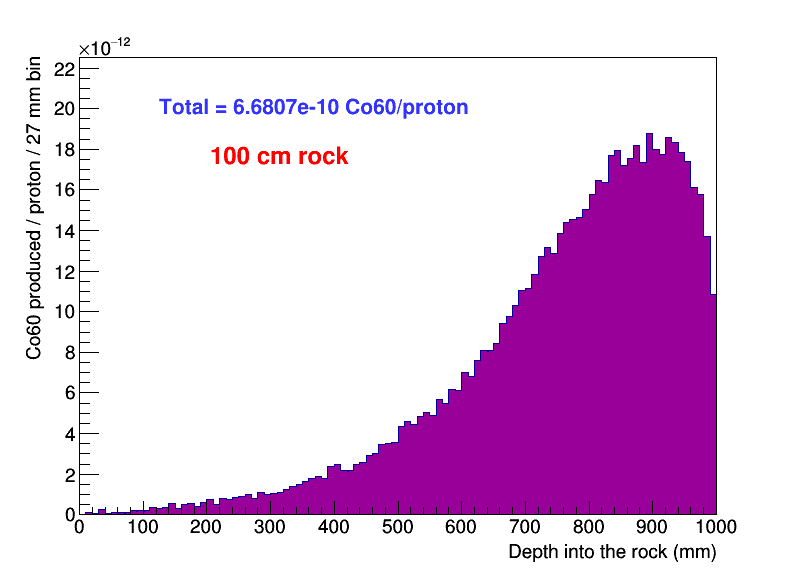} 
   \caption{The distribution of the production of $^{60}Co$ inside the layer of
     rock as a function of rock thickness. The total isotope
     production is 6.68 x 10$^{-10}$  $^{60}Co$ per proton.}
   \label{fig:Co60_vs_Z}
\end{figure}

The distribution of isotope production inside the layer of rock is not
uniform but varies with the depth inside the rock. Most of the
isotopes are produced in the proximity of the shielding box, mainly in
the first 20-30 cm, then the production drops to much lower values
at 100 cm depth inside the rock. Figure~\ref{fig:Co60_vs_Z} shows the
production of $^{60}Co$ as a function of depth and similar plots were
obtained for other isotopes. 

%%%%%%%%%%%%%%%%%%%%%%%  %%%%%%%%%%%%%%%%%%%%%%%%%%%%%%%%%%%%%

The isotope production is also non-uniform in the horizontal plane
with most of the isotopes produced in the region corresponding to the
backside of the target where the neutron flux is forward biased as
fig.~\ref{fig:008} has shown. In order to study the isotope production
distribution in the $XY$ plane after 5 years run, the 100 cm of rock 
were divided into 20 layers of rock of 5 cm thickness. Since the layer 
of rock considered in our studies is extending 100 cm on all sides of the 
shielding box, a significant amount of isotopes are produced on the 
lateral sides of the shielding and at the back end where the water pipes are
located. Also, a very large fraction of isotopes are produced at the
front end of the target where a space was left for the wobbler
magnets.  Again, the shielding geometry has changed, new calculations
will show markedly less activity in the forward direction.

\begin{figure}[t]
   \centering
   \includegraphics*[height=80mm, width=150mm]{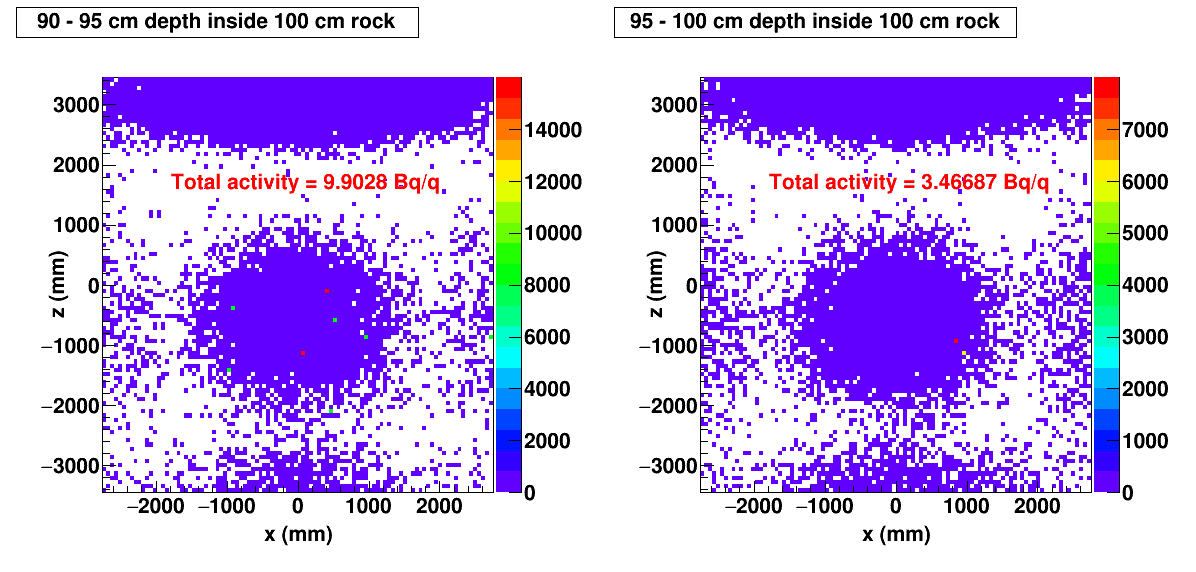} 
   \caption{The distribution of the production of all isotopes inside
     one of the slices of the rock. This (horizontal) slice is sitting above the shielding, the beam is entering from the top, so the very intense region represents backscattered neutrons escaping through the hole for beam in the shielding.  Revised shielding configuration will drastically reduce this activation.  The circular part is directly above the target, while the weaker downstream component represents neutrons escaping from the back of the target, along the water-cooling lines.}
   \label{fig:all_iso_slice}
\end{figure}

\begin{figure}[p]
   \centering
   \includegraphics*[height=170mm, width=150mm]{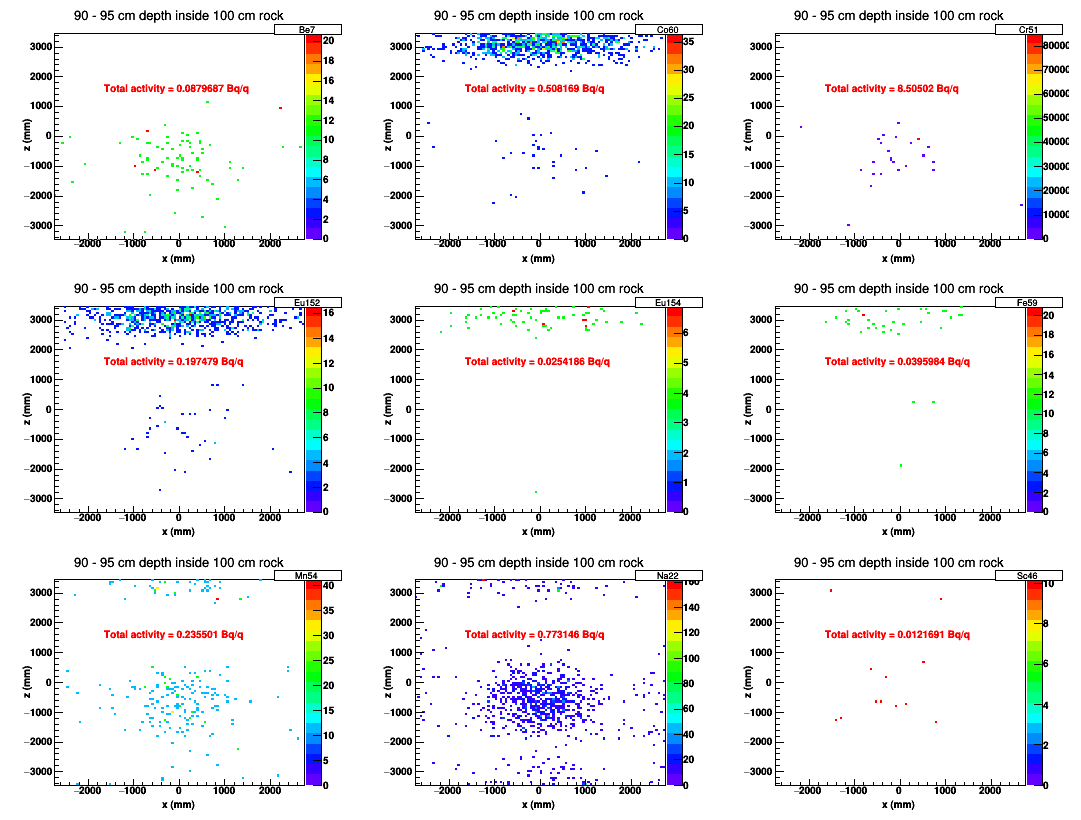} 
   \caption{The distribution of the production of individual isotopes (label in upper right box) inside
     one of the slices of the rock.}
   \label{fig:thirteen_iso_slice}
\vspace{1 cm}
\end{figure}

Figure~\ref{fig:all_iso_slice} shows the distribution of the isotope
production in the horizontal plane in the slices closer to the
shielding box where the isotope production reaches a peak.   
Figure~\ref{fig:thirteen_iso_slice} shows the individual 
production in slice 90-95 cm for most of the isotopes. The
contribution of $^{44}Ti$, $^{56}Co$, $^{57}Co$, and $^{58}Co$ is low
and it is not shown here.

%%%%%%%%%%%%%%%%%%%%%%%  ok!  %%%%%%%%%%%%%%%%%%%%%%%%%%%%%%%%%%%%%

The simulation gives the number of every isotope $i$ that is produced per
incident proton ($N_{iso}/N_{p}$). From the standpoint of isotope production
and consequently induced activity, we distinguish two different
periods of time. During the run period, the rate at which each isotope 
is produced depends on both the production and decay rate 
(Eq.~\ref{eq:beamon}) and the solution of this equation is
given by Eq.~\ref{eq:beamon_solution}. 

\begin{equation}
\frac{dN_{i}}{dt}(t)=\frac{N_{iso}I}{N_{p}e} - \lambda_{i}N_{i}(t)
\label{eq:beamon}
\end{equation}

\begin{equation}
N_{i}(t) = \frac{N_{iso}I}{N_{p}e\lambda_{i}} (1 - exp(-\lambda_{i}t))
\label{eq:beamon_solution}
\end{equation}

After the beam is switched off, the amount of which each isotope
$N_{i}$ is present in the rock depends only on the decay rate of that particular isotope. Considering 
the time $t_{1}$ when the beam is switched on and the time
$t_{2}$ the period since the beam was switched off, the amount of each
isotope $N_{i}$ is given by Eq.~\ref{eq:beamoff}. 

\begin{equation}
N_{i}(t) = \frac{N_{iso}I}{N_{p}e\lambda_{i}} (1 - exp(-\lambda_{i}t_{1})) exp(-\lambda_{i} t_{2})
\label{eq:beamoff}
\end{equation}

Each isotope will give a different induced rock activity depending on
its decay constant (Eq.~\ref{eq:oneisotope_activity}) and the total
activity in the rock during the beam on and the beam off period is
given by the summation of the activities of all isotopes produced 
(Eq.~\ref{eq:allisotopes_activity}). 

\begin{equation}
A_{i}(t) = \lambda_{i} N_{i} (t)
\label{eq:oneisotope_activity}
\end{equation}

\begin{equation}
A(t) = \sum_{i}{A_{i}(t)}
\label{eq:allisotopes_activity}
\end{equation}

\begin{figure}[!t]
   \centering
   \includegraphics*[height=73mm, width=90mm]{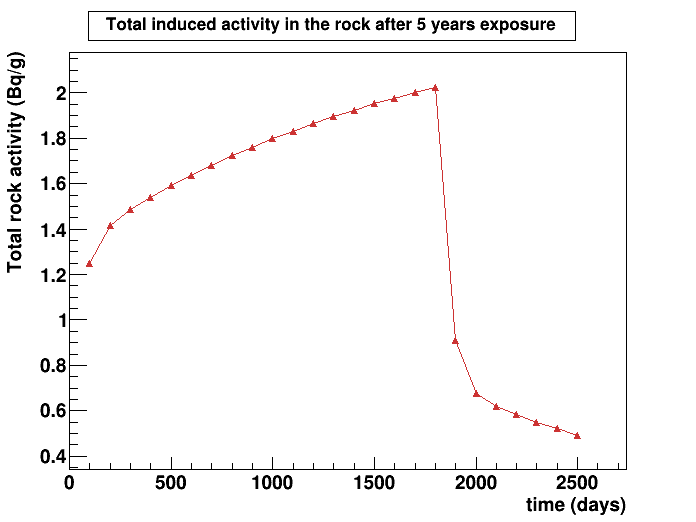} 
   \caption{Total induced activity inside the rock after five
     years exposure. The total activity assumes a uniform isotope
     production inside the 100 cm thick rock.}
   \label{fig:activity}
\end{figure}

The total induced activity in the rock for the five years run period
and after beam switch off is shown in Fig.~\ref{fig:activity}. 
The activity peak value after five years run assuming a uniform
isotope distribution inside the layer rock with dimensions $ 5512.7
\times 1000 \times 6912.7$ mm is 2 Bq/g. As discussed previously, a major
contribution to the rock activity comes from the region where the space for
magnet is left as all the backscattered neutrons are escaping and hit
the cavern ceiling. With the revised upstream shielding configuration only a fraction
of these will reach the cavern ceiling.

%%%%%%%%%%%%%%%%%%%%%%%  ok so far!  %%%%%%%%%%%%%%%%%%%%%%%%%%%%%%%%%%%%%
\subsubsection{Current Status and Future Work}
The GEANT4 tools have been well benchmarked now, and are believed to 
be reliable predictors of shielding efficiency and isotope production.
It remains to refine the shielding configurations, both in terms of materials
and thicknesses, as well as ensuring that ``holes" upstream and downstream
are adequately plugged.  The end result must be a configuration
that meets the 0.1 Bq/gm requirement.

We have also not yet begun serious assessments of shielding requirements in the cyclotron
vault and along the beam line to the target.  These are more difficult to assess, because
the source terms are not certain.  They are dependent on beam loss, which we are
committed to keep as low as possible.  The usual procedure is to plan shielding for the 
maximum likely exposure to personnel that might occur.  However, in our case the sensitive
element is the rock walls, not personnel (that will be excluded from the area during operation
in any event).  What is important in rock activation is not peak exposure but the integrated dose
over the lifetime of the experiment.  Developing the methods and criteria for
shielding design for these areas will be an interesting project.

\subsubsection{Risk assessment and Risk Mitigation}

\textbf{Risk:  Inaccuracies in GEANT4 calculations}

\emph{Mitigation: Further benchmarking studies, and allowing adequate
margins of safety in case of underestimates in isotope production}

\textbf{Risk:  Rock composition different from expected}

\emph{Mitigation:  In May 2015 a more extensive set of rock samples were collected, 
with specific emphasis on the target and the cyclotron vaults.  These will
be analyzed through reactor neutron activation, and possibly exposure to fast neutrons if a suitable
source can be identified.}

\clearpage
\subsection{Acquiring and Handling FLiBe \label{sec:flibe}}

A key element of our neutrino source is the sleeve consisting of a lithium fluoride-beryllium fluoride (LiF-BeF$_2$) salt mixture, often called FLiBe.    Traditionally, interest in FLiBe has been associated with molten-salt reactors, which were developed in the 1950s and 1960s.  A great deal of operational and practical information about FLiBe was accumulated during these efforts.  Since 2011 a private company called Flibe Energy has been updating this reactor technology and plans to produce reactor-grade FLiBe, which has been unavailable for many years.  We are collaborating with this company on this project.

Their development plan complements our own needs very well, since they intend to resolve the two challenges associated with the use of FLiBe.  The first is the need for isotopic separation of lithium in order to generate the highly-depleted lithium (HDLi) needed both for their reactors and for our experiment.  The second is the hazard associated with the beryllium content in the salt.  Because of the need for isotopic separation, lithium dominates the cost of the salt, but after production beryllium dominates the handling considerations of the salt.  Flibe Energy is also supplying expert advice on issue of understanding hazardous by-products that are produced during the run.

\begin{figure}[t]
\centering
\includegraphics[width=4.5in]{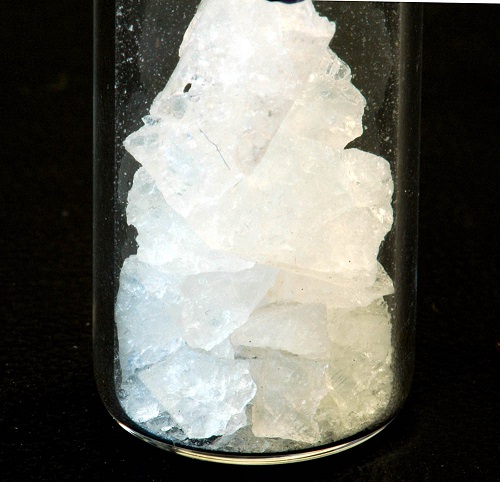}
\caption{\footnotesize Photo of a beaker of purified FLiBe salt crystals (credit: B. C. Kelleher, published in Creative Commons.)
\label{FLiBe}}
\end{figure}

Isotopic separation of lithium was originally developed in the 1950s in order to produce enriched lithium in support of the thermonuclear weapons program.  The rejected stream of lithium from this process was only depleted to about 1\% lithium-6 and large quantities were recently sold to commercial customers.  It was too impure to be considered for our experiment.  Much smaller quantities of lithium depleted to approximately 1000 ppm lithium-6 was also produced in special runs in order to support the Molten-Salt Reactor Program.  A small amount of this material remains although some of it has been transferred to the Czech Republic.  Greater purity than 1000 ppm would be greatly desired for this experiment.  The legacy separation technique to produce all these material streams used mercury which leaked into the surrounding environment and has led to a large cleanup problem, thus the traditional lithium separation technique can no longer be feasibly entertained as an option for future needs.

Lithium depleted to 1000 ppm lithium-6 is used in hydroxide form as a chemical buffer in existing pressurized-water nuclear reactors, but this lithium comes from sources in Russia and China and is likely produced via the mercury-based process.  The Department of Energy is concerned about this vulnerability to a foreign supply of depleted lithium \cite{reister:lithium}, but has not initiated any efforts to develop a
new and environmentally acceptable technique for lithium isotope
separation. A review of the possible methods is given in 
Ref.~\cite{ault:lithium}.

Flibe Energy is developing a new lithium separation technique in cooperation with the University of Utah that appears very promising and does not have any undesirable environmental side effects.  They have recently joined our development team and will produce depleted lithium according to our specifications.  They also have a close relationship with Materion, the manufacturer of all US-origin beryllium materials, and anticipate formulating the FLiBe needed for the experiment and developing the safe handling guidelines for this unique material.

Lastly, the experts at Flibe Energy have a thorough knowledge of the hazardous by-products associated with using FLiBe.  They are advising us so that the design can address these issues.  The rate of tritium production is well understood and methods to remove the tritium gas are already included in the design.  We are now in the process of working with Flibe Energy to identify and mitigate any other hazardous materials that will or potentially may be produced within the target.

\subsubsection{Risk Assessment \&  Risk Mitigation}

\textbf{Risk: Volatility of the FLiBe Market }   The primary element that has great fluctuations in availability is the $^7$Li.

\emph{Mitigation: Work with a US-based company to acquire the FLiBe}

\textbf{Risk: Handling the FLiBe}  The primary issue is handling the beryllium.

\emph{Mitigation: We will ship an approved container to Flibe Energy.  As an experienced company, they can safely handle the material.}

\textbf{Risk: Hazardous By-products}  Hazardous by-products must be properly handled and removed.

\emph{Mitigation: We are working with Flibe Energy to understand all potential chemical hazards and to mitigate them.}

\subsection{Preliminary Mechanical Design of a Target Prototype}

\begin{figure}[!t]
\centering
\includegraphics[width=5.5in]{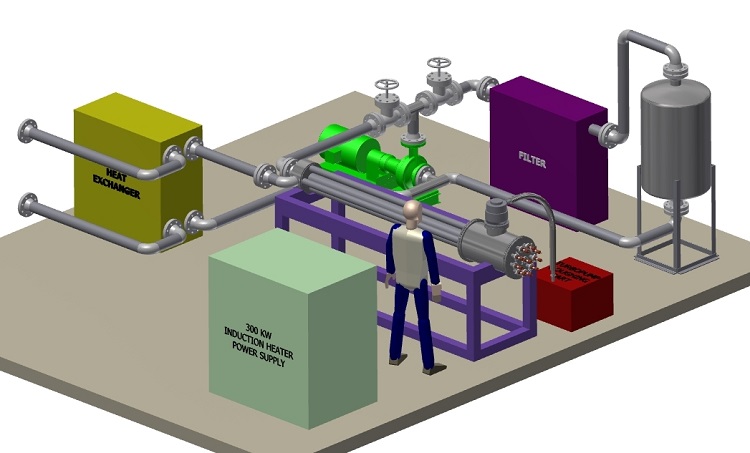}
\caption{\footnotesize A view of the target prototype test stand 
\label{targprot1}}
\end{figure}

\begin{figure}[!h]
\centering
\includegraphics[width=5.5in]{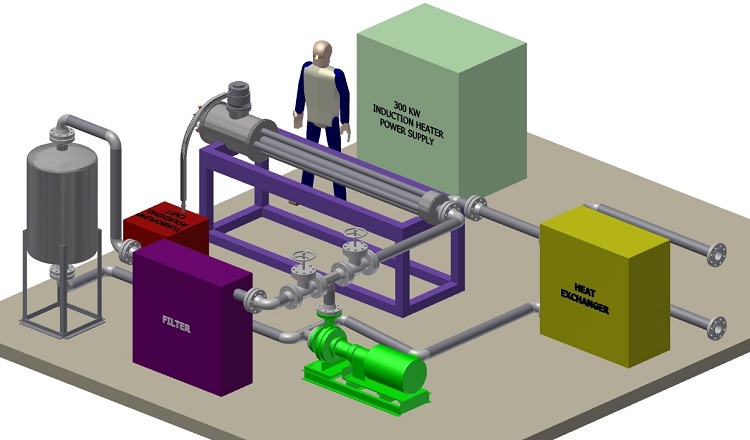}
\caption{\footnotesize Another view of the target prototype test stand 
\label{targprot2}}
\end{figure}

A conceptual design for the IsoDAR target prototype test stand is shown in Fig.~\ref{targprot1} and Fig.~\ref{targprot2}.  The prototype is intended to be a full scale test of the ability of the beryllium target vessel to remove the $\geq$300 kW of heat deposited in the front face of the vessel.  The reason for running the test at full scale is because of a lack of a simulant fluid that is scalable to full scale in thermal and mechanical properties.  One method of heating the target is to use induction coils and an induction coil power supply.  Other methods are also being explored, such as electron beam heating.  The filtration system and water reservoir are in a parallel loop running at a lower flow rate with the pump that feeds the higher flow target cooling loop.  The test stand will be able to vary the static pressure in the system, the velocity of the water striking the inner face of the beryllium vessel, and the inlet temperature of the water.  The goal is to finalize the parameters of the cooling system and verify the CFD simulation that happens before the prototype test.  Erosion of the beryllium will also be studied to estimate the working life of the target torpedo.

\begin{figure}[h]
\centering
\includegraphics[width=5.5in]{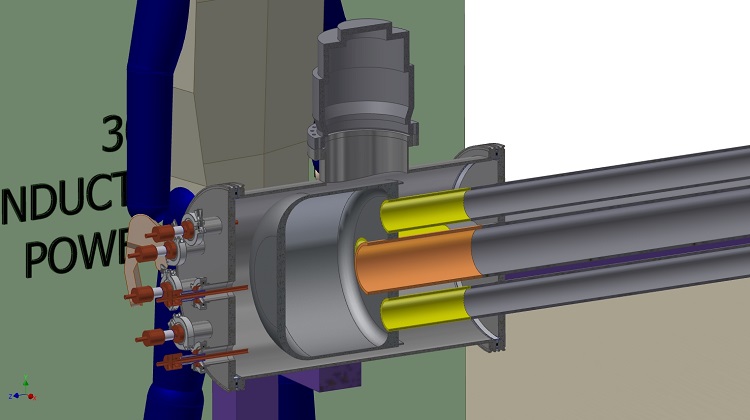}
\caption{\footnotesize Close-up section view of the beryllium vessel inside a vacuum chamber
\label{targprot3}}
\end{figure}

The beryllium vessel is enclosed in a vacuum chamber as shown in Fig.~\ref{targprot3}.  The actual target vessel is in vacuum and this prototype vacuum vessel serves several roles.  It creates the possibility of measuring heat transfer to the rest of the target through thermal radiation from the surface of the target facing the beam.  It protects the beryllium from any elevated temperature corrosion that would happen in air.  Finally, the vacuum can serves as secondary containment of the beryllium in the event of a beryllium vessel failure.  The feedthroughs shown in the figure are rated for 35 kW at 20 kHz and are commercially available for induction furnaces.  Coil design will be done by a company specializing in induction coils such that only the surface exposed to the beam sees as close an approximation as achievable to the  level and depth of energy deposition created by the proton beam.

Commercial sources for items such as the heat exchanger, pump and filtration system are being identified with the help of water system engineers from Fermilab to ensure that the components satisfy the radiation containment requirements.

%\clearpage
\section{Controls}
Controls are, without a question, one of the most important aspects 
of a complicated machine like the IsoDAR neutrino source. 
However, the concepts and hardware for successful control systems
are readily available and thus we do not consider a detailed treatment
of the control systems part of the Conceptual Design Report (CDR). 
Only a coarse structure shall be provided with references to 
subsections above in which some of the controls of individual 
components have been discussed already.
In general, our approach to controls will be that of an integrated 
plan covering all aspects from the ion source through the target,
including data acquisition and diagnostics (where applicable) as 
well as safety and interlocks.

In the cost appendix, we will give a first estimate of cost that 
follows the structure of subsections in this section.

Both, the details of the control system and the costs, will be 
refined by the time of the Preliminary Design Report (PDR).

\subsection{High Level Controls}
The high level controls include the main computer systems, 
data storage, and networking and communication. As a possible 
candidate for the high level control system, we are considering 
EPICS.

\subsection{Low Level Controls}
This is the hardware oriented side of the control system, 
divided into subsystems corresponding to the different parts of 
the machine. In the case of diagnostics this also includes beam 
diagnostic hardware like Faraday cups, emittance scanners, probes,
and other devices. The low level controls include everything
from the Programmable Logical Controllers (PLC's) down to the individual 
sensors and devices.

\subsubsection{Safety and Interlocks}
This includes beam loss monitoring (radiation monitors), machine 
protection, and interlock systems.

\subsubsection{Interface Controls}
This includes the controls for the systems described in 
\secref{sec:interface} (cooling water system and electrical power
distribution) that function as interfaces to the Conventional Facilities.

\subsubsection{Front End Controls}
This includes the ion source and the low energy beam transport (LEBT) 
controls, data acquisition, and diagnostics.
See also \secref{sec:frontend}.

\subsubsection{Cyclotron Controls}
This includes the central region, RF, general cyclotron, and
extraction system controls, data acquisition, and diagnostics.
See also \secref{sec:cyclotron}.

\subsubsection{MEBT Controls}
This includes the Medium Energy Beam Transport (MEBT) controls, 
data acquisition and diagnostics. See also \secref{sec:mebt}

\subsubsection{Target Controls}
This includes target controls, data acquisition and diagnostics.

\clearpage
\section{Interface to Conventional Facilities\label{sec:interface}}
As will be discussed in more detail in \secref{sec:conventional},
the conventional facilities will be the topic of a separate CDR,
because they need to be developed in close collaboration with the 
host site.
However, there are several subsystems that are part of the technical 
design presented here, but are also directly connected to the 
conventional facilities. These are listed in the cost-appendix and 
consequently need to be part of this CDR and the work breakdown structure
as well. We call these subsystems the ``Interface to the Conventional Facilities''
and discuss them briefly here.

\subsection{Power Distribution}
This interface to the provided main power line includes several cabinets 
for breaking out the power to the separate subsystems like ion source,
LEBT, cyclotron main magnet, cyclotron RF, MEBT, target, etc. 
The cabinets will contain breakers, distribution systems, and
interlocks for machine and personnel protection. As these systems are 
well-understood and straight-forward, we will postpone a detailed 
discussion to the Preliminary Design Report (PDR).

\subsection{Cooling Water System}
The rudiments of the cooling water system are discussed in 
\secref{water} and a schematic is shown in \figref{fig:coolingschem}.
Bringing the cooling water of the secondary loop from the underground stream
to the cyclotron vault will be part of the separate CDR. The interface 
between the subsystems in need of cooling and the secondary loop
will be the primary cooling loops and the heat exchangers.

\subsection{Additional Interfaces}
Depending on the results of ongoing and proposed studies in collaboration
with the host site, we may in the future identify other interface systems 
necessary to build the bridge between the technical facility and the host 
site. These could be in the areas of, e.g., 
``Radiation Protection'' and ``Ventilation''.

% All below was moved to chapter 5, some of it was re-written.
%\input{Ch4.5-radprot.tex}
%\clearpage
%\input{Ch4.6-civil_construction_installation.tex}
%\clearpage
%\input{Ch4.7-utilities.tex}
%\clearpage
%\input{Ch4.8-operation.tex}
%\clearpage
%\input{Ch4.9-decom.tex}

\clearpage 
\chapter{Conventional Facilities \label{sec:conventional}}

The topic of this Conceptual Design Report is the Technical Facilities
for isoDAR$@$KamLAND, and it is beyond the scope of this
document to provide the design details for the Conventional
Facilities.  This will be the subject of a separate CDR.    

With this
said, the Conventional Facilities are rather unconventional in the
world of particle physics. Also,  design choices for the Technical
Facility have been driven by
aspects of the Conventional Facilities. Similarly, vice versa:
design choices for the Conventional Facilities are linked ot very
specific needs of the Technical Facility.   For these three reasons,
we provide an overview of the Conventional Facilities issues here to provide
the necessary context.   Specifically we discuss installation issues,
the required utilities, and radiation protections issues.

\section{Space Constraints and Civil Construction}

The designs of the subcomponents of the technical facility are
strongly affected by the constraints of running in the underground
environment.    The purpose of this section is to explain how we have
informed ourselves of the space constraints and how we plan to address civil
construction.    

\begin{figure}[t]
\centering
\includegraphics[width=4.5in]{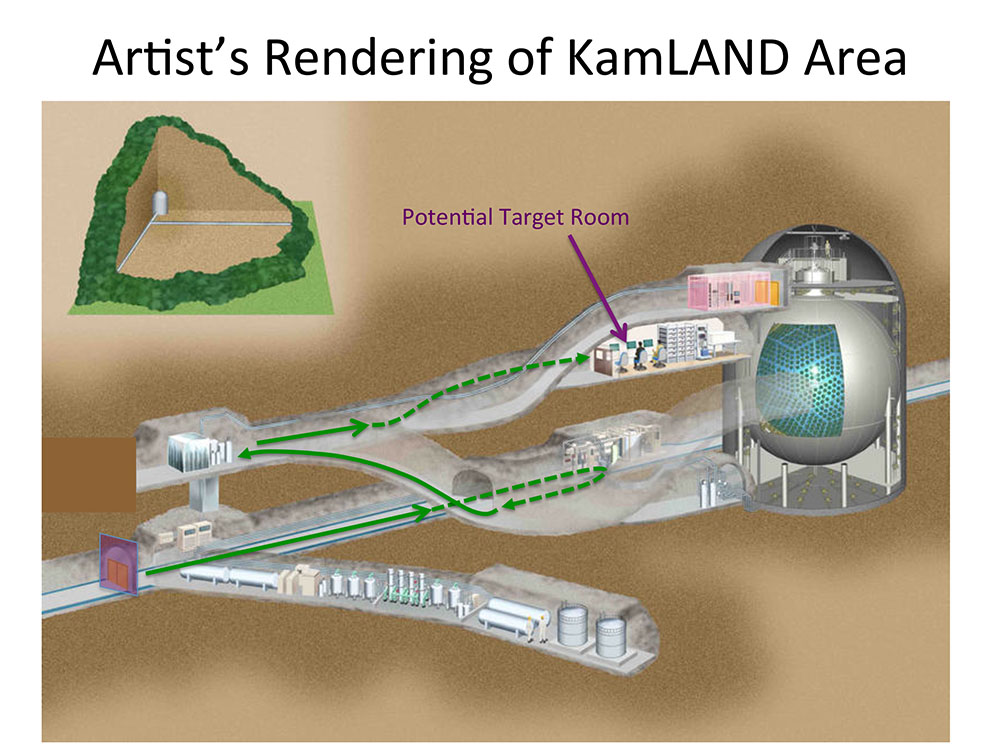}
\caption{\footnotesize An artist's rendering of the underground area around the KamLAND detector.
\label{artist01}}
\end{figure}

\subsection{Laser Mapping of the Present KamLAND Space}

In order to inform ourselves of the space constraints,  we employed
extensive laser mapping of the present KamLAND Space.   The data set is very rich and allows us to
investigate important questions.
One example is to study the question of
whether the rock underneath the cyclotron, which is above a lower
drift, is strong enough to support the cyclotron.  Another 
is the maximum size objects can be and still be moved through the
KamLAND-region tunnel system.  The success of this approach may lead
us to augment the data set in the future, as only the tunnels around the
detector and not the detector hall itself were scanned. 

The mining company responsible for the underground areas of KamLAND provided 2D AutoCAD drawings of the tunnels around the detector.  These drawings were initially used to estimate the location of the cyclotron and target in the tunnels around the detector, and the access to bring in components for assembly.  The only other representation of the tunnels around the detector that was available was an artist rendering, shown in Fig.~\ref{artist01}.  In an effort to get a more realistic picture of the tunnels and access drifts, in April of 2013 a Japanese company scanned areas of the KamLAND mine with a FARO laser scanner and produced 24GB of data.  The ultimate goal of the scan was to produce a 3D CAD model of the KamLAND tunnel that could be used in Autodesk Inventor to develop the cyclotron and target CAD models.

\begin{figure}[H]
\centering
\includegraphics[width=4.in]{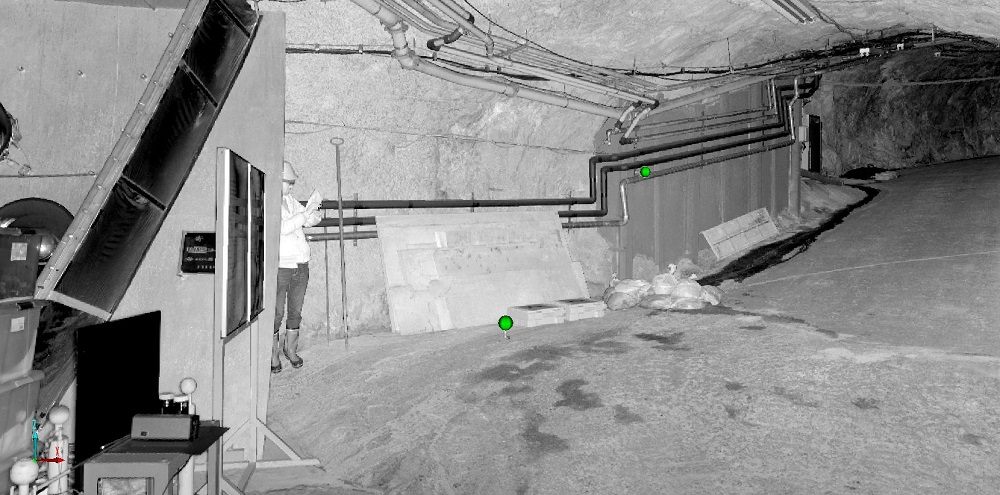}\\
\vspace{0.1in}
\includegraphics[width=4.in]{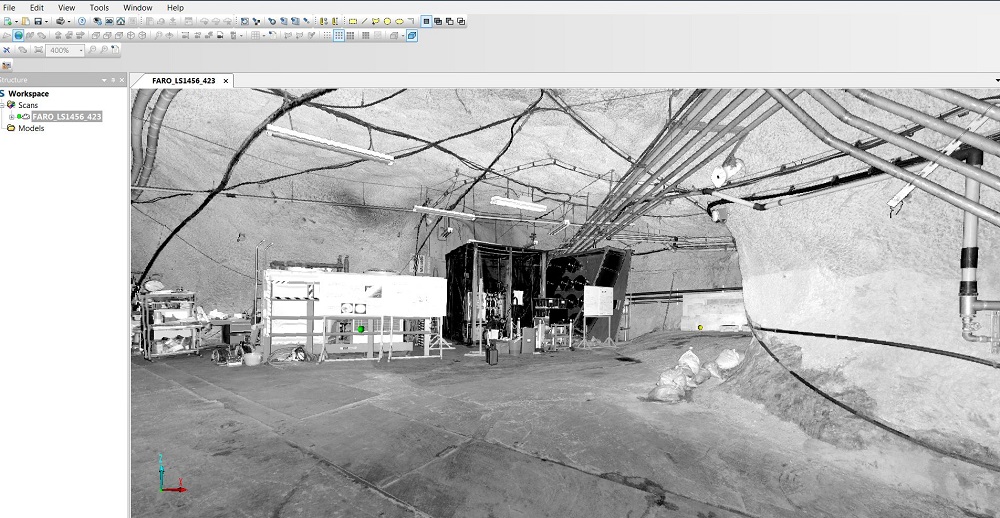}
\vspace{0.1in}
\includegraphics[width=4.in]{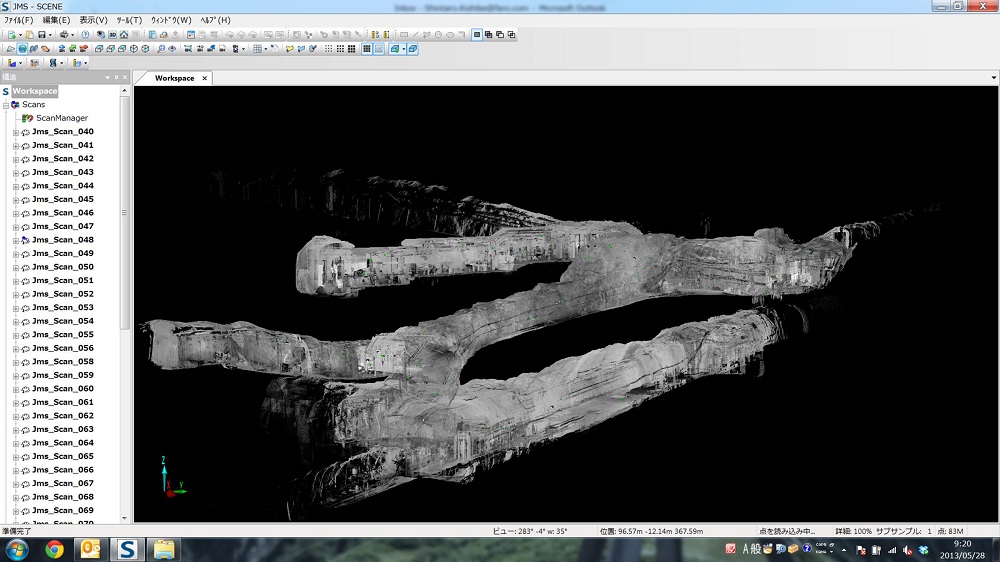}
\caption{\footnotesize Top and middle:  Two example scenes from the KamLAND tunnel as viewed in the SceneLT software.  The green dots are targets used to align the coordinate systems of different sets of data.   Bottom: Screen shot of Scene software viewing the point cloud from outside the tunnel from a vantage point somewhere in the rock.
\label{scan-00}}
\end{figure}

As seen in Fig.~\ref{scan-00} (top, middle), when the FARO data is viewed with SceneLT software, the laser scan looks almost like a photograph of the area.  Laser scanning is accomplished by putting targets in the scene to be scanned and collecting reflected laser light from all the surfaces in the scene.  The scanner creates its own coordinate system and records the location of the point of reflection from the surface into a table of data.  This table of data is called a "point cloud".  The targets in the scene must be visible from different locations as the scanner is moved through the scene so that adjacent areas of geometry can be transformed into a common coordinate system.
Fig.~\ref{scan-00} (bottom) is a screen shot from the Scene software that shows what the point cloud data looks like from a point outside the tunnel.  

\begin{figure}[H]
\centering
\includegraphics[width=4.5in]{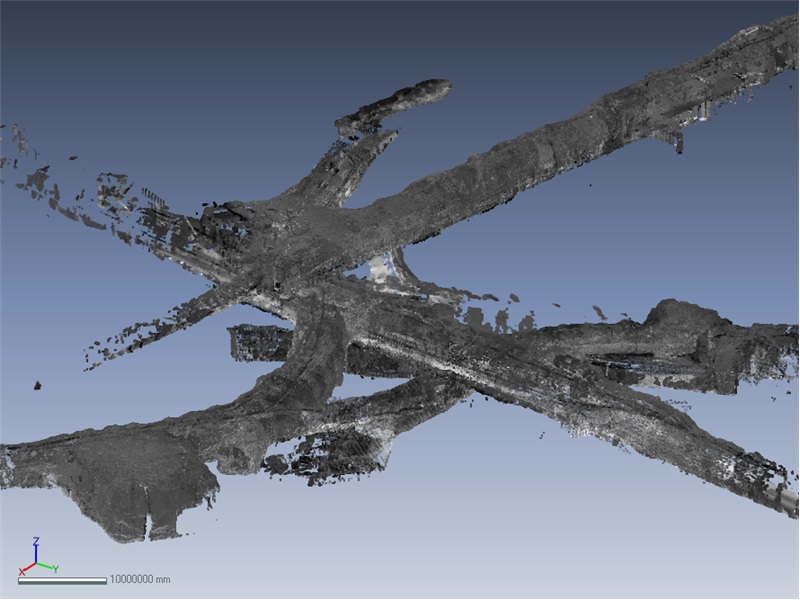}\\
\vspace{0.2in}
\includegraphics[width=4.5in]{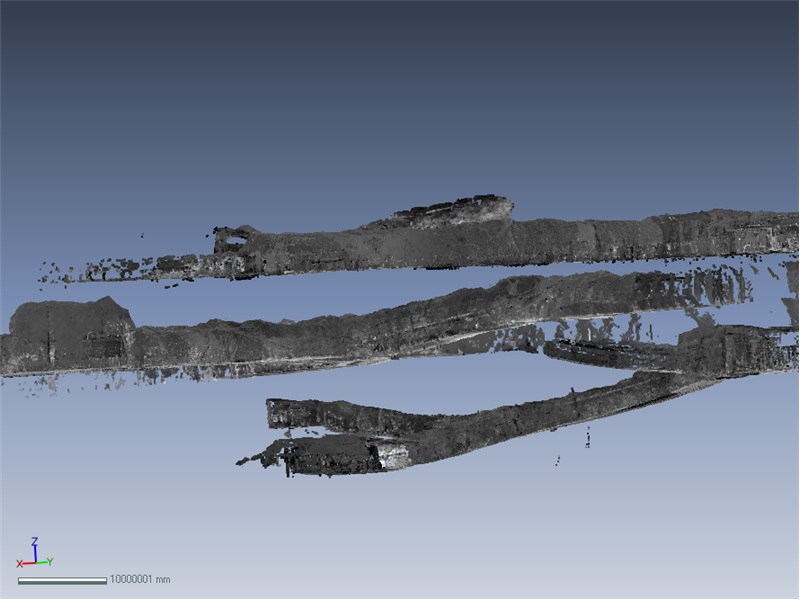}
\caption{\footnotesize 
Two examples of viewing the point cloud from outside the tunnel using GEOMAGIC.
\label{CaptureScreen12}}
\end{figure}

Fig.~\ref{CaptureScreen12} shows two views of the data from another piece of software, GEOMAGIC, used to process the point cloud.  The SceneLT software is used to view the data, but does not produce files that are importable to CAD software.  MIT purchased a license of GEOMAGIC to process the tunnel data and convert it into Inventor solids.  The views of the data from outside the tunnel compare favorably to the artist's rendering and show how much of the tunnel area was scanned.

\newpage

\begin{figure}[H]
\centering
%\begin{minipage}{.45\linewidth}
\includegraphics[width=5.5in]{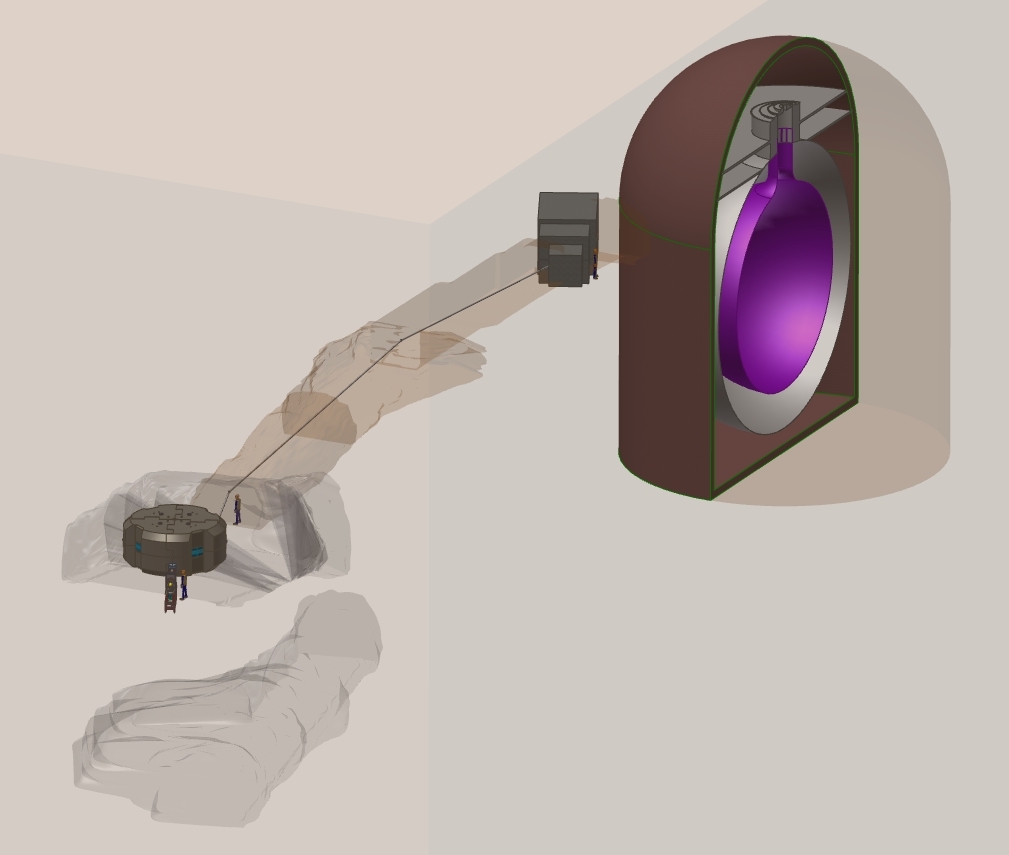}\\
%\end{minipage}
\vspace{0.2in}
%\hspace{.05\linewidth}
%\begin{minipage}{.45\linewidth}
\hspace{-0.2in}\includegraphics[width=5.5in]{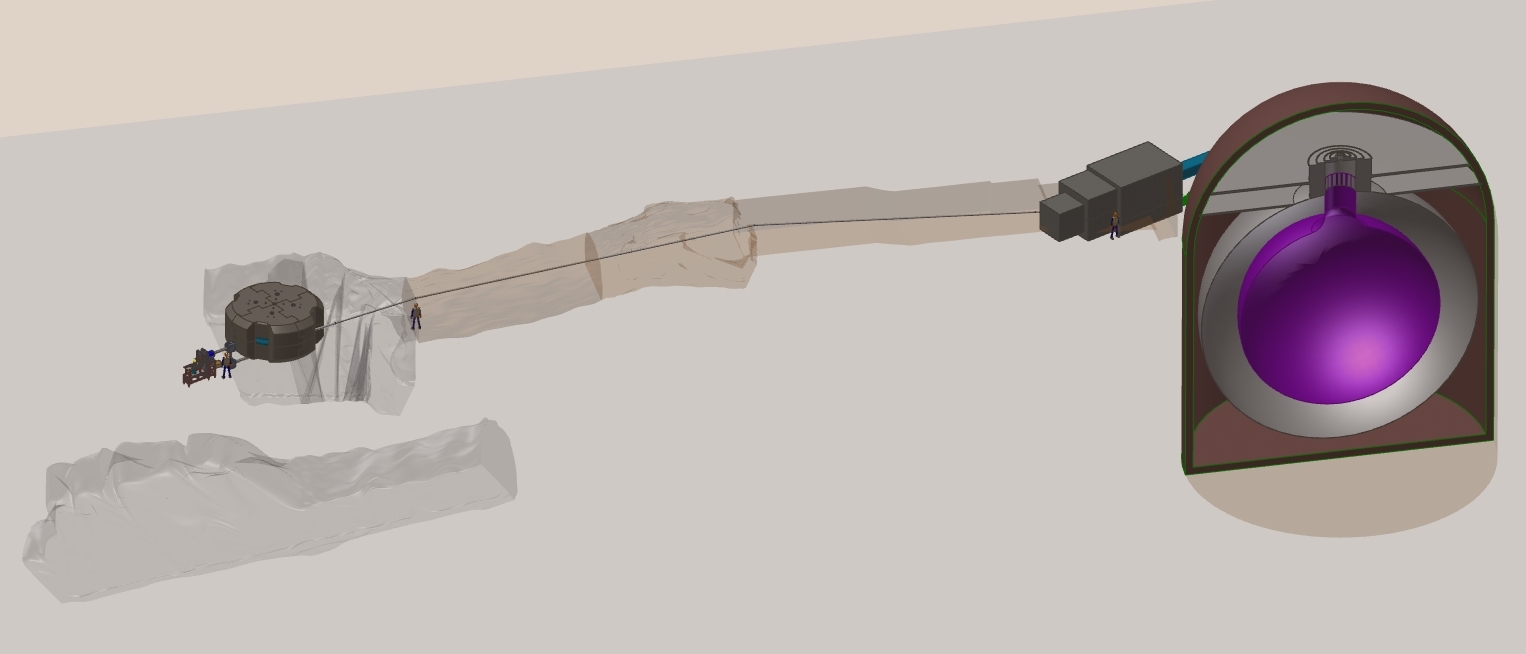}
%\end{minipage}
\caption{\footnotesize Example views of the tunnel model in Inventor showing the cyclotron, target and KamLAND detector.
\label{KAMLANDAssembly-2-l006}}
\end{figure}

\begin{figure}[H]
\centering
\includegraphics[width=6.5in]{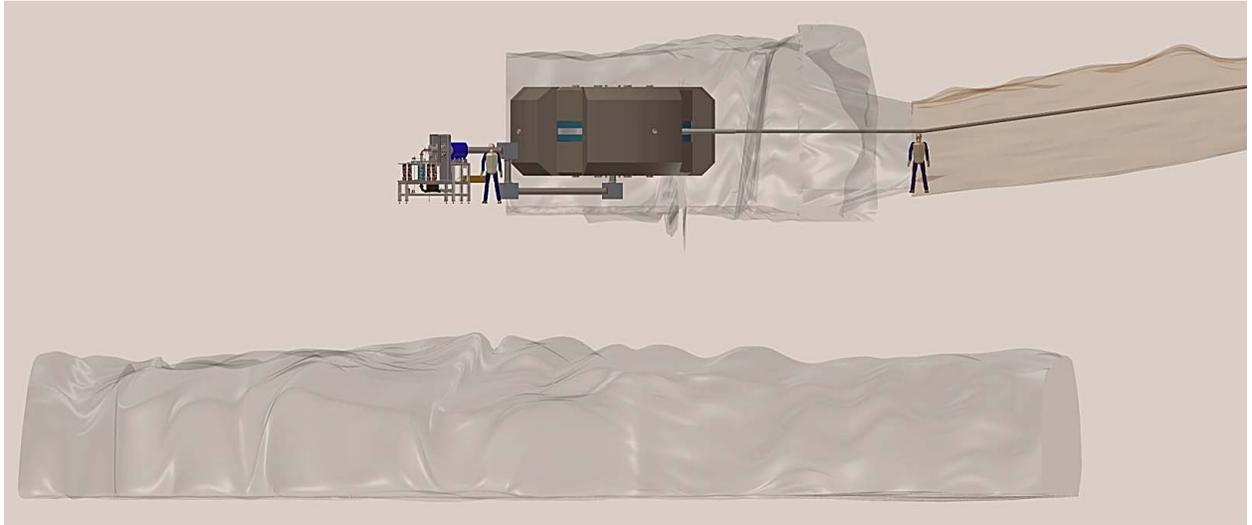}\\
\vspace{-1.5in}
\includegraphics[width=6.5in]{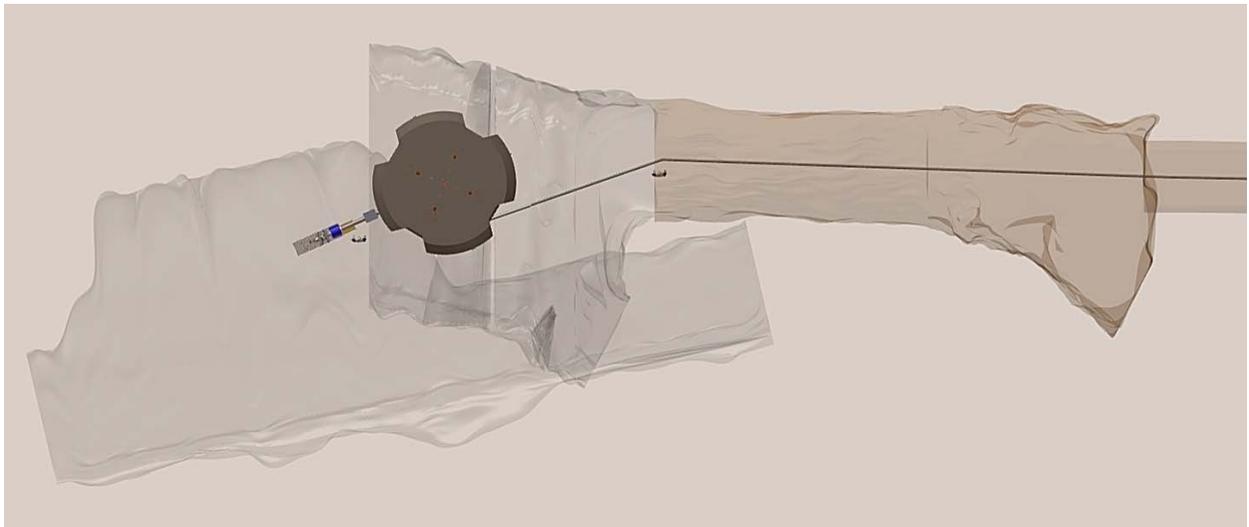}
\vspace{-.5in}
\caption{\footnotesize Elevantion view (top) and plan view (bottom) of the KamLAND tunnels around and below the cyclotron in Inventor.
\label{KAMLANDelevview-01}}
\end{figure}

The present data set is now under analysis.  Certain
clean-up of images must be performed manually and is yet to be
complete.  This includes removal of images of temporary items that were in the
tunnels.  The rough rock geometry also sometimes causes extraneous
surfaces in the CAD model which are not present in reality and must be
manually eliminated in the CAD model.

Fig.~\ref{KAMLANDAssembly-2-l006}  show examples of what the CAD model looks like in Inventor.  The rock has been rendered transparent and the tunnels are the shaded regions around the cyclotron, beam line and target.  The target shown is a previous version somewhat smaller than the current version.  The size increase was necessitated by shielding considerations.

Fig.~\ref{KAMLANDelevview-01} (top) shows an elevation view of the cyclotron chamber and the chamber below it.  Gravity is vertical in this picture.  Fig.~\ref{KAMLANDelevview-01} (bottom) shows the plan view.  In the location the cyclotron is shown, it is not right over the lower chamber, but overlaps it.  Once the 3D solid model is finished, 2D projections can be made and overlaid on the 2D CAD drawings we got from the mine.  Preliminary work using these projections has shown mismatches between the 2D mine-provided drawings and our 3D geometry that need to be understood.  Ultimately, the 3D scan geometry must be considered the true representation of the tunnel.

\subsection{Enlarging Spaces Using Well-Understood
  Non-blasting Methods \label{noblast}}

Even at this early stage of the design, it seems very likely that some excavation will be
required.    This is particularly true in the region of the target, to
assure that the necessary shielding can be installed.
We understand that 
conventional mining techniques using explosives are not acceptable for
this enlargement because of the need to prevent shock waves that can
damage the detector.   As an alternative, we propose, for
collaboration consideration,  a well-developed
``gentle'' mining technique known as ``rock-splitting mortar.''

Consultation with a US engineering firm confirms the maturity of the
technique, and its feasibility for the specific tasks at hand
\cite{daigh}. The technique involves drilling a series of holes,
$\sim$5 cm diameter, to the depth of the 
desired excavation, and in a carefully prescribed pattern to cover the
face of the volume to be removed.  
Most (but not all) of these holes are filled with a special mortar, which expands on hardening, 
exerting extreme pressure on the surrounding rock---but over
a period of hours, so there is no shock involved---causing the entire
area of the drilled face to fracture for easy removal.  
The unfilled holes provide space for the rock to relieve pressure into, enabling the fracturing.
The cost of this procedure is somewhat more than conventional blasting, however it does provide 
a technique for rock removal that should be acceptable to the KamLAND collaboration.

\subsection{Areas where Rock Removal may be required}

While detailed plans still need to be developed, there are certain areas where providing more space can already be seen to be beneficial for the experiment.  
The drifts around the KamLAND detector were originally used for the excavation of the cavern housing the Kamiokande and now the KamLAND detectors.  After the cavern was completed, these drifts were repurposed for support and access to the detector installed in the cavern.  Additionally, several areas were enlarged for utilities and service areas necessary for the operation.

The equipment required for the IsoDAR experiment have also some specialized needs for space that may be difficult to accommodate within the existing and available spaces.  Below we identify the principal needs:

{\bf Target:}  The size of the target shielding enclosure is calculated to prevent an unacceptable flux of neutrons from reaching the rock walls.  As seen in the Radiation Protection section, there is a very stringent requirement on allowed activation of rocks or other infrastructure that must remain following decommissioning of the experiment.  Meeting this requirement has led to the specification of the shielding surrounding the target, which has been optimized for maximum effectiveness with minimum material.  Still, the outer dimensions of the target shielding are larger than the drift identified as the location for the target.  A few cubic meters of rock will need to be removed to accommodate the target.

{\bf Target Maintenance:}  As seen in the target section above, spent targets will be retracted from the shielding assembly from the downstream end.  There must be sufficient space behind the target to bring in a cart with a shielded casket to accept the spent target, and to then transport this casket to a suitable repository for storage.  

Two options are available for the storage of these caskets.

a)  One option would be to prepare an area in the vicinity of the target, designated as a target storage area. This might be a new drift running away from the target, with a number of recessed vaults drilled into the rock sufficiently large and deep to each hold a single spent-target ``torpedo'' with or without its casket.  The recesses would need to be deep enough to be able to insert the torpedo, and emplace a plug on the surface suitable to attenuate the gamma radiation from the target.  These vaults must also be sufficiently far away from the KamLAND detector that high-energy gamma radiation has no chance of penetrating through the rock to reach the sensitive volume of the detector.  This target storage area could also contain fresh target assemblies, so that the entire process of replacing a spent target with a new one can take place without the need for transporting these ``torpedos'' long distances.

b)  An alternate is to plan on transporting the caskets out of the KamLAND area, either for storage in another area of the mine, or to be driven to the surface, and shipped to a designated area elsewhere in Japan for storage of high-level radioactive waste. 
Consulting with KEK and JPARC on their procedures for storing their highly-radioactive spent elements could point to a viable solution. 

Transport of the casket from the area behind the target will require sufficient space for clearance around the target shielding, best would be on the side away from the KamLAND detector.
Removal of the casket out of the KamLAND area will require identification of a route by the MEBT, and may involve removal of a section of the MEBT to allow passage of the cart with the casket and the driving tractor.  The most logical path for the cart is to pass from the back of the target alongside the target on the side away from the KamLAND detector.  However the path must cross the MEBT to pass between this line and the cyclotron, hence the need to remove a portion of the MEBT.  It is unlikely that the beam height would be high enough for the cart to pass underneath it.
This plan may also require further enlargement of the MEBT passage to provide clearance.

{\bf MEBT route}  The path of the MEBT follows the natural contour of the drifts, which involves a rise, then about half-way to the target a drop.  The total elevation difference between cyclotron and target is about  2 M, however the maximum elevation difference along the path is almost 5 to 6 M.  For ease of transport of components for construction and maintenance of the target, as well as for installation and operation of the beam line, it would be desirable to explore the possibility of removal of rock to provide as much as possible a level surface between the cyclotron and the target.

{\bf Cyclotron vault}  The large cavern designated for the cyclotron contains a decommissioned water-purification plant.  The ceiling height is about 4 meters, however this might not be sufficient for the installation and maintenance of the cyclotron, and for the installation of needed shielding.  In addition, a substantial staging area is needed for the construction of the cyclotron, it may be difficult to develop a staging plan within the available floor-space.  While a detailed space-requirement plan has not yet been developed, it should be kept in mind that enlarging the cyclotron cavern may in all likelihood be a desirable option.

{\bf Power substation}  The present substation for the 1.5 MW transformers and power-distribution equipment occupies a space that is sized just for this equipment.  The 3.4 MW required for IsoDAR will necessitate a larger space for the additional equipment.  This could either be accommodated by an enlargement of the existing area, or identification of a new location within the mine that is still close enough to the experiment to allow reasonable cable runs.

\subsection{Understanding Constraints Outside of the KamLAND Space}

A substantial region of tunnels exists between the main adit and the
beginning of the KamLAND space.    This space cannot be modified.   As this is outside the control of
KamLAND, we were not able to arrange laser mapping of this space.   We
may do so in the future.   At present we have relied on careful
measurements of the space.  We have found one region with opening too
small to fit the coil for the cyclotron.   It is this stricture which
drives the decision to construct the coil in two segments, or,
alternatively, to produce the coil undergrounds.

\section{Utilities}

\subsection{Electrical Power \label{power}}

The installed power requirement for the IsoDAR system is approximately 3.4
MW. This covers, with some reserve capacity, 
the 600 kilowatts of beam power with an RF efficiency of 50 percent as well as 
the power for magnets, vacuum systems, water
pumps and air conditioning.
Present installed power at KamLAND, which is approximately 1.5 MW, 
only covers the existing operations. 
Hence arrangements must be made for bringing in power to the site.  

The mining company owns and operates 
ten hydroelectric generating plants located on the various rivers within a ~10 km radius of the main adit (mine entrance) with total  generating capacity of 37 MW.   Because of the current power crisis in Japan these power plants have been put under the control of the local power utility company, to support the local grid.  As a result, obtaining power necessary for IsoDAR will require negotiations with the local power company. 

The power company will provide the electricity
at the site of its main trunk line, which runs along the main highway by the river.  
The mine entrance is about 2 kilometers from the highway, along a small secondary road.    Power could be brought to the mine adit via overhead power lines.  A 5 - 6 kV underground-certified cable from the adit to the transformer banks at the KamLAND site, about a 3 km run will be required.   

%A new transformer station for IsoDAR load, and a new station must be built. Space beyond the existing transformer bay must be identified.

Note that the vault containing the current KamLAND power-distribution transformers will not be large enough to accommodate the additional transformers to power the IsoDAR systems.  In addition to the new transformers themselves, the existing cavern must be enlarged.

\subsection{Water Cooling \label{water}}

\begin{figure}[!t]
\begin{center}
\includegraphics[width=5.5in]{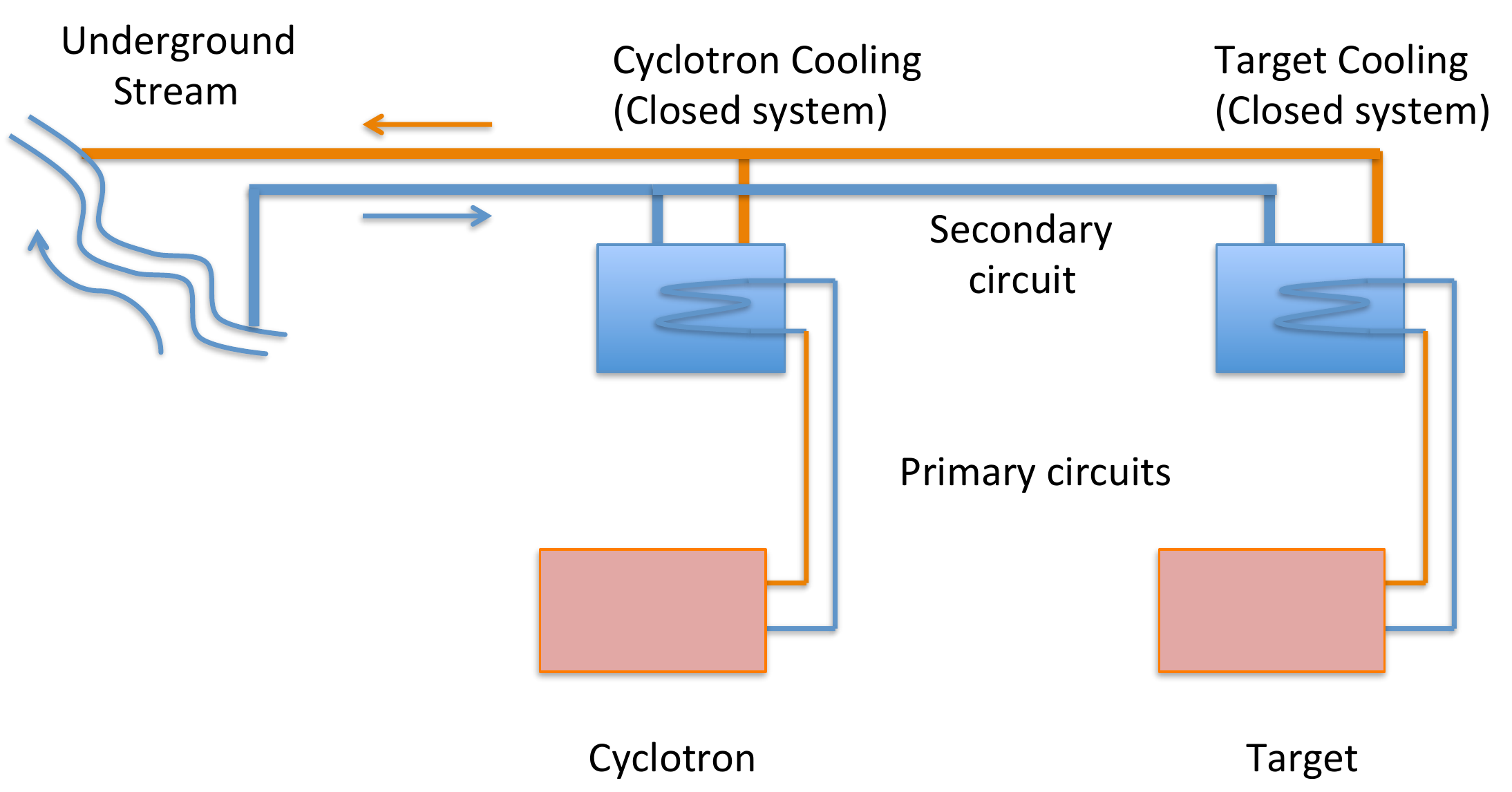}
\end{center}
%\vspace{-1.0in}
\caption{\footnotesize  Schematic showing the two primary cooling loops and one secondary loop.
\label{fig:coolingschem}}
\end{figure}

IsoDAR will need to have at least two primary cooling circuits for
which the water must be contained and purified.
The first primary system will cool the cyclotron and beam
line components.  In this system, the
water will become slightly activated, but must be able to withstand 
the high voltages of the cyclotron RF system.  
The second
primary system is dedicated to cooling the target.  In this case, the
cooling water is in direct contact with the
beam (the Bragg peak is deposited in the water) and so will become
activated. The design for containment, purification and filtration of this 
water must be very carefully thought through because of this activation. 
This second primary system will be an integral part of the target.
Both of these primary systems will go through heat-exchangers 
to transfer heat to a secondary system which connects to the outside environment. 
The cooling system is shown schematically in Fig. ~\ref{fig:coolingschem}

\begin{figure}[!t]
\begin{center}
\includegraphics[width=5.5in]{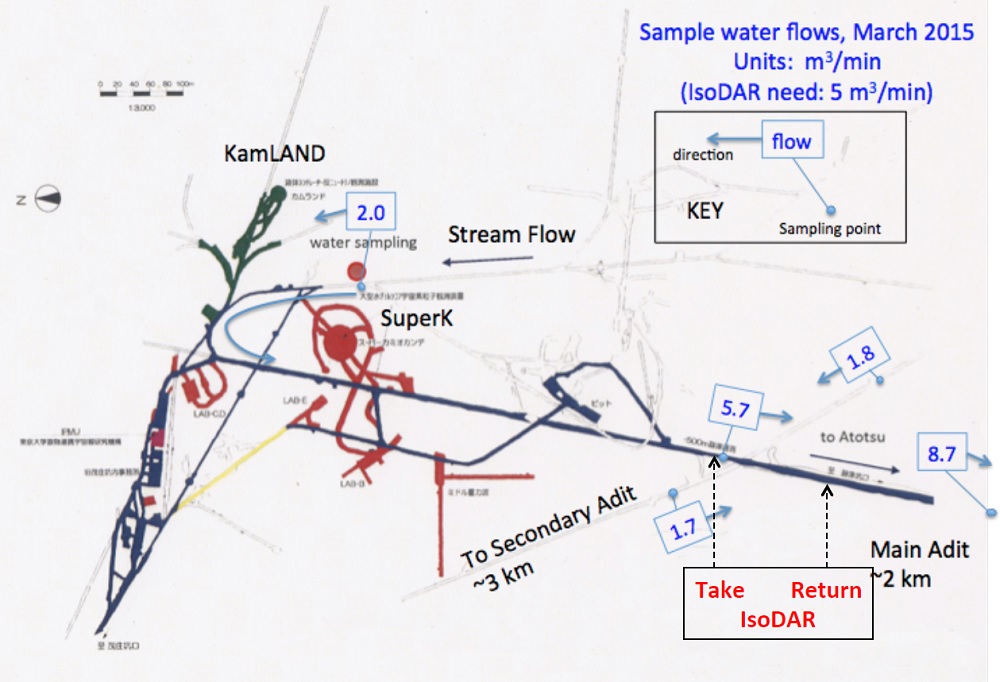}
\end{center}
%\vspace{-1.0in}
\caption{\footnotesize  Path of underground stream, following underground drifts.  It flows from right to left above SuperK, loops around below SuperK, and flows through the access drift to the Atotsu entrance (adit).  The sampling point for the data in Fig. ~\ref{fig:StreamFlow} is the red dot about midway between KamLAND and SuperK.  At typical low flow months the rate is inadequate to supply the 5 m$^3$/min needed for IsoDAR.  Below SuperK, however, the flow rate has increased due to addition of tributaries.  The points where IsoDAR could take and return water are shown.
\label{fig:UndergroundStream}}
\end{figure}

\begin{figure}[!t]
\begin{center}
\includegraphics[width=5.5in]{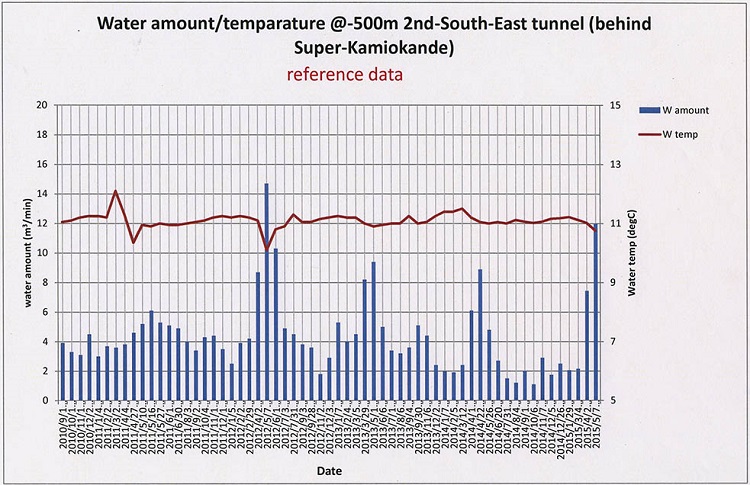}
\end{center}
\caption{Flow of underground stream, measured monthly, for last 5 years.  Seasonal variations are seen clearly.  2014 dry seasons have seen substantially reduced flow, possibly due to excavations for Japan's gravity-wave experiment, that has opened up new pathways for water to flow underground.  The sampling point is the red dot between KamLAND and SuperK in Fig.
\protect\ref{fig:UndergroundStream}.
\label{fig:StreamFlow}}
\end{figure}

The secondary system will require 5 m$^3$/min of water flow for cooling, for a temperature rise of 10$^\circ$C or less.
An underground stream runs through the area, and currently provides sufficient cooling for 
both KamLAND and SuperK.  Fig. ~\ref{fig:UndergroundStream} shows the course of this stream,
and indicates flow volumes at different locations.  Fig. ~\ref{fig:StreamFlow} shows the volume of water flow, measured every month for the last five years, sampled at a point approximately halfway between KamLAND and SuperK.  One can clearly see seasonal variations.  In particular, at this sampling point the water flow would not be adequate to service all three needs:  KamLAND, SuperK and IsoDAR.  However, this stream receives flow from tributaries as it continues, and at a point a few hundred meters downstream from SuperK the flow is already up to 5.7 m$^3$/min.  It is almost 9 m$^3$/min at the mine entrance.  So even in low-flow months, by taking water from below SuperK and returning it a bit beyond, the stream can provide adequate cooling.  

As a contingency, should the flow during a particularly dry period fall below the 5 m$^3$/min mark, the water will still serve to cool IsoDAR, but the temperature rise will be above 10$^\circ$C.  The water will be diluted from the downstream tributaries, but if the temperature at the adit is still above the legal limit for discharge to the local river, one can install a small evaporative cooling tower, probably mounted above the present air-handling building at the mine entrance.

Note that with this solution, the run of 25 cm diameter intake and return pipes for the water will need to be less than 1 km each way.  This is a better alternative than making the secondary IsoDAR cooling circuit also a closed loop, which requires running these pipes to a cooling tower at the adit, a distance of about 3 km each way.

%We have been informed that there is an underground stream with a flow
%that appears to be sufficient to provide adequate water for the
%secondary system, although this needs to be verified.  
%If not, then provision will need to be made to bring water in from the outside river.

%Following the heat exchanger, the secondary water will need to be
%discharged.  We will work with the collaboration to determine the best
%solution.    It can either be transported back along the route to the
%external environment, or discharged back into the underground stream.
%However, we understand that this stream also is a source of
%water for SuperK cooling needs, and so its temperature cannot be elevated
%beyond a certain degree.  This may require that the discharge of
%IsoDAR water takes place beyond the intake point of SuperK.  This could
%be addressed through additional lengths of piping.

\subsection{Ventilation}

As discussed in the Radiation Protection section, a careful study is needed of activation of air in the vicinity of the high neutron-flux areas.  If these studies reveal that activation of air is a concern, then mitigation measures will be necessary.  An evaluation of air-flow patterns in the vicinity of KamLAND will be needed, and a plan for limiting air exchanges and containing potentially activated air until sufficient time has elapsed for the short-lived activities produced to decay.  

In any event, air conditioning equipment should be provided to ensure that temperature in the areas containing technical equipment are maintained at a safe level for operation of the equipment.

% This was moved to this section post-KamLAND meeting.
\section{Radiation Protection}

All radiation protection policies and guidelines must be in compliance
with the Japanese law. In accordance with this, measures will be taken
to maintain radiation exposure as low as reasonable achievable through
shielding design and administrative controls. The primary method of
radiation control will be through physical design features such as
shielding and barriers to radiological areas. Occupational exposure of
personnel is most likely to occur in the vicinity of experimental
areas. Effectiveness of the shielding will be actively monitored by
radiation instruments located in the control room and by frequent
area-surveys performed by the health physics personnel. 

Our collaborators at RIKEN have been very valuable in establishing a clear
understanding of the
shielding and radiation protection guidelines.    Standards and
regulations for installations in Japan follow guidelines established
by IAEA Document RS-G-1.7 ``Safety Guide.''    In this section, we
describe a first iteration on the conceptual design for the radiation
protection, with the plan to expand on this in the CDR for the
Conventional Facility.

\subsection{Personnel Protection}

Personnel protection requires minimization of radiation doses to personnel, where doses are kept as low as reasonably attainable (ALARA), but in no case above limit mandated by Japanese radiation safety regulations.  Protection is needed both for prompt radiation (when the beam is on) and residual radiation from activated material.

\subsubsection{Prompt Radiation}

There are several sources of prompt radiation.
During operation of the cyclotron, radiation results from x-rays
associated with high voltages in the ion-source platform and cyclotron
RF systems.
Fast neutrons and gamma rays result from nuclear reactions of the accelerated ions through beam interactions in the cyclotron, transport line and target systems.  

All areas where radiation exist must be contained inside ``Controlled
Areas for Radiation Protection.''  Depending on the level of
radiation, these areas may require complete exclusion of personnel
while running, or access limitations commensurate with the radiation fields present.

%\subsubsubsection{Residual Radiation}

\subsubsection{Residual Radiation}

When the beam is off, radiation fields remain in areas where beam
losses have occurred, as well as in the vicinity of the target and
target-service areas.  These radiation fields result from radioactive
decay and are primarily gammas with some high-energy betas.  They are the
result of activation from beam particles or secondary neutrons induced
by beam interactions.  The greatest part of the radioisotopes produced
will be contained in solid materials, but activated liquids will also
be present in the primary coolant circuits.  
There will possibly also be gaseous products,
generally very short-lived, from 
activation of air in the vicinity of neutron fluxes from the experiment.

A careful assessment will be performed on the need for addressing activation of 
air in the vicinity of high-neutron fluxes.  Should it be established that this is required, the matter will be 
managed by careful engineering of ventilation systems, involving slowing of 
air exchanges to ensure that short-lived gaseous activities will have decayed prior 
to mixing of air from within the experimental area with the external environment.

Activated liquids are managed by very careful engineering and
maintenance of primary cooling circuits to ensure absence of leaks,
and shielding of conduits and reservoirs.  Particular attention to
shielding should take place in the vicinity of pumps, heat-exchangers
and valves,  where access is required for maintenance personnel. 

Similarly, shielding of areas containing activated solids, such as
magnets, beam pipes, and components exposed to primary protons or
neutrons (especially the target) should be shielded so that personnel
requiring access for maintenance are suitably protected.  Engineering
of the accelerator and 60 MeV/amu beam 
transport should ensure beam losses in these areas are sufficiently
low  to allow for hands-on maintenance.  
Experience at PSI has indicated that this requires maintaining
beam-losses above 30 MeV to less than 200 watts in the accelerator vault, 
and less than 1 watt per meter in the 60 MeV/amu beam transport areas.  

Target maintenance requires specialized remote-handling systems and proper storage of highly-activated spent components.  These systems must be carefully designed and constructed to totally prevent any contamination from uncontrolled spread of activated material.

\subsubsection{Environmental Protection \label{environ}}

While minimizing radiation doses to operations and maintenance personnel is a key driver of the shielding design in the KamLAND area, an equally important aspect is preservation of the low-radiation environment of the underground Kamioka mine site both for standard radiation safety requirements and to keep the site viable for the deployment of future sensitive experiments.

Basic Japanese requirements have been clearly specified in a
communication with Y. Uwamino, Director of RIKEN Safety Management
Group, as given in IAEA-RS-G-1.7.  Briefly stated,
artificially-produced radioactivity of the isotopes we are likely to
produce that are at levels above 0.1 Bq/gm must be located inside a
Controlled Area.  The second requirement we must meet is that when the IsoDAR experiment is
completed, and fully de-commissioned, there must be no Controlled
Areas remaining due to artificially-produced radioactivity.

In the Shielding section above we evaluated the environment near the target after the end of the IsoDAR decommissioning period. During IsoDAR operations, some neutrons will penetrate into the rock walls of the target hall and activation will occur. After approximately a 1-year cool-down period for the area immediately surrounding the target hall, dismantling of the target and surrounding shielding can begin. (Access to the KamLAND detector and its support facilities should be completely available during the cool-down period.)

This process is likely to take perhaps another year or more
year. When this process is complete, then a further waiting period may be advisable prior to 
performing the critical assay to
establish that no residual radioactivity above the mandated limit 
exists in the surrounding rocks.
By that time, radioactive isotopes that have half-lives shorter
than a few months will have decayed away, and will be of no
consequence in the assessment of residual radioactivity. The isotopes
of concern will be those with long half-lives, in particular $^{60}$Co,
$^{152}$Eu, $^{154}$Eu and $^{22}$Na. It is then important to determine the concentration of the 
progenitor isotopes to these activities, and the integrated neutron flux.

During two visits to KamLAND site, in 2014 and 2015, rock samples were collected from
various locations around the KamLAND Control Room and accelerator and
transport areas. The first samples were returned to MIT and have been
irradiated at the MIT reactor in fluxes of both thermal and
high-energy (up to 5 MeV) that are several orders of magnitude above the levels estimated for the integrated flux of neutrons likely to reach the rock surfaces. Preliminary assays have been performed of the noted trace elements, and the samples have now been transferred to the Lawrence Berkeley National Lab where low-background counting facilities exist that can follow the decay of the very long-lived components produced in the irradiation. The first results have identified on a preliminary basis the amounts of cobalt and europium in the rocks, forming, as we have seen in a previous section, the basis for establishing activation limits from different shielding configurations.

Further analyses on these rock samples will refine the rock environment, leading to more accurate assessments of shielding needs.

\subsection{Interlocks \label{interlock}}

It is anticipated that the boundary of the exclusion area during
operation of the accelerator will include all access drifts to the
KamLAND area, so that during Beam On conditions no personnel access
will be possible to the detector or area above the detector.     After
Beam Off is established, the wait-time to access the detector is
expected to be very short.  The time frame will be dominated by a
radiation survey that assures allowable levels.  We estimate less than
30 minutes for this procedure.

\clearpage 
% !TEX root = ./CDR.tex

\chapter{Simulations \label{sec:simulations}}

\section{Introduction}
Careful simulations are the bread and butter of a successful particle
accelerator design. This includes the front-end (ion source, LEBT), the
accelerators (cyclotron), and beam transport lines (MEBT). 
In this chapter, we discuss
the simulation software we use to design the different parts of IsoDAR
(cf. \secref{sec:software}) and present results of simulations we 
performed to test the feasibility of our conceptual design as well as
simulations we did for test experiments like the BCS test beam line
in Vancouver in order to benchmark our simulation tools and better
understand the experimental results of the tests.

\section{Simulation Software \label{sec:software}}

In order to better understand the theory behind the simulations, we
introduce briefly the different programs and offer short explanations
on the main principles behind them.

\subsection{Ion Source: IGUN/KOBRA-INP/IBSimu}

Several codes exist for the simulation of ion beam extraction from 
plasma ion sources like non-resonant ECR (VIS) and multicusp (MIST-1):
We are using IGUN \cite{becker:igun1, becker:igun2, becker:igun3}, 
KOBRA-INP \cite{spaedke:kobra}, and in the future IBSimu \cite{kalvas:ibsimu1}.

IGUN  is a numerical simulation
code for ion beam formation at a plasma boundary. IGUN is a 2D code in RZ 
coordinates, exploiting the fact that most extraction systems are rotationally
symmetric and in many cases, the extracted beams are as well. The problem is 
set up in IGUN by the user through an input file containing the basic parameters 
(INPUT1), axial values of the magnetic field (INPUT3), boundary points of the 
electrodes, and runtime variables (INPUT5) \cite{becker:igun3}.

IGUN then solves the problem by iterating multiple times over the following 
steps until a stable solution has been reached \cite{becker:igun2}:
\begin{enumerate}
    \item Solve Poisson's equation for the potential map. This is done on a
          rectangular mesh using a finite element method and a successive 
          over-relaxation solver (SOR).
    \item Solve the equations of motion for the macro-particles using
          ray-tracing with a fourth order Runge-Kutta procedure.
    \item Allocate space charge to the potential map. This is done 
          in two steps: 
          Inside the plasma sheath, the space charge is calculated analytically
          using a 1D sheath model. Outside the sheath the space 
          charge density is generated from the particle trajectories obtained
          in step 2 multiplied with a compensation term according to the 
          thermal distribution of electrons. 
\end{enumerate}

The magnetic field data (if specified) is given as $\mathrm{B}_z(z)$ at $r=0$ 
and the field values for $r > 0$ are calculated through radial expansion up 
to order 6.

The ion source group in Catania did the modeling of the ion beam extraction
with the 3D extraction simulation software KOBRA-INP.
The principles are similar to IGUN, but KOBRA-INP can do the simulation
in 3D. Some results of these simulations are reported in \secref{sec:lebt_sim}
where the KOBRA-INP results were used as the starting conditions of the 
BCS test-beamline simulations with WARP.

Finally, in the future, we will also use a new simulation code for 
ion beam extraction, developed by the ion source group in Jyv\"ask\"yl\"a, 
called IBSimu. Like KOBRA-INP, IBSimu has the
capability of 2D (R-Z symmetric systems) and 3D treatment of the problem
and offers a useful interface by being a self-contained C++ library that
can be embedded in any self-written code.

\subsection{Low Energy Beam Transport: WARP \label{sec:warp}}
WARP \cite{grote:warp1, grote:warp2, friedman:warp} is a particle-in-cell (PIC)
code originally developed for the simulation of intense ion beams used in 
heavy-ion driven inertial confinement fusion (HIF).

In PIC, macro-particles representing a number of real particles are advanced in
time using the Lorentz equation of motion. Electrostatic self-fields of the 
particles are calculated self-consistently on a mesh. Particles are advanced
through a combination of the \emph{Leapfrog} and \emph{isochronous Leapfrog}
methods \cite{friedman:leapfrog1}.

\subsubsection{Main Packages (python class name in \emph{italic})}
The main package supervising the simulations is called \emph{top}.

For advancing particles and solving for the electrostatic fields (external
plus self-fields) WARP has three different geometry modes:

\begin{itemize}
    \item WARP3d (particles: \emph{w3d}, 
          field solver: \emph{f3d}). The full three dimensional
          model.
    \item WARPxy (particles: \emph{wxy}, 
          field solver: \emph{fxy}). A transverse slice model.
    \item WARPrz (particles: \emph{wrz}, 
          field solver: \emph{frz}). A cylindrically symmetric model.
\end{itemize}

The separation of particle pusher and field solver enables the user to combine 
3D pusher and RZ field solver for cases where the electrodes are RZ symmetric,
but a 3D treatment of the beam is necessary.

In addition to the PIC model, there is also an envelope solver (package: 
\emph{env}) based on a matrix formalism. The envelope solver has not been 
used and will not be described further.

\subsubsection{Particle Loading}
Particles are loaded and injected separately. This is practical since the same
initial distribution loaded can be used with different injection methods or in
different geometry modes in subsequent simulation runs.
WARP provides a variety of built-in initial transversal particle distributions: 
Gaussian, Semi-Gaussian or K-V (uniform). Longitudinal distributions can be 
cigar shaped or uniform. WARP also lets the user specify arbitrary transversal
and longitudinal particle distributions. The loaded beam can then be injected
in one of two ways: \emph{Constant current}, or \emph{space charge limited}.

\subsubsection{Fields and Lattice Elements}
WARP has four built in lattice elements: \emph{quadrupoles, dipoles, bends}, and
\emph{acceleration gaps}. Lattice segments can be periodically extended.
Special attention should be given to the bend, which is not a lattice element 
per se, but rather a region of coordinate transformation. In this region,
specified by a start point, end point and bending radius, Frenet-Serret 
coordinates are used to transform the bend into a straight line. These elements
are used in superposition with all forms of dipole elements to keep the 
design trajectory straight.

In addition to the built-in lattice elements, WARP lets the user set electrostatic
or magnetostatic external fields in a number of ways:
\begin{itemize}
    \item Hard-edged multipole description 
    \item Axially varying multipole description
    \item Gridded elements
\end{itemize}
For the first two types of descriptions, WARP calculates the fields at the 
particle position through multipole expansion. Gridded elements are given on a
3D grid and are interpolated to the particle position as needed.

The user can also generate electrode geometries, specify their voltages, and
have WARP calculate the electrostatic fields directly through a finite elements method.

In addition, there are supplemental python scripts for generating realistic
models of common beam optics elements like electrostatic quadrupoles or 
solenoids.

\subsubsection{Fieldsolvers}
In 3D and in transverse slice mode, the self field of the beam is calculated on
a Cartesian mesh. WARP has a variety of field solvers at its disposal (fast 
Fourier transforms - FFT, capacity matrices, successive over relaxation - SOR, and
the multigrid solver - an extension of the SOR solver which utilizes additional
coarser grids to speed up the calculations). Because we need to
resolve rather complex structures in some of the electrodes, the SOR and
multigrid solver are used for the simulations in this report.
In WARP the variable \emph{fstype} sets the fieldsolver with 3 being the SOR
solver and 7 the multigrid solver.

\subsubsection{Space Charge Compensation}
We are using WARP primarily for the simulation of the LEBT, where the
beam currents are high and the energies are low.
In high intensity, low energy beams, it is important to include space charge
effects as well as make some assumptions about space charge compensation. 
The way this is done in the WARP simulations is described in more detail in 
a recent paper by our group \cite{winklehner:bcs_tests}.
Space charge compensation arises from the interaction of beam ions with the 
residual gas molecules. There are several processes that can occur, but the
two dominant ones are charge-exchange and ionization. Together, they lead to 
low energy secondary ions and electrons inside the beam envelope. Because of the 
positive beam potential, electrons are attracted and positive ions are repelled. 
This can lead to a significant lowering of the beam potential. 
In the LEBT simulations, space charge compensation is calculated dynamically 
at each time step. The model is based on the work of Gabovich et al. 
\cite{gabovich:spacecharge1, gabovich:spacecharge2, soloshenko:spacecharge1,
soloshenko:spacecharge2, soloshenko:spacecharge3} and was recently reviewed 
and updated \cite{winklehner:scc2, winklehner:phd}. In the model, the energy
balance of the secondary electrons is used to calculate a steady-state value
of the space charge compensation factor $\mathrm{f}_\mathrm{e}$. It depends on
the neutral gas density (beam line pressure), the beam currents, the beam
energy, and the beam size. It also relies on knowing the cross sections for
total secondary ion and secondary electron production through the aforementioned
processes. This model 
is a `best case' approximation, where the beam is uniform and round and all
species have the same radius. No collective effects, plasma oscillations, 
and non-linearities are taken into account. In many cases, however, it works 
remarkably well.

As mentioned, one important parameter in the compensation 
estimation is the pressure along the beam line, because it determines the 
density of neutral gas molecules available for secondary ion and electron
production through residual gas ionization and charge-exchange processes. 
Simulations using the code MOLFLOW \cite{kersevan1991molflow} 
can be performed to estimate the pressure.

\subsection{Cyclotron without Space Charge: OPERA}

VectorFields OPERA \cite{opera:online} is a well-established multiphysics 
suite of programs. We are mainly using TOSCA, a Finite Elements Method (FEM) 
solver for electrostatic and magnetostatic problems. The Post-Processor 
has a ray-tracing module that can be used to track particles through the
calculated combined electrostatic and magnetostatic fields using a 
Leapfrog method. Through user programming, time varying RF fields can be 
simulated by applying a $\sin$ or $\cos$ variation to the field calculated
from the maximum dee voltage. As OPERA is such a widely known program, we
will not go into further detail here. Similarly, results of OPERA 
calculations are not presented in the simulation results section of this 
chapter, but directly included with the design studies in the respective 
subsections of \secref{sec:cyclotron}.

\subsection{Cyclotron with Space Charge: \opal}
For this discussion we briefly introduce \opalcycl, one of the four flavors of the parallel open source tool \opal.\ Detailed information about the models used in \opal\ and applications to high intensity problems can be found in 
\cite{yang:cyclotron_sim,
      bi:cyclotron_sim,
      adelmann:poisson}.

%1.Coordinates frame, basic equation of particle motions, time integrator, external field interpolation and treatment.
%2.Brief introduction of PIC methods, such as FFT, Green function, CIC scheme.
\subsubsection{Governing Equation}
In the cyclotron under consideration, the collision between particles can be neglected because the typical bunch densities are low.
In time domain, the general equations of motion of charged particle in electromagnetic fields can be expressed by
\begin{equation}\label{eq:motion}
  \frac{d \bs{p}(t)}{dt}  = q\left(c\mbox{\boldmath$\beta$}\times \bs{B} + \bs{E}\right), \nonumber \\
\end{equation}
where $m_0, q,\gamma$ are rest mass, charge and the relativistic factor. With $\bs{p}=m_0 c \gamma \mbox{\boldmath$\beta$}$ we denote the momentum of a particle, 
$c$ is the speed of light, and $\mbox{\boldmath$\beta$}=(\beta_x, \beta_y, \beta_z)$ is the normalized velocity vector. In general the time ($t$) and position ($\bs{x}$) dependent electric and magnetic vector fields are
written in abbreviated form as \textbf{B} and \textbf{E}.

If $\bs{p}$ is normalized by $m_0c$, 
Eq.\,(\ref{eq:motion}) can be written in Cartesian coordinates as 
\begin{eqnarray}\label{eq:motion2}
  \frac{dp_x}{dt} & = & \frac{q}{m_0c}E_x + \frac{q}{\gamma m_0}(p_y B_z - p_z B_y),    \nonumber \\
  \frac{dp_y}{dt} & = & \frac{q}{m_0c}E_y + \frac{q}{\gamma m_0}(p_z B_x - p_x B_z),   \\
  \frac{dp_z}{dt} & = & \frac{q}{m_0c}E_z + \frac{q}{\gamma m_0}(p_x B_y - p_y B_x).    \nonumber 
\end{eqnarray}
The evolution of the beam's distribution function $ f(\bs {x},c\mbox{\boldmath$\beta$},t)$ can be expressed by a collisionless Vlasov equation:
\begin{equation}\label{eq:Vlasov}
  \frac{df}{dt}=\partial_t f + c\mbox{\boldmath$\beta$} \cdot \nabla_x f +q(\bs{E}+ c\mbox{\boldmath$\beta$}\times\bs{B})\cdot \nabla_{c\mbox{\boldmath$\beta$}} f  =  0, 
\end{equation}
where $\bs{E}$ and $\bs{B}$ include both external applied fields, space charge fields and other collective effects such as wake fields
%\begin{eqnarray}\label{eq:Allfield}
%  \bs{E} & = & \bs{E_{\RM{ext}}}+\bs{E_{\RM{sc}}}+\bs{E_{\RM{wake}}}, \nonumber\\    
%  \bs{B} & = & \bs{B_{\RM{ext}}}+\bs{B_{\RM{sc}}}+\bs{B_{\RM{wake}}}.
%\end{eqnarray}
\begin{eqnarray}\label{eq:Allfield}
  \bs{E} & = & \bs{E_{\RM{ext}}}+\bs{E_{\RM{sc}}}, \nonumber\\    
  \bs{B} & = & \bs{B_{\RM{ext}}}+\bs{B_{\RM{sc}}}.
\end{eqnarray}
In order to model a cyclotron, the external electromagnetic fields are given by measurement or by numerical calculations. 
%Wake fields depend to first order on the structure of the vacuum chamber and are neglected in this study. 

% In the working area of a cyclotron, the boundary conditions are open on both the radial and azimuthal directions, and 
% on vertical direction it is smooth and flat, therefore $\bs{E_{wake}}$ and $\bs{B_{wake}}$ are negligible when compare with space charge fields.
\subsubsection{Self-fields}
The space charge fields can be obtained
by a quasi-static approximation. In this approach, the relative motion of the particles is non-relativistic in the beam rest frame, so the self-induced magnetic field is practically absent and the electric field can be computed by solving Poisson's equation
\begin{equation}\label{eq:Poisson}
  \nabla^{2} \phi(\bs{x}) = - \frac{\rho(\bs{x})}{\varepsilon_0},
\end{equation}
where $\phi$ and $\rho$ are the electrostatic potential and the spatial charge density in the beam rest frame. The electric field can then be calculated by
\begin{equation}\label{eq:Efield}
  \bs{E}=-\nabla\phi,
\end{equation}
and back transformed to yield both the electric and the magnetic fields, in the lab frame, required in Eq.\,(\ref{eq:Allfield}) by means of a Lorentz transformation.
Because of the large gap in our cyclotron, the contribution of image charges and currents are minor effects compared to space charges \cite{baartman:cyclotrons}, and hence it is a good approximation to use open boundary conditions. 

The combination of Eq.\,(\ref{eq:Vlasov}) and Eq.\,(\ref{eq:Poisson}) constitutes the Vlasov-Poisson system. 
In the content followed, the method of how to solve these equations in cyclotrons using PIC methods.

\subsubsection{External fields}
With respect to the external magnetic field two possible situations can be considered: 
in the first situation, the real field map is available on the median plane of the existing cyclotron machine using measurement equipment.
%In view of the narrow gaps of magnets, 
In most cases concerning cyclotrons, the vertical field, $B_z$, is measured on the median plane ($z=0$) only.
Since the magnetic field outside the median plane is required to compute trajectories with $z \neq 0$, the field needs to be expanded in the $Z$ direction. 
According to the approach given by Gordon and Taivassalo 
\cite{gordon:cyclotron_extraction}, 
by using a magnetic potential and measured $B_z$ on the median plane
at the point $(r,\theta, z)$ in cylindrical polar coordinates, the 3$rd$ order field can be written as    
\begin{eqnarray}\label{eq:Bfield}
  B_r(r,\theta, z) & = & z\frac{\partial B_z}{\partial r}-\frac{1}{6}z^3 C_r, \nonumber\\    
  B_\theta(r,\theta, z) & = & \frac{z}{r}\frac{\partial B_z}{\partial \theta}-\frac{1}{6}\frac{z^3}{r} C_{\theta}, \\     
  B_z(r,\theta, z) & = & B_z-\frac{1}{2}z^2 C_z,  \nonumber    
\end{eqnarray}
where $B_z\equiv B_z(r, \theta, 0)$ and  
\begin{eqnarray}\label{eq:Bcoeff}
  C_r & = & \frac{\partial^3B_z}{\partial r^3} + \frac{1}{r}\frac{\partial^2 B_z}{\partial r^2} - \frac{1}{r^2}\frac{\partial B_z}{\partial r} 
        + \frac{1}{r^2}\frac{\partial^3 B_z}{\partial r \partial \theta^2} - 2\frac{1}{r^3}\frac{\partial^2 B_z}{\partial \theta^2}, \nonumber  \\    
  C_{\theta} & = & \frac{1}{r}\frac{\partial^2 B_z}{\partial r \partial \theta} + \frac{\partial^3 B_z}{\partial r^2 \partial \theta}
        + \frac{1}{r^2}\frac{\partial^3 B_z}{\partial \theta^3},  \\
  C_z & = & \frac{1}{r}\frac{\partial B_z}{\partial r} + \frac{\partial^2 B_z}{\partial r^2} + \frac{1}{r^2}\frac{\partial^2 B_z}{\partial \theta^2}. \nonumber
\end{eqnarray}

All the partial differential coefficients are computed on the median plane data by interpolation, using Lagrange's 5-point formula.

In the other situation, 3D field for the region of interest is calculated numerically by building a 3D model using commercial software 
during the design phase of a new cyclotron. In this case the calculated field will be more accurate, especially at large distances from the median plane i.e. a
full 3D field map can be calculated. For all calculations in this paper, we use the method by Gordon and Taivassalo 
\cite{gordon:cyclotron_extraction}.

Finally both the external fields and space charge fields are used to track particles for one time step using a 4$th$ order Runge-Kutta (RK) integrator, in which 
the fields are evaluated for four times in each time step. Space charge fields are assumed to be constant during one time step,
because their variation is typically much slower than that of external fields. 

For the radio frequency cavities we use a radial voltage profile along the cavity $V(r)$, the gap-with $g$ to correct for the transit time. For the time dependent field we get
 
\begin{equation}
\label{ }
 \Delta E_{\RM{rf}} = \frac{\sin\tau}{\tau} \Delta V(r) \cos[\omega_{\RM{rf}} t - \phi],
\end{equation}

with $F$ denoting the transit time factor $F=\frac{1}{2} \omega_{\RM{rf}}  \Delta t$, and
$ \Delta t$ the transit time
\begin{equation}
 \Delta t = \frac{g}{\beta c}.
\end{equation}

In additions, a voltage profile varying along radius will give a phase compression of the bunch, which is induced by an additional magnetic field component $B_{z}$ in the gap,
\begin{equation}
B_{z} \simeq \frac{1}{g \omega_{\RM{rf}} } \frac{d V(r)}{dr} \sin[\omega_{\RM{rf}} t - \phi].
\end{equation}
From this we can calculate a horizontal deflection $\alpha$  as
\begin{equation}
\alpha \simeq \frac{q}{m_{0} \beta \gamma c  \omega_{\RM{rf}}t} \frac{d V(r)}{dr} \sin[\omega_{\RM{rf}} t - \phi].
\end{equation}

\subsection{Transport to Target: TRANSPORT}
The TRANSPORT program has been a mainstay for beam line design and beam envelope calculations since 1963.  
Based on the Courant and Snyder matrix formulation 
\cite{courant:synchrotron} 
a beam line is represented by a series of [n x n] matrices, with specific representations for drifts, 
bending magnets, focusing magnets and other elements. 
The beam is represented by a [1 x n]  vector, the elements representing transverse positions, 
divergences and energy spread of the envelope.  
This vector is propagated through the line by sequential multiplication with each of the beam line element matrices.  
The result  of each multiplication shows how this element has transformed the beam envelope. 
For example, a drift section changes the beam size depending on initial divergence, but not the divergences themselves; 
a focusing element (that might be approximated by a thin lens) changes the divergence without changing the size. 
As developed by K.L. Brown, D.C. Carey and others 
\cite{carey:transport} the code can perform 
 optimization of the beam transported by varying parameters of the beam line elements, 
 to produce a layout that will transport an incident beam into a desired final beam vector.  
 
Development of the code has continued, adding more sophisticated features.  
For example, U. Rohrer at PSI \cite{rohrer:transport} 
describes implementation of the effect of 
space charge on the beam envelope, by approximating the space-charge forces 
as a series of thin defocusing lenses placed along the beam line.
	
The IsoDAR MEBT (Medium-Energy Beam Transport) line has been developed using this code, 
which has resulted in an acceptable beam envelope at the target with three sets of quadrupoles 
and three dipoles.  
However, it should be noted that this code is not capable of calculating beam halo, 
or other growth mechanisms in high-current beams that can lead to small, 
but highly-significant beam loss along the transport line.  
This is pointed out in \secref{sec:mebt_risk} and in
\secref{sec:sim_risk} (the risk sections of the respective 
chapters),
noting the requirement for using more sophisticated PIC codes for evaluating these effects.

\section{Simulation Results}
In this section we present preliminary simulations of the different parts
of the IsoDAR accelerator system and of the experimental tests we 
performed to demonstrate feasibility.

\begin{figure}[!t]
	\centering
	\includegraphics[width=0.8\textwidth]{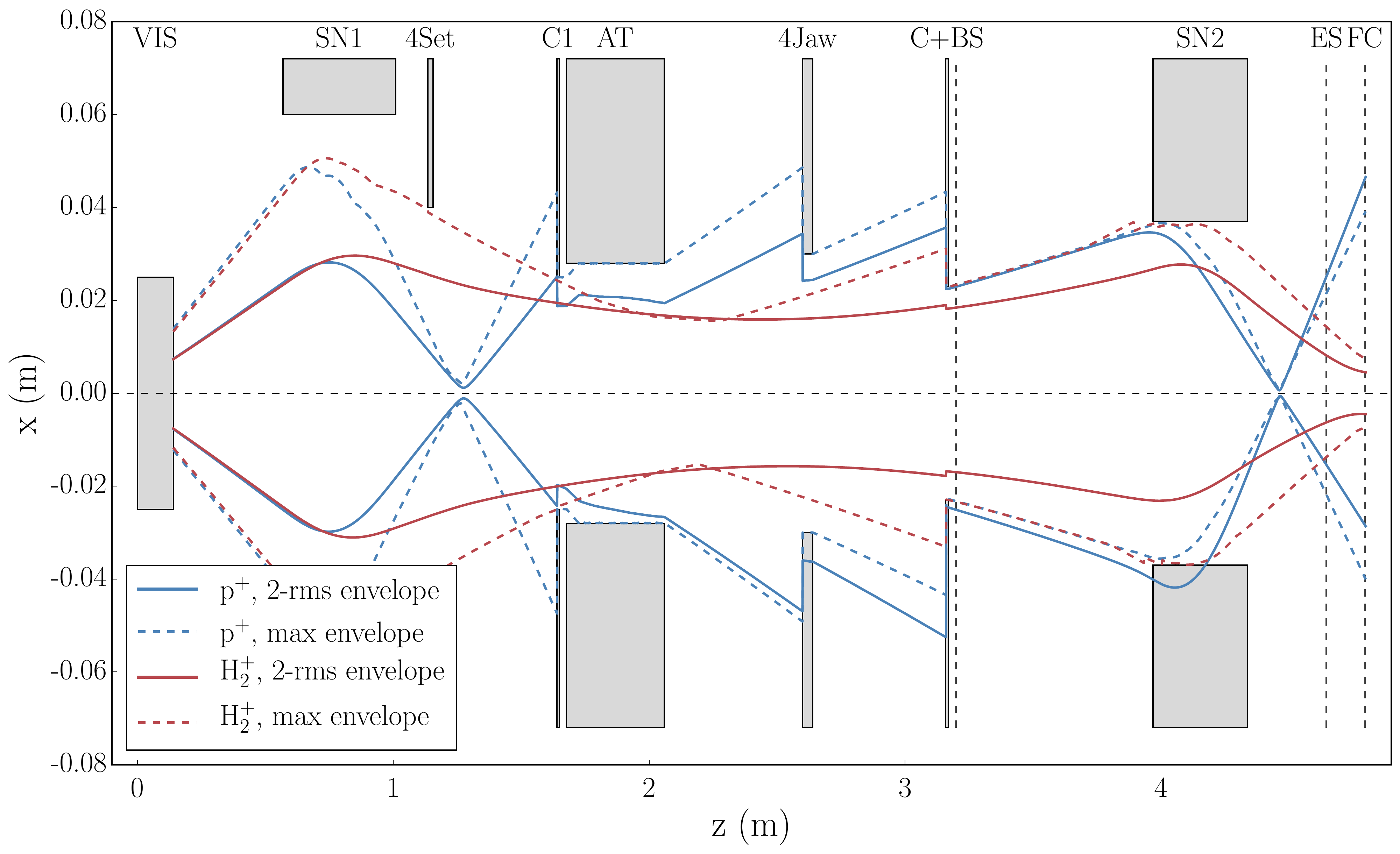}
	\caption{Schematic layout of the BCS test beam line with horizontal
	         beam envelopes for protons (blue) and \htp (red).
	         The important parts of the beam line are the ion source (VIS),
	         the two focusing solenoids (SN1, SN2) and the 
	         two diagnostic devices at the end (ES -- Electrostatic Emittance 
	         Scanner, FC -- Faraday Cup). The rest are various apertures and
	         slits.
	         With 353 A on SN1 and 230 A on SN2, \htp is focused at the 
	         location of the Faraday cup at the end of
             the beam line which roughly coincides with the entrance aperture 
             of the spiral inflector. \label{fig:SN2_230A_Env_horz}}
\end{figure}

\subsection{Ion Source and Low Energy Beam Transport \label{sec:lebt_sim}}

For the simulation of the conventional LEBT, WARP is used in transverse
XY slice mode (self fields of the DC ion beam are only considered on a
2D transversal grid) including a simple home-made space charge compensation model
(cf. \secref{sec:warp}). The injection tests in the summers of 2013 and 2014 at BCS 
in Vancouver provided a good opportunity to test the capabilities of WARP 
with our space charge compensation model and led to an extensive technical 
report \cite{winklehner:bcs_tests}. 
In the following subsection, we will summarize the results, 
followed by a brief section on the ion source extraction simulation 
of our new ion source MIST-1 using IGUN.

\subsubsection{Comparison of WARP LEBT Simulations with the BCS Test Beam Line
               \label{sec:warp_simulations}}

\begin{figure}[!t]
\centering
\begin{minipage}{.45\linewidth}
  \includegraphics[width=\linewidth]{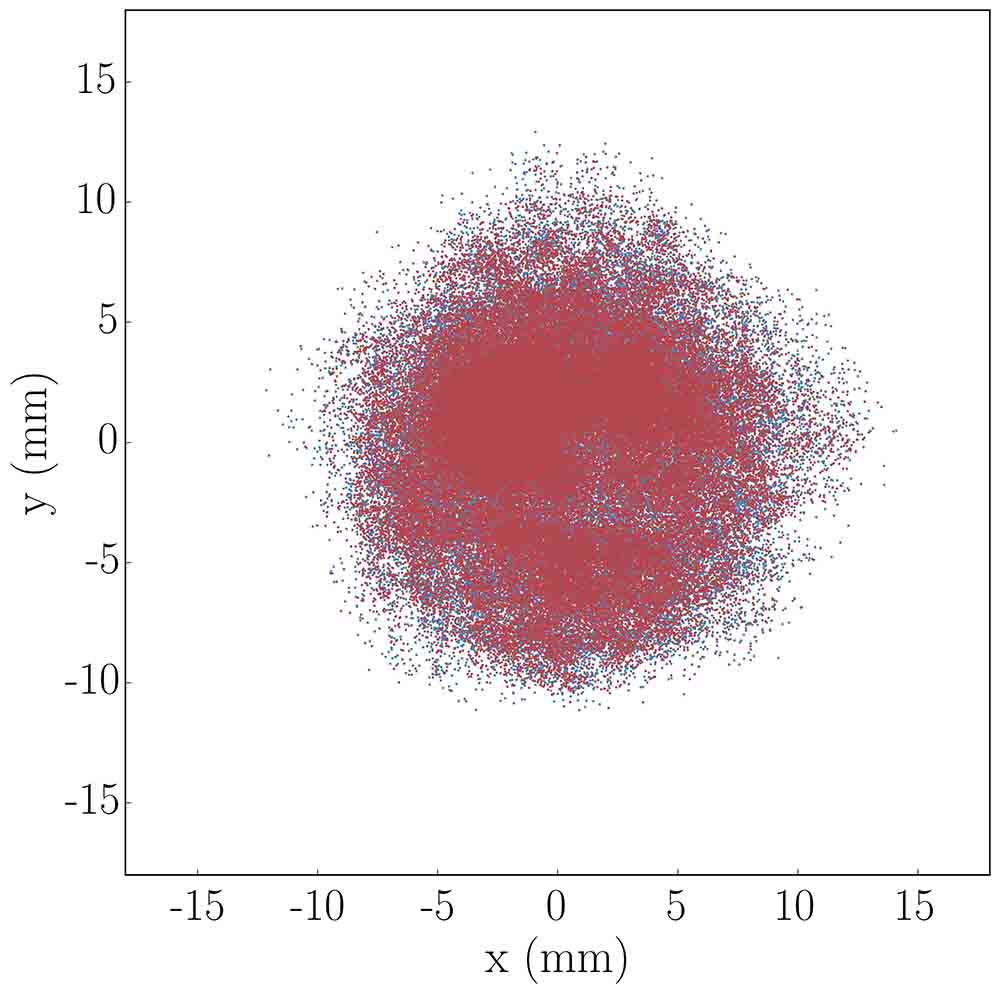}
\end{minipage}
\hspace{.05\linewidth}
\begin{minipage}{.45\linewidth}
  \includegraphics[width=\linewidth]{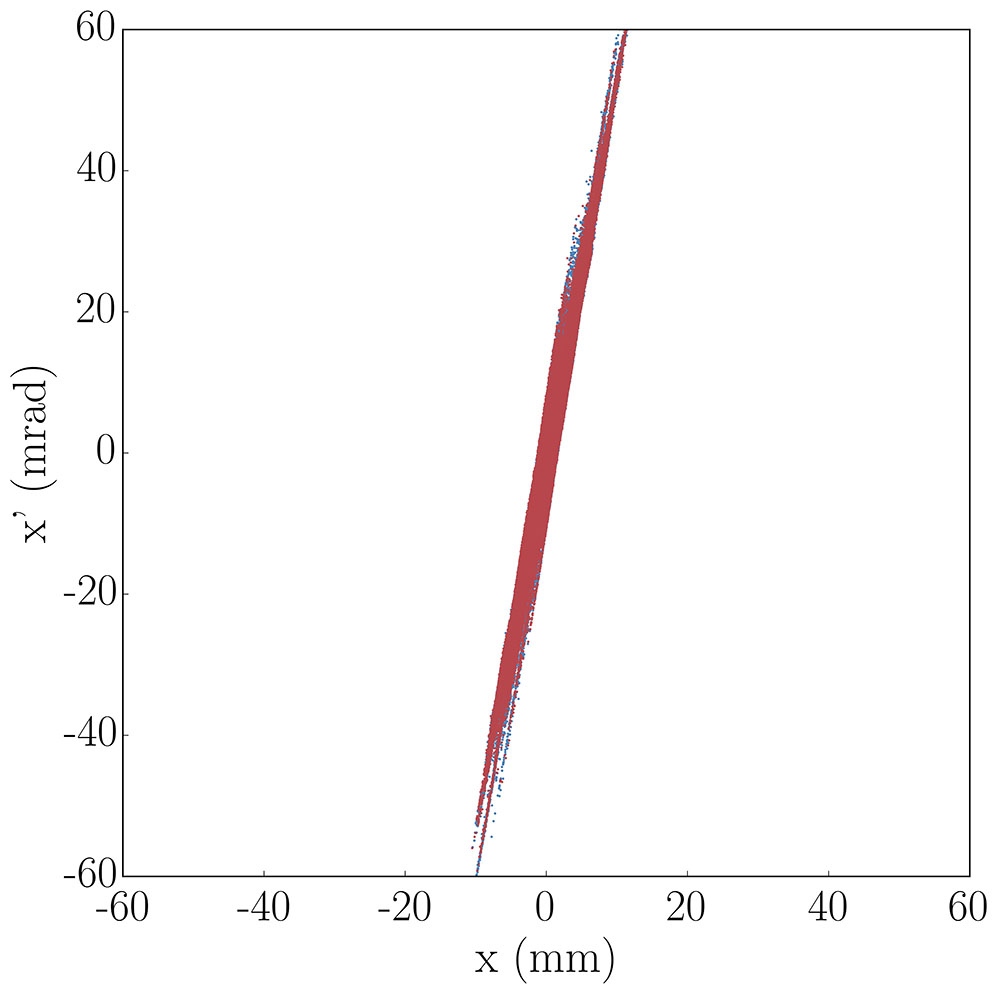}
\end{minipage}
\caption{Raw initial particle distributions used in the WARP simulations,
         obtained with KOBRA-INP. The red markers represent protons, the blue
         markers represent \htp. Left: Initial beam cross-section. Right:
         initial x-x' phase space. There are initial asymmetries, which can
         be attributed to numerical effects in the 3D simulations.}
\label{fig:particles_initial}
\end{figure}

\begin{table}[!b]
	\caption{The initial diameters and emittances for protons and 
             \htp obtained by KOBRA-INP extraction simulations and
             used in the WARP beam transport simulations. The 4-rms emittances
             include $\approx85\%$ of the beam.}
	\label{tab:particles_initial}
	\centering
    \vspace{5pt}
    \renewcommand{\arraystretch}{1.25}
		\begin{tabular}{lll}
            \hline
                                               & \textbf{protons}   & $\mathbf{\mathrm{H}_2^+}$ \\
            \hline \hline
            x--diameter (2-rms)                 & 15.2 mm            & 15.0 mm \\
            y--diameter (2-rms)                 & 15.4 mm            & 15.0 mm \\
            x-x'--emittance (4-rms, normalized) & 0.54 $\pi$-mm-mrad & 0.36 $\pi$-mm-mrad\\
            y-y'--emittance (4-rms, normalized) & 0.59 $\pi$-mm-mrad & 0.39 $\pi$-mm-mrad\\
            \hline
		\end{tabular}
\end{table}

\figref{fig:SN2_230A_Env_horz} shows the schematic layout of the test beam
line along with the resulting beam envelopes from simulating the injection
conditions during the measurement. It shold be noted that, due to space 
constraints at the test site, it was not possible to use a dipole magnet for 
separation of protons and \htp. This led to increased \htp ion beam emittance
(discussed below).

The initial particle distribution was obtained by the Catania group using the 
self-consistent 3D ion source extraction simulation software KOBRA-INP 
\cite{spaedke:kobra}. The beam profile and x-x' phase space 14 cm after the 
extraction aperture of the source are shown in Figure 
\ref{fig:particles_initial}. This is the starting point of the WARP simulations.
An extraction voltage of 60~kV was used to obtain this initial distribution and 
the longitudinal velocity component ($v_z$) of each particle is scaled 
appropriately in the WARP simulation if a different beam energy is desired (i.e.
if the experiment the simulation should be compared to was at a different 
extraction voltage). The initial 2-rms diameters and 4-rms normalized emittances 
are listed in Table \ref{tab:particles_initial}.
As a benchmark and to make sure the initial asymmetries were not causing the 
effects attributed to the beam line elements, a Gaussian beam with the same 
Twiss parameters was generated and occasionally used instead of the KOBRA-INP 
initial distribution. The results exhibited the same aberrations and effects
down the line.

The beam envelopes for optimal transport from ion source to the cyclotron are
shown in \figref{fig:SN2_230A_Env_horz}
The particle distributions at the end of these simulations 
were used as initial conditions for the spiral inflector simulations reported in
\secref{sec:injsc}. 
In \figref{fig:SN2_230A_Env_horz}, a tight focal point of 
protons early on can be seen, which leads to strong aberrations in the \htp
beam.  
\begin{figure}[!t]
    \centering
    \includegraphics[width=0.75\textwidth]
                    {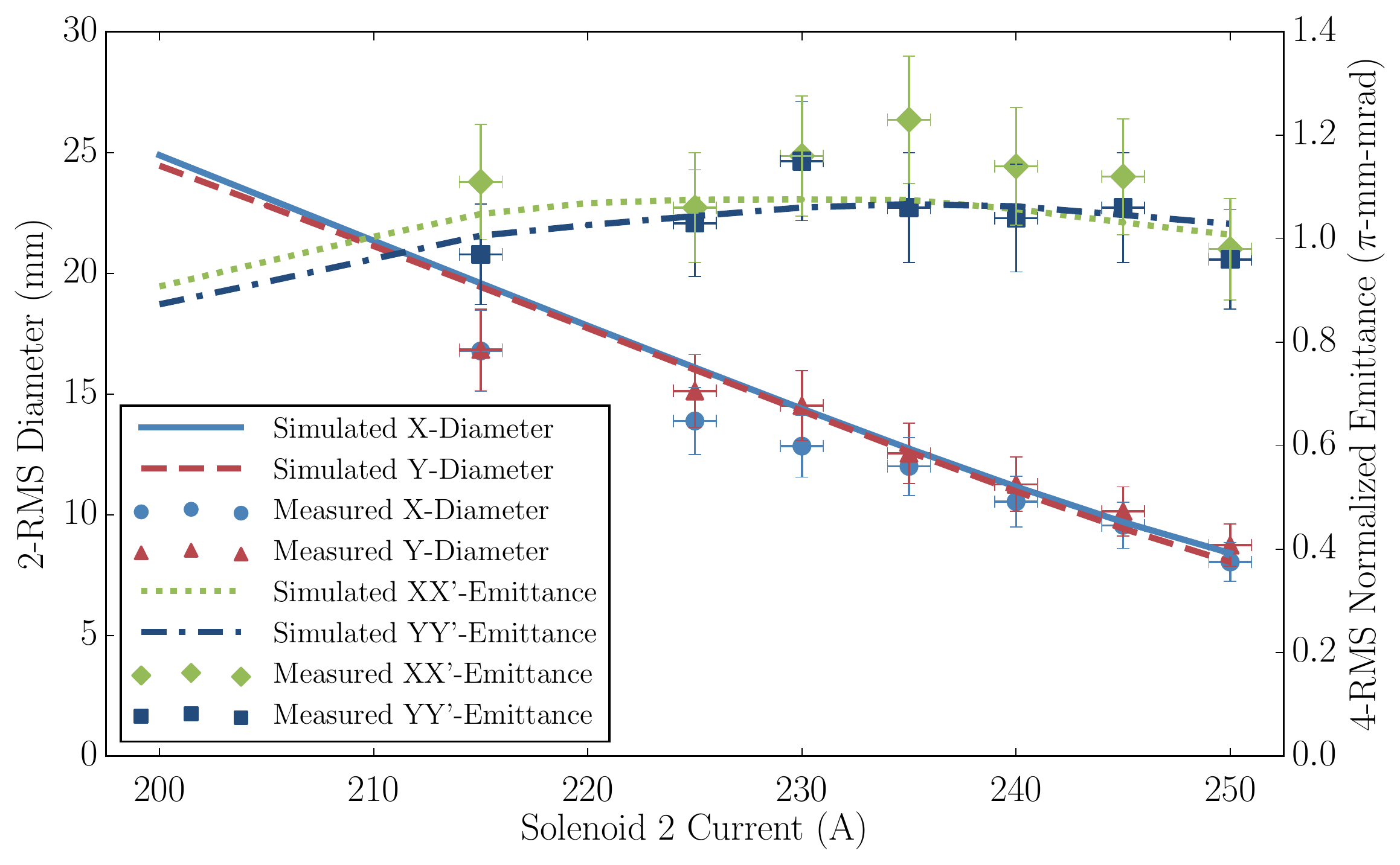}
    \caption{Emittance (4-rms normalized)
             and beam diameter as a function 
    		 of the current on SN2.
             It should be noted that, 
             there is an additional systematic error 
             of $+0.01 \mathrm{to} +0.18$~$\pi$-mm-mrad, due to the
             fact that many of the measured phase spaces used to 
             calculate the emittances did not cover the full beam
             area thus leading to artificially reduced emittance
             values (most notably at the beginning and end of the
             scan). The graph shows the values without the
             systematic errors and the simulated phase spaces,
             truncated with the same limits, yield good agreement
             in emittances and diameters.}
    \label{fig:emittance_diameter}
\end{figure}
The diameters and emittances of the \htp beam for different settings of 
SN2 are compared to the measured values in \figref{fig:emittance_diameter}. It should be noted that, in the measured phase spaces, part of the beam is outside of the axes limits and thus the calculated values are underestimated. In \figref{fig:emittance_diameter} the same limits were applied to the simulation results leading to good agreement. When using the full untruncated particle
distributions to calculate diameters and emittances from the simulations, the
diameters remain largely unchanged, while the emittances increase. The average
increase of emittances from truncated to untruncated phase space is 
$\approx 17.3\%$ for the horizontal phase spaces and $\approx15.4\%$ for the
vertical ones, with a maximum of 35.6\% and 31.4\% for SN2 at 200 A (largest beam). This is treated as a systematic error, but is not included in the 
errorbars in \figref{fig:emittance_diameter}.

\clearpage
\begin{figure}[!p]
    \centering
    \begin{minipage}{.45\linewidth}
        \centering
        \hspace{-25pt}
        \includegraphics[height=2.6in]{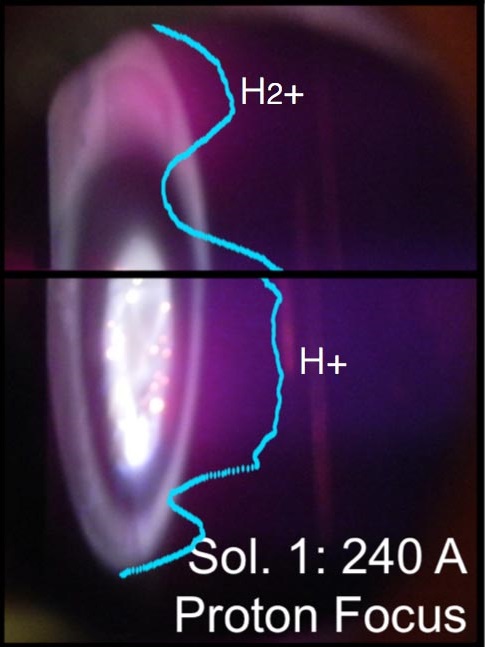}
    \end{minipage}
    \hspace{.05\linewidth}
    \begin{minipage}{.45\linewidth}
        \centering
        \hspace{-25pt}
        \includegraphics[height=2.6in]{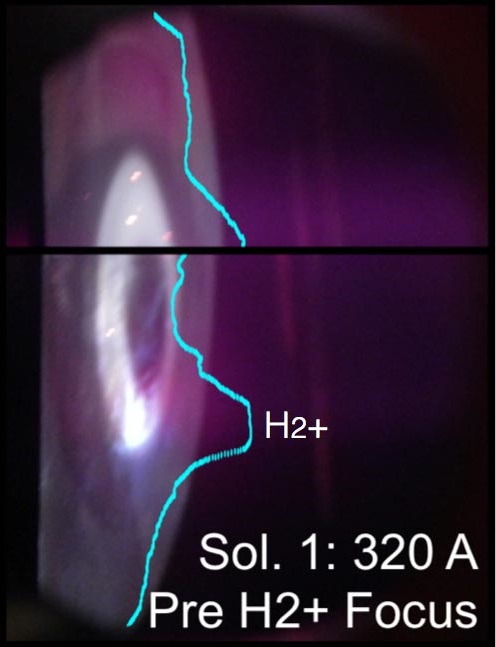}
    \end{minipage}
    \vspace{.05\linewidth}
    \begin{minipage}{.45\linewidth}
        \centering
        \includegraphics[height=2.6in]{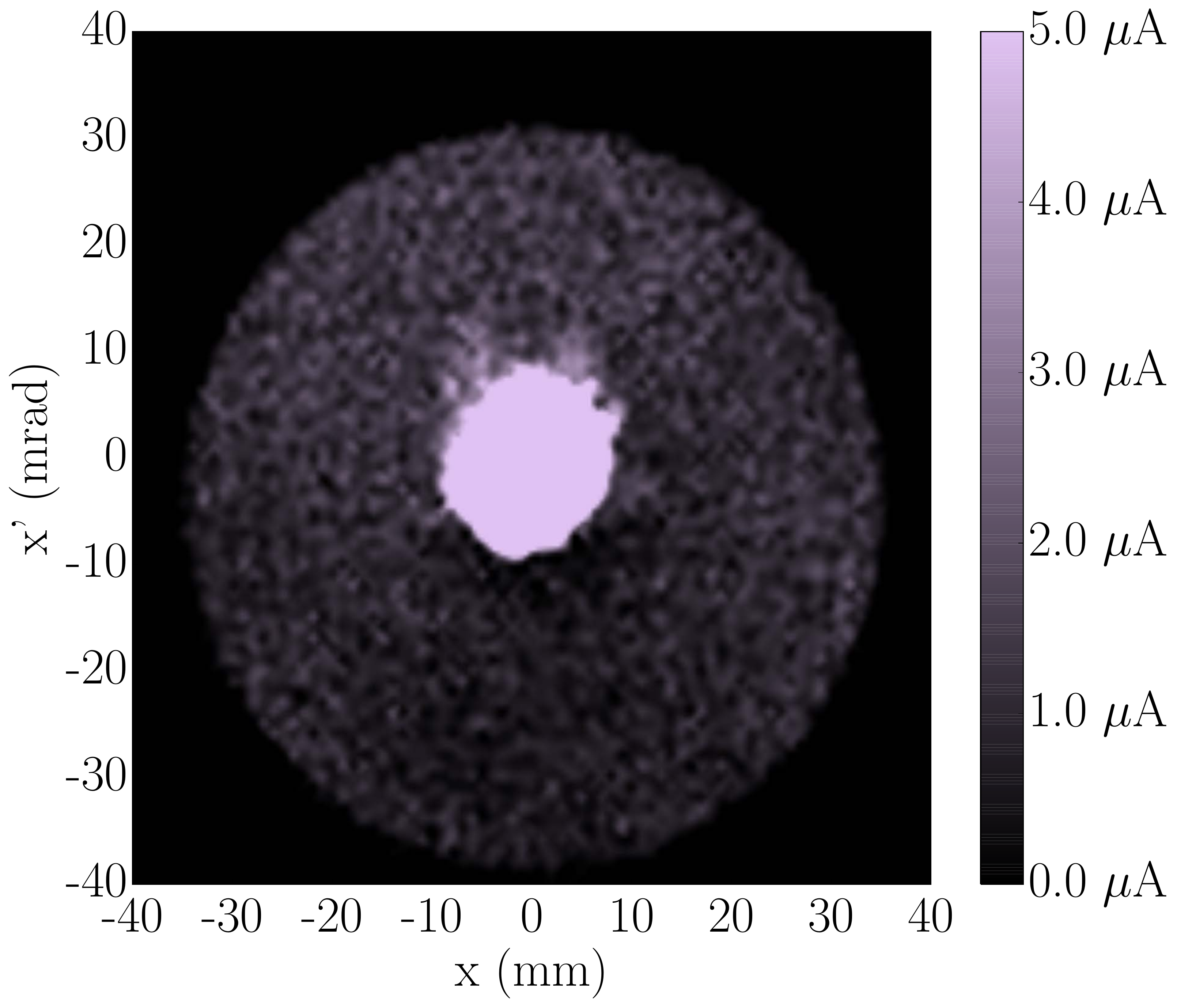}
    \end{minipage}
    \hspace{.05\linewidth}
    \begin{minipage}{.45\linewidth}
        \centering
        \includegraphics[height=2.6in]{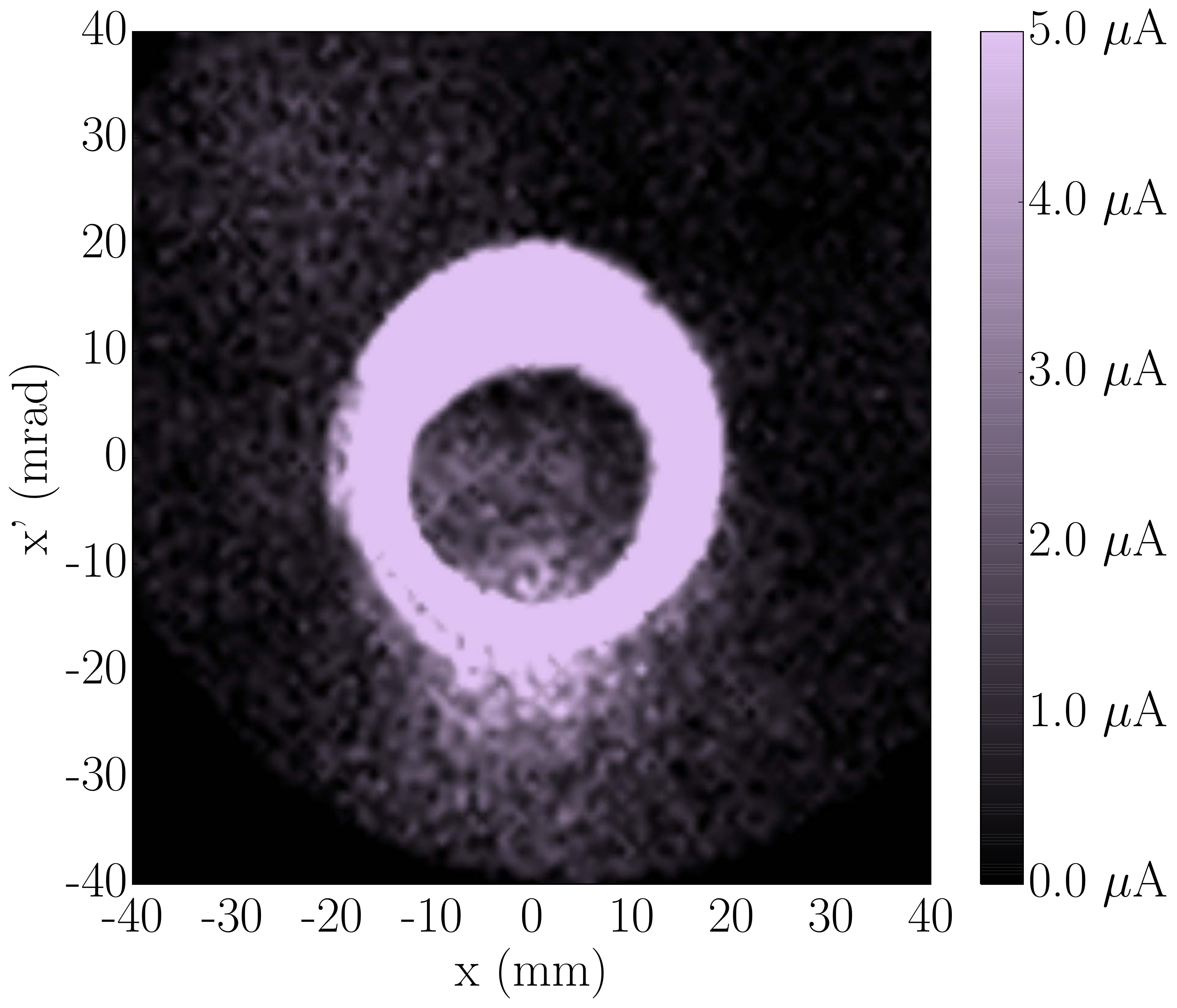}
    \end{minipage}
    \caption{\\
             \\
             \textbf{Top:} Transverse beam profiles observed on the 4-jaw slits           
            (4Jaw). The 
             beam enters the image from the right and strikes the face of two 
             horizontally separated water-cooled copper plates. Left: The 
             focused proton beam is seen as the central illuminated circle while 
             the H$_2^+$ ions form a concentric outer ring. Right: The bright 
             inner ring is the H$_2^+$ being brought into focus. By focusing the 
             H$_2^+$, the protons are over focused and blast out a hole in the 
             center of the beam, seen as the central intensity deficit. Note the 
             horizontal splice in the image corresponds to the small separation 
             between the water-cooled plates.
             \\
             \\
             \textbf{Bottom:} Colored current density plots of 
             the beam cross-sections at the 
             location of the 4-jaw slits.
             Left: 240 A on SN1, 
             Right: 320 A on SN1. The ring structure of the hollow \htp beam is
             clearly visible on the right.}
    \label{fig:SN1_320A_4Jaw_rastered}
\end{figure}

\clearpage
\begin{figure}[!t]
    \centering
    \begin{minipage}{.45\linewidth}
        \includegraphics[width=\linewidth]{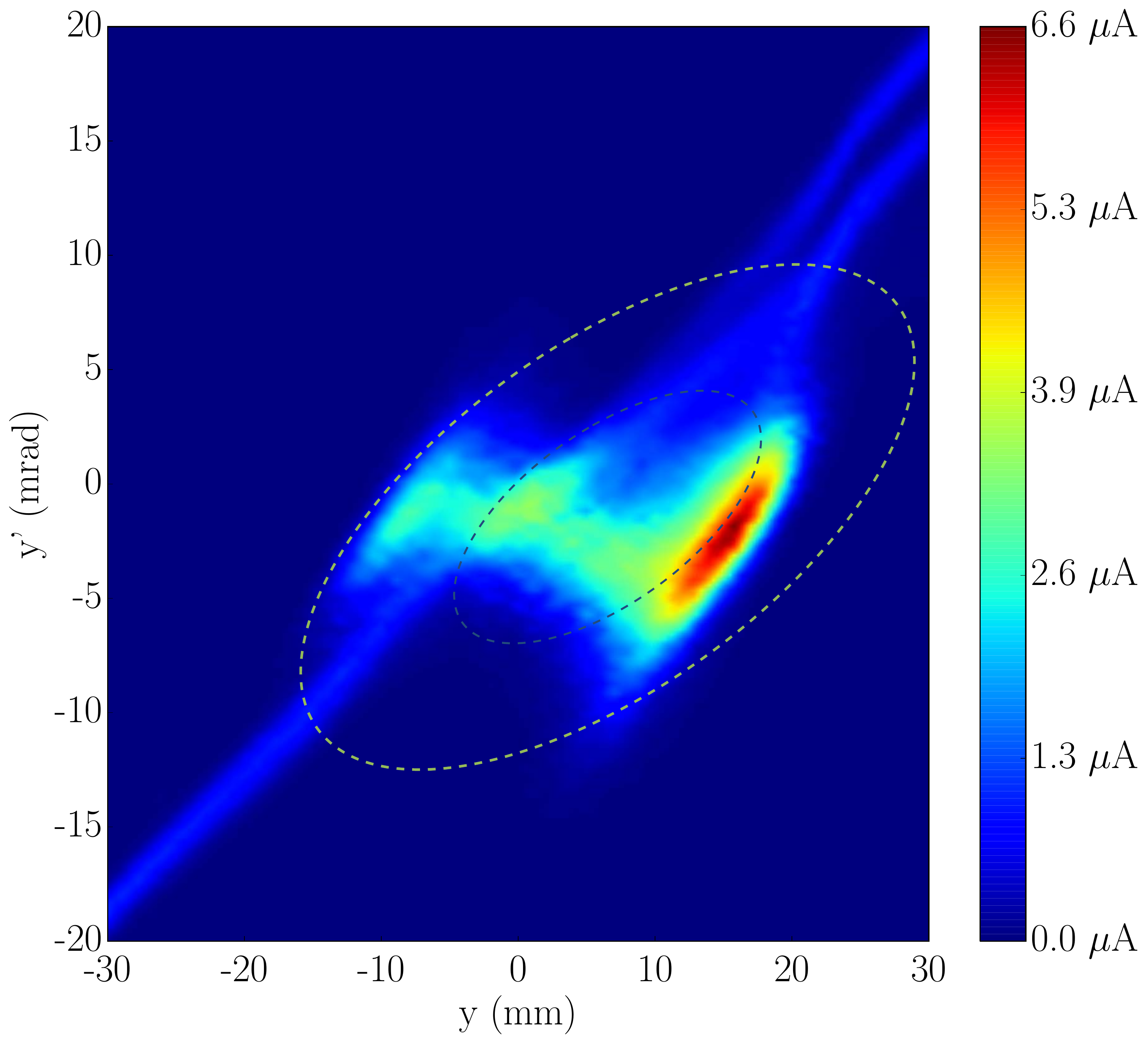}
    \end{minipage}
    \hspace{.05\linewidth}
    \begin{minipage}{.45\linewidth}
        \includegraphics[width=\linewidth]{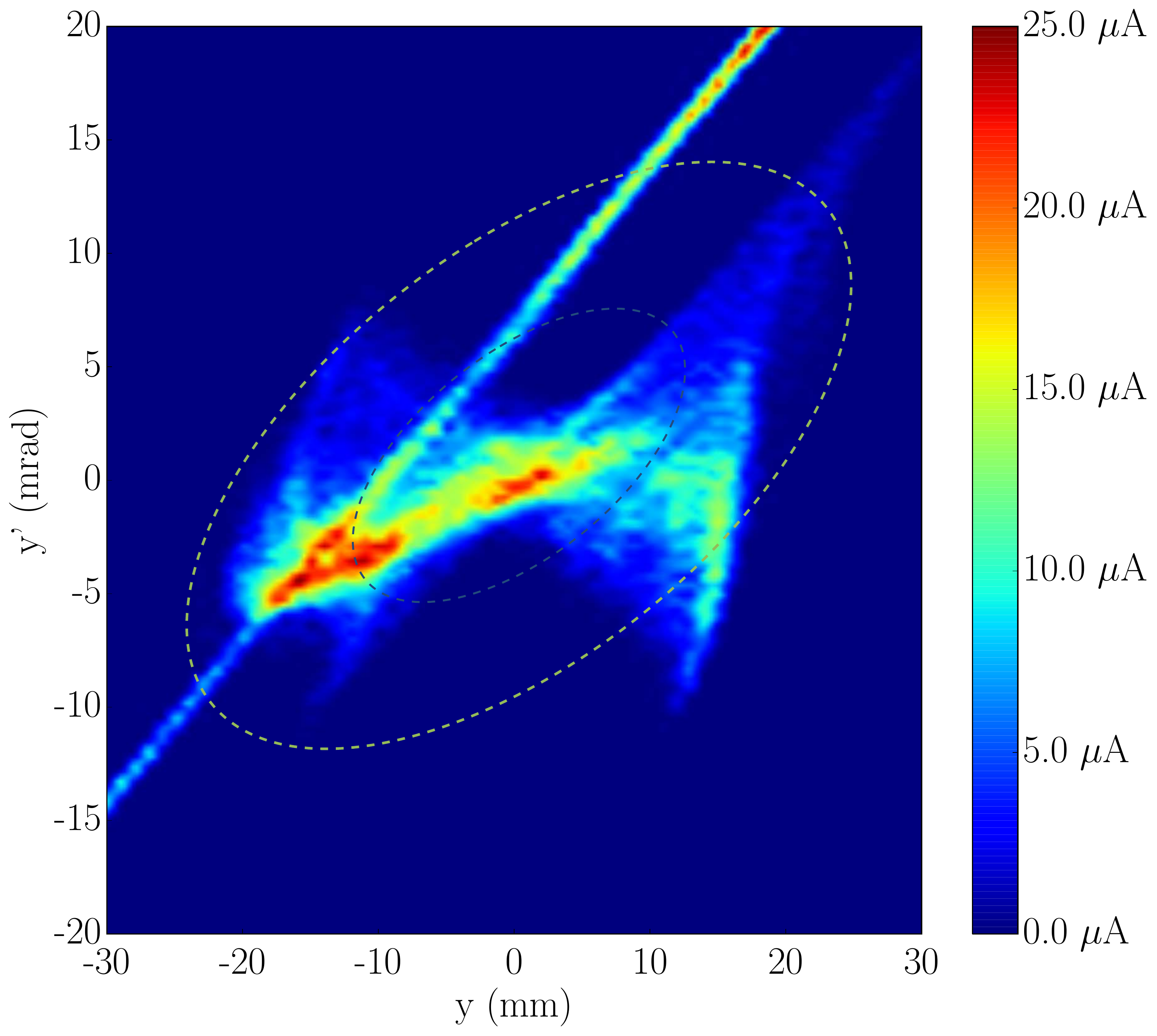}
    \end{minipage}
    \caption{Comparison of measured horizontal phase space (left) and current 
             density plot of simulated horizontal phase space (right) with
             SN1 at 320 A. The 'N'-shape is typical for a hollow beam
             and leads to significantly increased emittance. The beam energy 
             is 55 keV in this case. The different intensity scales arise from 
             the fact that the entrance slit of the emittance scanner was 
             smaller than the step size.}
    \label{fig:SN1_320A_phasespaces}
\end{figure}

The reason for the generally high emittance and some of the aberrations is
the aforementioned tight focus of protons early after the solenoid SN1. 
Space charge leads to a very strong potential at this location and pushes \htp
ions out of the center of the beam, creating a ``hollow beam''. This 
phenomenon is reproduced well in the simulations as can be seen in 
\figref{fig:SN1_320A_4Jaw_rastered} and \figref{fig:SN1_320A_phasespaces}. \figref{fig:SN1_320A_4Jaw_rastered} shows a comparison of photos taken of the surface of the fully closed 4-jaw slits (4Jaw in \figref{fig:SN2_230A_Env_horz})
and density plots of the corresponding particle distributions from the 
simulations. 
In \figref{fig:SN1_320A_phasespaces}, a y-y' phase space measured with the
electrostatic emittance scanner placed at position ``C+BS'' in 
\figref{fig:SN2_230A_Env_horz} is put side by side with the simulated phase
space at the same location. They agree well and show similar aberrations
typical for a hollow beam.

From the comparison of our simulations with the measurements at BCS in Vancouver,
we draw the following conclusions:
\vspace{-\topsep}
\begin{itemize}
  \item In WARP, we have a well developed tool for the simulation of 
        high intensity, low energy beams in a conventional LEBT like 
        the one currently planned as main option for IsoDAR.
  \item We can predict and reproduce effects that arise from space charge
        and can calculate space charge compensation dynamically along the
        beam line from parameters like beam size, beam intensity and
        residual gas pressure.
  \item An LEBT set-up with an analyzing dipole magnet will greatly reduce
        the \htp emittance, to an extent where we can safely increase the 
        ion source aperture to extract more current without surpassing the
        spiral inflector acceptance limit.
\end{itemize}

\clearpage
\subsubsection{Ion Source Extraction Simulation (MIST-1)}
\begin{figure}[!t]
\centering
\includegraphics[width=0.8\textwidth]
                {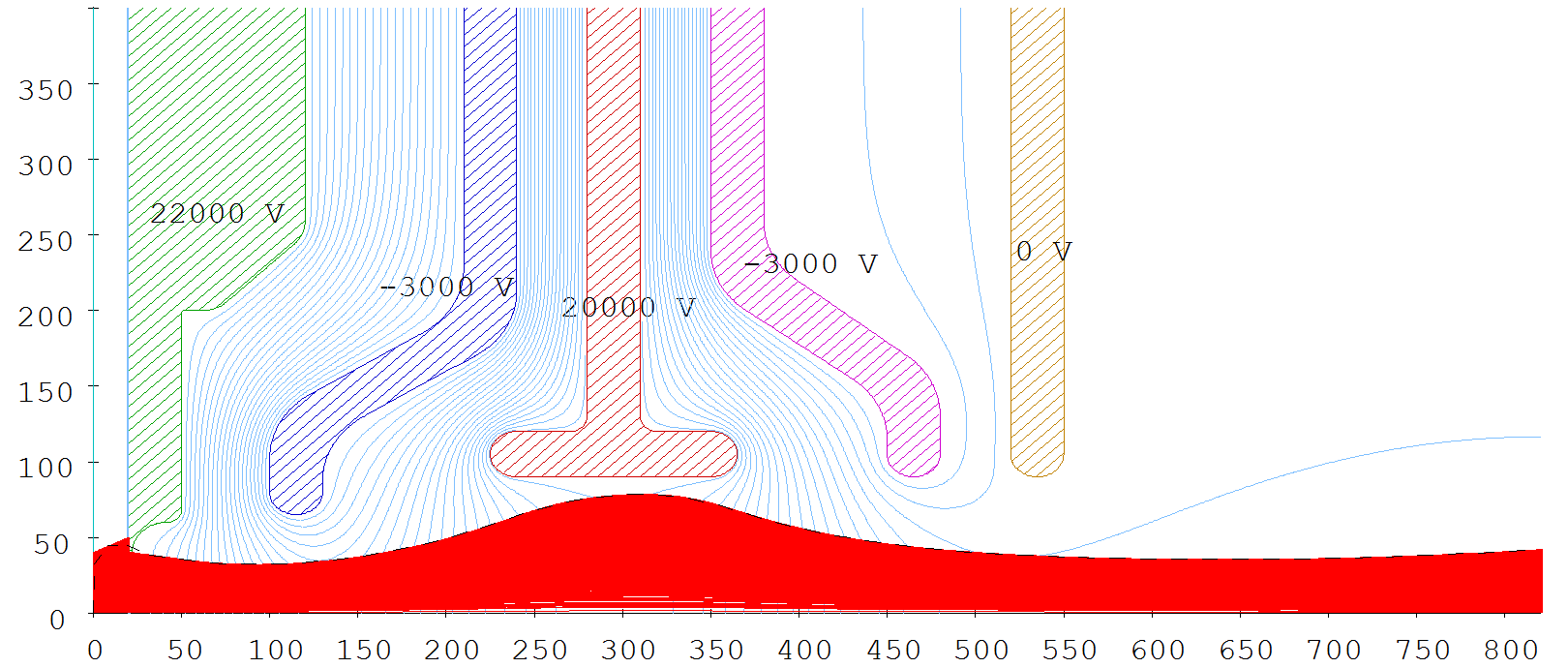}
	\caption{Particle trajectories for the preliminary extraction system 
	         of the MIT multicusp ion source MIST-1. This is a 50 mA \htp
	         beam extracted from 22 kV ion source high voltage potential.
	         \label{fig:IGUN-MIST1_trajectories}}
\end{figure}
\begin{figure}[!b]
    \centering
    \begin{minipage}{.45\linewidth}
        \includegraphics[width=0.8\linewidth]{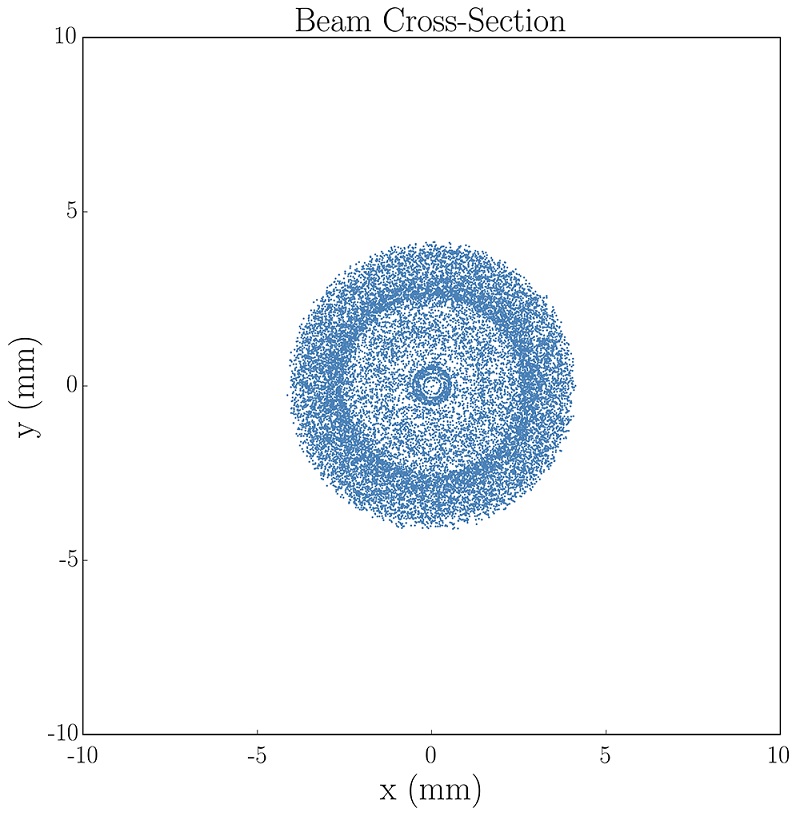}
    \end{minipage}
    \hspace{.05\linewidth}
    \begin{minipage}{.45\linewidth}
        \includegraphics[width=0.8\linewidth]{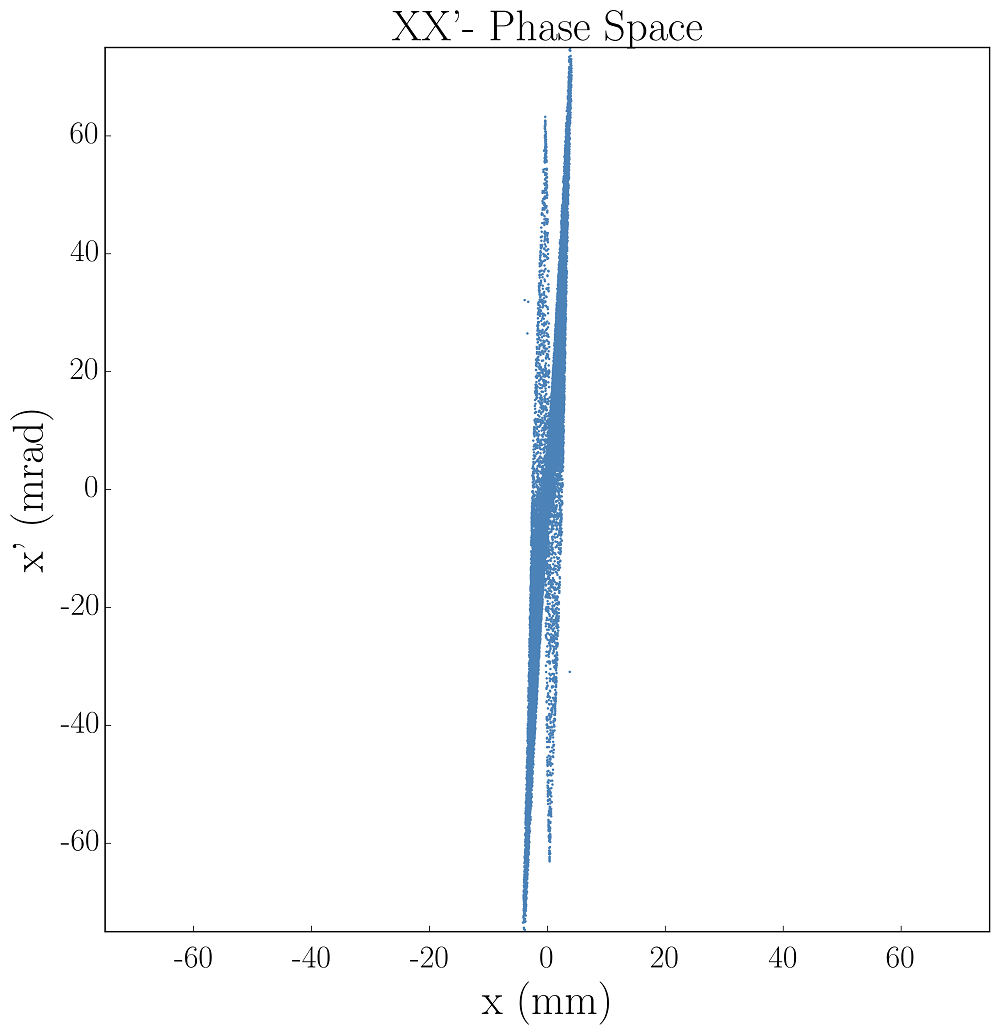}
    \end{minipage}
    \caption{Particle distributions for the preliminary extraction system 
	         of the MIT multicusp ion source MIST-1. This is a 50 mA \htp
	         beam extracted from 22 kV ion source high voltage potential.
	         Left: XY cross-section, Right: XX' phase space.}
    \label{fig:IGUN-MIST1_distributions}
\end{figure}

A preliminary design for the extraction system of our multicusp ion source 
MIST-1 can be seen in Figure \ref{fig:IGUN-MIST1_trajectories} along with ion trajectories. 
The simulation was performed with the well-established 2D ion source simulation software IGUN \cite{becker:igun1}.
The electrode system is a pentode system with the plasma electrode at $15-22$ 
kV, a puller at negative $1-5$ kV, an einzel lens at $13-20$ kV, an electron repeller at negative $1-5$ kV, and a grounded electrode.
It yields a 50 mA \htp ion beam with a 2-rms diameter of 8 mm and a normalized 
4-rms emittance of 0.6 $\pi$-mm-mrad at the end of the simulation (80 mm after extraction). The phase spaces (shown in \figref{fig:IGUN-MIST1_distributions}) exhibit aberrations due to the high fill-factor of the einzel lens part. 
Hence, the diameter of the electrode will be increased in future design 
iterations. This will reduce the emittance of the beam.

\clearpage
\subsection{Cyclotron Injection with Space-Charge \label{sec:injsc}}

\begin{figure}[!t]
\centering
\includegraphics[width=0.6\textwidth]
                {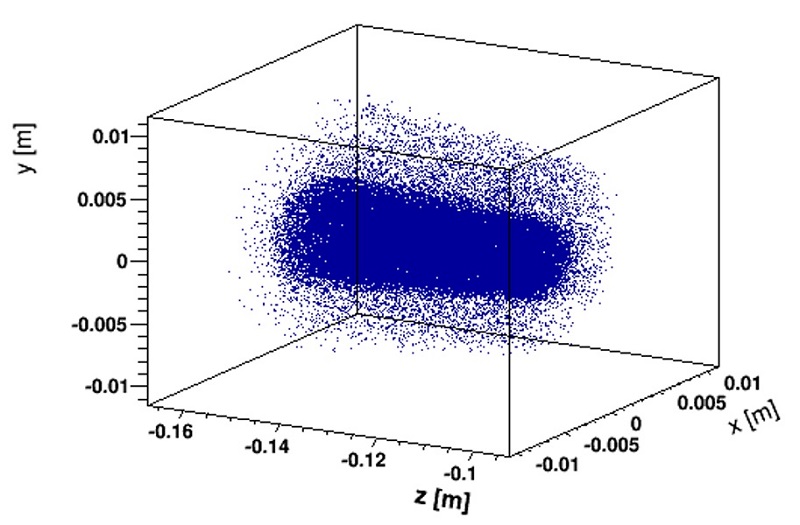}
	\caption{Initial particle distribution of the OPAL injection simulation
	         for the BCS 1 MeV cyclotron based on LEBT simulations with WARP.
	         A significant halo due to the influence of protons on the \htp 
	         beam can be seen. 
	         \label{fig:OPAL_Bunch}}
\end{figure}

In this subsection, we report on results of ion beam simulations
of the \htp beam entering the BCS 1 MeV test cyclotron through the spiral 
inflector using OPAL. For this purpose, \opal ~was upgraded with an improved 
class for geometry handling, an SAAMG fieldsolver, and routines for
arbitrary 3D rotation and translation to accommodate axial injection. 
A publication on this work and the first results of the spiral inflector 
simulations is being put together at the moment.
The results presented here are a work in progress and are to be considered 
highly preliminary. However, they already show that with the nominal beam 
current of 50 mA, significant reduction of injected beam can be expected.

The initial transversal beam distribution for these studies was obtained 
from the WARP LEBT simulation reported above (\secref{sec:warp_simulations}).
In \figref{fig:OPAL_Bunch}, the bunch created from this is shown.
The longitudinal distribution is projected forward using the particle angles
and has a length corresponding to a duty factor of 100\% (DC beam) at the 
cyclotron frequency.

\begin{figure}[!t]
\centering
\begin{minipage}{.45\linewidth}
  \includegraphics[width=\linewidth]{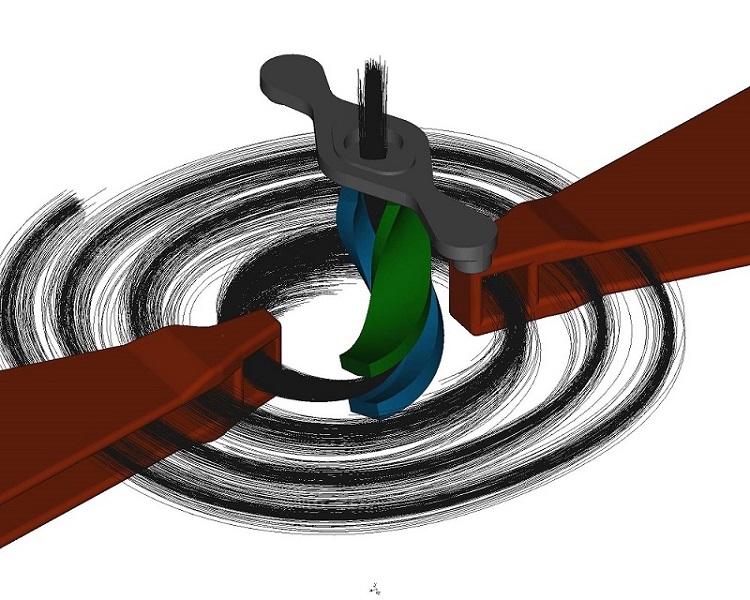}
\end{minipage}
\hspace{.05\linewidth}
\begin{minipage}{.45\linewidth}
  \includegraphics[width=\linewidth]{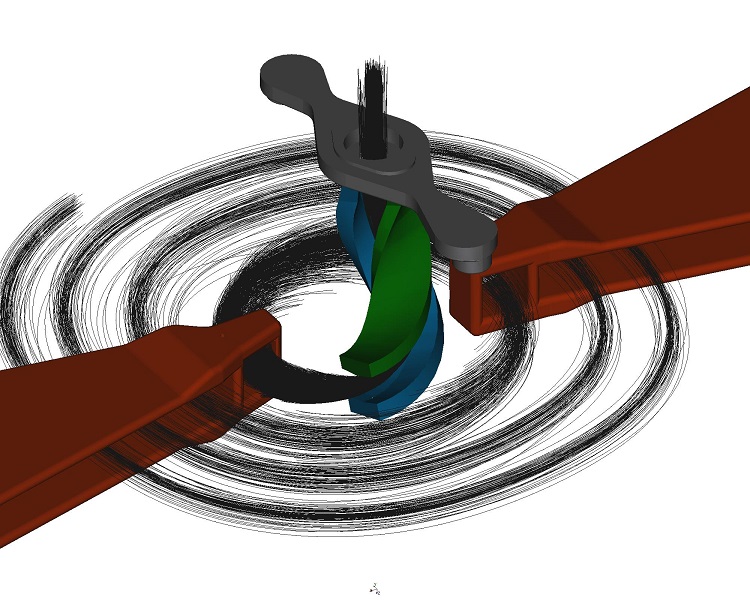}
\end{minipage}
\caption{Injection through the spiral inflector and the first four turns 
         of the BCS test cyclotron. The beam entering from the top is 
         identical except for the beam current. Left: 10 mA, Right 50 mA.
         It can be seen that the number of particles making it to the end
         is reduced for higher beam current.}
\label{fig:opal_injection_trajectories}
\end{figure}

\begin{figure}[!b]
\centering
\includegraphics[width=0.6\textwidth]
                {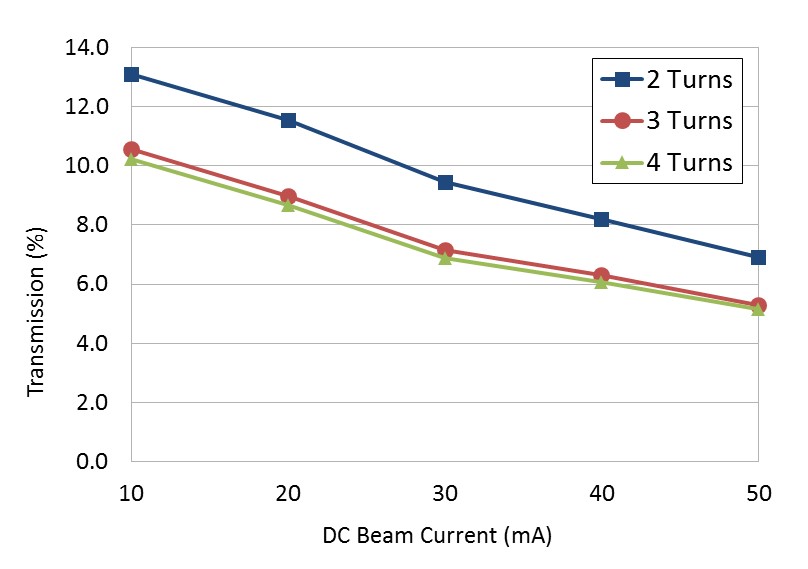}
	\caption{Transmission through the spiral inflector and the first few
	         turns of the BCS 1 MeV test cyclotron. It should be noted that
	         this simulation was performed to mimic the conditions (empirical
	         tuning of beam current into the cyclotron) during the
	         injection tests at BCS in Vancouver and no effort was made 
	         in the simulation to improve matching of the beam to the cyclotron. 
	         \label{fig:opal_transmission}}
\end{figure}

The trajectories through the spiral inflector and the first four turns are
shown in \figref{fig:opal_injection_trajectories} for a 10 mA and a 50 mA
beam. It can be seen that the number of ions that make it to the end (i.e.
are captured by the cyclotron RF bucket) is reduced for the higher current.
This is also plotted in \figref{fig:opal_transmission}, where the capture
efficiency of the beam goes down from 10\% to 5\% when increasing the beam
current from 10 mA to 50 mA. This is not surprising as space charge becomes
more and more important for higher currents and there is no space charge 
compensation in this region due to the presence of the strong electrostatic 
fields generated by the spiral inflector electrodes.
It should be noted that this beam is considered unmatched. During the tests in Vancouver, the objective was to obtain maximum \htp current at the end of the 
beam line and inject this into the cyclotron. Better performance can be expected 
in the simulation when the beam is carefully matched to the spiral inflector and
first turn acceleration.
Furthermore, the beam line at BCS in Vancouver did not have a analyzing dipole 
which lead to significant deterioration of the \htp beam as discussed in 
\secref{sec:lebt_sim}. As an example, in \figref{fig:OPAL_Bunch} a significant halo can be seen that was included in the transmission calculation.
Much better beam quality can be expected from a 
conventional LEBT in the full IsoDAR case. Similarly, the Versatile Ion Source 
(VIS) has a solenoid field at extraction, which is known to lead to higher 
emittance beams. Use of a multicusp source (MIST-1) with low plasma temperature 
will also lead to better beam quality at the entrance of the spiral inflector
and will lead to better transmission.

\clearpage
\subsection{Cyclotron Beam Dynamics with Space Charge \label{sec:cyclsc}}
The beam dynamics with space charge, not considering the spiral inflector, is described in detail in 
\cite[Section II]{yang:daedalus}. Later, in \secref{sec:injsc}
first attempt to describe a space charge dominated beam throughout a spiral inflector is given. 

\subsubsection{Stationary Beam Distribution}

\begin{figure}[t]
\centering
\includegraphics[width=0.45\textwidth]{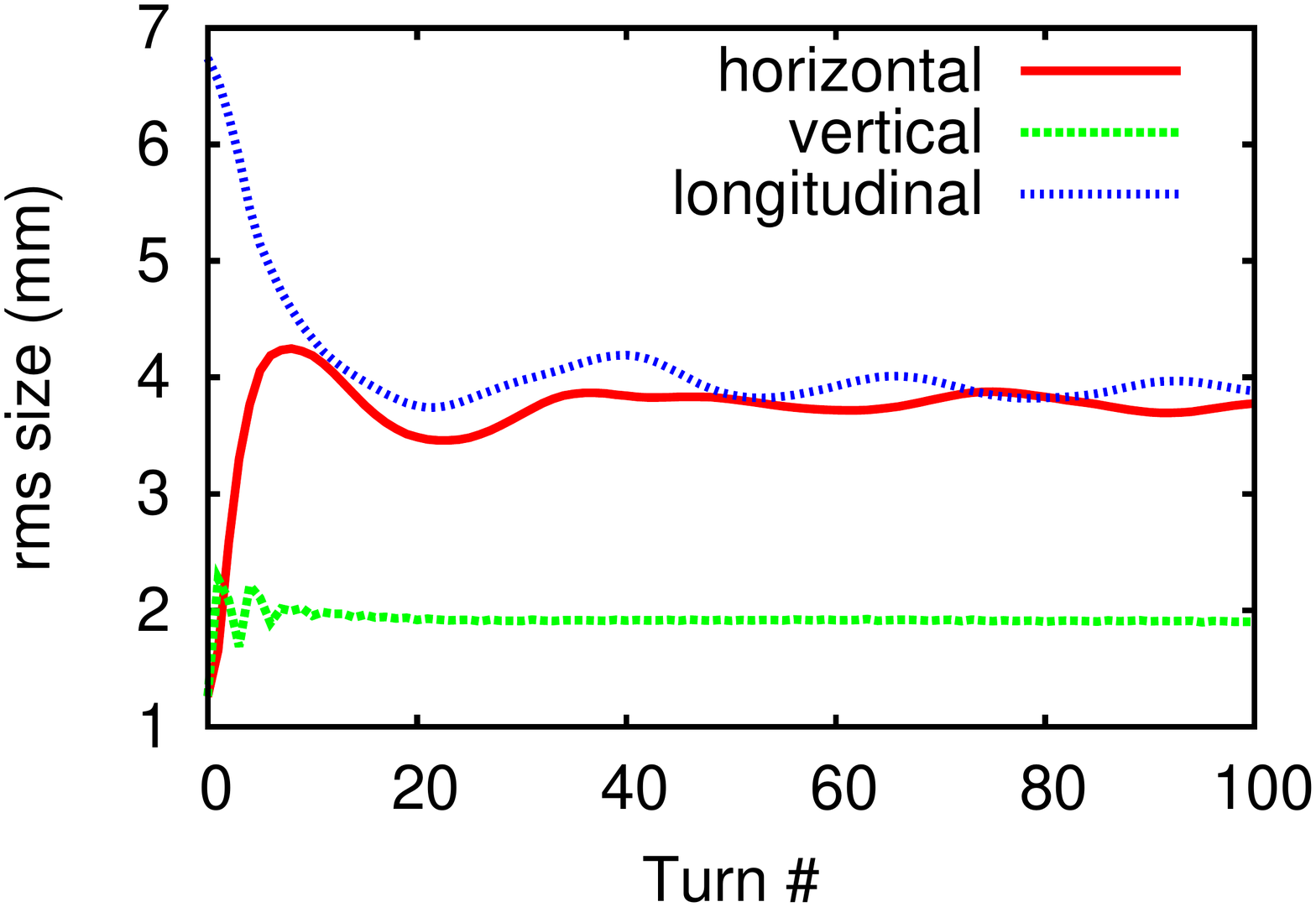}
\includegraphics[width=0.45\textwidth]{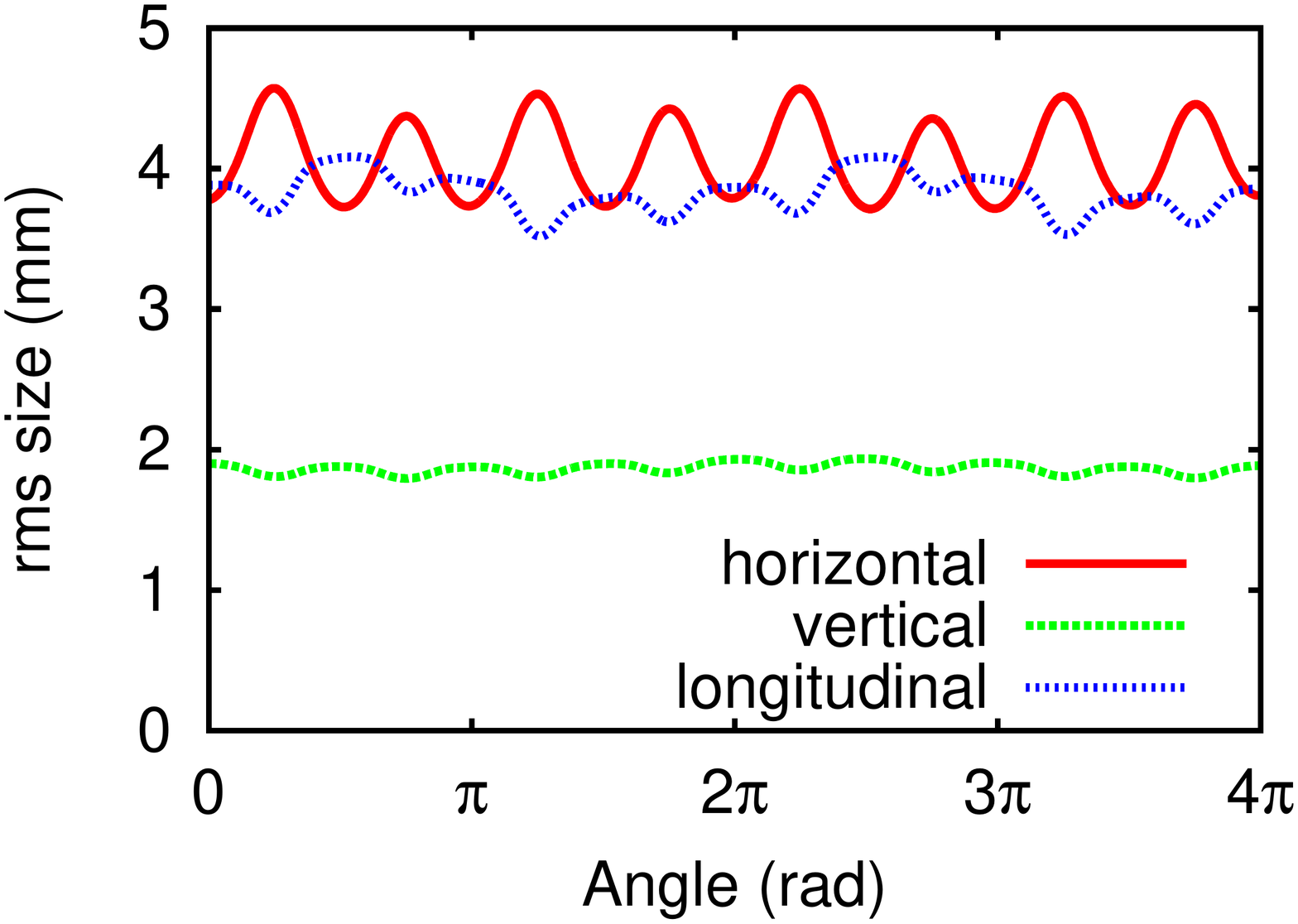}
\caption{The rms size snapshot at 0$^\circ$ azimuth in the 100 turns (left), and the rms envelop in the last two turns (right).\label{fig:statb}}
\end{figure}

The  space charge force combined with the  radial-longitudinal coupling motion develops a vortex motion inside the bunch
that can change the shape of the beam.\
If the  vortex motion  in the horizontal-longitudinal plane is strong enough, the beam will be approximately circular in this plane.\ This is experimentally verified \cite{humbel:private} at PSI. 

When a coasting beam is perfectly matched with the cyclotron,  
its beam size oscillates periodically with a frequency equal to the number of sectors.
Baumgarten  \cite{baumgarten:cyclotron} developed a theoretical 
model to compute the second moments of an ellipsoidal distribution 
($\sigma$ matrix) matched to a cyclotron,
in a linear approximation for a given beam current and emittance.
In practice, the beam does not necessarily maintain an uniform distribution. Hence there exists non-linear components, and that is why we have to 
use the 3D code \opal\, that  provides us with numerical methods to compute not only  the $\sigma$ matrix, 
but also realistic particle distributions. 

\subsubsection{Space charge effects during acceleration}

\begin{figure}[!t]
\centering
\includegraphics[width=0.51\textwidth]{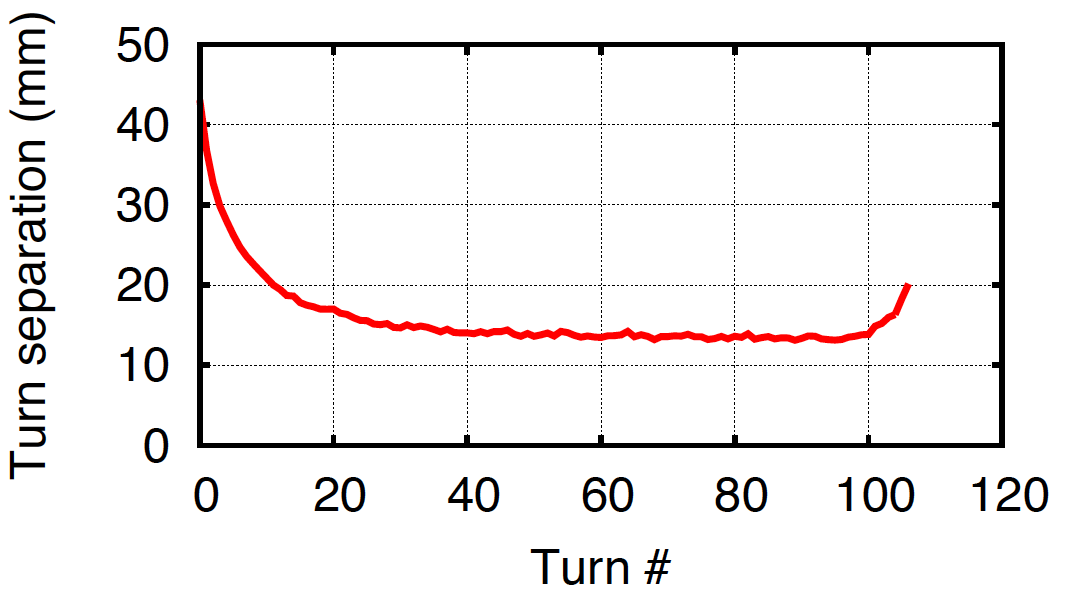}
\includegraphics[width=0.45\textwidth]{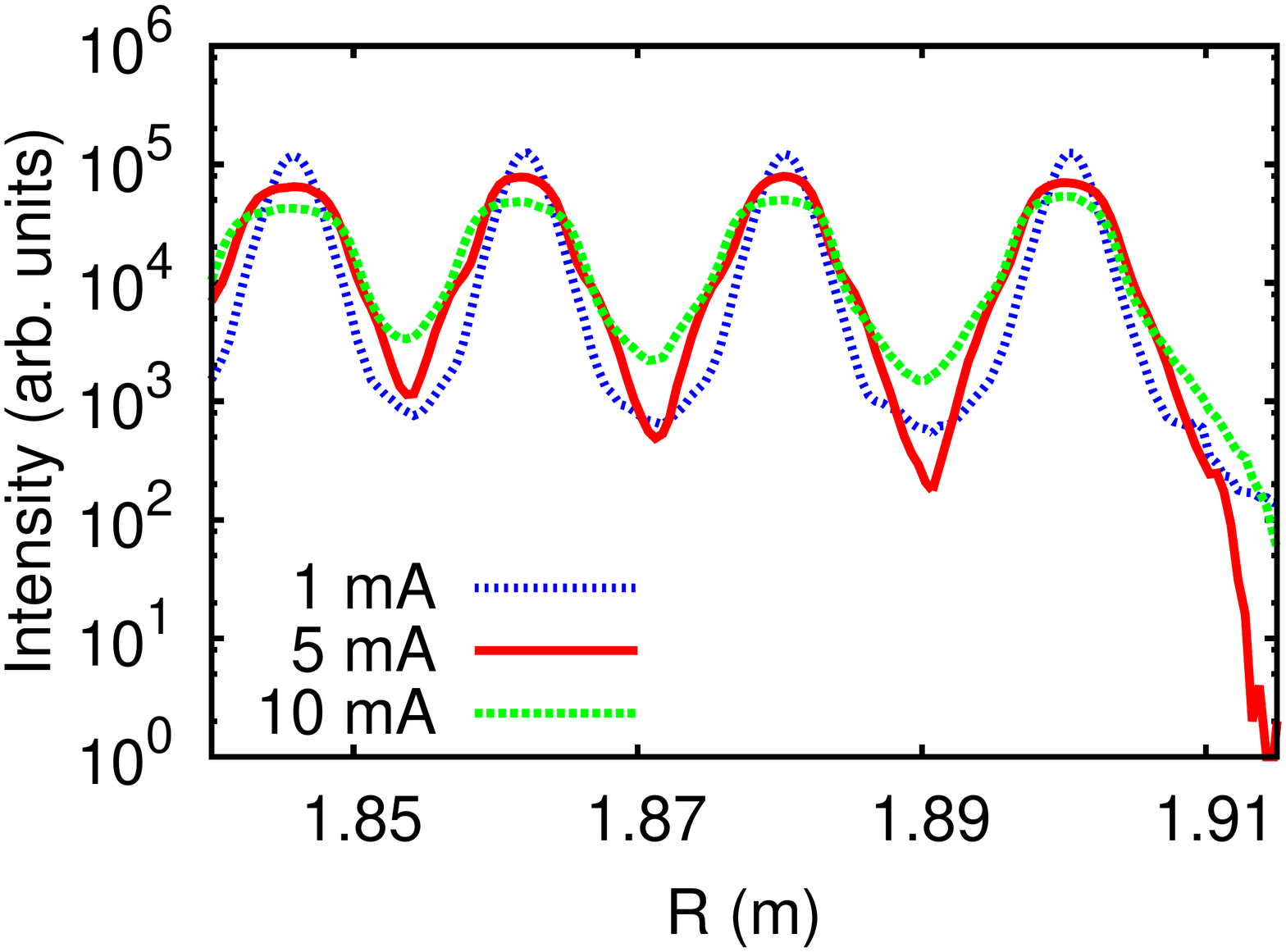}
\caption{Left, turn separation of the central particle in the beam. Right, the radial profile of the last 4 turns at the center on the valley for the different beam current with the initial phase width of 20$^{\circ}$. \label{fig:drdn}}
\end{figure}

 The simulations of the IsoDAR described in this paper start at the exit of the central region. 
Considering that both the space charge effects in the injection line and the transverse-longitudinal coupling motion in the spiral inflector inevitably  increase the emittance,
the initial normalized emittance at the exit of central region is set to three times larger than that of ion source, i.e., 0.6 $\pi$ mm-mrad.  %( $2\sigma$). 
The phase acceptance and initial energy spread are assumed to be $10^\circ-20^\circ$ and  0.4 \% respectively. %($2\sigma$).
In order to reduce the tail particles of the extracted beam,  four collimators are placed at around 1.9 MeV/amu to cut off about 10 \% of the halo particles. 
With the four collimators, the simulation is carried out for 1 mA, 5 mA  and 10 mA current respectively. 
To achieve the high statistics for halo particles,  $10^6$ macro particles are used the in the simulations
and a $64^3$ rectangular grid is used for space charge calculation.\\
The \htp beam needs 107 turns to reach the final energy. The phase slipping during the acceleration is within $\pm10^\circ$ in the first 100 turns. During the final turns it increased by $45^\circ$, 
caused by the sharp field drop, needed to increase the turn separation by passing the $\nu_r$=1 resonance.  
The total turn separation consists of two parts:
\begin{equation}\label{eq:dR}
\frac{dR}{dn} = \frac{dR}{dn}\mbox{(accel)}+\frac{dR}{dn}\mbox{(precession).}
\end{equation} 
The contribution of acceleration can be expressed by
\begin{equation}\label{eq:dRaccel}
\frac{dR}{dn}\mbox{(accel)} = R \times\frac{\Delta E } {E} \times\frac{\gamma}{\gamma+1}\times\frac{1}{\nu^2_r},
\end{equation} 
where $R$ denoting the radius, $\Delta E$ is the energy gain per turn, $E$ is the total energy of the
particles, and $\nu_r$ is the radial focusing frequency. 
The maximum turn separation of precession motion can be expressed by
\begin{equation} \label{eq:dRprecss}
\frac{dR}{dn}\mbox{(precession)}=2\times x_c\times \sin{\pi(1-\nu_r)},
\end{equation}
where $x_c$ is a coherent betatron oscillation amplitude of beam. 
Figure \ref{fig:drdn} (left) shows the turn separation obtained by particle tracking.
The separation of the last turn is $\sim$ 20 mm, in very good agreement  with Eq.\ (\ref{eq:dR}),
i.e. the contribution of acceleration and precession are 13 mm and 7 mm respectively. 

 In \opal\, a radial profile probe element enables one to compute the radial beam profile at the location of  the electrostatic deflector. 
 Figure \ref{fig:drdn} (right) shows the radial beam profile at the position of the electrostatic deflector during the last four turns.
A very conservative septum thickness of 0.5 mm keeps the beam loss at extraction to less than 120 W. 
In the DAE$\delta$ALUS  experiment, the IsoDAR will work at the a duty cycle of 20\%, hence the beam power at the deflector is less than 24 W.
The IsoDAR electron antineutrino source requires the IsoDAR to work at a duty cycle of 90\% ,
in which case the beam power will be less than 108 W. Based on the experience with the PSI Injector II cyclotron, and the sensitivity of the beam losses to variation of injection conditions we presented in \cite[Section II]{yang:daedalus},
we conclude that the most crucial part of the IsoDAR, the extraction is feasible.

\subsection{Transport to Target}

As a starting point for the MEBT beam calculations, we assume the \htp beam emerges from the cyclotron with the characteristics summarized in Table \ref{BeamParameters}. The average current is 5 mA (effectively 10 mA of protons), but the longitudinal bunch size is about 12 mm.  As bunches are separated by about 3.1 m (extraction radius is 2 meters, and there are 4 bunches per turn), the peak proton current is about 2.6 A. The energy spread of about 0.6$\%$ will cause the bunch to elongate, diluting this peak current; at the location of the target bunches are about 30 cm long.  For the purpose of beam instrumentation, the beam remains effectively bunched, so that excellent signal strength can be expected for capacitive or other pickups that rely on time-varying signals.

\begin{table}[!b]
	\caption{Assumed initial beam parameters for MEBT design    
	         \protect\cite{abs:daedalus}.}
	\label{BeamParameters}
	\centering
	\renewcommand{\arraystretch}{1.25}
    \begin{tabular}{ll} 
    	\hline
		\textbf{Parameter} & \textbf{Value} \\ 
		\hline\hline
		Beam from cyclotron & \htp 60 MeV/A 5 mA \\
		Horizontal bunch size & 12 mm \\
		Horizontal divergence & 3.2 mrad \\
		Horizontal emittance & 19 $\pi$ mm-mrad \\
		Vertical bunch size & 12 mm \\
		Vertical divergence & 2.1 mrad \\
		Vertical emittance  &  13 $\pi$ mm-mrad \\
		Longitudinal bunch length & 12 mm \\
		Energy spread & 0.6$\%$ \\
		Beam transported & Protons \\
		Average current & 10 mA \\
		Peak current & 2.6 amps\\ 
		\hline
	\end{tabular}
\end{table}

\subsubsection{Beam Envelope Calculations}

\begin{figure}[!t]
\begin{center}
\includegraphics[width=5.5in]{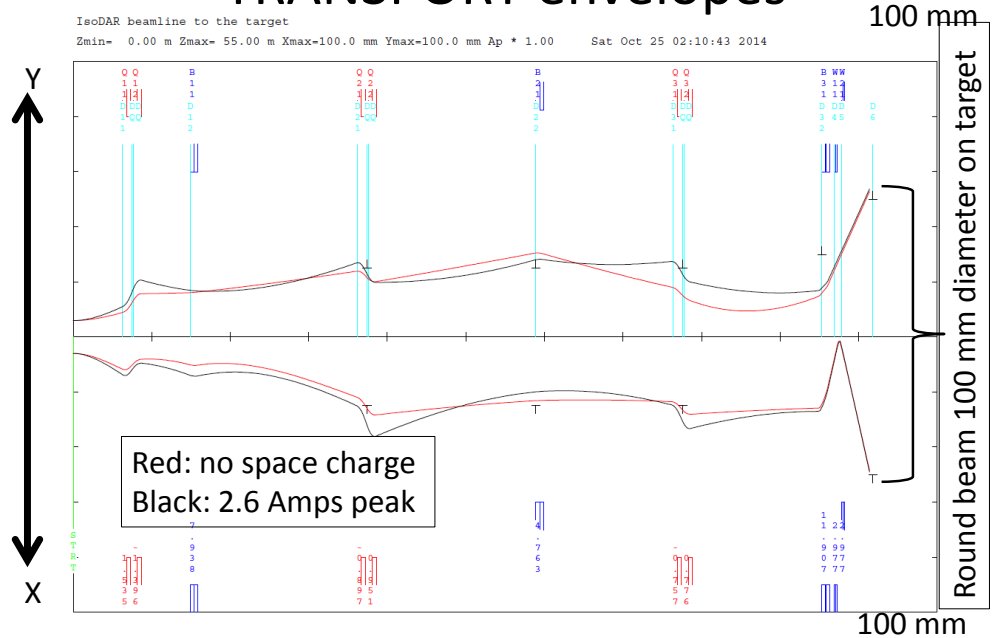}
\end{center}
%\vspace{-1.0in}
\caption{\footnotesize  Beam envelopes calculated by TRANSPORT (ref. W. Kleeven)
\label{fig:transport}}
\end{figure}

As indicated before, and shown in \figref{fig:BeamLineSchem}, 
beam is transported through three sets of quadrupole doublets, and 
three dipole bends. The beam envelopes, calculated with TRANSPORT, 
are shown in \figref{fig:transport}.

The red envelope shows horizontal (x) and vertical (y) envelopes for the beam with no space charge, while the black traces show the effect of space-charge for the peak current of 2.6 amps.  This is an overestimate, as this peak current will be diluted as the bunch spreads out due to the energy spread of the beam.  In any event, one can see in the figure that the effect of this high space charge on the beam envelope is not really significant.
	
However, space charge may still play a role in halo generation, a study
which remains to be undertaken.

\section{Risk Assessment and Mitigation \label{sec:sim_risk}}

\textbf{Risk: Start-to-end simulations not yet complete.}
The individual pieces for a successful start-to-end simulation are
available, but the exercise of complete start-to-end simulations is
not complete. This start-to-end simulation will allow us to 
precisely estimate transmission and losses and guide the realization
process.

\emph{\textbf{Mitigation:} 
A dedicated small cluster of 128-256 core will be very helpful 
in order to avoid long waiting times for beam dynamics jobs.  One post doc and two PhD students are necessary to build, based on the present
model fragments, a sound start-to-end simulation.}

\textbf{Risk: Halo prediction.} For the operational phase, on-line
model with halo prediction capabilities will  be very helpful. 

\emph{\textbf{Mitigation:} 
A PhD. project underway at PSI could be
augmented with one post doctoral researcher and one PhD student to accomplish this task.}

\section{Current Status and Future Work}

Coupling the two codes \opal\ and WARP, will provide us with a start-to-end simulation of the IsoDAR cyclotron. The LEBT where neutralisation 
is very important will be handled by WARP. The subsequent beam transport through the cyclotron and the beam delivery line can be done
by \opal. We mention the newly added capability of simulating the spiral inflector, using a Poisson solver \cite{adelmann:poisson} that takes complicated boundaries into account. 

In order to further aid the design we are able to perform multi-objective optimisation within the full design process using \opal\ together with optPilot 
\cite{ineichen:parallel_opt,
      ineichen:solver}. 
This will be very important in the fine-tuning of the design but 
also in the later phases of commissioning and operation.

\clearpage
\chapter{Conclusion}

This conceptual design report has described the
technical facility for the IsoDAR
  electron-antineutrino source at KamLAND. 
We have presented a strong science case that continues to be
developed.   We provided an 
explanation for  the general approach using cyclotrons as drivers for
the neutrino source, in comparison to alternative designs.    This was
followed by the detailed
information on the conceptual design of the subsystems, with
particular
emphasis on the
cyclotron and target, which are the two most complex systems.    We
provide a short discussion of the 
conventional facility, and note 
that a separate CDR providing
further details on this is planned.    Lastly, we reviewed the state
of our simulation packages.

At each point in the conceptual design discussion, we have listed the risks that have been identified
and our approach to mitigating or retiring these risks.  In particular, we discuss alternative
designs and, in some cases, continue to pursue secondary designs in
order to mitigate risk.
A substantial portion of the conceptual design is based on
three years of experimental efforts, which reduces the overall risk.    We have also enlisted
substantial industry experience on the cyclotron design and on the
 FLiBe handling and acquisition, which again helps to reduce the risks.

Associated appendices on costs and project management are available
upon request.

We conclude that IsoDAR has a strong physics case, a strong conceptual design, and strong backup using detailed simulations. Having benchmarked the progress with this Conceptual Design Report, we will now proceed to the next level in the design and engineering process incorporating the feedback gained from the response to this report.

\clearpage
\printbibliography

\end{document}